# On Attitude Recovery of Spacecraft using Nonlinear Control

Siamak Tafazoli

A Thesis
in
The Department
of
Electrical and Computer Engineering

Presented in Partial Fulfillment of the Requirements
for the Degree of Doctor of Philosophy at
Concordia University
Montréal, Québec, Canada

April 2005

**CONCORDIA UNIVERSITY**

**SCHOOL OF GRADUATE STUDIES**

This is to certify that the thesis prepared

By: **Siamak Tafazoli**

Entitled: **On Attitude Recovery of Spacecraft using Nonlinear Control**

and submitted in partial fulfillment of the requirements for the degree of

DOCTOR OF PHILOSOPHY (Electrical and Computer Engineering)

complies with the regulations of the University and meets the accepted standards with respect to originality and quality.

Signed by the final examining committee:

______________________ Chair
Dr. C. Mulligan

______________________ External Examiner
Dr. D. Staley

______________________ External to Program
Dr. K. Demirli

______________________ Examiner
Dr. M.R. Soleymani

______________________ Examiner
Dr. S. Hashtrudi Zad

______________________ Thesis Supervisor
Dr. K. Khorasani

Approved by ______________________
Dr. M. Mehmet-Ali, Graduate Program Director

April 28, 2005

______________________
Dr. N. Esmail
Dean of Engineering and Computer Science

# ABSTRACT

On Attitude Recovery of Spacecraft using Nonlinear Control

Siamak Tafazoli
Concordia University, 2005

The general objective of this Ph.D. thesis is to study the dynamics and control of rigid and flexible spacecraft supported by a high-fidelity numerical simulation environment. The attitude control system is one of the critical technology elements, a significant cost driver on most spacecraft programs and hence a key area of study and research. The thesis will focus on the different aspects of the problem including the modeling of flexible dynamics and the design and implementation of a suitable subclass of an attitude control system. We have named this the Attitude Recovery System which would be initiated autonomously in an emergency situation such as a thruster misfire, a micro-meteorite collision, a mechanical wheel malfunction, or any other large disturbing torques. The system proposed could equally well be used in the normal attitude acquisition phase beginning right after the launcher tip-off, which could typically impart unwanted angular body rates to the spacecraft.

The demand for greater attitude pointing precision, attitude maneuvering or recovery with the increased use of lightweight and flexible materials necessitates the consideration of flexible dynamics in the control strategy. These highly nonlinear dynamics which increase the order of the system are extremely difficult to model with high degree of accuracy. A general model for attitude and flexible dynamics for a class of spacecraft is hence derived in detail based on the so-called *hybrid coordinates* approach. The spacecraft considered has a star topology with a rigid central bus and flexible plate-type appendages. Given that the flexible spacecraft is under-actuated, the input-output feedback linearization technique is specifically used to partition the system into two distinct parts, namely an external linear system and an internal unobservable nonlinear system. A general internal/zero dynamics theorem for a class of nonlinear systems is proved and then applied to a flexible spacecraft which results in a linear asymptotically stable zero dynamics. The overall closed-loop stability of the flexible spacecraft is also analyzed rigorously and shown to be locally asymptotically stable using the Lyapunov theory. The robustness of the controller against modeling and parametric uncertainties is examined through extensive numerical simulations. Overall, the feedback linearization control scheme has been proven to be feasible and efficient for the attitude recovery of a spacecraft and has also become front and center in other application areas in the recent years.

Finally, in order to validate the theoretical work performed in the thesis, a high-fidelity, user-friendly modeling and control software environment was developed and which has been validated using a set of special simulation cases. These cases are based on analytical results or energy and angular momentum conservation principles. Once validated, the simulator is then used for the investigation of the attitude recovery of spacecraft and simulation results are presented to demonstrate the capability and advantages of the proposed modeling and control techniques for attitude recovery under severe malfunctions affecting the spacecraft.

The present thesis has made the following contributions to the spacecraft dynamics, simulation and control literature:

- The important topic of *automated* attitude recovery (e.g. without any human intervention) of rigid and specially, flexible spacecraft has been studied in this thesis. This is a topic which has not been addressed specifically by the aerospace literature.

- A generic attitude and flexible dynamics formulation for a class of spacecraft having a star topology consisting of a rigid central bus with *n* flexible plate-type appendages, has been fully detailed.

- A high-fidelity simulator which includes the spacecraft's dynamics, environment, and control system was designed and implemented and the many simulation test cases results can be used by other researchers to validate their own simulators.

- A robust and asymptotically stable controller based on feedback linearization has been successfully developed and implemented for the attitude recovery of rigid and flexible spacecraft.

- A new approach based on quaternion addition for producing the control error signal has been proposed. The proposed new approach effectively deals with the existence of dual equilibrium points.

- A design criterion for a class of flexible spacecraft has been established which needs to be satisfied so that the *zero dynamics* of the system is asymptotically stable. Under this condition, the overall closed-loop system was also proven to be asymptotically stable using the Lyapunov theorem.

- A new explicit analytical solution for the construction of the *normal form* and the *zero dynamics* of a general class of nonlinear systems was developed and applied specifically to the flexible spacecraft system.

# ACKNOWLEDGEMENTS

I would like to acknowledge and sincerely thank my thesis advisor, Professor K. Khorasani, for his kind support and guidance throughout my Ph.D. program. I wish to express my profound gratitude to Dr. Staley for his valuable feedback and constructive criticism of the first version of my thesis, which improved the quality of the final version. I would also like to extend a special thanks to the other members of my examining committee for reviewing my thesis work and providing valuable comments.

# DEDICATION

*I dedicate this work to my family for their unwavering support.*

# TABLE OF CONTENTS

# LIST OF ACRONYMS

| | |
|---|---|
| *n*D | *n* Dimensional |
| ABM | Adams-Bashforth-Moulton |
| ACS | Attitude Control System |
| AOCS | Attitude and Orbit Control System |
| ARM | Attitude Recovery Maneuver |
| ARS | Attitude Recovery System |
| CMC | Coupled Modal Control |
| CMG | Control Moment Gyro |
| COM | Center Of Mass |
| CPU | Central Processing Unit |
| DAC | Disturbance Accommodation Control |
| DARTS | Dynamics Algorithms for Real-Time Simulation |
| DC | Direct Current |
| DCAP | Dynamics and Control Analysis Package |
| DLL | Dynamically Linked Library |
| Dshell | DARTS Shell |
| EA | Eigenstructure Assignment |
| FEM | Finite Element Model |
| FLC | Feedback Linearization Control/Controller |
| FPA | Flexible Plate-type Appendage |
| FSC | Frequency Shaping Control |
| FSD | Flexible Spacecraft Dynamics |
| GPS | Global Positioning System |
| IMSC | Independent Modal-Space Control |
| ISS | International Space Station |
| LQG | Linear-Quadratic-Gaussian |
| LSS | Large Space Structure |
| LTI | Linear Time Invariant |
| LTR | Loop Transfer Recovery |
| LTV | Linear Time Varying |
| LVLH | Local Vertical Local Horizontal |
| MESS | Model Error Sensitivity Suppression |
| MIMO | Multi-Input Multi-Output |
| MSS | MultiSatSim |
| NASA | National Aeronautics and Space Administration |
| ODE | Ordinary Differential Equation |
| OFC | Output Feedback Control |
| Open-SESSAME | Open-Source, Extensible Spacecraft Simulation And Modeling Environment |
| PA | Pole Allocation |
| PCM | Positivity Control Method |
| PCT | Perturbation Control Technique |
| PDE | Partial Differential Equation |
| PID | Proportional-Integral-Derivative |
| RAM | Random Access Memory |
| RK | Runge-Kutta |
| SB | Spin Bearing |
| SCOLE | Spacecraft Control Laboratory Experiment |
| SMCC | Spin Motor Commanded Current |
| TWC | Traveling Wave Control |
| UDC | Uniform Damping Control |
| XIPS | Xenon Ion Propulsion System |
| XPOP | X-axis Perpendicular to Orbital Plane |

# LIST OF FIGURES

# LIST OF TABLES

# LIST OF APPENDICES

# CHAPTER 1

## INTRODUCTION AND BACKGROUND

Designing and building space systems is challenging and aerospace engineers need to get it right the first time since product recalls are extremely costly and almost never done. The exception being for example, the shuttle missions to repair the very expensive Hubble telescope. Furthermore, actual ground tests are also incredibly costly, possibly dangerous and often just plain impossible. Virtual prototyping and high-fidelity simulation is used by spacecraft manufacturers where physical testing is difficult or impossible to prove out critical aspects of their designs, such as deployments, separations and spacecraft attitude control. With the shrinking budgets, the aerospace community is focusing on small and/or lightweight spacecraft to satisfy a wide range of applications. New and emerging designs and development processes are helping to make spacecraft considerably less costly to build, however, matching improvements in the technologies are also required. Many future spacecraft are expensive and have a high public profile (e.g. the unsuccessful NASA Mars probes: Polar Lander and Climate Orbiter) and therefore instability and failure are not acceptable.

The objective of this Ph.D. thesis is to address an ever more important and specific problem of *automated* spacecraft attitude recovery (i.e. without human intervention). This is aligned with the current trend for more autonomous spacecraft (i.e. satellite and planetary probes) which in essence would take the human operators out of the loop, by adding enough on-board intelligence and technologies to cut the human link for at least part of the mission and increase the utility and functionality of the spacecraft. There are good indications that future spacecraft will become more and more autonomous in all aspects and will eventually include an on-board subsystem dedicated solely to spacecraft health monitoring and failure detection and recovery (e.g. NASA's Deep Space I spacecraft) with very minimal human intervention, if any. These tasks are presently mainly done by the ground segment (i.e. mission operators and engineering team on ground) but we are already starting to see more tasks being pushed on-board (e.g. Pegasus launcher tip-off rate recovery and autonomous sun-pointing for Scisat satellite in case of a bad injection). Presently, the most common approach to failure recovery is to put the spacecraft in a *safe hold mode* while the ground segment investigates the cause of the failure/fault. However, a safe hold mode can last up to several months which represents a significant loss of scientific or communications data for that time period. If instead, an independent system was available to address the fault immediately then this loss of precious data could be avoided. Of course, the addition of a new system involves design time and ultimately extra money and hence there would have be a trade-off analysis performed.

In this light, the present thesis would be addressing the automated attitude recovery problem, where a fault in ACS would be autonomously detected, isolated and a recovery and reconfiguration would be initiated. For this purpose, the Attitude Recovery System (ARS) could be even considered to be separate from ACS (e.g. the ARS software could run on a different dedicated processor and an independent set of sensors and actuators might even be envisaged to exist for the attitude recovery system).

The thesis will focus on the different aspects of the automated attitude recovery problem including the system dynamics modeling and simulation, and the design and implementation of a suitable controller. The attitude recovery system would be used when a malfunction affects the attitude of the spacecraft and throws it into a spin. Some examples of such malfunctions or failures are: a thruster misfire due to actuator fault or operator error, a severe failure of a mechanical wheel, or a debris/micro-meteorite impact, all of which impart a torque on the spacecraft and their magnitude, potentially significant, are not known *a priori*.

The demand for greater attitude pointing precision, attitude maneuvering or attitude recovery with the increased use of lightweight and flexible materials necessitates the consideration of flexible dynamics in the control strategy. Hence, the automated attitude recovery problem is investigated here mainly for a flexible spacecraft whose dynamics are highly nonlinear, increase the order of the system and are extremely difficult to model with high degree of accuracy. The dynamics formulation method used is the *hybrid coordinates* approach which is based on the Lagrangian formulation. The proposed control scheme is based on the Feedback Linearization Control (FLC) technique which has proven to be feasible and efficient and has become front and center in the control research arena in the recent years. Feedback linearization control basically refers to the approach where the nonlinear differential equations of the plant are linearized by appropriate state space coordinate transformation and nonlinear control transformation schemes. In other words, exact linearization is achieved by the change of the representation of a nonlinear system into a linear one, which is controllable and then a stabilizing controller is designed that guarantees desired performance specifications.

In the next few sections, a literature review is presented for spacecraft (i) dynamics; (ii) control; (iii) simulation; and (iv) AOCS failures.

## 1.1 Literature Review

The attitude control field is about half a century old. Its beginning can be traced back to the late 1950s with most of the work being from 1957 onwards. October 4$^{th}$, 1957 marked the historic launching of the first human made artificial satellite into Earth orbit, Sputnik I. The launch of this small satellite ushered in new political, military, and most importantly technological and scientific developments. The attitude dynamics and control field became and has remained a topic of great interest for many researchers. The actual number of journal articles, conference papers and technical reports published in the area of spacecraft dynamics and control is in the hundreds and is quite overwhelming. There are also a great number of relevant books in print today.

The literature review given in the following paragraphs is by no means comprehensive and exhaustive but it presents the key references to the published papers and other review articles. This review is deemed sufficiently complete to familiarize a new comer with this intriguing and challenging field. One thing is for certain, and that is attitude dynamics and control is here to stay!

First, a brief paragraph is provided on the most researched category of spacecraft, namely the rigid spacecraft, following which flexible spacecraft which are of key interest in this thesis are examined.

### 1.1.1 Rigid Spacecraft Dynamics and Control

Rigid attitude control of spacecraft was reviewed early on by Meyer [1]-[2] and his is a seminal investigation of the attitude control problem. The subject matter is now so mature that several excellent books in the field have appeared which cover the rigid attitude dynamics and control problem [3]-[6]. There are hundreds of articles that deal with the rigid attitude control using a variety of control techniques from the classical Proportional-Integral-Derivative (PID) to adaptive control. Also recently, many intelligent-based attitude control approaches have also been investigated by researchers using such techniques as Neural Networks [7], Fuzzy Logic [8], combined Neuro-Fuzzy approach [9] and combined Genetic Algorithm and Neural Networks [10]. Recently, Fortuna et al. [11] present a comparative study between four different control strategies for attitude control of rigid spacecraft, namely (i) PD plus hypercomplex multilayer perceptron neural network; (ii) PD plus hypercomplex radial basis function neural network; (iii) classical PD; and (iv) adaptive control. Fortuna et al. [11] conclude that in both cases where neural networks is used, the results show an improvement with respect to the simple PD and adaptive controllers.

### 1.1.2 Flexible Spacecraft Dynamics and Control

Dynamics and control design for flexible spacecraft, which is sometimes referred to in the literature as Large Space Structure (LSS), has received much attention in the past three decades. One of the initial reports by National Aeronautics and Space Administration (NASA) [12] shed light on a problem that was experienced with several early spacecraft. The attitude of these spacecraft were noticed to act strangely without any recognizable reason at first. However, further investigation by NASA and reported in [12] identified the culprit to be the structural flexibility which was interacting with the ACS in a variety of ways. This report along with a couple of other papers [13]-[14] cite useful references and compiles a review of this problem for that early time period. Since those early days, a large number of papers have appeared in the field. In 1974, Modi [15] reviewed the state of the art on attitude dynamics of spacecraft with flexible appendages, and this review is still relevant today. Also more recently, Junkins and Kim have published a book that cover the more established areas of dynamics and control of flexible space structures with many useful references [16].

On the control side, Meirovitch et al. [17] present a comparison of control techniques for large flexible space systems. These control techniques are grouped based on two broad approaches, namely (i) coupled control which does not place any requirements on the number of actuators, provided controllability condition is satisfied and (ii) independent modal-space control which requires that the number of actuators be equal to the number of controlled modes. Meirovitch et al. [17] conclude that the independent modal-space control method is superior to coupled control since it offers a large choice of control techniques (both linear and nonlinear control), is easier to design, exhibits virtually no instability characteristics, and minimizes the input energy. Some of the other reviews relevant to the topic of control of flexible spacecraft are given by Balas [18], Croopnick et al. [19], Meirovitch and Oz [20] and Robinson [21].

Meirovitch [22] presents a fairly comprehensive literature review from a structural control point of view which encompasses many space applications. The following sections are reviews of some of this relevant literature on the control of flexible spacecraft or LSS grouped by control strategy. These papers are not specifically related to the topic of attitude recovery but to the more general topic of flexible spacecraft/LSS attitude control which includes vibration suppression, attitude pointing, and attitude slew maneuvering. This part of the literature review which also includes seminal papers in each control area is given here for completeness and for the interested new comer to the field. Note that some of these seminal papers are not in the area of spacecraft/LSS attitude control but originated the ideas which were used in the subsequent work by spacecraft control researchers.

#### 1.1.2.1 Classical Control

The early study of control systems involved heavily the transform techniques. Basically, the dynamics of the system to be controlled would be expressed by linear differential equations and transformed to *s*-domain (or frequency domain) using Laplace transformation. This transformation allows relating the system output to the input via a transfer function, which is a ratio of two polynomials in *s*. When the system involves more than one input and output (MIMO), then the inputs and outputs can be related by a matrix of transfer functions known as the transfer matrix. One of the main concerns with any control system is stability which, in the case of classical control, is investigated by examining the poles of the closed-loop transfer function. Various graphical techniques were developed allowing determination of the closed-loop poles by working with the open-loop transfer function. Widely used techniques, taught at the undergraduate level, are the root-locus method, Nyquist plots, Bode diagrams and Nichols plots. The implication is that the system is characterized by a single input and a single output (SISO), and hence by a single transfer function. In the case of multi-input multi-output systems, the situation is complicated by the fact that one needs to consider the transfer functions relating every input to every output. Moreover, the approach is limited to linear systems with constant coefficients. However, one of the advantages of classical control is that it offers good physical insight. Classical control uses time-domain and frequency-domain performance criteria such as rise time, maximum overshoot, settling time, steady-state error, gain margin, phase margin, maximum magnitude and bandwidth. The theory described briefly above which is concerned with transform techniques for the design of feedback control systems has come to be known as classical control.

Proportional-Integral-Derivative (PID):

Hughes and Abdel-Rahman [23] study the linear attitude control of flexible spacecraft using a classical PID feedback control law. They assume that the spacecraft is composed of a rigid central bus to which are attached one or more flexible appendage (modeled as a series of second order systems) and that there is at least one axis about which the spacecraft's attitude motion is uncoupled from the motion about the other two axes. The principal result of the paper is that if the controller is unconditionally stable (with respect to gain), then appendage flexibility can not destabilize it. The PID control law for flexible spacecraft is also considered by Meirovitch and Sharony [24].

#### 1.1.2.2 Modern Control

With the advent of the space age in the 1950s, the interest turned to a new class of control techniques encompassing new performance criteria, such as minimum time and minimum fuel. These performance criteria have led to nonlinear control laws. Other performance criteria attracting a lot of attention are quadratic performance criteria, leading to linear control laws. The approach consisting of the minimization of a performance index is known as optimal control and falls in the domain of modern control. Major advances in optimal control are Bellman's dynamic programming and Pontryagin's minimum principle, where the latter can be regarded as a generalization of the fundamental theorem of calculus of variation. A control approach developed in the late 1960s is known as pole allocation, and can be considered as part of modern control. Modern control should not be regarded as replacing classical control but complementing it, as a mix of both approaches together can enhance the understanding of the problem. Modern control is essentially a time-domain approach, based on the state-space description of the behavior of the dynamical systems. Over the years, many new control techniques have been devised and need to be classified within the modern control domain. Feedback linearization control is one such technique dating back to the 1980s. Other more recent approaches are based on neural networks and fuzzy logic. The followings are the relevant modern control techniques found in the literature, and covered alphabetically here, which have found applications in flexible spacecraft attitude control arena.

Coupled Modal Control (CMC):

The term modal control can be interpreted in a broad sense as the control of a given number of spacecraft flexible modes. CMC represents a family of techniques whose underlying principle is to induce asymptotic stability by changing the open-loop eigenvalues so that the motion of the corresponding modes is damped out. Basic investigation on CMC was undertaken in [25]-[29] as discussed further below in connection with the pole allocation method. The subject of CMC in conjunction with structures is considered in [30] where Breakwell investigates the problem of optimal feedback slewing maneuvers of flexible Spacecraft from one position to another while leaving an arbitrary number of bending modes inactive at the end of the maneuver. Wu et al. [31] look at control of large flexible space structures using pole placement design techniques and includes the CMC method. Meirovitch et al. [17] undertake a comparative study of CMC and independent modal-space control where the basic difference between the two approaches lies in the manner in which the feedback controls are designed. The two approaches are compared qualitatively from design and computational viewpoints, and quantitatively through a performance index and *spillover* effects. Spillover refers to the phenomenon in which the energy intended to go into damping the *controlled modeled modes* is pumped elsewhere, for example into the *uncontrolled modeled modes* which must be avoided. Note that all the infinite number of flexible modes of a system can be broken down to *modeled* and *unmodeled* modes, and the modeled modes can be further broken down to *controlled modeled modes* or *uncontrolled modeled modes*. In conclusion, Meirovitch et al. [17] conclude that independent modal-space control method to possess advantages over CMC, as it permits easier design and implementation and requires less computational effort and control energy for implementation. However, some of their conclusions were based on the computational and implementation techniques available in the early 1980's and should be now reassessed, as much advance has occurred in the computing and software field in past two decades.

Decentralized Control:

Decentralized control for flexible spacecraft is concerned with the placement of actuators and sensors away from the rigid central body (e.g. on the flexible structure) and hence having better measurements and control authority for the control of the flexible parts. West-Vukovich et al. [32] consider the decentralized control problem for space structure using point force actuators and point displacement/displacement-rate sensors. They show that a solution to the decentralized control problem exists if it exists for the centralized control problem and their proposed decentralized controller also eliminates the spillover into the unmodeled high-frequency flexible modes. Young [33] advances a design framework for the decentralized control of space structures integrating finite element modeling and control design. Decentralized control is also used by Siljak [34] and Meirovitch and France [35].

Disturbance Accommodation Control (DAC):

The theory of DAC is based on the premises that disturbances are inputs which are not known accurately *a priori*, not accurately predictable, and not directly measurable in real-time [36]-[37]. DAC theory is a collection of non-statistical modeling and controller design techniques for the class of multi-variable control problems in which it is required to maintain set-point regulation or servo-tracking in the face of a broad range of uncertain, multi-variable and persistently acting disturbances. Moreover, DAC theory permits that control performance to be realized in such a way as to optimally exploit any useful energy or other effects which may be associated with the disturbances. Chun et al. [38] investigate the problem of large angle maneuvering of a flexible spacecraft, where the disturbance-accommodating feedback control tracks a desired output state which is obtained from an open-loop solution for the linear model. They develop closed form solutions of the feedback gains, state trajectory, and residual mode response for the tracking/disturbance-accommodating controller. Some numerical simulations presented show that DAC outperforms the simple linear tracking controller which does not use information about the disturbances. The DAC approach is also used by Meirovitch and Sharony [24] and [39], and Sharony and Meirovitch [40]-[41].

Eigenstructure Assignment/Pole Allocation (EA/PA):

The goal of linear feedback control is to place the closed-loop poles/eigenvalues on the left half of the complex plane so as to ensure asymptotic stability of the closed-loop system. The pole allocation approach consists of selecting first the closed-loop poles associated with the modes to be controlled and then computing the control gains required to produce these poles. Because this amounts to controlling a system by controlling its modes, this approach is known as modal control. The algorithm for producing the control gains is known as Pole Allocation, Pole Placement, or Eigenstructure Assignment. A comprehensive treatment of pole allocation can be found in the book by Porter and Crossley [29]. A design approach using pole placement techniques for a class of flexible space structures is developed by Wu et al. [31]. They examine numerical problems of the pole placement algorithm when employed towards high dimensional systems with very low-frequency eigenvalues. This control strategy is also considered for flexible spacecraft/LSS control by Kida et al. [42] and Tseng and Mahn [43].

Frequency Shaping Control (FSC):

Gupta [44] extends the Linear-Quadratic-Gaussian (LQG) method for feedback control design (covered below) to include frequency-shaped weighting matrices in the quadratic cost functional, and hence combining classical design requirements with automated computational procedures of modern control. The approach is also used by Chun et al. [45] for slew maneuvering a flexible spacecraft where a frequency-shaped open-loop rigid body solution was presented and a frequency-shaped perturbation control scheme was defined for maintaining the closed-loop rigid and flexible body system response in the vicinity of the rigid body nominal solution. They also found that the use of control smoothing in both the rigid body nominal solution and the perturbation feedback controller greatly reduces the flexible excitations.

Independent Modal-Space Control (IMSC):

This method is based on the idea of coordinate transformations, whereby the system is decoupled into a set of independent second-order systems in terms of the modal coordinates and as a result, a stable controller can be devised for each second-order system independently, no matter how large the model is. The control gains thus obtained lead to modal control forces which are abstract forces corresponding to the modal coordinates. This procedure not only guarantees controllability but also insures that no spillover into the uncontrolled modeled modes occurs, provided that the system in not under-actuated (i.e. the number of actuators used is equal to the order of the discretized system). Gould and Murray-Lasso [46] devise a Laplace transform version of independent modal-space control for distributed systems. Implementation is carried out on the assumption that the output and the forcing functions have negligible eigenfunction content beyond a finite number. Takahashi et al. [47] introduce some basic ideas in the independent modal-space control method. Meirovitch et al. [48] develop the independent modal space control for spinning flexible spacecraft. An IMSC dual-level control approach for controlling rigid-body and elastic motions of a space structure is presented by Meirovitch and Oz [49]. The IMSC method is also considered by VanLandingham and Meirovitch [50], Meirovitch and Oz [51], Meirovitch et al. [17], Baruh and Silverberg [52], McLaren and Slater [53], Quinn and Meirovitch [54] and Meirovitch and Quinn [55].

Linear-Quadratic-Gaussian/Loop Transfer Recovery (LQG/LTR):

Doyle and Stein [56] developed the LQG/LTR method. The LQG/LTR approach provides a means of including robustness-to-uncertainties in the control design process itself. Also, since it operates in the frequency domain, it extends the basic frequency domain design guidelines, such as bandwidth, crossover frequency, etc. from a scalar system to a multivariable system. It uses a low frequency design model of the plant and a high frequency characterization of the modeling errors. This method, which characterizes unstructured uncertainty with singular-value bounds, appears to be particularly well suited for the control of large flexible spacecraft due to the considerable modeling inaccuracy that inherently exists in the mathematical models. The method is applied to the synthesis of an attitude control system for a space structure by Joshi et al. [57] and Sundararajan et al. [58]. Sundararajan et al. [58] conclude that this method is useful for alleviating spillover effects common to LSS control problems modeled from finite element data. A modified version of the LTR robust control is presented by Blelloch and Mingori [59].

Model Error Sensitivity Suppression (MESS):

Sesak et al. [60] propose the MESS method to solve the problems of sensitivity to modeling errors and the use of low-order state estimators. The method is based on the decentralized control concept, covered earlier, which consists of a collection of subsystem estimators and controllers, each independently charged with a subset of the system states. The key controller concept is the penalization in the performance index of any control action that excites modeled states other than those for which the subsystem controller is charged, thereby inhibiting control spillover. These performance index modification concepts have utility with or without a decentralized controller. Sesak et al. [60] provide proofs demonstrating that performance indices can be modified to reduce control and observation spillover arbitrarily while preserving stability. Sesak and Likins [61] consider the capability of MESS to handle the truncation of known vibration modes. Sesak [62] advances a suppressed mode damping scheme in conjunction with MESS based on direct feedback of sensor signals. Sesak and Halstenberg [63] apply the MESS method to the decentralization of attitude and vibration control in a flexible space structure.

Multi-Objective Optimization:

Hale et al. [64]-[66] consider the problem of optimal integrated structure control design for a maneuvering space structure by means of reduced-order model. The problem is formulated as a multi-objective control problem involving the deflection, velocity and force spent, and the necessary condition for optimality is stated in the form of a maximum principle. Bodden and Junkins [67] develop an eigenspace optimization approach for the design of feedback controllers for the maneuvering and vibration suppression in flexible structures. Lisowski and Hale [68] use reduced-order models to study the unified problem of synthesizing a control and structural configuration for open-loop rotational maneuvering of flexible structures. Onoda and Haftka [69] present an approach to the simultaneous optimal design of a structure and control system for flexible spacecraft whereby the combined cost of the structure and the control system is minimized subject to a constraint on the magnitude of the response of the structure.

Optimal Projection/Maximum Entropy:

Bernstein and Hyland [70] present a review of optimal projection/maximum entropy approach which could be used to design low-order controllers for high-order systems with parameter uncertainties. The basic concepts are introduced by Hyland [71] where the active control of a large flexible spacecraft is considered. Hyland notes that difficulties arising from inherent inaccuracies in structural modeling and from the high dimensionality of the dynamical system force a re-examination of the problem of optimal control of LSS within the context of stochastic system theory. Recognizing that a complete probabilistic description of modal parameters can never be provided, one first identifies the minimum set of *a priori* data needed to preserve any measure of modeling fidelity. Acknowledging this data as available, a complete probability assignment is then induced by a maximum entropy principle. The resulting mean-square optimization problem is consequently reduced to the solution of a modified Riccati equation of particularly simple form.

Output Feedback Control (OFC):

Frequently the designer of control systems does not have a complete set of state variables directly available for feedback purposes. Moreover, one may wish to generate the control signals directly by taking linear combinations of the available output variables instead of first reconstructing the state via a Kalman filter. If the original system is time invariant and the linear transformations are also constrained to be time invariant, the design problem then reduces to choosing an appropriate matrix of feedback gains which leads to an optimization problem. A technique for determining optimal linear constant output feedback controls is devised by Levine and Athans [72]. This feedback control law does not require state estimation. Canavin [73] and Balas [74] suggest controlling the vibration of a flexible spacecraft by means of multivariable output feedback. The approach of Levine and Athan [72], Canavin [73] and Balas [74] is known as direct feedback control, which implies that sensor outputs are multiplied by a gain matrix to produce actuator commands. Elliott et al. [75] demonstrate that the output feedback controller suggested in [73] is capable of high performance. Lin et al. [76] investigate four output feedback control methods for possible application to vibration control in space structures. An output feedback control design capable of increasing the damping ratio and the frequency of selected modes in space structures is proposed by Lin and Lin [77]. The OFC is also discussed in Sesak [62], Joshi et al. [57], von Flotow and Schafer [78] and McLaren and Slater [53].

Perturbation Control Technique (PCT):

Auburn [79] investigates a perturbation approach to determine the closed-loop eigenvalues and eigenvectors resulting from a low-authority controller which imparts small forces to the structure. Sharony and Meirovitch [40]-[41] are concerned with the control of perturbations experienced by a flexible spacecraft during a minimum-time maneuver. The controller is divided into an optimal finite-time linear quadratic regulator and a disturbance accommodating control. Meirovitch and Sharony [35] present a general review of the problem of flexible space structure maneuvering and control with emphasis on a perturbation technique. PCT are also used by Quinn and Meirovitch [54] and Meirovitch and Quinn [55].

Positivity Control Method (PCM):

Iwens et al. [80] discuss a technique for designing robust controllers for flexible space structures based on positivity of operators. There are basic conditions for stability based on positivity. A multi-input multi-output system including controller and plant transfer matrices is asymptotically stable if at least one of the transfer matrices is strictly positive real and the other is positive real. This stability concepts need not be restricted to transfer matrices but can include nonlinear and time-variant operators defined on a vector valued Hilbert functional space. Benhabib et al. [81] propose a technique for the design of stable and robust control systems for space structures based on positivity of transfer matrices.

Benhabib [82] extends the concept of positivity presented in [81] to discrete-time systems. McLaren and Slater [53] investigate robust multivariable control of space structures designed on the basis of a reduced-order model using positivity concepts. Three control design approaches are compared: (i) ordinary multivariable control; (ii) independent modal-space control; and (iii) decentralized direct output feedback control.

Traveling Wave Control (TWC):

Vaughan [83] examines wave propagation concepts to control the bending vibration in beams. Meirovitch and Bennighof [84] are concerned with the use of the Independent Modal-Space Control method to control traveling waves in a structure. Von Flotow [85] describes the elastic motion in a space structure in terms of traveling waves. To achieve control, the energy is shunted into unimportant portions of the structure or is absorbed by an active wave absorber. Von Flotow and Schafer [78] describe the experimental implementation of the wave absorbers introduced in [85]. They discuss the similarity between wave absorbing compensator and direct velocity output feedback.

Uniform Damping Control (UDC):

Silverberg [86] introduces the idea of uniform damping control of flexible spacecraft where the control torques are generally a function of the displacement and the spatial derivatives of the displacements. He also shows that the uniform damping control solution represents a first-order approximation to a special globally optimal control problem. The UDC control approach exhibits three distinctly attractive features, namely (i) the associated uniform damping control law is independent of spacecraft stiffness; (ii) the associated control forces are proportional to the spacecraft mass matrix; and (iii) the UDC law in centralized. Since the mass matrix is time invariant, these control formulations are not affected during the maneuver and are therefore robust. Meirovitch and Quinn [87] make use of UDC for the problem of slewing a large structure in space and suppressing any vibration at the same time, but treating each aspect independently. Also they found that the optimal actuator locations are the points of maximum mass both for maneuver and vibration control. The method is also used by Meirovitch and France [35] and Quinn and Meirovitch [54].

Feedback Linearization Control (FLC):

There are several excellent books covering the general topic of Feedback Linearization Control [88]-[90]. Essentially, feedback linearization as a control technique is referred to the approach where the nonlinear differential equations of the plant are linearized by an appropriate coordinate and control law transformations. In other words, exact linearization is achieved by the change of the representation of a nonlinear system into a linear one, which is controllable and then a stabilizing controller is designed that guarantees the desired performance specifications. There exist several methods of deriving feedback linearization with the most common three being:

(i) State linearization which is a nonlinear change of coordinates by a diffeomorphism.

(ii) Input-state feedback linearization (sometimes called state feedback linearization or just feedback linearization) consists of finding a nonlinear feedback and a state space diffeomorphism. This is a generalization of the pole placement for linear systems.

(iii) Input-output feedback linearization (sometimes referred to as partial feedback linearization) consists of a selection of the outputs that will make the system input-output feedback linearizable. It also requires a nonlinear feedback and a state space diffeomorphism. This is a generalization of the pole-zero cancellation technique of linear systems.

A particular problem that has received much attention from the research community was the design challenge posed by Taylor and Balakrishnan [91]. In the design challenge, the Spacecraft Control Laboratory Experiment (SCOLE) was set up by NASA Langley Research Center to provide a standard configuration to test control laws for investigators. The laboratory apparatus consisted of the Space Shuttle connected by a long (130 ft) flexible beam to a hexagonal antenna or reflector. The reflector and Shuttle were treated as rigid bodies. When the Shuttle maneuvered, the flexible modes of the beam were excited. Two force actuators and a torquer were provided to exert sufficient force to suppress unwanted vibration of the beam. Research related to both dynamics and control system design has been performed and reported in literature and SCOLE workshops at NASA Langley Research Center [55] and [92]-[99]. Of particular interest to this thesis, nonlinear invertiblity theory (e.g. feedback linearization) has been used in [100]-[101] for designing controller for a simplified spacecraft system. For such a system, adaptive control system and sliding mode controllers have been also designed to compensate for parameter uncertainty and disturbance torques acting on the spacecraft [102]-[103]. Azam et al. [104] treat the question of the large-angle attitude maneuver and vibration suppression of the SCOLE system based on input (Space Shuttle torque) - output (attitude angles) feedback linearization for attitude control and linear feedback for vibration suppression. An input-output feedback linearization control law is derived. This gives rise to decoupled linear attitude dynamics and allows independent control of attitude angles through PID control of the attitude errors. For the design of the controller only attitude angles, angular rates and tip elastic deflection components and torsion deflection at the tip of the beam are assumed to be measured by sensors. An observer is designed to estimate the elastic modes and their rates for use in the feedback linearization control loop, as the control requires knowledge of the complete state vector. Simulations of the system show that the large maneuvers of the spacecraft can be performed to follow precise attitude trajectories and elastic modes can be stabilized, however the paper does not prove the stability of the overall system including the controller and the observer.

Bennett et al. [105] develop a nonlinear adaptive control system design, based on the Lagrangian dynamical system and input-output feedback linearization, for the flexible multi-body model of SCOLE system. The approach to adaptive input-output feedback linearization is shown to be effective for a spacecraft with significant uncertainty in the appendage flexure dynamics. Also, it is noted that the importance of integrated design including low authority damping enhancement for robust slewing and pointing control underlies the potential significance of nonlinear dynamics in understanding and compensating for control-structure interactions in flexible multi-body systems. The approach taken in [105] is different from that of [104] in that the control derivation is based on the explicit structure of the dynamics arising from Lagrange's equations which simplifies the structure of the nonlinear control law.

The work by Monaco and Stornelli [106] shows that the method of feedback linearization can be successfully applied to perform a large angle maneuver for a spacecraft with Flexible Beam-type Appendages (FBA). More recently, Lin and Lin [107]-[109] present a nonlinear feedback control law which is able to perform three-axis attitude reorientation of a flexible spacecraft with a rigid bus and two flexible beam-type appendages. The attitude feedback control was derived with the method of input-output feedback linearization and the formulation of stable attitude error dynamics, which eliminates the error vector projected from the four-dimensional Quaternion space of the distance between the present and the final attitudes. Lyapunov stability analysis was applied and a set of appendage vibration feedback gains were determined from the analysis to ensure the stability of the attitude control system, however the rate of the Lyapunov function was not shown rigorously to be always negative and the zero dynamics of the closed-loop system was not analyzed. Also, the vibration feedback gains found are not constant but rather time varying and are in sort adding damping to the structures. Their control method requires only three torque actuators located in the rigid bus to perform attitude maneuvers. Lin and Lin [108] look at a different approach to the same system as in [107] and [109] by considering input-output feedback linearization and sliding mode method to select the gains. Here again, the appendage stabilization in the form of adaptive damping was derived from the procedure of Lyapunov stability and was a part of the attitude feedback control law. Papers [106]-[109] have focused on the problem of rest-to-rest slew maneuvers, subject to near zero angular rate initial conditions for spacecraft with flexible beam-type appendages using feedback linearization control.

Indeed, all of the attitude control literature presented in the previous sections has addressed:

(i) attitude regulation or pointing problem about an equilibrium point;
(ii) attitude tracking problem; and
(iii) rest-to-rest slew maneuvers

for rigid or flexible spacecraft using different control strategies. However, there is a lack of research on the important problem of automated attitude recovery, in case of a severe malfunction causing significant initial angular rate perturbations which has not been addressed directly in the literature. This is an important problem to address as many if not all spacecraft in orbit today, do not have an automated recovery procedure in case of a malfunction affecting the attitude of the spacecraft, and rely on ground intervention for the recovery procedures.

In this thesis, the problem of automated attitude recovery for a spacecraft with generic Flexible Plate-type Appendages (FPA) is addressed including the analysis of the zero dynamics and the overall closed-loop stability. It is important to also note that flexible plate-type appendages are more generic structures than the flexible beam-type appendages usually found in the academic literature.

### 1.1.3 Spacecraft Simulation

In order to design and analyze ACS for spacecraft, it is important to be able to simulate the dynamics of the spacecraft. Hardware-in-the-loop simulations of a spacecraft, such as air-bearing spacecraft simulators, are not only expensive to build, but they cannot provide the full experience of micro-gravity. Moreover, a low torque environment is often the key to the success of high-precision systems, but duplicating such environments on ground to validate control concepts is quite difficult [110]. Also many future spacecraft are large in dimension and lightweight and will not be able to support their own weight in a gravity field and hence no ground testing will be possible. An alternative is to develop a purely software simulator. For an academic setting, it is preferable that the simulator be inexpensive and the source code readily available so that it may be adapted for different requirements.

For rigid spacecraft simulation, there exist several simulators such as MultiSatSim (MSS), developed by Princeton Satellite Systems [111]; FreeFlyer developed by A.I. solutions [112]; and the Open-Source, Extensible Spacecraft Simulation And Modeling Environment (Open-SESSAME), which is a collection of open-source libraries and toolkits [113]. However, for flexible spacecraft simulation, the existing few simulators which are mentioned below, either have limited functionality, thus limiting their usefulness; or the simulators are proprietary, thus making their internal operations inaccessible; and they're simply too expensive for an academic setting.

The Dynamics Algorithms for Real-Time Simulation (DARTS) Shell (Dshell) is a spacecraft simulator built by NASA [114]. Dshell is an interface to (i) the DARTS computational engine, which solves equations for the flexible dynamics of a spacecraft, and (ii) a set of hardware model libraries. It also has the advantage of being portable from desktop computer workstations to real-time, hardware-in-the-loop simulation environments. However, the disadvantage is that Dshell is not easily accessible. The Dynamics and Control Analysis Package (DCAP) is a spacecraft simulator developed by ESA [115]. DCAP is a tool for designing and verifying the dynamics and control performance of coupled rigid and flexible structural systems. Basically, the DCAP model consists of the spacecraft, sensors, actuators, and the controller. The simulator is menu-driven, and has 3D modeling and animation capability. Unfortunately, access to the source code is difficult, if not impossible. Alternative to the two government-based simulators just described, there exists a few commercial flexible spacecraft simulation software, such as the Dynawiz XSV provided by Concurrent-Dynamics International [116]. Dynawiz XSV is a multi-body simulator for both rigid and flexible body dynamics. The simulation parameters are setup from the DOS command line, and the flexible dynamics engine is interfaced through SIMULINK. The disadvantage of Dyanwiz XSV is that the simulator is not user-friendly and the source code to the dynamics engine is not readily available. MSC Software Corporation has developed ADAMS, a simulator for a spacecraft with flexible panels and antennas. It includes functionalities such as attitude stabilization during the deployment of flexible solar panels, and includes a control system integrated with mechanical equations [117]. The big disadvantage with the ADAMS simulator is its cost. Finally, Open Channel Software offers the Flexible Spacecraft Dynamics (FSD) simulator, which is from the NASA COSMIC collection [118], but it was also not easy to obtain. Consequently, since there does not exist any inexpensive flexible spacecraft simulator with easy access to its source code, a software simulator was developed for this thesis.

### 1.1.4 Spacecraft Failures

A spacecraft failure investigation was conducted where about 150 spacecraft failures were analyzed and classified in order to identify the critical spacecraft systems that failed over the last 25 years. The spacecraft failures were classified in 8 different categories. The breakdown of these failures by spacecraft subsystem is displayed in Figure 1.1.

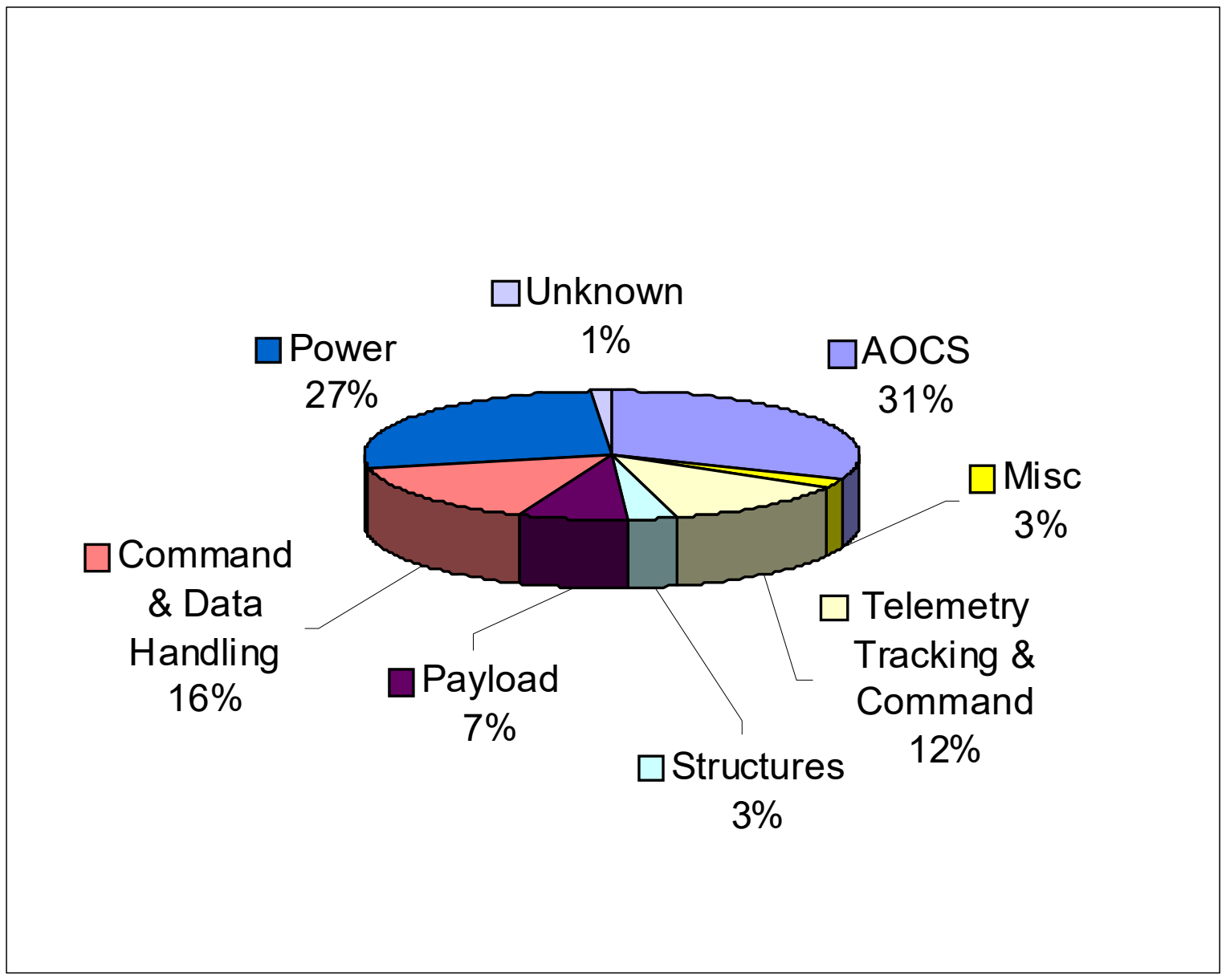

Figure 1.1: Spacecraft subsystem failure statistics

AOCS failures represent about one third of the total failures affecting spacecraft. Figure 1.2 shows the sources of the AOCS failures.

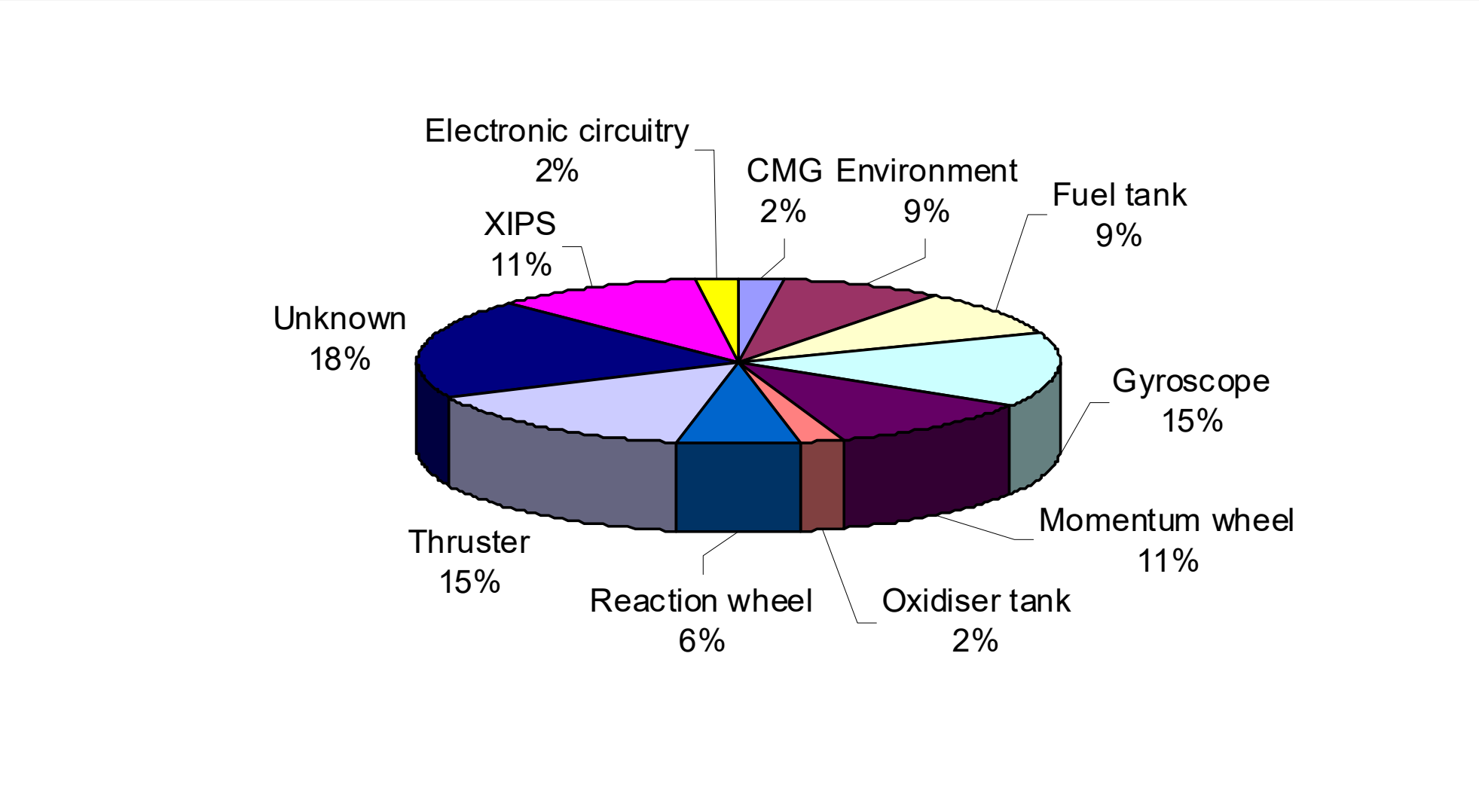

Figure 1.2: Sources of failures for the attitude and orbit control subsystem

The next three subsections addresses three of the main AOCS failure possibilities that could severely affect the attitude of the spacecraft, and hence requiring an attitude recovery maneuver.

Thruster Failures:

An important part of the Attitude and Orbit Control System (AOCS) is the propulsion subsystem which is used for orbit and/or attitude control. Thrusters and fuel tank failures represent about 25% of all AOCS failures. Table 1.1 summarizes some recent on-orbit thruster assembly failures, ordered by their failure dates.

On-Orbit Collisions:

The first ever credited collision of two catalogued space objects was detected in low Earth orbit on 24 July 1996. After a year of successful operation of the CERISE satellite, the operators at the mission control center in Guilford, Surrey, observed a sudden change to the attitude of the microsatellite as it was thrown into a rapid end-over-end tumble in its 700 km polar Earth orbit. After initial investigations, a collision with a space debris was suspected and subsequent observations and analysis by the UK Space Track and NASA confirmed the on-orbit collision. The debris appears to have impacted the gravity gradient boom, which extended 6m from the rigid bus, in a head-on course at an impact speed of over 14 km/sec. Eventually, the ground control engineers were able to recover the spacecraft's attitude, by uploading a new control algorithm [119].

Even though on-orbit collisions have a low probability, they can happen, and especially with the steady increase of the number of space debris over the years, this is becoming a more serious problem that will require attention. An automated attitude recovery system could be utilized effectively for on-orbit collisions.

Table 1.1: Summary of on-orbit failures for thruster assemblies

| Spacecraft | Failure Description | Date of Failure | Mission Degradation? | Loss of Mission? | References |
|---|---|---|---|---|---|
| JCSat-1B (JCSat-5) | JSAT claims that the satellite experienced an anomaly in one of its thrusters. Preparations were made to operate the satellite using its remaining functional thrusters. Services provided by JCSAT-1B were restored on 18 January. | 17 January 2005 | No | No | [120]-[121] |
| EchoStar VIII | During 2002, two of the thrusters on EchoStar VIII experienced anomalous events and are not currently in use. During March 2003, an additional thruster on EchoStar VIII experienced an anomalous event and is not currently in use. | March 2003 | No | No | [120],[122] |
| Space Shuttles | Extrusion of a polytetrafluoroethylene (PTFE) pilot seal located in the Space Shuttle Orbiter Primary Reaction Control Subsystem (PRCS) thruster fuel valve has been implicated in 68 ground and on-orbit fuel valve failures. | December 2001 | No | No | [122] |
| EchoStar V | One of the thrusters on EchoStar V experienced an anomalous event resulting in a temporary interruption of service. | August 2001 | No | No | [120],[124] |
| EchoStar VI | EchoStar operators believe that the satellite was hit by one or more micrometeorites, in its attitude control system. More specifically EchoStar VI has a hole causing a propellant leak in or near one of the thrusters. When the thruster was fired, ground control observed unexpected movement of the spacecraft. | April 2001 | No | No | [125]-[126] |
| AO-40 (Phase 3D) | It was determined that a plugged valve vent on the 400N thruster had prevented proper functioning of the burn valves and had probably allowed build-up of fuel pressure in the cooling coils around the motor bell housing. These coils apparently ruptured and in the process damaged one or both of the burn valves. During cycling of the pressurization valve the following day, one component of the fuel apparently escaped from the damaged burn valve at the motor housing and mixed with residual second fuel component in the motor, creating a localized explosion. | 1 December 2000 | Yes | No | [120],[127] |
| Hot Bird 2 | Eutelsat's Hotbird 2 went into safe hold mode during an unexpected anamoly. The reason for the anomaly is unknown, however some recent suggestions link it to the problem several satellite have with leaking thrusters supplied by Daimler Chrysler Aerospace. | 21 March 2000 | No | No | [120],[128] |

Mechanical Wheel Failures:

Mechanical wheel and mechanical gyroscope failures represent about 35% of all AOCS failures, but wheel failures often happen to be more critical than gyroscope failures. For example, in several occasions, engineers found alternative solutions to control spacecraft without affecting the scientific or telecommunication operations even when all gyroscopes failed (e.g. BeppoSAX and ERS 2 spacecraft).

On the other hand, mechanical wheel failures often degrade a spacecraft permanently. The failure of one or more wheels forces an overuse of remaining wheels and/or thrusters which decrease the spacecraft life. Mechanical wheel failures also result in a temporary interruption of the spacecraft operation when switching from normal mode to safe hold mode. This represents a loss of scientific data or communications that can extend over several weeks. This has occurred for many spacecraft, such as Echostar V, Far Ultraviolet Spectroscopic Explorer (FUSE), Radarsat I, etc.

Table 1.2 summarizes some on-orbit failures of momentum/reaction wheels and control moment gyros, ordered by their failure dates. A more detailed summary and causes of failure on 3 spacecraft, which are reported in the literature, are presented in the next 3 sub-subsections.

Table 1.2: Summary of on-orbit failures for momentum/reaction wheels and control moment gyros

| Spacecraft | Failure Description | Date of Failure | Mission Degradation? | Loss of Mission? | References |
|---|---|---|---|---|---|
| FUSE | On November 25, 2001, the *x*-axis reaction wheel stopped. Science operations continued using the three remaining operable wheels. On December 10, 2001, the *y*-axis wheel also stopped. Spacecraft was temporarily put in safe hold mode with solar arrays pointed towards the Sun to maintain power to the spacecraft's systems. Spacecraft was recovered by (i) adjusting the coordinate system of the satellite so that the skew wheel controls a new axis, which is halfway between the old *x* and *y* axes, and (ii) adapting the magnetorquers so that they can provide control on the third axis.<br><br>FUSE experienced a significant anomaly that temporarily suspended science operations on 27 December 2004. For reasons that are still under investigation, one of the spacecraft's two remaining reaction wheels stalled placing the satellite into safe hold mode until now. | 27 December 2004 | Yes | Yes | [129] |
| Radarsat-1 | In September 1999, the primary pitch momentum wheel started suffering from excessive friction and temperature, consequently, control was shifted to a back-up wheel. This back-up pitch momentum wheel also developed similar problems and was taken offline on 27 November, 2002. Canadian Space Agency (CSA) developed new pointing procedures that eliminates the pitch momentum wheel, relying instead on roll and yaw reaction wheels and magnetorquers to accurately point the spacecraft. | December 2002 | Yes | No | [130] |
| International Space Station | On June 8$^{th}$ of 2002, one of the Control Moment Gyros (CMG) suffered a bearing failure and is no longer operational. | 8 June 2002 | Yes | No | [131] |
| EchoStar V | EchoStar V experienced the loss of one of its three momentum wheels. Two momentum wheels are utilized during normal operations and a spare wheel was switched in at the time. A second momentum wheel experienced an anomaly in December 2003 and was switched out resulting in operation of the spacecraft in a modified mode utilizing thrusters to maintain spacecraft pointing. | July 2001 | No | No | [132] |
| GOES-9 | An emergency outage occurred on the GOES-9 weather satellite. Following that, it was noted that the active momentum wheel on-board GOES-9 became dangerously hot, apparently resulting from of a high current drawn by the wheel. Momentum wheel 2 was switched off on January 2003. | 7 July 1998 | Yes | Yes | [133] |
| ANIK E1 and E2 | The momentum wheel on both Anik E1 and E2 failed due to, most probably, a dielectric bulk charging followed by a discharge disabling key circuitry. ANIK E1 was out of service for just a few hours, but the attitude recovery of ANIK E2 took many months, and the cost of rescue, loss of revenue and reduced lifetime is estimated at tens of millions of dollars. | January 1994 | Yes | No | [134] |
| GPS BI-05 | Wheel no. 2 and no. 3 stopped completely with full motor voltage applied. Spacecraft attitude recovery for satisfactory operations was not achieved and the satellite mission was ended. | December 1983 | Yes | Yes | [135] |

**Far Ultraviolet Spectroscopic Explorer (FUSE) Reaction Wheel Failures**

Overall, three reaction wheels failed in November 2001, December 2001 and December 2004. The first two reaction wheels failed in November and December 2001 (within a two-week period), which put the spacecraft into safe hold mode for 3 months. Then, the third reaction wheel failed in December 2004 and the spacecraft remains in safe hold mode until now.

A formal investigation was conducted to determine the first FUSE reaction wheel anomaly. The formal investigation about the wheel torque anomaly studied 5 possible sources of failures and the events leading to the wheel shutdown. FUSE experienced a gradual increase of torque over 12 hours on day 214 of year 2000. Three days later, FUSE spacecraft was put into safe hold mode when "a torque event in the pitch wheel resulted in a decrease in the wheel speed of 175 rpm in a time of as little as 1 or 2 seconds"[129]. The source of the problem was identified to be a *mechanical touchdown caused by air bubbles*.

The investigation also considered a *sudden catastrophic failure* hypothesis that was rejected, but this failure type could happen on other spacecraft. It is also a possibility that must be considered in the failures of the 2 reaction wheels that permanently failed in 2001 and 2004 on FUSE. The sudden catastrophic failure that was studied is a *contamination in reaction wheel bearings*. The authors describe the failure as follows: "Contamination in the bearings typically will not result in a gradual increase in torque. It can eventually lead to bearing failure, which would be similar in symptoms to the eventual catastrophic failure due to lubricant starvation" [129].

**International Space Station (ISS) Control Moment Gyro (CMG) Failure**

A total failure of Control Moment Gyro (CMG) occurred on the International Space Station (ISS). A formal investigation was also conducted to determine the possible causes of failure. The incident happened as follows: "Several hours after [shuttle] docking, the flight control team noted a flag that indicated the accelerometer on CMG1 had measured 0.5 Gs. The CMG1 SMCC [Spin Motor Commanded Current] was elevated and one spin bearing temperature was beginning to rise. GN&C [Guidance, Navigation & Control] personnel determined that the CMG would be monitored but no immediate action would be taken. The event repeated itself twice during the next 16 hours. Then, following a 30-minute period with no ground communication, telemetry showed that the CMG1 SMCC was at its maximum value, 1.6 Amps, and the wheel-speed was decreasing. After several seconds of maximum current and wheel-speed reduction, Fault Detection and Isolation software declared the CMG failed and automatically shutdown the spin motor" [131].

The most probable factors leading to the failure are the combination of a change in pre-load on the bearing from thermal contraction and induced ball skidding from high gimbal rates produced during the same period on CMG1 spin bearing #2. However, due to either less lubrication or more extensive ball skid, the event caused a catastrophic failure [131]. The ball skidding phenomena happens when one or more balls in bearing ride on opposite race from the others which adds stress to the retainer. However, if properly lubricated, it should not cause damage.

Again, the investigation studied a sudden catastrophic failure hypothesis that was rejected. This sudden catastrophic failure is called *Thermal Clamping*. The failure can be described as follows: "The SB [Spin Bearing] design was conducted with the assumption that the thermal power distribution between the inner and outer races was 70%/30%. During LVLH [Local Vertical Local Horizontal] flight periods, the heater cycles the outer race temperatures with little thermal transfer to the inner race. This creates a temperature differential within the bearing as the outer race is warmer and inner race is cooler, causing the gap between the bearing to increase slightly. However, in the XPOP flight attitude, the CMGs are heated evenly and both races maintain approximately the same temperature. As the ISS transitions from XPOP [X-axis Perpendicular to Orbital Plane] to LVLH, the temperatures drop dramatically as the assembly sheds thermal energy. The outer races conduct thermal energy to the inner and outer gimbals with large surface areas to radiate heat. The inner race transfers the heat down to the spin axis into the momentum wheel that acts as a heat sink. It has been shown that with a large enough temperature differential, the outer race could clamp down on the balls and cause the bearings to seize. As the bearings rub on the outer race, heating from friction will increase the temperature of the outer race and it will expand. With initial damage to bearings, failure would occur quickly after the first instance" [131].

**GPS Reaction Wheel Failure**

GPS Satellite Navstar 5 experienced anomalies in reaction wheels no. 2 and no. 3. After few months of intermittent problems, both reaction wheels failed within 4 minutes and could not be recovered. The wheels stopped completely even though full motor voltage was applied. Telemetry also indicated that the failed wheels drew current when no acceleration was observed.

The possible causes of failures in wheel driver electronics, as specified by the author, are: improper switch of the H bridge, thus bypassing one of the motor windings and maintaining constant DC current on one winding. The suspicious areas within the wheel assembly electrical circuit are the open phase and the shorted windings. The suspicious areas within the mechanical wheel assembly included bearing degradation from loss of lubrication and the change in preload [135]. Another reason explaining the failures involves the operations conducted to clear the intermittent problems that have been encountered before the critical failures. During these operations, the wheels were driven with full motor voltage for more than a few hours. In November 1983, the wheel driver circuit designer recommended to run the wheels at 2900 rpm for a week. It was the first time the circuit board attempted this operation. This may have degraded the electrical and mechanical parts of the wheel.

However, the author concluded that the catastrophic failures result from a combination of more than one failure: "The cause for the failure of two wheel systems on the same satellite within 4 min is still unresolved. … Even considering each wheel as totally separate, no single failure for either wheel can account for all of the observations on either wheel alone"[135].

Even though no serious catastrophic wheel failure has been reported in the literature, sources (e.g. thermal clamping) do exist that could lead to such failure and hence it must be considered as a real possibility that could sooner or later happen on a present or a future spacecraft. Since a sudden catastrophic failure would most probably not be detected quickly enough by ground, an automated attitude recovery system could become useful. For planetary spacecraft, the situation is even worse due to the longer communication delays.

The possible spacecraft anomalies described in the previous sections (i.e. thruster malfunction, on-orbit collision and wheel failure) are probable events and when they do occur, they can seriously jeopardize the mission. Also not specifically addressed above, are many anomalies or failures caused by incorrect operational procedures or operator errors which may have a significant effect on the attitude of the spacecraft, and could necessitate an automated attitude recovery maneuver. As mentioned previously, the particular malfunctions that we are considering in this thesis, are those that impart a large disturbance torque on the spacecraft resulting in tumbling. For example in a manned spacecraft, such motion if left uncontrolled would create a hazardous environment for the crew who would experience angular accelerations. The structural integrity of the spacecraft (specially its appendages) can also be jeopardized by prolonged or fast tumbling of the spacecraft. But perhaps the single most important cause for alarm for many spacecraft, in case of loss of attitude control, is the danger to the power subsystem and depletion of the batteries since the solar panels might point away from the Sun.

## 1.2 Objectives and Scope

The fundamental overall objective of this thesis is to contribute to the field of dynamics, simulation and control of flexible spacecraft. The specific focus of this research will be on the design of an automated attitude recovery system for a common class of rigid and flexible spacecraft.

Attitude recovery, in a general sense, can be defined as reducing the spacecraft's angular kinetic and elastic potential energies while achieving a reference/target attitude. It is assumed that the spacecraft is equipped with appropriate ACS actuators which can deliver high authority control torques about the spacecraft body axes. At the present time, such actuators could be either thrusters or Control Moment Gyros (CMGs). Also, the spacecraft would need to be equipped with appropriate attitude sensors (e.g. gyros) and attitude determination algorithms to deliver the required attitude state to the attitude recovery system. It is further assumed that the effects of the environmental torques (i.e. gravity gradient, solar pressure, aerodynamic and magnetic torques) on the spacecraft are negligible as they are small (in the order of $10^{-6}$ to $10^{-3}$ Nm) compared to the control torques applied during the attitude recovery maneuvers which are usually much larger. In this thesis, we will be dealing with extreme cases (i.e. large initial disturbance torques) to validate the controller proposed and the attitude recovery torques are between 1 Nm and 50 Nm. Even at 1 Nm level, the control torques are at least about 1000 times larger than the environmental torques. However, these environmental disturbances are modeled and included in the software simulator.

The scope of the thesis can be summarized in the four major points below:

1. Mathematical formulation of a generic star topology spacecraft with a rigid central bus and *n* flexible plate-type appendages, in a deployed configuration. The mathematical models shall be such that they lend themselves well to numerical simulation (e.g. analytical models as opposed to Finite Element Models (FEM)).

2. Design and implementation of a generic numerical simulation of the spacecraft described above (#1). The simulation needs to be of high fidelity and shall be implemented in such a way that is easy to be used for the purposes of testing and investigating the attitude recovery system (#3).

3. Investigate and propose a controller for the attitude recovery of the spacecraft in an emergency situation (e.g. thruster misfire). This high level requirement on the attitude recovery system translates itself to the following control system requirements:

a. After an unpredictable incidence, the controller needs to determine that it must react to the situation, for example the attitude recovery system can monitor the angular velocity of the spacecraft and if it exceeds a certain limit, it initiates a recovery procedure.
b. The usual recovery procedure would be to take the spacecraft from an undesired angular velocity to a safe-hold orientation with near-zero final angular velocity.
c. The attitude recovery system must also damp the vibrational motion of the appendages as it is performing the attitude recovery maneuver.

4. Formal mathematical analysis supported by numerical simulations for validation and verification of the closed-loop control system's stability and performance requirements.

## 1.3 System Performance Requirements

The system performance requirements for attitude recovery can be stated simply as the regulating (or zeroing) of the attitude angles or rates towards a specified reference, starting from an unknown initial condition. In the case of a flexible spacecraft, we must as well include the requirement of vibration suppression of the appendages. The recovery maneuver needs to be performed autonomously without ground intervention and in less than some specified time depending on a given mission. In order to make the present problem more challenging, we chose a rather short and arbitrary maneuver time limit of 600 seconds (10 minutes). One can assume that this is the maximum amount of time where battery power is available to the attitude recovery system before batteries go below their acceptable depth of discharge. This could be a worst case scenario where for example the spacecraft is in eclipse or its solar panels are moved away from the sun, due to the attitude failure, so that the batteries can not be continuously charged by the solar panels. Of course, the exact duration for which power could be available to the attitude recovery system would vary greatly upon many factors, including the type of malfunction/failure, the magnitude of the initial disturbance, the battery size, the spacecraft orbit, etc.

## 1.4 Methodology

The methodologies used in this thesis are described in the following sections:

Dynamics Formulation

The common way of modeling flexible spacecraft found in literature is by using the Lagrangian formulation based on energy. The Newton-Euler formulation based on angular momentum is used mostly for multi-body dynamics of robotic systems with many link and joint and is not used much in the flexible spacecraft modeling.

The Lagrangian approach is summarized as follows and was given the special name of *hybrid-coordinates* modeling [136]-[139]. The hybrid-coordinates approach separates the system into two distinct subsystems which are categorized as rigid body and flexible appendages. The flexible appendages are assumed to have a linearly elastic structure for which small deformation assumption must also hold valid. That is the deformations are such that elastic stresses remain proportional to deformations. The flexible appendages are then modeled using distributed or modal/generalized coordinates whereas discrete coordinates are used to describe the unconstrained motion of the rigid body. This approach was also used earlier by [140] but was not given a particular name. The symbolic derivation of the equations, if possible, provides the fastest possible numerical simulations, so that a complex spacecraft with many appendages can be simulated in real time on modest computers.

Hence, the Lagrangian approach will be also used here to model the dynamics of the flexible spacecraft. The topology of the spacecraft is restricted to the most commonly found in the industry and literature which is the star topology. The star topology includes a rigid bus and $n$ (usually $n < 10$) flexible appendages attached to the rigid bus at one end and free at the other. The flexible appendages are chosen to be plate-type which are again closest to the shapes of actual appendages found in spacecraft design. Deployable solar panels, antennas and radiators are more commonly rectangular or circular plate-type objects which are mostly flexible. Also the more realistic flexible plate-type appendages were chosen since they are less treated in the literature, due to the complexities of their mathematical models. There is a good deal of research and experimentation data [141] available for cantilever type plates with fixed boundary condition (no rotation allowed), but hardly any experimental data (e.g. flight data) for the same plates with rotational liberty at their boundary, as is the case for the system under consideration.

Another reason why flexible plate-type appendages were not as popular as flexible beam-type appendages were due to the fact that the computational power needed to simulate plate dynamics is higher. However, due to the advent of affordable high speed processors and low cost computer memory at the present day, the required computational power for simulating a complex spacecraft with several appendages is not a major constraint, anymore. Indeed, this simulation can be performed on a high end laptop!

Simulation

The flexible spacecraft simulator was developed to aid in the analysis of the control system proposed. This numerical simulation served in-lieu of a terrestrial experimental setup for the 3 Degree Of Freedom (DOF) attitude motion which is extremely difficult and expensive to build properly. Motion tables with 2 DOF do exist but are expensive to acquire in a university setting, and 3 DOF motion tables are extremely rare. Also many future spacecraft are large in dimension and lightweight and will not be able to support their own weight in a gravity field and hence no ground testing will be possible. As such, effort must be put towards making this numerical simulation of high fidelity so as to closely represent the flexible dynamics of the typical spacecraft described previously.

The validation of the simulator was an important step in this thesis and its validation was done in several steps:

- Test and validation of individual subroutines (e.g. spacecraft inertia calculation subroutine) which implements the basic building blocks of the overall system.

- Validation of the overall dynamic system based on the following criteria
  - Verification of the system's conservation of energy
  - Verification of the system's conservation of angular momentum

- Verification of the special cases where analytical solutions may be obtained for both rigid and flexible cases.

Control

The control strategy is based on feedback linearization control which has proven to be a very powerful technique in handling nonlinear systems. The approach was to investigate the input-output linearization technique and to analyze its merits and drawbacks for the present problem. The relevant analytical results investigated are stability analysis and robustness. The controller is designed to be asymptotically stable and have sufficient robustness to handle inevitable model/parameter uncertainties.

## 1.5 Thesis Outline

The thesis is organized as follows: The present chapter serves as an introduction outlining the problem of attitude recovery and providing a relevant literature survey of the spacecraft attitude dynamics, simulation and control fields. Chapter 2 provides the necessary background for the understanding of the dynamics of the problem and formulating the rigid and flexible spacecraft dynamics. Chapter 3 presents the high-fidelity dynamics and control simulator developed to serve as the test-bed for verification of the control strategies and for analysis of the system closed-loop dynamics. The important topic of validation of this simulator is detailed for both rigid and flexible cases. Chapters 4 and 5 provide the control formulation and simulation results for the attitude recovery of the rigid and flexible spacecraft, respectively. Chapter 6 contains all the necessary stability analysis for the rigid and flexible spacecraft and provides a derivation of a useful general formulation for obtaining the *normal form* of a class of nonlinear systems. Chapter 7 provides the thesis conclusions and directions for future work. Two relevant appendices are also included to complement the main body of the thesis.

## 1.6 Thesis Contributions

The present thesis has made the following contributions to the spacecraft dynamics, simulation and control literature:

- Firstly, the important topic of *automated* attitude recovery (e.g. without any human intervention) of rigid and specially, flexible spacecraft has been studied in this thesis. This is a topic which has not been addressed specifically by the aerospace academic literature. The literature has been mainly devoted to (i) attitude regulation or pointing problem about an equilibrium point; (ii) attitude tracking problem; and (iii) rest-to-rest slew maneuvers. The important difference in attitude recovery is the fact that the initial conditions are not known *a priori* and could be quite large, disturbing the attitude of the spacecraft far from its equilibrium point.

- Development of a generic attitude and flexible dynamics formulation for a class of spacecraft having a star topology consisting of a rigid central bus to which $n$ flexible plate-type appendages are attached. These appendages can be attached in any orientation and could have different boundary conditions. The existing formulations found in the literature are either generic and do not provide detailed expressions or use flexible beam-type appendages which is a simpler case. Also, in order to have closed-form expressions for all the terms in the formulation presented in this thesis, a number of non-trivial integrals of beam eigenfunction mode shapes were derived analytically.

- A high-fidelity simulator which includes the spacecraft's dynamics, environment, and control system was designed and implemented and can be configured to be used for different spacecraft configurations and control algorithms. Many validation test cases were devised, studied and documented which add confidence in regards to the correctness of this simulator's outputs. These results can be used by other researchers to validate their own simulators. This simulator was an invaluable tool for the realization of this thesis and will play a central role for future work in this area. Also, to the best of the author's knowledge, the use of fundamental physical conservation concepts (i.e. conservation of energy and momentum) for validation of complex spacecraft simulations is novel.

- A nonlinear controller based on feedback linearization was successfully developed and implemented for the attitude recovery of rigid and flexible spacecraft. The robustness of this controller to model/parameter uncertainties was shown using many simulation test cases.

- A new approach based on quaternion addition for producing the control error signal has been proposed which effectively deals with the existence of dual equilibrium points. This approach only uses the vector part of the quaternions to construct an error signal which can then be regulated by the controller.

- A design criterion for a class of flexible spacecraft has been established which needs to be satisfied so that the *zero dynamics* of the system is asymptotically stable. Under this condition, the overall closed-loop system was also proven to be asymptotically stable using the Lyapunov theorem.

- The usual difficulty in transforming a nonlinear system into its *normal form*, has to do with finding the appropriate diffeomorphisms which requires the analytical solution of a set of partial differential equations. In this thesis, a new explicit analytical solution for the construction of *normal form* and the *zero dynamics* of a general class of nonlinear systems has been developed and applied specifically to the flexible spacecraft system.

# CHAPTER 2

## SPACECRAFT DYNAMICS

In this chapter, the dynamic model governing the general 3-dimensional attitude-flexural motion of a spacecraft with flexible plate-type appendages are derived using the Lagrangian formulation. Plate type appendages were selected as opposed to beam type appendages since many existing and proposed spacecraft consist of flexible appendages whose behavior can be best approximated by modeling them as plates. However, the choice of plates as the model for flexible appendages makes the mathematical modeling more complex and increases the system's dimension. This is in fact the principal reason why most works in this area have used beams to model the appendage flexibility.

The necessary background material on attitude representation and kinematics using Euler angles, quaternions and frames of reference are first presented to allow for better appreciation of the formulations to follow in Sections 2.2 and 2.3. The basic orbital notions are also briefly reviewed, but it is to be noted that these do not play an important role in the present thesis, nevertheless they were considered and implemented in the software simulator for completeness. Note that all vector quantities are identified by an arrow and matrices are underlined.

## 2.1 Basics of Attitude Dynamics

The next four subsections will serve as a brief introduction to the terminology and basic concepts necessary to better understand the orbital and attitude dynamics of spacecraft. The orbital dynamics of spacecraft is not of direct interest in this thesis and has only been implemented in the software simulation for completeness.

### 2.1.1 Frames of Reference

To determine the spacecraft's attitude with respect to another body, a reference system must be used. While any reference system will provide correct answers if it is applied consistently, it is desirable to choose a convenient and intuitive set of reference frames. An appropriate reference frame simplifies calculations and reduces the obscuring of physical theory by numerous algebraic transformations. Transformations between reference systems will be necessary in all spacecraft design, simulation and control activities. The following six frames of reference were defined and utilized through out the thesis. It is important to note that the *x*, *y*, and *z* notation in each frame of reference is independent of the other reference frames.

Ecliptic Reference Frame

The ecliptic is the plane of the orbit of the Earth around the Sun. As the Sun is the center of attraction, it is the also the center of the inertial ecliptic reference frame ($x_S, y_S, z_S$). The *x*-axis is in the plane of the ecliptic and points to a reference point among the stars known as the direction of vernal equinox ($\Upsilon$) and also referred to as the first point of Aries. The symbol designates the ram due to fact that the direction of the vernal equinox pointed to the constellation of Aries during Christ's lifetime. Due to procession, however, the vernal equinox now points to the constellation Pisces. The *y*-axis is also in the plane of the ecliptic, at right angles to the *x*-axis and the z-axis is orthogonal to the plane of the ecliptic, and through the center of the Sun (Figure 2.1). This reference frame was not used in this thesis.

Celestial Reference Frame

This is the inertial celestial frame ($X, Y, Z$) with its origin located at the center of the Earth. Its $x$-axis points to the direction of the vernal equinox in the equatorial plane, $z$-axis is in the same direction as the Earth's rotation axis, and its $y$-axis completes the right-hand triad. This frame is sometimes referred to as the non-rotating Earth-centered frame, the Geocentric inertial frame or the Galilean frame (Figure 2.2). In the present thesis, this reference frame was used for the dynamics formulation, simulation and control.

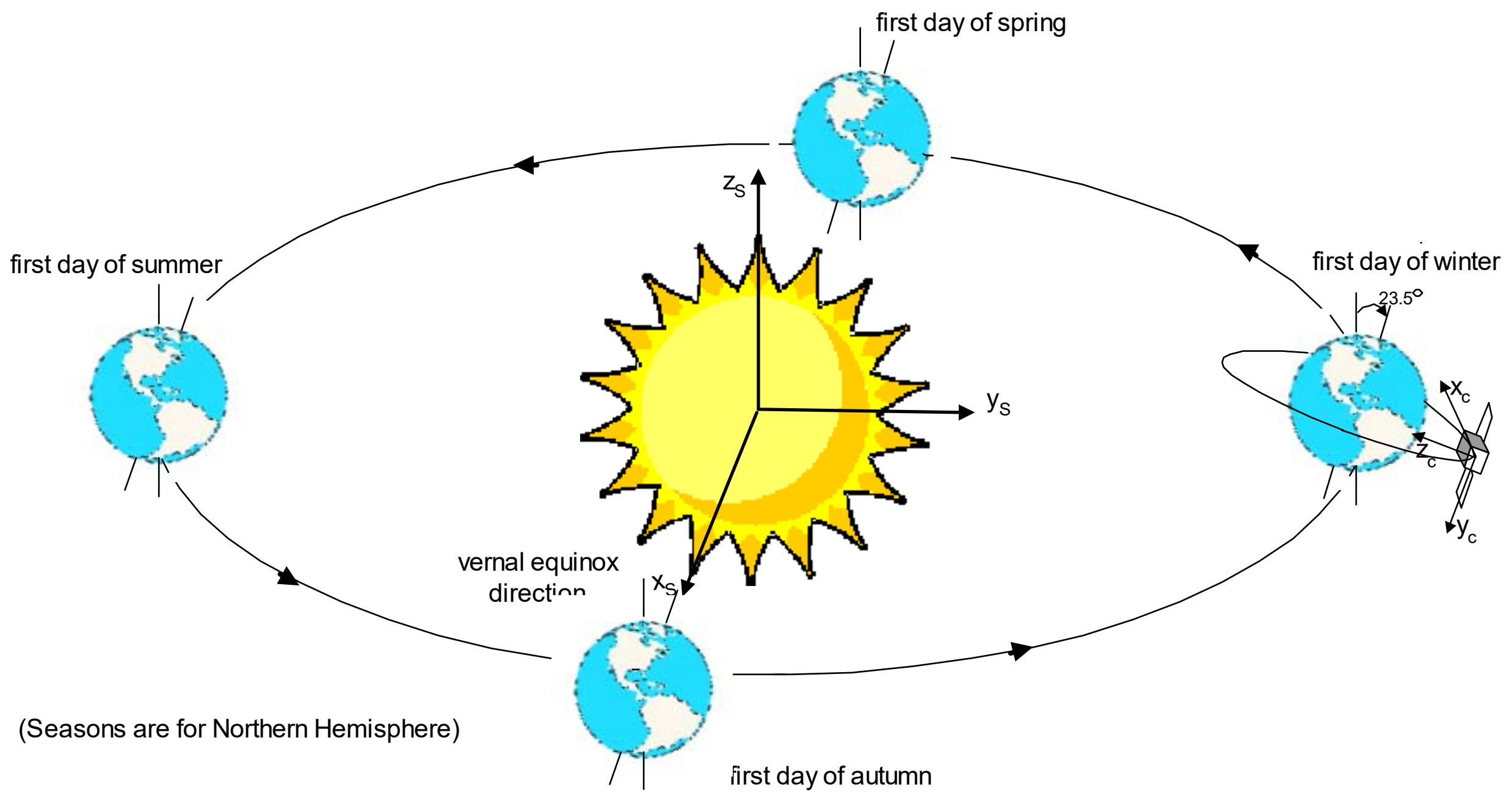

Figure 2.1: Ecliptic reference frame

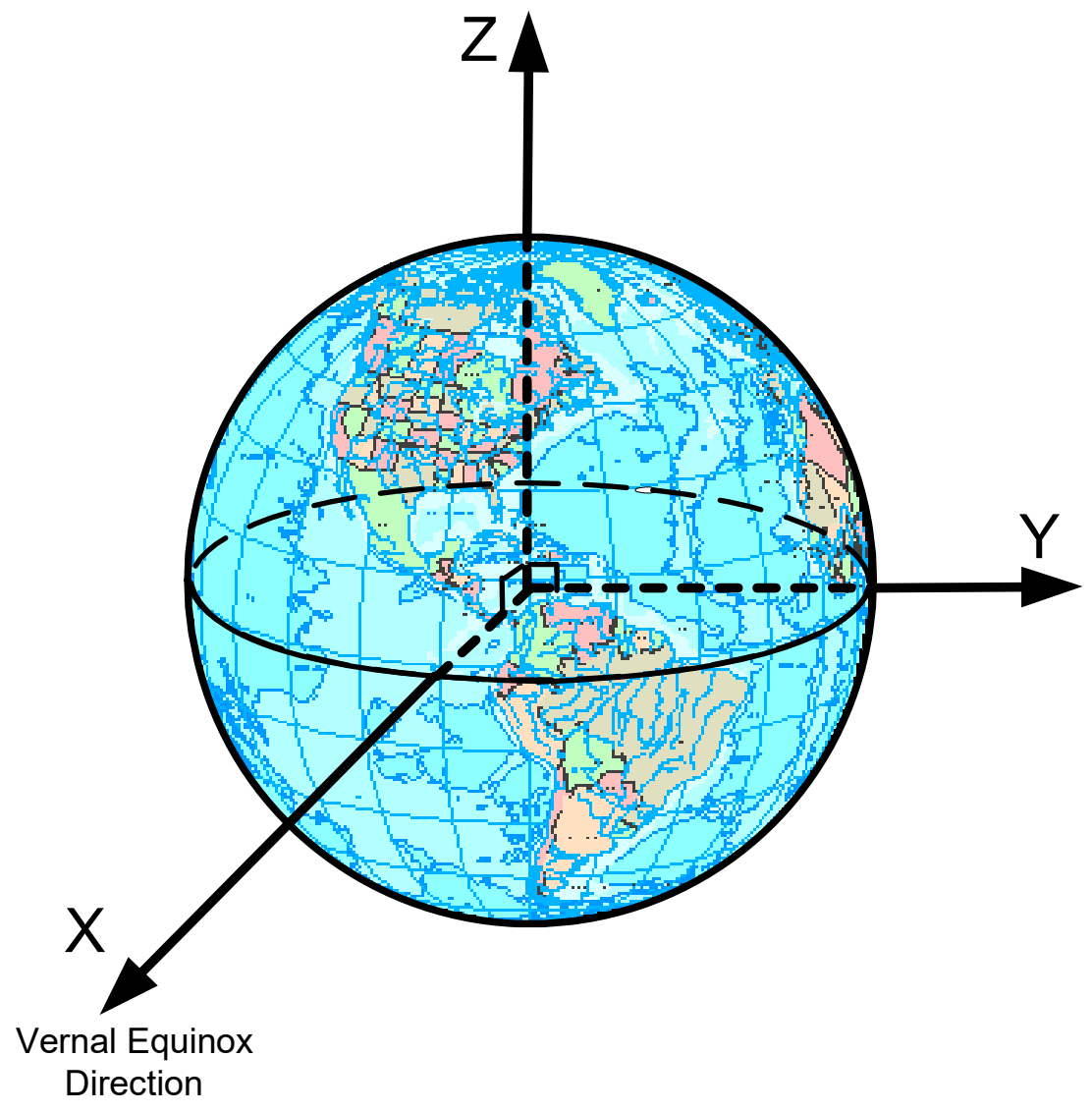

Figure 2.2: Celestial reference frame

Earth Reference Frame

This is the non-inertial Earth-centered frame ($x_E, y_E, z_E$) that has its origin located at the center of the Earth. Its *x*-axis points through the Greenwich Meridian in the equatorial plane, z-axis is in the same direction as the Earth's rotation axis, and its *y*-axis completes the right-hand triad (Figure 2.3). This reference frame was only used in the simulation to calculate the Greenwich Hour Angle which is an input to the magnetic field model.

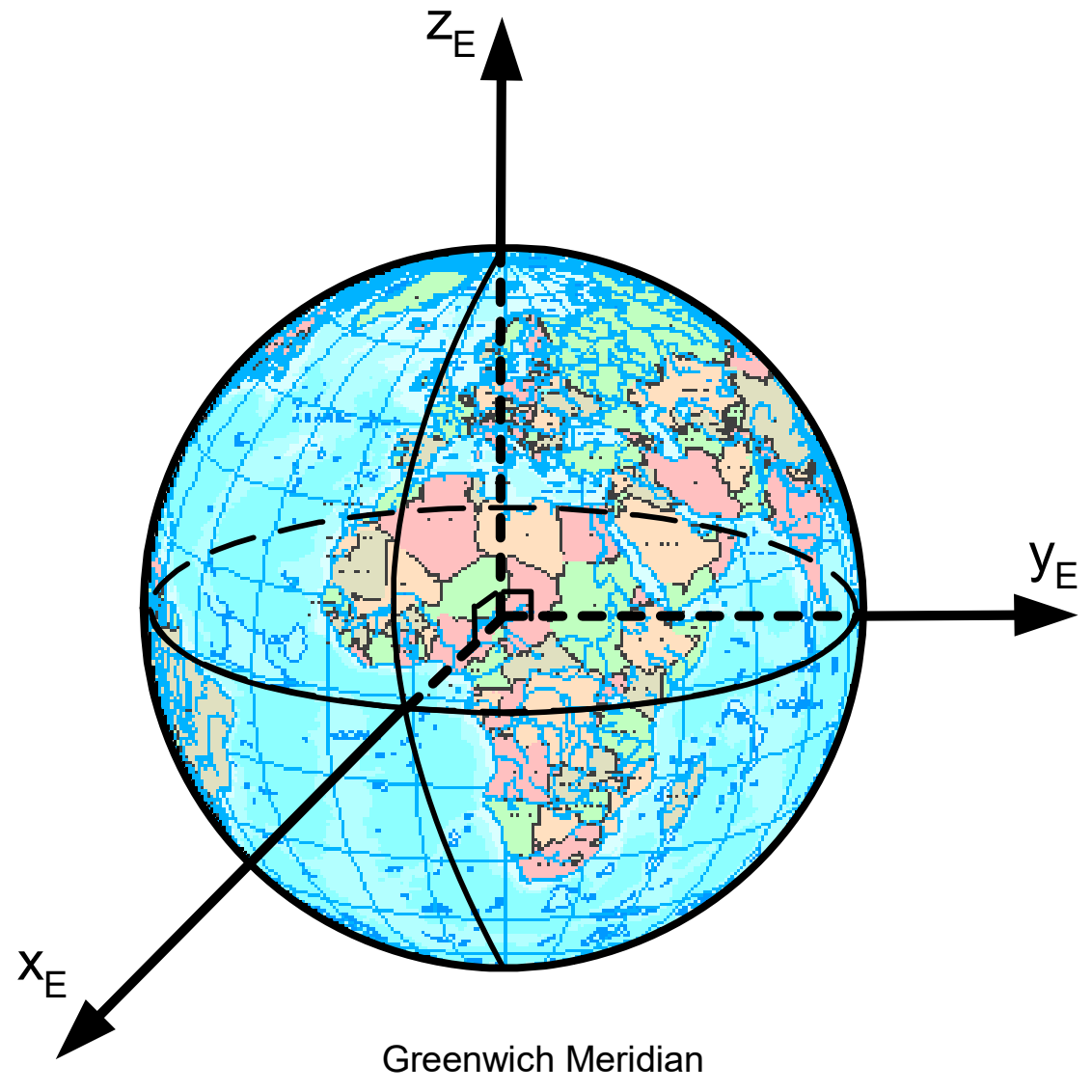

Figure 2.3: Earth reference frame

## Orbital Reference Frame

The non-inertial orbital reference frame ($x_o, y_o, z_o$) is the well-known Earth-pointing reference system where the *x*-axis (Roll) is along the spacecraft's velocity vector, the z-axis (Yaw) points towards the center of the Earth and the *y*-axis (Pitch) is along the orbit normal which completes the right-hand triad (Figure 2.4). This reference frame was not used in this thesis.

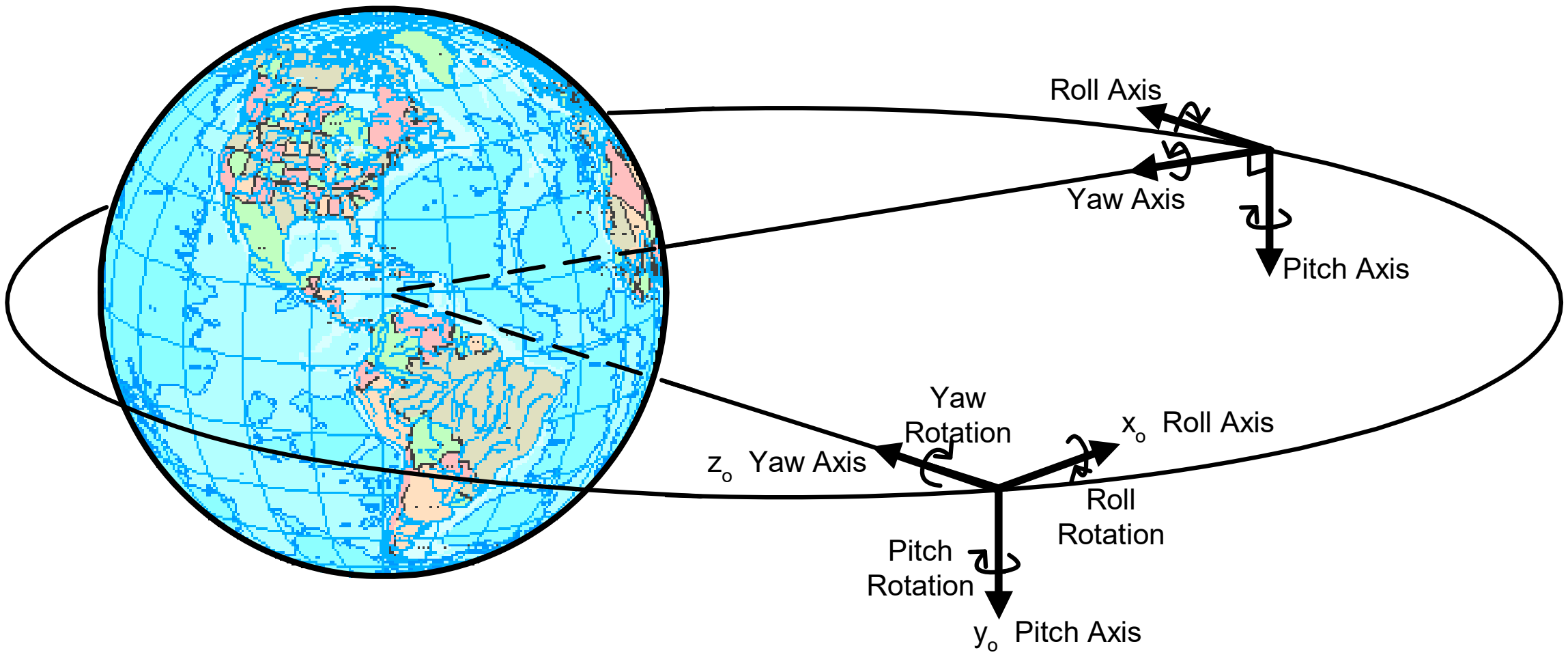

Figure 2.4: Orbital reference frame

## Spacecraft/Body Reference Frame

The non-inertial spacecraft frame ($x_c, y_c, z_c$) is a body-fixed frame of reference. Its origin is located at the spacecraft Center Of Mass (COM) and its orientation is based on the spacecraft geometry. For example, the *y*-axis could be along an appendage and the *x* and *z*-axis along two other orthogonal directions (Figure 2.5) completing a right-hand triad. In the present thesis, this reference frame was used for the dynamics formulation, simulation and control.

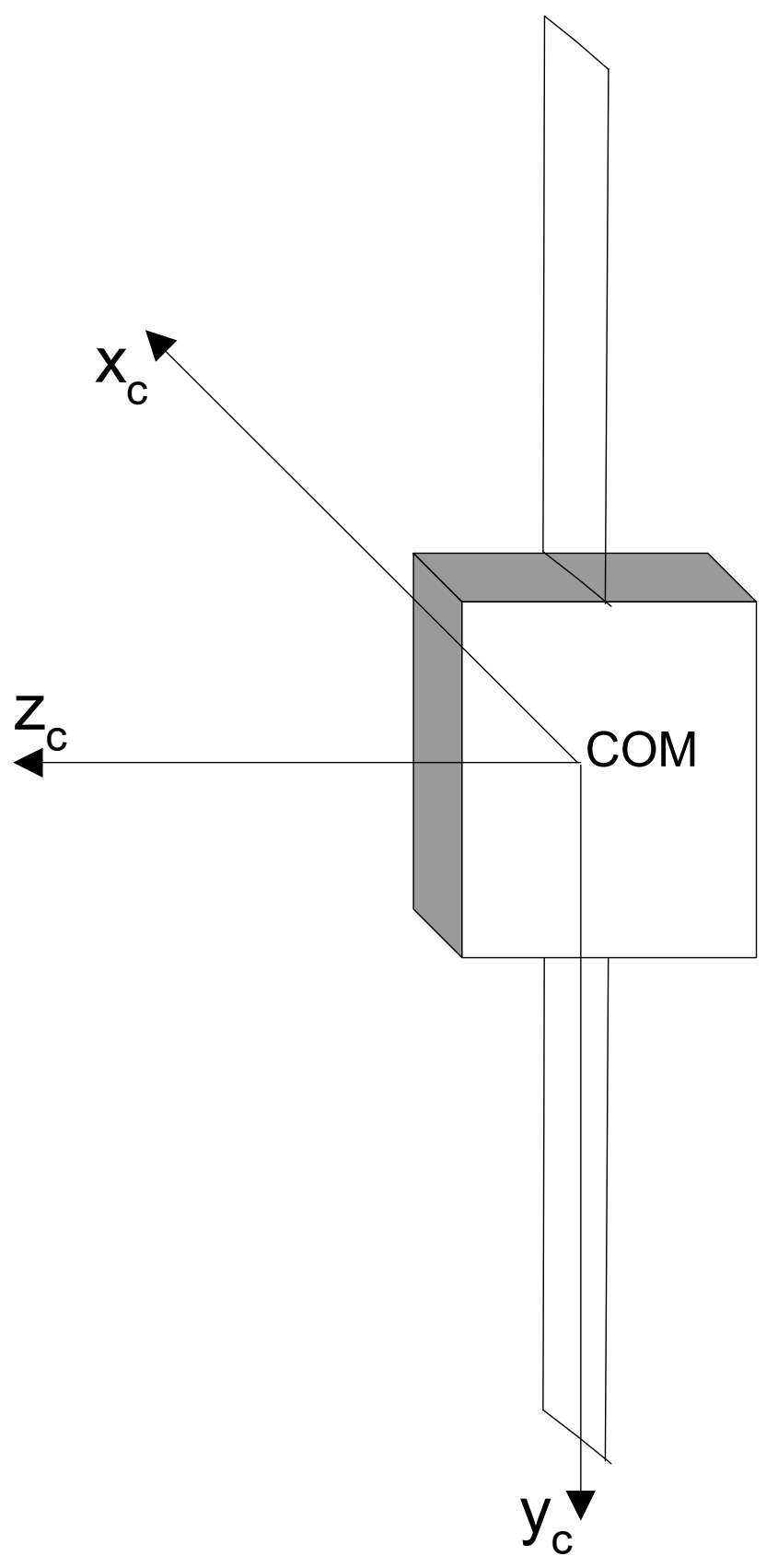

Figure 2.5: Spacecraft/Body reference frame

Appendage Reference Frame

For simplicity, a set of plate axes ($x_i, y_i, z_i$) is defined for the $i^{th}$ appendage such that $x_i, y_i$ are along the width and length of the $i^{th}$ appendage and $z_i$ complete the right-hand triad. The axes have a constant orientation with respect to the spacecraft axes ($x_c, y_c, z_c$) and their origin is located at the point of attachment of the appendage to the rigid bus, given by vector $\vec{d}_i$ (Figure 2.6). In the present thesis, this reference frame was used for the dynamics formulation and simulation.

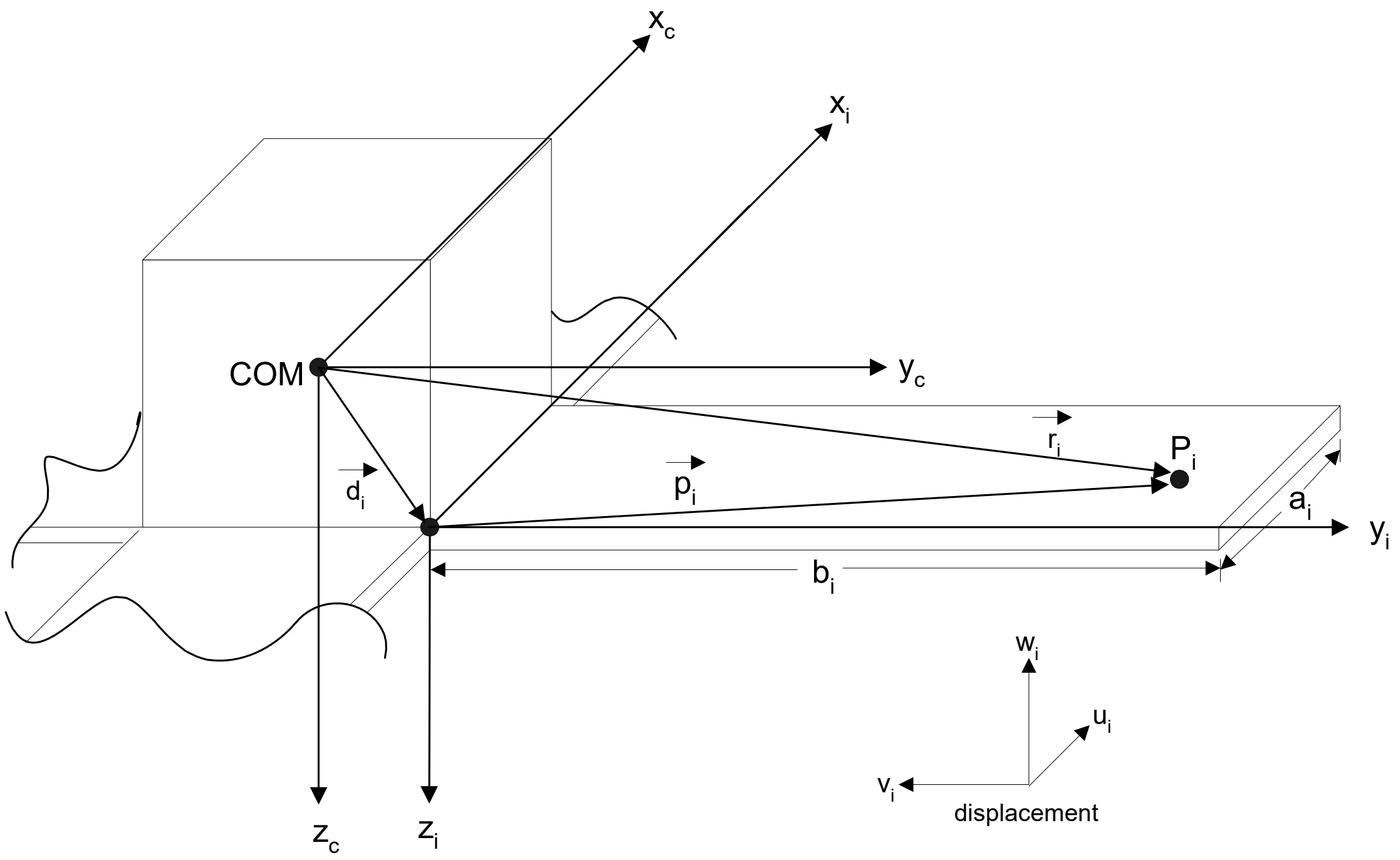

Figure 2.6: Spacecraft configuration and appendage reference frame

### 2.1.2 Orbits

There are multiple ways of defining an orbital motion depending on the chosen set of parameters to form a mathematical model. These models will all have seven degrees of freedom to describe the position (3 DOF) and velocity (3 DOF) of the spacecraft at a given instant of time (1 DOF) which in turn uniquely defines an orbit. The set of parameters commonly used are called the Keplerian orbital elements, named after Johannes Kepler and his three laws (Figure 2.7) [6]. These Keplerian elements are defined below:

1. Semi-major Axis ($a$): Semi-major axis is half the long axis of an ellipse.

2. Eccentricity ($e$): Eccentricity is the common parameter associated with every conic section and for elliptical orbits could be regarded as a measure of how much the ellipse deviates from a circle. It is the distance from the center of the ellipse to focus divided by the semi-major axis. For a circular orbit, $e = 0$.

3. Inclination ($i$): Inclination is the angular distance of the orbital plane from the equatorial plane, stated in degrees. For example an inclination, $i = 0^\circ$, means that the spacecraft is orbiting Earth in the equatorial plane and in the same direction as the Earth's rotation, where as $i = 180^\circ$ refers to a retrograde equatorial orbit. A polar orbit, in which the spacecraft passes over the north and south poles is specified by an inclination $i = 90^\circ$.

4. Right Ascension of the Ascending Node ($\Omega$): This is the angle measured at the center of the Earth in the equatorial plane from the vernal equinox eastward to the *ascending node*. The ascending node is the point in the orbit at which the spacecraft crosses the equatorial plane going from south to north.

5. Argument of Perigee ($\omega$): This is the angle measured at the center of the Earth in the orbital plane from the ascending node to *perigee*, going in the direction of the spacecraft's motion. The perigee is the point in the orbit where the spacecraft is closest to the Earth.

6. True Anomaly ($\nu_o$): This is the angle measured at the *barycenter* between the perigee and the spacecraft. The barycenter is the location of the center of mass between two objects. In this case, since the spacecraft's mass is negligible compared to Earth's mass, the barycenter can be considered to be at the center of the Earth.

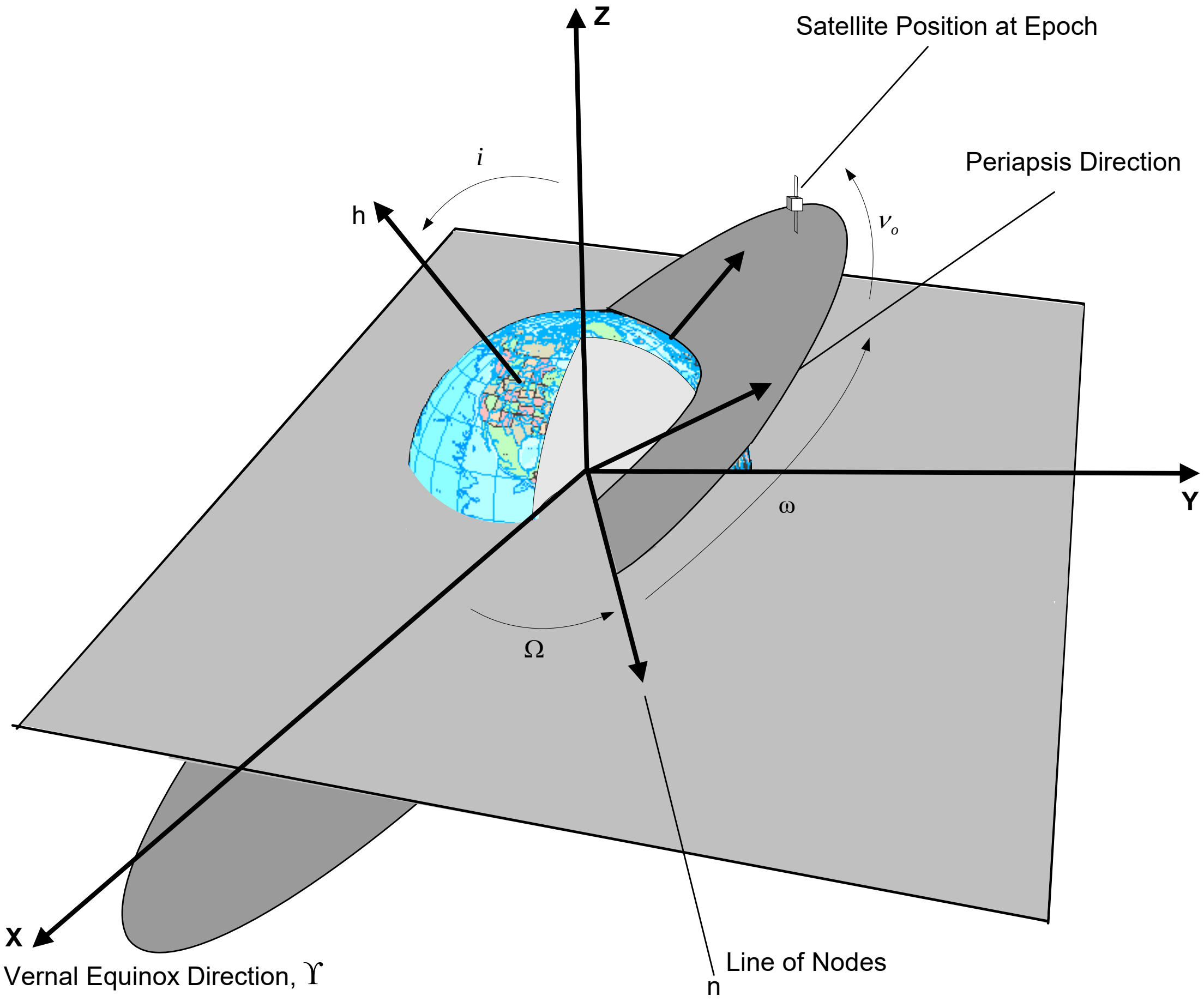

Figure 2.7: Keplerian orbital elements

Note that the first 6 Keplerian elements define the shape and orientation of an orbit at a given instant in time called the Epoch (seventh element).

### 2.1.3 Disturbance Environment

This section describes the environmental torques affecting the spacecraft in orbit. There are essentially four environmental torques acting on any given spacecraft, namely, the gravity gradient torque, the magnetic torque, the solar radiation torque and the aerodynamic torque. There were several assumptions made during the calculation of these torques which will be stated and briefly discussed. The effects of these environmental disturbance torques, for the purposes of the controller design and analysis, are neglected. This is due to the fact that these disturbances have small magnitudes (in the order of $10^{-6}$ to $10^{-3}$ Nm) and have long time scales (in the order of the orbital period; > 90min), compared to the control torques applied during the attitude recovery maneuvers which are between 1 Nm and 50 Nm and shorter than 10 minutes. Even at the 1 Nm level, the control torques are at least about 1000 times larger than the environmental torques. However, these disturbance forces are modeled and included in the high-fidelity spacecraft simulator.

It is to be noted that the internal disturbance torques are also to be considered in case there are any for a given spacecraft. Typical internal torques are those due to the antenna, solar arrays, or instrument scanner motion, or to the deployable booms and appendages. For this thesis, no internal torques were modelled in the simulator.

Please refer to Appendix C for a brief discussion of the external disturbance torques modeled in the simulator.

### 2.1.4 Attitude Representation and Kinematics

Attitude refers to the orientation of one reference frame with respect to another. There are several common ways of representing the attitude of an object, namely, Direction cosine matrix, Euler axis and angle, Gibbs vector, Cayley-Klein parameters, Euler angles (e.g. Roll, Pitch and Yaw), and Quaternions [6]. In this thesis, quaternions are used as the main attitude parameterization for dynamics formulation and control synthesis since quaternions avoid singularities (which are usually associated with Euler angles) due to the fourth redundant parameter. In fact, any three-parameter attitude representation has always a singularity and should be avoided in situation where large angular motions are present, as is the case in this thesis. The main drawback of attitude representation using quaternions however, is that the parameters do not have a convenient physical interpretation. Due to this fact, the more intuitive Euler angles are used as outputs in the simulator, only for visualization and data analysis. In the next three subsections, Euler axis and angle, quaternions and Euler angles are briefly introduced and for more details, Wertz [6] can be consulted.

Euler Axis and Angle

The simplest attitude representation is undoubtedly given by the Euler axis, $\vec{e} = [e_1 \quad e_2 \quad e_3]^T$, and the corresponding Euler angle, $\theta$. Euler axis is a principal unit vector about which the first reference frame rotates through the Euler angle to align with the second reference frame.

Quaternions

The quaternions, $\tilde{q}^T = \left[ q_0 \; \vec{q}^T \right] = \left[ q_0 \; q_1 \; q_2 \; q_3 \right]$, as the name suggests, are a set of 4 parameters related to the Euler axis and angle, as

$$\vec{q} = \vec{e} \sin\frac{\theta}{2}, \qquad q_0 = \cos\frac{\theta}{2} \tag{2.1}$$

The quaternion has unit length, $\tilde{q}^T \tilde{q} = 1$, which serves as a useful constraint on the four parameters (commonly referred to as the unitary condition). For example, the unitary condition could be used during numerical integration to check and maintain numerical accuracy by using quaternion normalization. It is interesting and useful to note that using quaternions, any 3D attitude motion is represented by a path on the surface of a 4D hypersphere.

Euler Angles

An intuitive way of describing the attitude is by using the Euler angles. In essence, the first reference frame is rotated through three angles about individual *x*, *y* and *z* axes to align with the second reference frame. There are several conventions for Euler angles depending on the axes and the order about which the successive rotations are carried out. Also, the transformations from the first frame to the second are not unique and there may be several rotation sequences that achieve the same final result. As was mentioned earlier, singularities are also a present feature of this attitude representation method when dealing with large angular rotations, however, with all its apparent drawbacks, it is the most widely used method for spacecraft attitude visualization. The three common attitude angles used in this thesis are the Euler angles, Roll, Pitch and Yaw, defined using the spacecraft reference frame.

Quaternions can be transformed to Euler angles using established methods in the literature, [6] and [142]. In this thesis, the transformation from quaternion to Euler angles is done through the intermediary of the rotation matrix, $\underline{R}$ (i.e. direction cosine matrix) which is obtained by the following formula

$$\underline{R} = (q_o^2 - \vec{q}^T\vec{q})\underline{i} + 2\vec{q}\vec{q}^T - 2q_o q^\times \qquad \text{where} \qquad q^\times = \begin{bmatrix} 0 & -q_3 & q_2 \\ q_3 & 0 & -q_1 \\ -q_2 & q_1 & 0 \end{bmatrix} \tag{2.2}$$

and where $\underline{i}$ is the $3\times3$ identity matrix.

The Euler angles are then obtained from the direction cosine matrix, $\underline{R}$. Basically, there are 12 different Euler rotation sequences possible [142] and implemented in the simulator where the user has the option of selecting a particular one. For all except one simulation case presented in this thesis, the (3,2,1) Euler rotation sequence was used. Also, it's important to note that the middle rotation is always constrained to $\pm 90^o$ to eliminate double values.

Attitude Kinematics

For every attitude representation, a set of kinematics equations of motion describes the rate of change of the attitude parameters due to the dynamics of the spacecraft. The kinematics equations for the quaternions are given as

$$\begin{aligned} \dot{q}_0 &= -\frac{1}{2}\vec{\omega}^T\vec{q} \\ \dot{\vec{q}} &= -\frac{1}{2}\underline{\Omega}\vec{q} + \frac{1}{2}q_0\vec{\omega} \end{aligned} \tag{2.3}$$

where $\vec{\omega}^T = \left[\omega_x \; \omega_y \; \omega_z\right]$ is the spacecraft body angular rate (representing the dynamics of the spacecraft) and

$$\underline{\Omega} = \begin{bmatrix} 0 & -\omega_z & \omega_y \\ \omega_z & 0 & -\omega_x \\ -\omega_y & \omega_x & 0 \end{bmatrix} \tag{2.4}$$

## 2.2 Rigid Attitude Dynamics Formulation

We can derive the general equations of rigid attitude motion considering a torque about the spacecraft center of mass. With reference to an absolute (non-rotating) coordinate system like the celestial reference frame, Newton's second law defines the relationship between torque and angular momentum

$$\vec{\tau} = \dot{\vec{H}} \tag{2.5}$$

where $\vec{\tau}$ is the torque and the spacecraft's total angular momentum, $\vec{H}$, is defined as

$$\vec{H} = \underline{I}_t \, \vec{\omega} \tag{2.6}$$

where $\underline{I}_t$ is the spacecraft total inertia tensor. However it is more convenient to express and work with attitude equations expressed in the rotating spacecraft reference frame, in which case, equation (2.5) becomes

$$\vec{\tau} = \dot{\vec{H}} + \vec{\omega} \times \vec{H} \tag{2.7}$$

The first term represents that part of $\dot{\vec{H}}$ due to the change in magnitude of the components of $\vec{H}$, and the cross-product term represents that part due to the changes in direction of the components of $\vec{H}$.

Using (2.6), equation (2.7) can be expanded to become

$$\vec{\tau} = \dot{\underline{I}}_t \vec{\omega} + \underline{I}_t \dot{\vec{\omega}} + \vec{\omega} \times (\underline{I}_t \vec{\omega} + \vec{h}_w) \tag{2.8}$$

which can be further simplified noting that a fully deployed rigid spacecraft has a non-varying inertia matrix (i.e. $\dot{\underline{I}}_t = \underline{0}$),

$$\vec{\tau} = \underline{I}_t \dot{\vec{\omega}} + \vec{\omega} \times (\underline{I}_t \vec{\omega} + \vec{h}_w) \tag{2.9}$$

Using (2.9), the state space form of the *rigid attitude dynamics equation* is obtained as

$$\dot{\vec{\omega}} = -\underline{I}_t^{-1} \left( \vec{\omega} \times \left( \underline{I}_t \vec{\omega} + \vec{h}_w \right) \right) + \underline{I}_t^{-1} \vec{\tau} \quad or \quad \dot{\vec{\omega}} = -\underline{I}_t^{-1} \underline{\Omega} \left( \underline{I}_t \vec{\omega} + \vec{h}_w \right) + \underline{I}_t^{-1} \vec{\tau} \tag{2.10}$$

where $\vec{h}_w$ is the internal angular momentum due to rotating mechanical wheel(s), if any.

Defining the state vector as $\vec{x} = \vec{\omega}$, the system's state equation, which is *3*-dimensional is

$$\begin{aligned} \dot{\vec{x}} &= \vec{f}(\vec{x}) + \underline{G}(\vec{x}) \vec{\tau} \\ \vec{y} &= \vec{h}(\vec{x}) = \begin{bmatrix} x_1 & x_2 & x_3 \end{bmatrix}^T = \begin{bmatrix} \omega_x & \omega_y & \omega_z \end{bmatrix}^T \end{aligned} \tag{2.11}$$

The $\vec{f}$ and $\underline{G}$ terms in (2.11) are obtained directly from (2.10).

Although the general attitude motion of a rigid spacecraft may be modeled by these equations, they have no general solution unless we specify the components of $\vec{\tau}$. Note that the equations are nonlinear and coupled. The coupling effect is most present when we are dealing with large angular rates, $\vec{\omega}$.

Equation (2.11) can also be expressed using quaternions, $\tilde{q}^T = \left[ q_0 \; \vec{q}^T \right] = \left[ q_0 \; q_1 \; q_2 \; q_3 \right]$, as opposed to the body angular rate, $\vec{\omega}$, and it's this representation that we have used in this thesis. To express (2.11) using quaternions, we start with the kinematics equations (2.3). Expanding and combining (2.3), we obtain

$$\begin{bmatrix}\dot{q}_0\\ \dot{q}_1\\ \dot{q}_2\\ \dot{q}_3\end{bmatrix}=\frac{1}{2}\begin{bmatrix}-\omega_x q_1-\omega_y q_2-\omega_z q_3\\ \omega_z q_2-\omega_y q_3+\omega_x q_0\\ -\omega_z q_1+\omega_x q_3+\omega_y q_0\\ \omega_y q_1-\omega_x q_2+\omega_z q_0\end{bmatrix}$$

$$=\frac{1}{2}\underbrace{\begin{bmatrix}0 & -\omega_x & -\omega_y & -\omega_z\\ \omega_x & 0 & \omega_z & -\omega_y\\ \omega_y & -\omega_z & 0 & \omega_x\\ \omega_z & \omega_y & -\omega_x & 0\end{bmatrix}}_{\underline{\Gamma}}\begin{bmatrix}q_0\\ q_1\\ q_2\\ q_3\end{bmatrix}=\frac{1}{2}\underbrace{\begin{bmatrix}q_0 & -q_1 & -q_2 & -q_3\\ q_1 & q_0 & -q_3 & q_2\\ q_2 & q_3 & q_0 & -q_1\\ q_3 & -q_2 & q_1 & q_0\end{bmatrix}}_{\underline{Q}}\begin{bmatrix}0\\ \omega_x\\ \omega_y\\ \omega_z\end{bmatrix} \tag{2.12}$$

Note that $\underline{\Gamma}$ is an orthogonal matrix and $\underline{Q}$ is an orthonormal matrix and hence the following equalities hold true

$$\underline{\Gamma}^{-1}=\frac{1}{\|\vec{\omega}\|^2}\underline{\Gamma}^T \quad and \quad \underline{Q}^{-1}=\underline{Q}^T \tag{2.13}$$

Next, we take the derivative of (2.12)

$$\begin{bmatrix}\ddot{q}_0\\ \ddot{\vec{q}}\end{bmatrix}=\frac{1}{2}\underline{\Gamma}\begin{bmatrix}\dot{q}_0\\ \dot{\vec{q}}\end{bmatrix}+\frac{1}{2}\underline{Q}\begin{bmatrix}0\\ \dot{\vec{\omega}}\end{bmatrix} \tag{2.14}$$

and using (2.12) we can re-write (2.14) as

$$\begin{bmatrix}\ddot{q}_0\\ \ddot{\vec{q}}\end{bmatrix}=\frac{1}{2}\underline{\Gamma}\left(\frac{1}{2}\underline{\Gamma}\begin{bmatrix}q_0\\ \vec{q}\end{bmatrix}\right)+\frac{1}{2}\underline{Q}\begin{bmatrix}0\\ \dot{\vec{\omega}}\end{bmatrix}=\frac{1}{4}\underline{\Gamma}^2\begin{bmatrix}q_0\\ \vec{q}\end{bmatrix}+\frac{1}{2}\underline{Q}\begin{bmatrix}0\\ \dot{\vec{\omega}}\end{bmatrix} \tag{2.15}$$

Now using (2.10) and (2.12), we can expand (2.15) to get

$$\begin{bmatrix}\ddot{q}_0\\ \ddot{\vec{q}}\end{bmatrix}=-\frac{1}{4}\|\vec{\omega}\|^2\begin{bmatrix}q_0\\ \vec{q}\end{bmatrix}+\frac{1}{2}\underline{Q}\begin{bmatrix}0\\ -\underline{I}_t^{-1}\left(\vec{\omega}\times(\underline{I}_t\vec{\omega}+\vec{h}_w)\right)+\underline{I}_t^{-1}\vec{\tau}\end{bmatrix} \tag{2.16}$$

Using (2.12) and (2.13), we can obtain $\vec{\omega}$ as a function of $\tilde{q}$ and $\dot{\tilde{q}}$

$$\begin{bmatrix} 0 \\ \omega_x \\ \omega_y \\ \omega_z \end{bmatrix} = 2\begin{bmatrix} q_0 & q_1 & q_2 & q_3 \\ -q_1 & q_0 & q_3 & -q_2 \\ -q_2 & -q_3 & q_0 & q_1 \\ -q_3 & q_2 & -q_1 & q_0 \end{bmatrix}\begin{bmatrix} \dot{q}_0 \\ \dot{q}_1 \\ \dot{q}_2 \\ \dot{q}_3 \end{bmatrix} = 2\underline{Q}^T\dot{\tilde{q}} \tag{2.17}$$

For example, $\omega_x(\tilde{q},\dot{\tilde{q}})$ is given as

$$\omega_x(\tilde{q},\dot{\tilde{q}}) = 2\left(q_0\dot{q}_1 + q_3\dot{q}_2 - q_2\dot{q}_3 - q_1\dot{q}_0\right) \tag{2.18}$$

and $\omega_y(\tilde{q},\dot{\tilde{q}})$ and $\omega_z(\tilde{q},\dot{\tilde{q}})$ can be obtained similarly from (2.17); and $\vec{\omega}^T(\tilde{q},\dot{\tilde{q}}) = \left[\omega_x(\tilde{q},\dot{\tilde{q}}) \quad \omega_y(\tilde{q},\dot{\tilde{q}}) \quad \omega_z(\tilde{q},\dot{\tilde{q}})\right]$.

Finally using (2.16) and (2.17), we can express the attitude motion of the rigid spacecraft as a function of $\tilde{q}$ and $\dot{\tilde{q}}$, that is

$$\ddot{\tilde{q}} = -\frac{1}{4}\left(\vec{\omega}^T(\tilde{q},\dot{\tilde{q}})\,\vec{\omega}(\tilde{q},\dot{\tilde{q}})\right)\tilde{q} - \frac{1}{2}\underline{Q}\begin{bmatrix} 0 \\ \underline{I}_t^{-1}\left(\vec{\omega}(\tilde{q},\dot{\tilde{q}})\times\left(\underline{I}_t\vec{\omega}(\tilde{q},\dot{\tilde{q}}) + \vec{h}_w\right)\right) \end{bmatrix} + \frac{1}{2}\underline{Q}\begin{bmatrix} 0 \\ \underline{I}_t^{-1}\vec{\tau} \end{bmatrix} \tag{2.19}$$

Defining the state vector as $\vec{x}^T = \left[x_1\ x_2\ x_3\ x_4\ x_5\ x_6\ x_7\ x_8\right] = \left[\tilde{q}^T\ \dot{\tilde{q}}^T\right]$, the system's state equation, which is *8*-dimensional becomes

$$\dot{\vec{x}} = \begin{bmatrix} x_5 \\ x_6 \\ x_7 \\ x_8 \\ f_{q_0}(\vec{x}) \\ \vec{f}_{\vec{q}}(\vec{x}) \end{bmatrix} + \begin{bmatrix} \vec{0} \\ \vec{0} \\ \vec{0} \\ \vec{0} \\ \vec{g}_{q_0}^T(\vec{x}) \\ \underline{G}_{\vec{q}}(\vec{x}) \end{bmatrix}\vec{\tau} \triangleq \vec{f}(\vec{x}) + \underline{G}(\vec{x})\,\vec{\tau}$$

$$\vec{y} = \vec{h}(\vec{x}) = \left[x_2 \quad x_3 \quad x_4\right]^T = \left[q_1 \quad q_2 \quad q_3\right]^T \tag{2.20}$$

where

$$f_{q_o} = -\frac{1}{4}\left(\vec{\omega}^T(\tilde{q},\dot{\tilde{q}})\,\vec{\omega}(\tilde{q},\dot{\tilde{q}})\right)q_0 + \frac{1}{2}\vec{q}^T \underline{I}_t^{-1}\left(\vec{\omega}(\tilde{q},\dot{\tilde{q}})\times\left(\underline{I}_t\vec{\omega}(\tilde{q},\dot{\tilde{q}}) + \vec{h}_w\right)\right)$$

$$\vec{f}_{\vec{q}} = -\frac{1}{4}\left(\vec{\omega}^T(\tilde{q},\dot{\tilde{q}})\,\vec{\omega}(\tilde{q},\dot{\tilde{q}})\right)\vec{q} - \frac{1}{2}\underline{Q}_{\vec{q}}\underline{I}_t^{-1}\left(\vec{\omega}(\tilde{q},\dot{\tilde{q}})\times\left(\underline{I}_t\vec{\omega}(\tilde{q},\dot{\tilde{q}}) + \vec{h}_w\right)\right)$$

$$\vec{g}_{q_0}^T = -\frac{1}{2}\vec{q}^T \underline{I}_t^{-1}$$

$$\underline{G}_{\vec{q}} = \frac{1}{2}\underline{Q}_{\vec{q}}\underline{I}_t^{-1}$$

and

$$\underline{Q}_{\vec{q}} = \begin{bmatrix} q_0 & -q_3 & q_2 \\ q_3 & q_0 & -q_1 \\ -q_2 & q_1 & q_0 \end{bmatrix}$$

Note that the formulation given by (2.20) is necessary for control synthesis (i.e. input-output feedback linearization) as will be described in Chapter 4. For numerical simulation implementation, the usual combination of the angular rates, $\vec{\omega}$, and quaternions, $\tilde{q}$, are used as the simulation states with the corresponding state equations (2.10) and (2.3).

## 2.3 Flexible Attitude Dynamics Formulation

The system under investigation consists of a rigid bus and *n* appendages attached to it (Figure 2.6) and is assumed to be in a fully deployed configuration. This is a star topology spacecraft which is by far the most common structural layout found for spacecraft to date. The spacecraft bus is assumed to be rigid and hence has a constant inertia matrix. It is assumed that the appendages' vibration has negligible effect on the location of center of mass which implies that the appendage mass is small compared to the rigid bus. The appendages are further assumed to be thin, long, and inextensible plates with homogeneous material properties. Structural damping is assumed to be of the viscous type with no modal coupling and hence the damping matrix is diagonal.

Appendix A contains all the detailed derivations, summarized here in the next few subsections. Also note that all motions are described in the spacecraft reference frame.

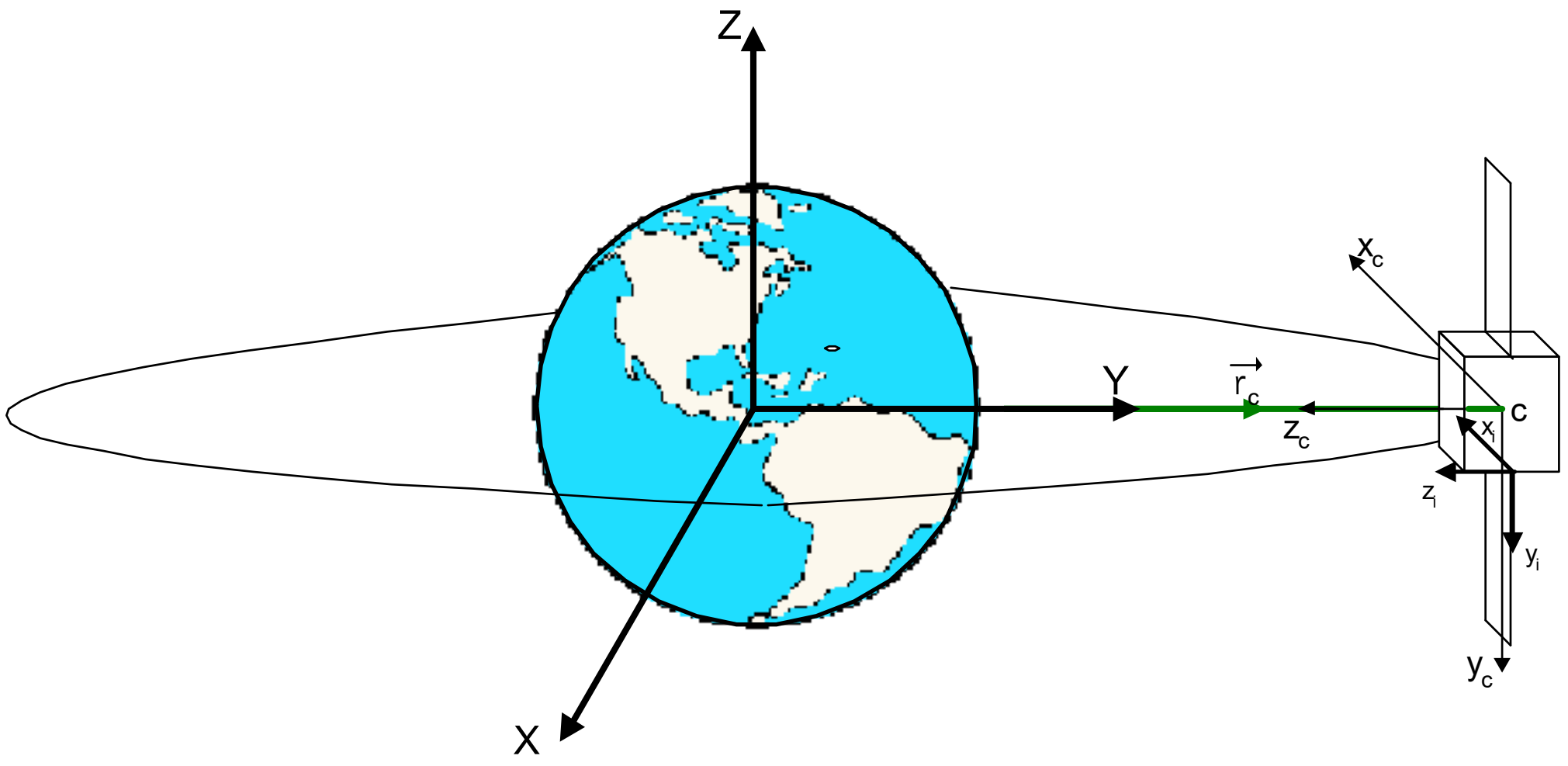

Figure 2.8: Spacecraft in a given orbit with relevant reference frames

<u>Appendage Dynamics</u>

The position of a point, $P_i$, on the appendage with respect to the appendage reference frame (Figure 2.6) is given by

$$\vec{r}_i = \vec{d}_i + \vec{p}_i \tag{2.21}$$

where the subscript *i* refers to the *i*^th^ appendage. Taking into account the displacement of this point where the deflection in the u direction is neglected, we have

$$\vec{r}_i^T = \begin{bmatrix} d_{ix} + x_i & d_{iy} + y_i - v_i & d_{iz} + w_i \end{bmatrix} \tag{2.22}$$

The instantaneous position of the spacecraft's center of mass is defined by the vector $\vec{r}_c$ (Figure 2.8) and hence the velocity of point $P_i$ is given by

$$\vec{v}_i = \dot{\vec{r}}_c + \dot{\vec{r}}_i + \vec{\omega}_i \times \vec{r}_i \tag{2.23}$$

and this is used to derive the rotational and vibrational kinetic energies of each appendage.

Spacecraft Kinetic Energy

The system total kinetic energy is the summation of the kinetic energies associated with the rigid bus rotation ($T_c$), and the appendage rotation and vibration ($T_i$), and the orbital motion of the system center of mass ($T_o$), that is

$$T = T_o + T_c + \sum_{i=1}^{n} T_i \tag{2.24}$$

The kinetic energy for the spacecraft can be derived to be

$$\begin{aligned} T &= \frac{1}{2} m_t v_c^2 + \frac{1}{2} \vec{\omega}^T \underline{I}_c \vec{\omega} + \frac{1}{2} \vec{\omega}^T \sum_{i=1}^{n} \underline{I}_i \ \vec{\omega} \\ &+ \sum_{i=1}^{n} \vec{\kappa}_i^T \vec{\omega} + \frac{1}{2} \sum_{i=1}^{n} \rho_i \int_{A_i} \dot{w}_i^2 \mathrm{d}A_i \end{aligned} \tag{2.25}$$

where $m_t$ is the total mass of the spacecraft, $v_c$ is the velocity of center of mass, $\vec{\omega}^T = \left[ \omega_x \ \omega_y \ \omega_z \right]$ is the spacecraft angular rate, $\underline{I}_i$ is the inertia tensor of the $i^{th}$ appendage, $\underline{I}_c$ is the inertia tensor of the rigid bus, $\vec{\kappa}_i$ is the $i^{th}$ appendage angular momentum due to flexibility, $\rho_i$ is the linear density of the $i^{th}$ appendage, the index $i$ ranges from 1 to $n$ and where $n$ is the total number of appendages.

Attitude Dynamics Equation

The attitude dynamics equation is given by

$$\frac{\mathrm{d}}{\mathrm{d}t}\left( \frac{\partial T}{\partial \vec{\omega}} + \vec{h}_w \right) + \underline{\Omega}\left( \frac{\partial T}{\partial \vec{\omega}} + \vec{h}_w \right) = \vec{\tau} \tag{2.26}$$

where $\vec{\tau}$ is the attitude recovery control torque and is obtained using thrusters or control moment gyros.

The final form of the *flexible attitude dynamics equation* is

$$\underline{I}_t\dot{\vec{\omega}} + \dot{\underline{I}}_t\vec{\omega} + \underline{\Omega}\left(\underline{I}_t\vec{\omega} + \vec{\kappa}_t + \vec{h}_w\right) + \dot{\vec{\kappa}}_t = \vec{\tau} \tag{2.27}$$

where

$$\underline{I}_t = \underline{I}_c + \sum_{i=1}^{n}\underline{I}_i \quad \dot{\underline{I}}_t = \sum_{i=1}^{n}\dot{\underline{I}}_i \quad \vec{\kappa}_t = \sum_{i=1}^{n}\vec{\kappa}_i \quad \dot{\vec{\kappa}}_t = \sum_{i=1}^{n}\dot{\vec{\kappa}}_i = \dot{\vec{\kappa}}_{\dot{\vec{\chi}}} + \dot{\underline{\kappa}}_{\ddot{\vec{\chi}}}^{T}\ddot{\vec{\chi}} \tag{2.28}$$

and $\vec{\chi}$ is the generalized coordinates vector for the appendages, described in the next section.

<u>Appendage Discretization</u>

The Raleigh-Ritz or assumed mode method is used to discretize the continuous appendage model. The deflection in the negative z-direction of the appendage reference frame (Figure 2.6) for the $i^{th}$ appendage is given by

$$w_i = w_i\left(x_i, y_i, t\right) = \sum_{r=1}^{p}\sum_{s=1}^{q}\chi_{i_{rs}}\left(t\right)\phi_r\left(x_i\right)\,\psi_s\left(y_i\right) \tag{2.29}$$

where $\chi_{i_{rs}}$ is a non-dimensional generalized displacement coordinate (temporal), $\phi_r$ and $\psi_s$ are the two families of admissible functions (i.e. mode shapes) satisfying the geometric boundary conditions (spatial), the indices $r, s$ range from 1 to $p, q$, and $p$ and $q$ are the number of discretized modes along $x$ and $y$ appendage axes, respectively.

In order to make the problem more specific and to be able to find the mode shape integrals analytically, cantilevered rectangular appendages were considered as these are the common shapes of solar arrays and antennas found on most spacecraft. However, note that the attitude and vibrational equations of motion derived in this thesis are generic and can accommodate any geometrically shaped appendage. In the case of an arbitrary shaped appendage, the mode shape integrals are more difficult to obtain analytically and one would have to resort to numerical techniques or finite element methods to obtain them.

The boundary conditions for the cantilevered appendage are taken to be clamped for $(x,0)$, and free for $(x,b)$, $(0,y)$ and $(a,y)$ (refer to Figure 2.6). The mode shape eigenfunctions for the free-free, $\phi_r(x)$, and clamped-free, $\psi_s(y)$, beams are given by (2.30)-(2.31). Note that the first two modes of the free-free beam (i.e. $\phi_1$ and $\phi_2$) are for the rigid body translation and rotation, respectively.

$$\phi_1(x) = 1$$

$$\phi_2(x) = \sqrt{12}\left(\frac{1}{2} - \frac{x}{a}\right)$$

$$\text{for } r = 3, 4, 5, \ldots, p \qquad (2.30)$$

$$\phi_r(x) = \left\{\left[\cosh\left(\lambda_r \frac{x}{a}\right) + \cos\left(\lambda_r \frac{x}{a}\right)\right] - \sigma_r\left[\sinh\left(\lambda_r \frac{x}{a}\right) + \sin\left(\lambda_r \frac{x}{a}\right)\right]\right\}$$

$$\text{for } s = 1, 2, 3, \ldots, q$$

$$\psi_s(y) = \left\{\left[\cosh\left(\lambda_s \frac{y}{b}\right) - \cos\left(\lambda_s \frac{y}{b}\right)\right] - \sigma_s\left[\sinh\left(\lambda_s \frac{y}{b}\right) - \sin\left(\lambda_s \frac{y}{b}\right)\right]\right\} \qquad (2.31)$$

where $\sigma_r$ and $\sigma_s$ are given by

$$\sigma_r = \frac{\cosh(\lambda_r) - \cos(\lambda_r)}{\sinh(\lambda_r) - \sin(\lambda_r)} \qquad (2.32)$$

$$\sigma_s = \frac{\sinh(\lambda_s) - \sin(\lambda_s)}{\cosh(\lambda_s) + \cos(\lambda_s)} \qquad (2.33)$$

and, $\lambda_r$ and $\lambda_s$ are the roots of

$$1 - \cosh(\lambda_r)\cos(\lambda_r) = 0 \qquad (2.34)$$

$$1 + \cosh(\lambda_s)\cos(\lambda_s) = 0 \qquad (2.35)$$

By using this discretization of the deflections, $w$, it will be possible to transform the set of Partial Deferential Equations (PDEs) describing the flexural motion of the appendages into a set of more easily manageable Ordinary Differential Equations (ODEs) which are also better suited for numerical implementation. These PDEs result from the Lagrangian formulation of the flexural/vibrational motions as will be described shortly, after giving the equations for the spacecraft potential energy.

<u>Spacecraft Potential/Elastic Energy</u>

The total potential energy is obtained by summing the orbital and flexible potential energies

$$K = K_{\text{orbital}} + K_{\text{flexible}} \tag{2.36}$$

The orbital potential energy is not a function of the generalized displacement, $\vec{\chi}$, and will not contribute to the present formulation. The potential energy stored in $n$ deflected appendages is given by the following equation

$$K_{\text{flexible}} = \frac{1}{2}\sum_{i=1}^{n} J_i \int_0^{b_i}\int_0^{a_i} \left[ \left(\frac{\partial^2 w_i}{\partial x_i^2}\right)^2 + \left(\frac{\partial^2 w_i}{\partial y_i^2}\right)^2 + 2\gamma_i \frac{\partial^2 w_i}{\partial x_i^2}\frac{\partial^2 w_i}{\partial y_i^2} + 2(1-\gamma_i)\left(\frac{\partial^2 w_i}{\partial x_i \partial y_i}\right)^2 \right] \mathrm{d}x_i \mathrm{d}y_i \tag{2.37}$$

where $J_i$ is the flexural rigidity and is defined by

$$J_i = \frac{E_i \, h_i^3}{12 \, \left(1\text{-}\gamma_i^{\,2}\right)} \tag{2.38}$$

$h_i$ is the thickness of the $i^{th}$ appendage, and $\gamma_i$ and $E_i$ are the Poisson ratio and the modulus of elasticity of the material, respectively.

The final form of the potential energy stored in the deflected appendages is

$$K_{\text{flexible}} = \frac{1}{2}\sum_{i=1}^{n} J_i \, \vec{\chi}_i^T \left[ \underline{M}_{\phi''\psi} + \underline{M}_{\phi\psi''} + 2\gamma_i \, \underline{M}_{\phi''\psi''} + 2(1-\gamma_i)\underline{M}_{\phi'\psi'} \right] \vec{\chi}_i \tag{2.39}$$

where $\vec{\chi}_i^T = [\chi_{i_{11}}\ \chi_{i_{12}} \cdots \chi_{i_{pq}}] = [\chi_1\ \chi_2 \cdots \chi_{pq}]$ and $\underline{M}_{\phi''\psi}, \underline{M}_{\phi\psi''}, \underline{M}_{\phi''\psi''}$ and $\underline{M}_{\phi'\psi'}$ are square matrices of dimension $pq$ obtained from integrals of mode shapes (2.30) and (2.31) and their spatial derivatives (i.e. $\phi', \phi'', \psi'$ and $\psi''$). As an example, the closed-form expression for matrix $\underline{M}_{\phi'\psi'}$ is detailed below

$$\begin{aligned}\underline{M}_{\phi'\psi'} &= \underline{M}_{\phi'\phi'} \otimes \underline{M}_{\psi'\psi'} \\ &= \int_0^a \vec{\phi}'\vec{\phi}'^T \mathrm{d}x \otimes \int_0^b \vec{\psi}'\vec{\psi}'^T \mathrm{d}y \\ &= \int_0^a \frac{\mathrm{d}\vec{\phi}}{\mathrm{d}x}\left(\frac{\mathrm{d}\vec{\phi}}{\mathrm{d}x}\right)^T \mathrm{d}x \otimes \int_0^b \frac{\mathrm{d}\vec{\psi}}{\mathrm{d}y}\frac{\mathrm{d}\vec{\psi}}{\mathrm{d}y}^T \mathrm{d}y\end{aligned} \tag{2.40}$$

where $\left(\frac{\mathrm{d}\vec{\phi}}{\mathrm{d}x}\right)^T = \vec{\phi}'^T = \left[\phi_1'\ \phi_2' \ldots \phi_p'\right]$, $\left(\frac{\mathrm{d}\vec{\psi}}{\mathrm{d}y}\right)^T = \vec{\psi}'^T = \left[\psi_1'\ \psi_2' \ldots \psi_q'\right]$, $\otimes$ is the tensor or Kronecker product and

$$\underline{M}_{\phi'\phi'} = \begin{cases} \dfrac{12}{a} & (2,2) \\ \dfrac{4\sqrt{3}}{\mathrm{a}}[(-1)^i + 1] & (2,i) \text{ and } (i,2),\ i \ge 3 \\ \dfrac{\lambda_i\sigma_i}{a}[6 + \lambda_i\sigma_i] & i = j,\ i \ge 3, j \ge 3 \\ \dfrac{4\lambda_j\lambda_i(\sigma_j\lambda_i^3 - \sigma_i\lambda_j^3)}{a\left(\lambda_i^4 - \lambda_j^4\right)}[(-1)^{i+j} + 1] & i \ne j,\ i \ge 3, j \ge 3 \\ 0 & \text{otherwise} \end{cases} \tag{2.41}$$

$$\underline{M}_{\psi'\psi'} = \begin{cases} \dfrac{\sigma_i\lambda_i}{b}(2 + \sigma_i\lambda_i) & i = j \\ \dfrac{4\lambda_j\lambda_i}{b\left(\lambda_i^4 - \lambda_j^4\right)}[(-1)^{i+j}(\sigma_j\lambda_i^3 - \sigma_i\lambda_j^3) - \lambda_j\lambda_i\left(\sigma_i\lambda_i - \sigma_j\lambda_j\right)] & i \ne j \end{cases} \tag{2.42}$$

The other integrals of mode shapes found in (2.39) and (2.46) are detailed in Appendix B. Using the notation

$$\underline{V}_i = J_i \left[ \underline{M}_{\phi''\psi} + \underline{M}_{\phi\psi''} + 2\gamma_i \ \underline{M}_{\phi''\psi''} + 2\left(1-\gamma_i\right) \underline{M}_{\phi'\psi'} \right] \tag{2.43}$$

the potential energy stored in the appendages may be expressed as

$$K_{\text{flexible}} = \frac{1}{2} \sum_{i=1}^{n} \vec{\chi}_i^T \underline{V}_i \ \vec{\chi}_i \tag{2.44}$$

<u>Flexural/Vibrational Equations</u>

The vibrational equations of motion are obtained by using the conventional form of Lagrange's equation, and the expressions for the kinetic and potential energies of the system given in (2.25) and (2.44)

$$\frac{\mathrm{d}}{\mathrm{d}t}\left( \frac{\partial T}{\partial \dot{\chi}_{i_{rs}}} \right) - \frac{\partial T}{\partial \chi_{i_{rs}}} + \frac{\partial K}{\partial \chi_{i_{rs}}} = Q_{i_{rs}} \tag{2.45}$$

The terms $Q_{i_{rs}}$ are the generalized forces due to environmental disturbances and are set to zero for our present purposes. Substituting the kinetic and potential energy equations in the Lagrange's equation, we obtain the final form of the vibrational equation for the $i^{th}$ appendage (the subscript $i$ is dropped for clarity of presentation):

$$\begin{aligned}
&\underline{M}_{\phi\psi} \ddot{\vec{\chi}} + \left( d_z \underline{M}_{\phi\psi'_\eta} \vec{\chi} + d_y \vec{m}_{\phi\psi} + \vec{m}_{y\phi\psi} \right) \dot{\omega}_x - \left( d_x \vec{m}_{\phi\psi} + \vec{m}_{x\phi\psi} \right) \dot{\omega}_y - \left( d_x \underline{M}_{\phi\psi'_\eta} + \underline{M}_{x\phi\psi'_\eta} \right) \vec{\chi} \dot{\omega}_z = \\
&\begin{bmatrix}
-\dfrac{\left(\underline{V} + \underline{V}^T\right)}{2\rho} - \omega_x^2 \left( d_y \underline{M}_{\phi\psi'_\eta} - \underline{M}_{\phi\psi} + \underline{M}_{y\phi\psi'_\eta} \right) + \\
\omega_y^2 \underline{M}_{\phi\psi} - \omega_z^2 \left( d_y \underline{M}_{\phi\psi'_\eta} + \underline{M}_{y\phi\psi'_\eta} \right) + \\
\omega_x \omega_y \left( d_x \underline{M}_{\phi\psi'_\eta} + \underline{M}_{x\phi\psi'_\eta} \right) + \omega_y \omega_z \left( d_z \underline{M}_{\phi\psi'_\eta} \right) \\
\
\end{bmatrix} \vec{\chi} +
\begin{bmatrix}
\omega_x^2 d_z \vec{m}_{\phi\psi} + \omega_y^2 d_z \vec{m}_{\phi\psi} - \\
\omega_x \omega_z \left( d_x \vec{m}_{\phi\psi} + \vec{m}_{x\phi\psi} \right) - \\
\omega_y \omega_z \left( d_y \vec{m}_{\phi\psi} + \vec{m}_{y\phi\psi} \right)
\end{bmatrix}
\end{aligned} \tag{2.46}$$

The above equation may be rewritten equivalently as

$$\underline{M}_{\phi\psi} \ddot{\vec{\chi}} + \vec{\kappa}_{x\ddot{\chi}} \ \dot{\omega}_x + \vec{\kappa}_{y\ddot{\chi}} \ \dot{\omega}_y + \vec{\kappa}_{z\ddot{\chi}} \ \dot{\omega}_z = \underline{K} \ \vec{\chi} - \underline{D} \ \dot{\vec{\chi}} + \vec{C} \tag{2.47}$$

where

$$\underline{K} = \begin{bmatrix} -\frac{\left(\underline{V}+\underline{V}^T\right)}{2\rho} + \omega_x^2\left(-d_y\underline{M}_{\phi\psi'_\eta} + \underline{M}_{\phi\psi} - \underline{M}_{y\phi\psi'_\eta}\right) + \\ \omega_y^2\underline{M}_{\phi\psi} + \omega_z^2\left(-d_y\underline{M}_{\phi\psi'_\eta} - \underline{M}_{y\phi\psi'_\eta}\right) + \\ \omega_x\omega_y\left(d_x\underline{M}_{\phi\psi'_\eta} + \underline{M}_{x\phi\psi'_\eta}\right) + \omega_y\omega_z\left(d_z\underline{M}_{\phi\psi'_\eta}\right) \\ \end{bmatrix} \tag{2.48}$$

$$\vec{C} = \omega_x^2 d_z\vec{m}_{\phi\psi} + \omega_y^2 d_z\vec{m}_{\phi\psi} - \omega_x\omega_z\left(d_x\vec{m}_{\phi\psi} + \vec{m}_{x\phi\psi}\right) - \omega_y\omega_z\left(d_y\vec{m}_{\phi\psi} + \vec{m}_{y\phi\psi}\right) \tag{2.49}$$

and

$$\begin{aligned} \dot{\vec{\kappa}}_{x\ddot{\chi}} &= d_z\underline{M}_{\phi\psi'_\eta}\vec{\chi} + d_y\vec{m}_{\phi\psi} + \vec{m}_{y\phi\psi} \\ \dot{\vec{\kappa}}_{y\ddot{\chi}} &= -d_x\vec{m}_{\phi\psi} - \vec{m}_{x\phi\psi} \\ \dot{\vec{\kappa}}_{z\ddot{\chi}} &= -d_x\underline{M}_{\phi\psi'_\eta}\vec{\chi} - \underline{M}_{x\phi\psi'_\eta}\vec{\chi} \end{aligned} \tag{2.50}$$

Note that since there is some damping in any flexible structure, we have added a damping term to (2.47). We have assumed viscous damping and no modal coupling as far as damping is concerned which results in a diagonal damping matrix, $\underline{D}$, with entries $\xi$ for the damping parameter. Finally, the vibrational equations of the $i^{th}$ appendage using matrix form can be written as

$$\begin{bmatrix} \dot{\underline{\kappa}}_{\ddot{\chi}} & \underline{M}_{\phi\psi} \end{bmatrix} \begin{bmatrix} \dot{\vec{\omega}} \\ \ddot{\vec{\chi}} \end{bmatrix} = \begin{bmatrix} \underline{K}\,\vec{\chi} - \underline{D}\,\dot{\vec{\chi}} + \vec{C} \end{bmatrix} \tag{2.51}$$

where $\dot{\underline{\kappa}}_{\ddot{\chi}} = \begin{bmatrix} \dot{\vec{\kappa}}_{x\ddot{\chi}} & \dot{\vec{\kappa}}_{y\ddot{\chi}} & \dot{\vec{\kappa}}_{z\ddot{\chi}} \end{bmatrix}$.

The complete set of attitude and flexural equations of the system, (2.27) and (2.51), can be combined using (2.28), and rewritten as

$$\begin{bmatrix} \underline{I}_t & \dot{\underline{\kappa}}_{\ddot{\chi}}^T \\ \dot{\underline{\kappa}}_{\ddot{\chi}} & \underline{M}_{\phi\psi} \end{bmatrix} \begin{bmatrix} \dot{\vec{\omega}} \\ \ddot{\vec{\chi}} \end{bmatrix} = \begin{bmatrix} -\left(\dot{\underline{I}}_t\vec{\omega} + \Omega\left(\underline{I}_t\vec{\omega} + \vec{\kappa}_t + \vec{h}_w\right) + \dot{\vec{\kappa}}_{\dot{\chi}}\right) \\ \underline{K}\,\vec{\chi} - \underline{D}\,\dot{\vec{\chi}} + \vec{C} \end{bmatrix} + \begin{bmatrix} \vec{\tau} \\ 0 \end{bmatrix} \tag{2.52}$$

$$\begin{bmatrix} \dot{\vec{\omega}} \\ \ddot{\vec{\chi}} \end{bmatrix} = \begin{bmatrix} \underline{I}_t & \underline{\dot{\kappa}}_{\ddot{\vec{\chi}}}^T \\ \underline{\dot{\kappa}}_{\ddot{\vec{\chi}}} & \underline{M}_{\phi\psi} \end{bmatrix}^{-1} \begin{bmatrix} -\left( \underline{\dot{I}}_t \vec{\omega} + \underline{\Omega}\left( \underline{I}_t \vec{\omega} + \vec{\kappa}_t + \vec{h}_w \right) + \vec{\kappa}_{\dot{\vec{\chi}}} \right) \\ \underline{K}\,\vec{\chi} - \underline{D}\,\dot{\vec{\chi}} + \vec{C} \end{bmatrix} + \begin{bmatrix} \underline{I}_t & \underline{\dot{\kappa}}_{\ddot{\vec{\chi}}}^T \\ \underline{\dot{\kappa}}_{\ddot{\vec{\chi}}} & \underline{M}_{\phi\psi} \end{bmatrix}^{-1} \begin{bmatrix} \vec{\tau} \\ 0 \end{bmatrix} \tag{2.53}$$

The mass matrix inversion in (2.53) can be obtained by using the block matrix inversion formula [127], and hence we can decouple the system of equations

$$\begin{aligned} \begin{bmatrix} \dot{\vec{\omega}} \\ \ddot{\vec{\chi}} \end{bmatrix} = & \begin{bmatrix} \underline{F}_{11}^{-1} & -\underline{I}_t^{-1} \underline{\dot{\kappa}}_{\ddot{\vec{\chi}}}^T \underline{F}_{22}^{-1} \\ -\underline{F}_{22}^{-1} \underline{\dot{\kappa}}_{\ddot{\vec{\chi}}} \underline{I}_t^{-1} & \underline{F}_{22}^{-1} \end{bmatrix} \begin{bmatrix} -\left( \underline{\dot{I}}_t \vec{\omega} + \underline{\Omega}\left( \underline{I}_t \vec{\omega} + \vec{\kappa}_t + \vec{h}_w \right) + \vec{\kappa}_{\dot{\vec{\chi}}} \right) \\ \underline{K}\,\vec{\chi} - \underline{D}\,\dot{\vec{\chi}} + \vec{C} \end{bmatrix} + \\ & \begin{bmatrix} \underline{F}_{11}^{-1} & -\underline{I}_t^{-1} \underline{\dot{\kappa}}_{\ddot{\vec{\chi}}}^T \underline{F}_{22}^{-1} \\ -\underline{F}_{22}^{-1} \underline{\dot{\kappa}}_{\ddot{\vec{\chi}}} \underline{I}_t^{-1} & \underline{F}_{22}^{-1} \end{bmatrix} \begin{bmatrix} \vec{\tau} \\ 0 \end{bmatrix} \end{aligned} \tag{2.54}$$

$$\begin{bmatrix} \dot{\vec{\omega}} \\ \ddot{\vec{\chi}} \end{bmatrix} = \begin{bmatrix} -\underline{F}_{11}^{-1} \vec{l}_{\vec{\omega}} - \underline{I}_t^{-1} \underline{\dot{\kappa}}_{\ddot{\vec{\chi}}}^T \underline{F}_{22}^{-1} \vec{l}_{\vec{\chi}} \\ -\underline{F}_{22}^{-1} \underline{\dot{\kappa}}_{\ddot{\vec{\chi}}} \underline{I}_t^{-1} \vec{l}_{\vec{\omega}} + \underline{F}_{22}^{-1} \vec{l}_{\vec{\chi}} \end{bmatrix} + \begin{bmatrix} \underline{F}_{11}^{-1} \\ -\underline{F}_{22}^{-1} \underline{\dot{\kappa}}_{\ddot{\vec{\chi}}} \underline{I}_t^{-1} \end{bmatrix} \vec{\tau} \tag{2.55}$$

where

$$\begin{aligned} \underline{F}_{11} &= \underline{I}_t - \underline{\dot{\kappa}}_{\ddot{\vec{\chi}}}^T \underline{M}_{\phi\psi}^{-1} \underline{\dot{\kappa}}_{\ddot{\vec{\chi}}} \\ \underline{F}_{22} &= \underline{M}_{\phi\psi} - \underline{\dot{\kappa}}_{\ddot{\vec{\chi}}} \underline{I}_t^{-1} \underline{\dot{\kappa}}_{\ddot{\vec{\chi}}}^T \\ \vec{l}_{\vec{\omega}} &= \underline{\dot{I}}_t \vec{\omega} + \underline{\Omega}\left( \underline{I}_t \vec{\omega} + \vec{\kappa}_t + \vec{h}_w \right) + \vec{\kappa}_{\dot{\vec{\chi}}} \\ \vec{l}_{\vec{\chi}} &= \underline{K}\,\vec{\chi} - \underline{D}\,\dot{\vec{\chi}} + \vec{C} \end{aligned} \tag{2.56}$$

By defining $\vec{\delta}^T = \left[ \vec{\chi}^T \ \dot{\vec{\chi}}^T \right]$, (2.55) can be expressed in a concise state space form as,

$$\begin{bmatrix} \dot{\vec{\omega}} \\ \dot{\vec{\delta}} \end{bmatrix} = \begin{bmatrix} \vec{f}_{\vec{\omega}}(\vec{\omega}, \vec{\delta}) \\ \vec{f}_{\vec{\delta}}(\vec{\omega}, \vec{\delta}) \end{bmatrix} + \begin{bmatrix} \underline{G}_{\vec{\omega}}(\vec{\delta}) \\ \underline{G}_{\vec{\delta}}(\vec{\delta}) \end{bmatrix} \vec{\tau} \triangleq \vec{f}(\vec{x}) + \underline{G}(\vec{x})\,\vec{\tau} \tag{2.57}$$

where the $\vec{f}$ and $\underline{G}$ terms can be obtained directly from (2.55) and the state and output vectors are given as

$$\vec{x} = [\vec{\omega}^T \quad \vec{\delta}^T]^T = [\omega_x \; \omega_y \; \omega_z \; \chi_1 \; \chi_2 \cdots \chi_{npq} \; \dot{\chi}_1 \; \dot{\chi}_2 \cdots \dot{\chi}_{npq}]^T$$
$$\vec{y} = \vec{h}(\vec{x}) = \begin{bmatrix} x_1 & x_2 & x_3 \end{bmatrix}^T = \begin{bmatrix} \omega_x & \omega_y & \omega_z \end{bmatrix}^T \tag{2.58}$$

This is a $m$-dimensional system with $m = 3 + 2npq$, with $n, p$ and $q$ defined previously.

Equation (2.57) can also be expressed using quaternions, $\tilde{q}$ *and* $\dot{\tilde{q}}$, as opposed to the body angular rate, $\vec{\omega}$, as was shown for the rigid case in Section 2.2, hence

$$\dot{\vec{x}} = \begin{bmatrix} x_5 \\ x_6 \\ x_7 \\ x_8 \\ f_{q_0}(\tilde{q}, \dot{\tilde{q}}, \vec{\chi}, \dot{\vec{\chi}}) \\ \vec{f}_{\vec{q}}(\tilde{q}, \dot{\tilde{q}}, \vec{\chi}, \dot{\vec{\chi}}) \\ x_{npq+9} \\ \vdots \\ x_{2npq+8} \\ \vec{f}_{\vec{\chi}}(\tilde{q}, \dot{\tilde{q}}, \vec{\chi}, \dot{\vec{\chi}}) \end{bmatrix} + \begin{bmatrix} \vec{0} \\ \vec{0} \\ \vec{0} \\ \vec{0} \\ \vec{g}_{q_0}^T(\tilde{q}, \vec{\chi}) \\ \underline{G}_{\vec{q}}(\tilde{q}, \vec{\chi}) \\ \vec{0} \\ \vdots \\ \vec{0} \\ \underline{G}_{\vec{\chi}}(\tilde{q}, \vec{\chi}) \end{bmatrix} \vec{\tau} \triangleq \vec{f}(\vec{x}) + \underline{G}(\vec{x})\vec{\tau} \tag{2.59}$$

where the state and output vectors will now be

$$\vec{x} = [q_0 \; q_1 \; q_2 \; q_3 \; \dot{q}_0 \; \dot{q}_1 \; \dot{q}_2 \; \dot{q}_3 \; \chi_1 \; \chi_2 \cdots \chi_{npq} \; \dot{\chi}_1 \; \dot{\chi}_2 \cdots \dot{\chi}_{npq}]^T$$
$$\vec{y} = \vec{h}(\vec{x}) = \begin{bmatrix} x_2 & x_3 & x_4 \end{bmatrix}^T = \begin{bmatrix} q_1 & q_2 & q_3 \end{bmatrix}^T \tag{2.60}$$

In this case the system is $m$-dimensional with $m = 8 + 2npq$, where $n, p, q$ were defined previously. Also, in (2.57) to (2.60), $\vec{\tau}$ is the $3\times1$ control input vector, $\vec{y}$ is the $3\times1$ vector of system outputs (i.e. angular rates or quaternion vector), $\vec{f}$ and $\vec{h}$ are $m\times1$ and $3\times1$, respectively, smooth vector fields, and $\underline{G}$ is a $m\times3$ matrix whose columns are $m\times1$ smooth vector fields $\vec{g}_j$, $j = 1, 2, 3$, that is $\underline{G} = \begin{bmatrix} \vec{g}_1 \; \vec{g}_2 \; \vec{g}_3 \end{bmatrix}$. The number of inputs and outputs is selected to be 3 so that we have a square system. The resulting system equations which are $m^{th}$ order, nonlinear and highly coupled, are now in the state space representation, ideal for the synthesis of the controller.

## 2.4 Concluding Remarks

This chapter summarized the dynamic formulation for both rigid and a class of flexible spacecraft. The 3D rigid dynamics equations were found to be nonlinear and coupled through the angular body rates and this coupling is more prominent for large angular motions. The flexible dynamics derivations were based on the hybrid coordinates approach and also resulted in highly nonlinear terms due to the addition of flexible appendages. There is indeed a strong coupling effect between the attitude and the elastic motion of the appendages. The formulation presented is applicable to star topology spacecraft and it is analytical (without resorting to any FEM models) and in closed-form.

# CHAPTER 3

## NUMERICAL SPACECRAFT SIMULATOR

Hardware-in-the-loop spacecraft attitude simulators, such as air-bearing simulators, are not only expensive to build, but they cannot provide the full experience of micro-gravity. Therefore a high fidelity numerical spacecraft simulator which includes both the rigid and flexible dynamics, as well as the orbital mechanics and the environment of the spacecraft, was developed and is presented in this chapter. It was decided to invest in developing an appropriate simulation environment from the ground up, since no suitable off-the-shelf simulator was available. Once validated, this simulation environment becomes an indispensable tool to test controllers for different types of attitude and orbital control systems and problems. In this thesis, this simulation environment was used primarily to test attitude recovery controllers.

Sections 3.1 to 3.3 describe the overall simulator as far as the environment, the architecture and the visualization capabilities are concerned. The validation cases for the rigid and flexible dynamics are presented in Section 3.3.

## 3.1 Simulator Environment

The flexible spacecraft simulator was developed using a standard PC running the Microsoft Windows operating system. The PC was based on an Intel Pentium IV, 1.80 GHz Central Processing Unit (CPU) with 256MB of RAM. The simulator can also run on a laptop, but the execution time may increase depending on the CPU speed and RAM available.

As for the software environment, Simulink Version 4.1 was used to build the modular spacecraft simulation by interconnecting a series of Simulink models; Matlab version 6.1 was used to set up the simulation parameters, automate simulation runs, generate output plots and archive the test results; and finally Microsoft Visual C/C++ version 6.0 was used to develop, debug and validate the different models. Matlab/Simulink was chosen for its strong numerical performance, necessary to solve the flexible dynamic equations. Particularly, Matlab was selected for its strength in matrix manipulations and numerical integration techniques. However, since Matlab is an interpreted language, it can be slow in comparison to compiled code, particularly when complex mathematical formulations are involved. Consequently, the simulator has been implemented to interface with compiled C model codes via S-Functions. An S-Function is essentially a Dynamically Linked Library (DLL) created using the Microsoft Visual C/C++ compiler, which is then called through a Simulink block. Furthermore, Matlab has the advantage of having a large user-base, and provides access to useful built-in toolboxes, including many native tools to visualize and analyze data, and control systems toolboxes to build dynamic controllers.

Before running a simulation case, a Matlab script is run to setup and initialize all the necessary and relevant parameters. The simulation can then be started and manipulated through the Simulink graphical user interface. Alternatively, for a series of simulation cases, it is preferable and more convenient to use a script to initialize, run and archive the output data for all the cases automatically without user intervention at every step. This way a large number of cases can be run over night without operator supervision.

## 3.2 Simulator Architecture

The top-level architecture for the simulator is shown in Figure 3.1. The system consists of the following high-level models which themselves contain many lower level models not described explicitly here:

- *Spacecraft Environment*: models the environmental factors that affect the spacecraft (e.g. magnetic field, sun flux); and models the spacecraft disturbances (i.e. gravity gradient torque, atmospheric drag force and torque, solar radiation force and torque, and magnetic torque).

- *Spacecraft Structure*: sets up the Flexible Plate-type Appendages' (FPAs) dimensions and characteristics (e.g. appendage length and width, number of modes, modulus of elasticity) and computes the appendage deflections.

- *Spacecraft Dynamics*: simulates the spacecraft orbital mechanics (not specifically discussed in this thesis) and, rigid or flexible attitude dynamics; and includes the spacecraft reference frame rotations and Euler angle calculations.

- *Spacecraft Attitude Recovery System*: models the spacecraft Attitude Recovery System (ARS) controller.

Each of the above-mentioned models has been built by connecting various Simulink built-in and user-defined (S-function) blocks.

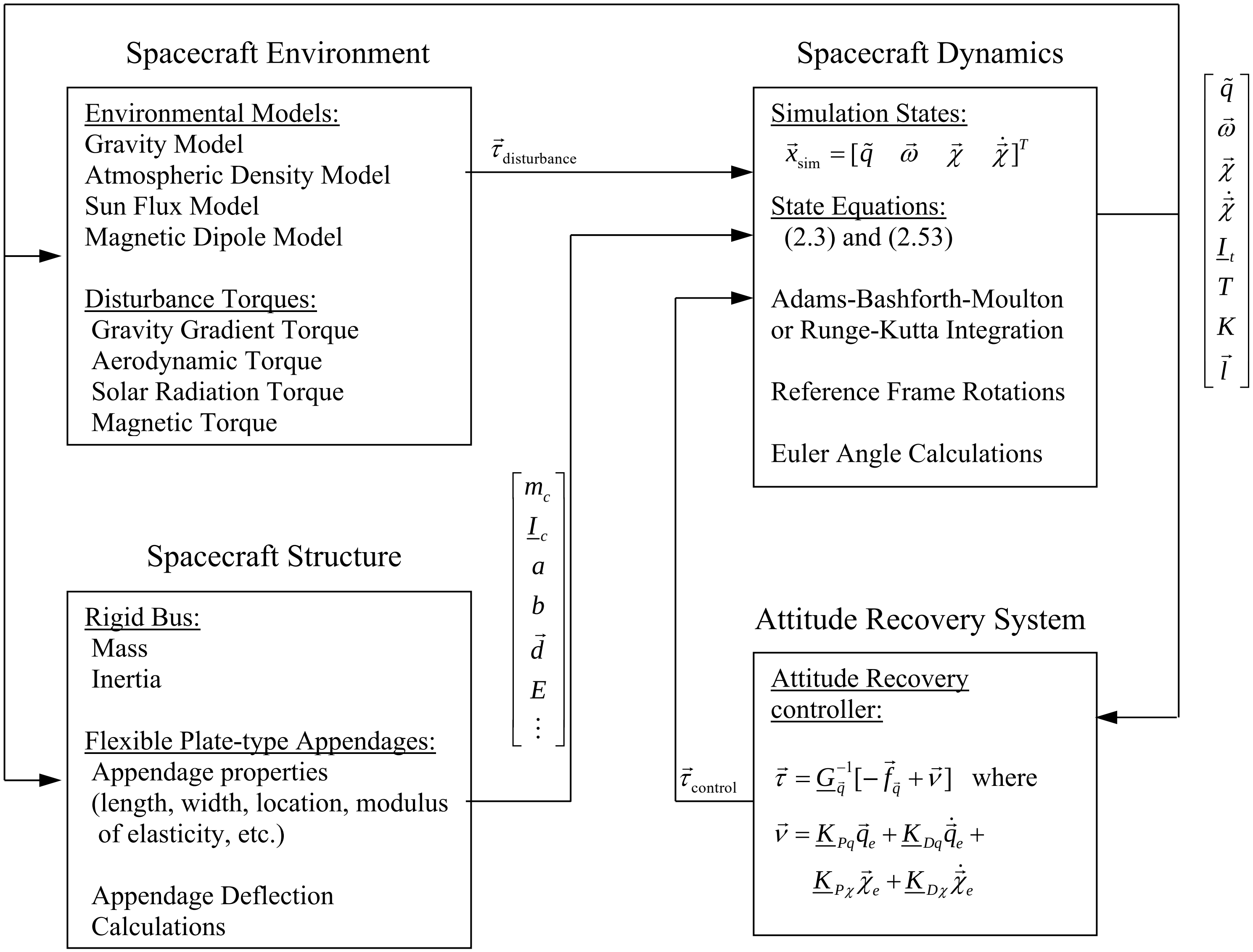

Figure 3.1: Simulator top-level architecture

As shown in Figure 3.1, the Spacecraft Environment, the Spacecraft Structure and the Spacecraft Attitude Recovery System models all feed into the Spacecraft Dynamics model. Note that the torques fed into the Spacecraft Dynamics model include both the disturbance torques from the Spacecraft Environment and the Spacecraft Attitude Recovery System torques. Switches have been included for the user to control whether or not to include the disturbance or attitude recovery torques. These switches may be turned on or off via software flags found in the simulator parameter setup m-file.

The Spacecraft Dynamics model outputs the quaternions, angular velocity, appendages' general displacements and velocities, total inertia, kinetic and potential energies and the angular momentum. The quaternions, angular velocity and general displacements and velocities are fed into the Spacecraft Attitude Recovery System, which in turn computes the control torque and feeds it back to the Spacecraft Dynamics model. The appendage's general displacements are also fed into the Spacecraft Structure model such that the appendages' deflections may be computed.

As indicated is Figure 3.1, the simulation states used for numerical integration are $\vec{x}_{\text{sim}} = [\tilde{q} \quad \vec{\omega} \quad \vec{\chi} \quad \dot{\vec{\chi}}]^T$ with the corresponding state equations (2.3) and (2.53).

## 3.3 Simulation Visualization

There are several options for visualizing the output of the simulator. The simplest way is to associate a numerical display to a scalar, vector or matrix parameter which is placed on the Simulink models. This however does not capture the time evolution of the parameter and one might rather visualize a time-series plot for different parameters which can also be easily obtained using a real-time *scope* available in Simulink. The scope is placed and connected to a particular parameter before the start of the simulation and is active while the simulation is running. These same time-series plots along with any other dynamical 2D or 3D plots, for the purposes of analysis and documentation, can be obtained after a simulation run. These plots are obtained using output data selected for archival and dumped to the Matlab workspace or hard disk. The standard *plot* command in Matlab can be used to generate these plots with many useful graphical features.

However, these plots can't really capture the true physical attitude and flexible motion of the spacecraft, and hence to give a more intuitive "feel" of how the attitude of spacecraft is being recovered, a 3D animation model is included in the Output block. This animation model allows the user to view the attitude of the spacecraft and the appendage deflections by displaying a window for the spacecraft attitude visualization, and one window for each appendage. Currently, the animation model can display up to two appendages, however, a future improvement will be to incorporate all displays into one, and to modify the animation block to display more appendages.

Figure 3.2 to Figure 3.4 show the output of the 3D animation block for a spacecraft with two appendages, where appendage 1 and 2 represent the right and left appendages, respectively. The simulation is for a spacecraft with 4 vibration modes in the *x* and the *y* direction of each appendage.

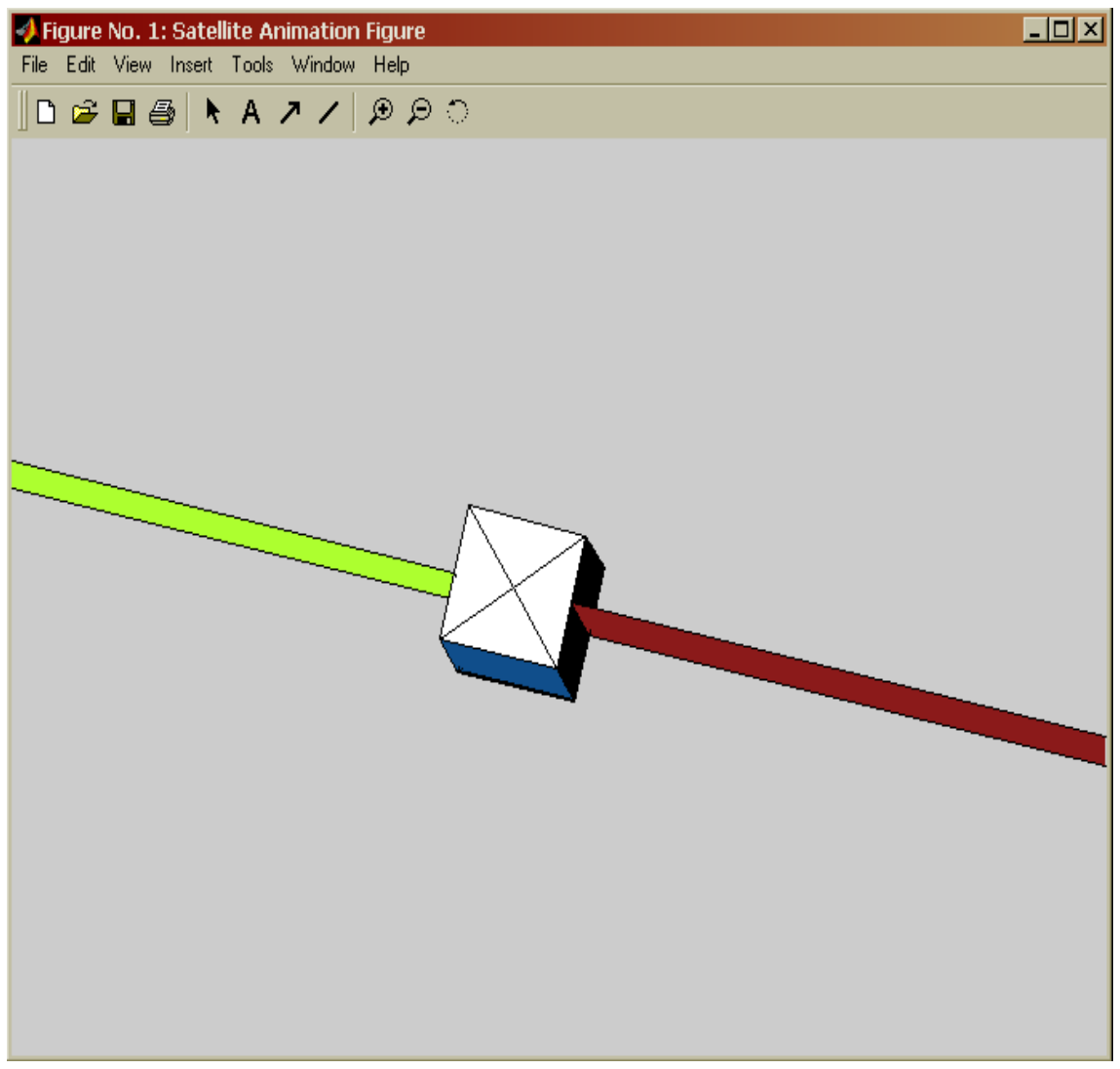

Figure 3.2: Spacecraft attitude animation window

To facilitate visualizing the spacecraft attitude motion, the animation figure has been color coded as follows:

| Spacecraft Part | Color |
|---|---|
| Front | White (with an X) |
| Back | Black |
| | |
| Top | White |
| Bottom | Blue |
| | |
| Left appendage | Lime |
| Right appendage | Red |

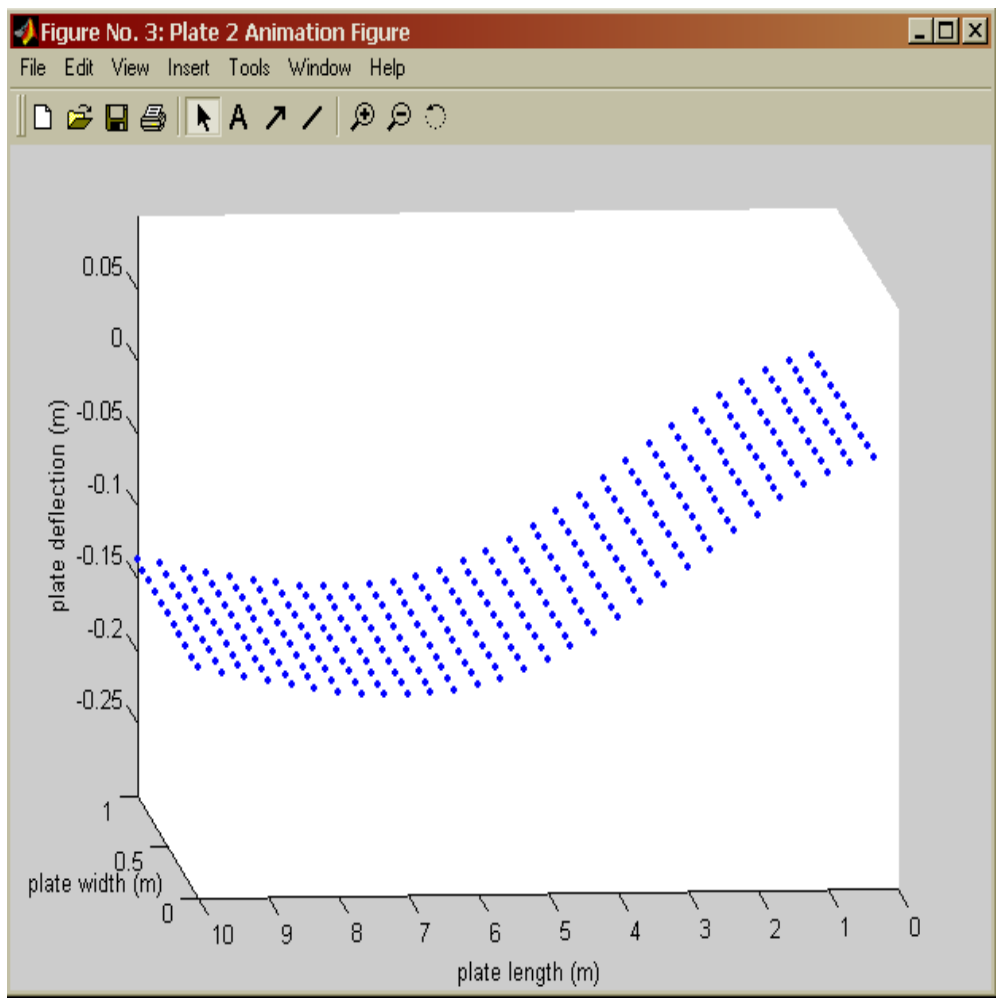

Figure 3.3: Left appendage animation window

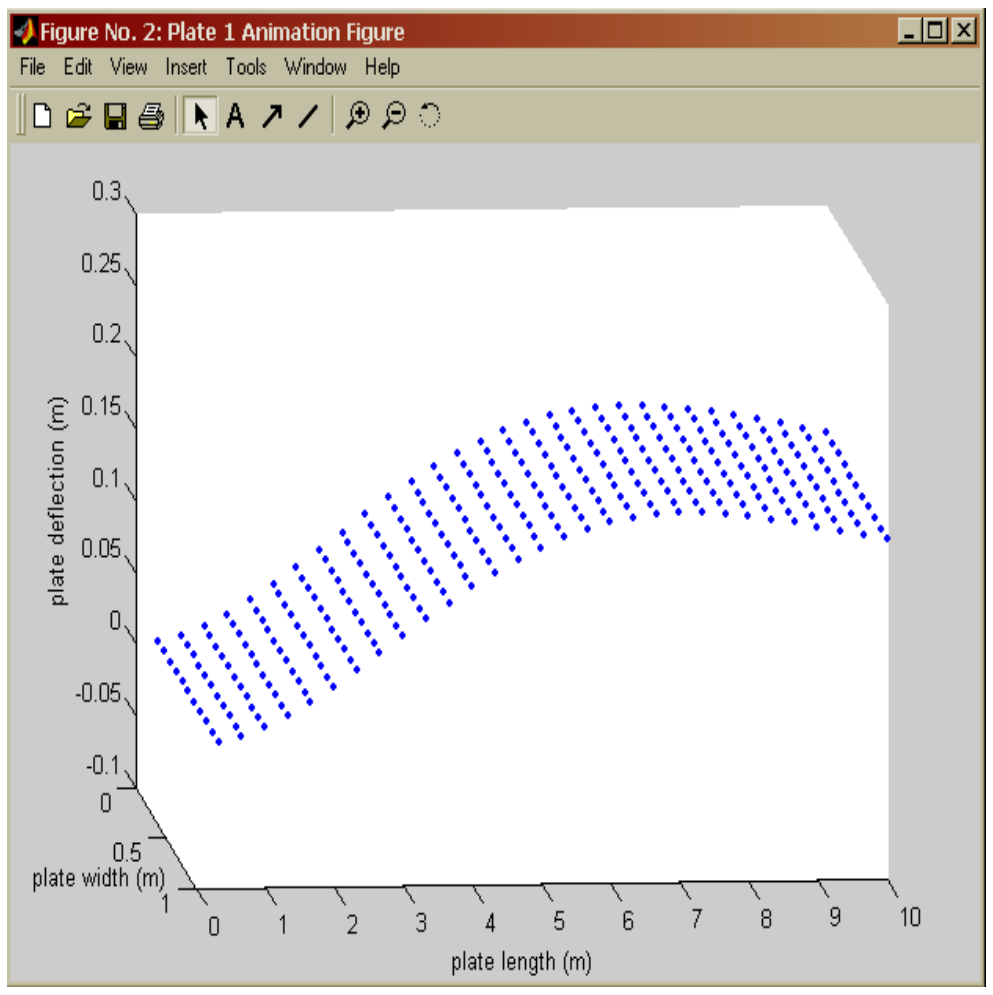

Figure 3.4: Right appendage animation window

The 3D animation model may be controlled via its block parameter dialog box as shown in Figure 3.5. Using the block parameter dialog box, the user may control which figure windows to view using the checkboxes located at the bottom of the dialog box. The block parameters have been set to read from the values set in the simulator parameter setup file, thus they need not be modified here. A description of the usage of each of the parameters is shown in Table 3.1 below.

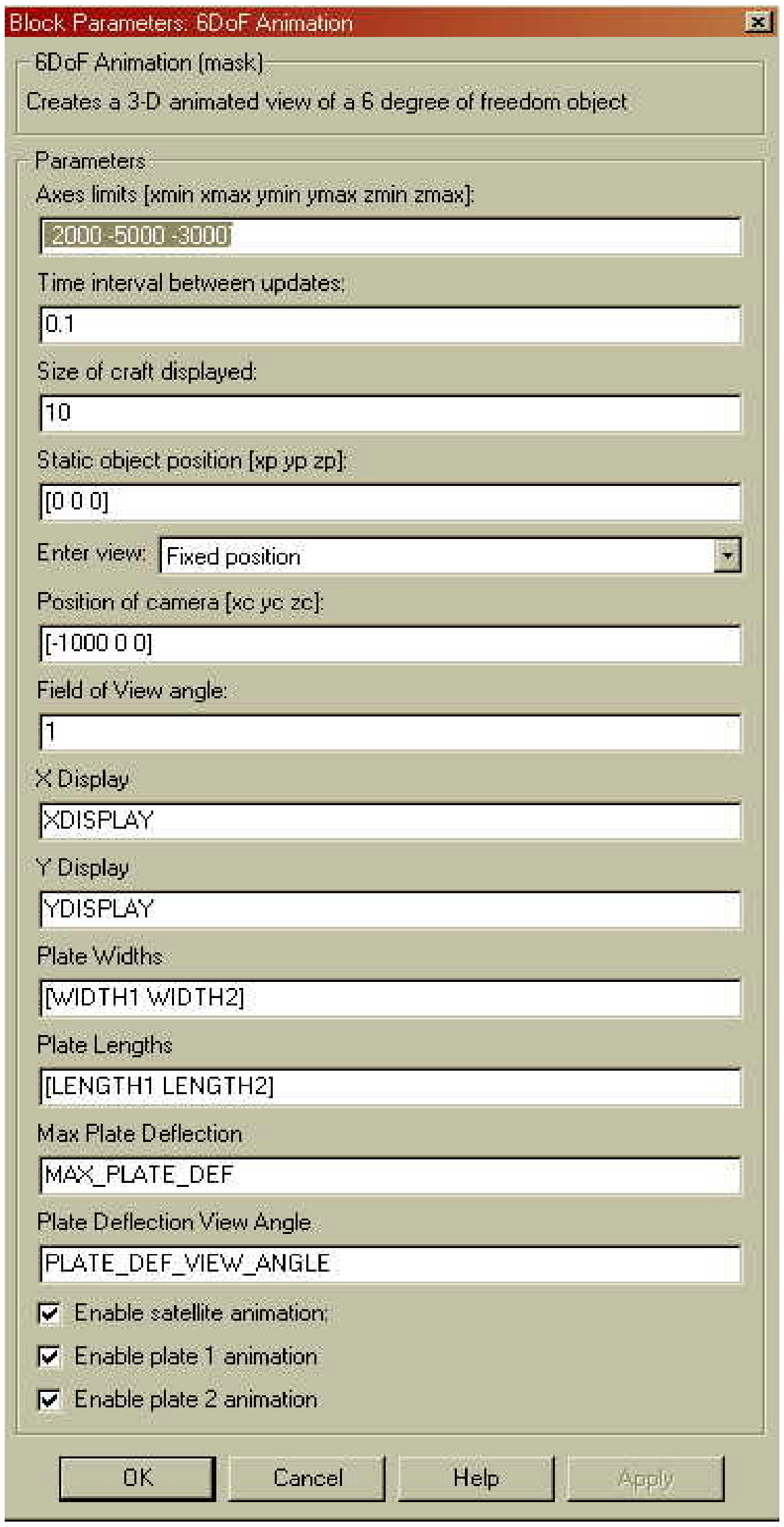

Figure 3.5: 3D animation model parameter dialog box

Table 3.1: 3D animation model parameter descriptions

| ***Parameter*** | ***Description*** |
|---|---|
| XDISPLAY | Number of points displayed for each appendage in the *x*-direction |
| YDISPLAY | Number of points displayed for each appendage in the *y*-direction |
| WIDTH1, WIDTH2 | Width of appendages 1 and 2 respectively. |
| LENGTH1, LENGTH2 | Length of appendages 1 and 2 respectively. |
| MAX_PLATE_DEF | Maximum appendage deflection (z-axis limit). |
| PLATE_DEF_VIEW_ANGLE | Appendage vertical elevation view angle (in degrees). |

## 3.4 Simulator Validation

The validation of the simulator was an important step in this thesis and was done in several steps:

(i) Test and validation of individual subroutines (e.g. inertia calculation subroutine) which implements the basic building blocks of the overall system;

(ii) Verification of the special cases where analytical solutions may be obtained for the overall spacecraft (both rigid and flexible case); and

(iii) Validation of the entire dynamic system based on the following criteria
   a. Verification of the spacecraft's conservation of energy; and
   b. Verification of the spacecraft's conservation of angular momentum.

Step (i) above was undertaken as the last step of the development stage of individual simulation models implemented in C-language subroutines. Every C-language subroutine was debugged for syntax and logic and then finally the mathematical model was verified individually using appropriate test cases, very similar to what was done for the overall system in step (ii), as will be detailed in the next two sections.

The important quaternion unitary condition was also verified for every simulation run and it was found to hold within acceptable numerical accuracy of about 1e-12 to 1e-9. This result was achieved without the need of a quaternion normalization step during the numerical integration.

The integration techniques and their relevant parameters used for the simulator are shown in Table 3.2. Note that the variable-step Adams-Bashforth-Moulton (ABM) solver is preferred over the fixed-step Runge-Kutta (RK) solver as it may be more efficient if using stringent error tolerances for solving our computationally intensive set of Ordinary Differential Equations (ODEs). In this thesis, most simulation cases where executed using the numerical parameters listed in Table 3.2. However, for two of the validation cases (i.e. cases 7 and 8), a much lower relative tolerance of $1\times10^{-12}$ had to be used as will be explained later.

Table 3.2: Integration techniques and common settings used in the simulator

| | |
|---|---|
| *Solver*: | Fixed-Step, ODE4 (RK) |
| *Fixed Step Size*: | 0.01 |
| *Mode*: | auto |
| *Solver*: | Variable-Step, ODE113 (ABM) |
| *Max Step Size*: | 0.01 |
| *Min Step Size*: | 0.0001 |
| *Initial Step Size*: | 0.01 |
| *Relative Tolerance*: | 1e-5 |
| *Absolute Tolerance*: | auto |
| *Refine Output Factor*: | 1 |

For all the test cases studied, the errors were computed as follows

$$\text{Absolute Error} = \text{Expected Value - Simulated Value}$$
$$\text{Relative Error (\%)} = \frac{\text{Expected Value - Simulated Value}}{\text{Expected Value}}\times 100$$

where the expected values are obtained using the analytical solution for each case and the simulated values are the results obtained using the spacecraft simulator. Note that in the discussions in this chapter, the words *expected* and *analytical* results or value are interchangeable. In the next two subsections, the results of the validation test cases are detailed for both rigid and flexible spacecrafts.

### 3.4.1 Rigid Spacecraft Validation

All test cases for the rigid spacecraft have been validated using both the RK and ABM integration techniques, however, only the results obtained from the more accurate ABM solver are presented here. The test cases were validated for 10000 seconds of simulation which is much longer than the attitude recovery maneuver time considered in this thesis (i.e. 600 seconds as per our performance requirements of Section 1.3). However, for cases 7 and 8, only 1000 seconds of simulation time was used since these two cases involve very large angular speeds even at 1000 seconds.

The *real* simulation execution time was determined using the Matlab *tic* and *toc* commands. The relationship between the simulation time and real execution time was determined to be polynomial for all cases, as shown in Table 3.3 and Figure 3.6 for validation case 4.

Table 3.3: Results for simulation and execution times

| Simulation Time (sec) | Execution Time (sec) |
|---|---|
| 10 | 0.021 |
| 100 | 0.17 |
| 1000 | 4.146 |
| 2000 | 14.24 |
| 3000 | 30.504 |
| 4000 | 52.527 |
| 5000 | 82.49 |
| 6000 | 116.179 |
| 7000 | 155.637 |
| 8000 | 202.416 |
| 9000 | 260.87 |
| 10000 | 312.396 |

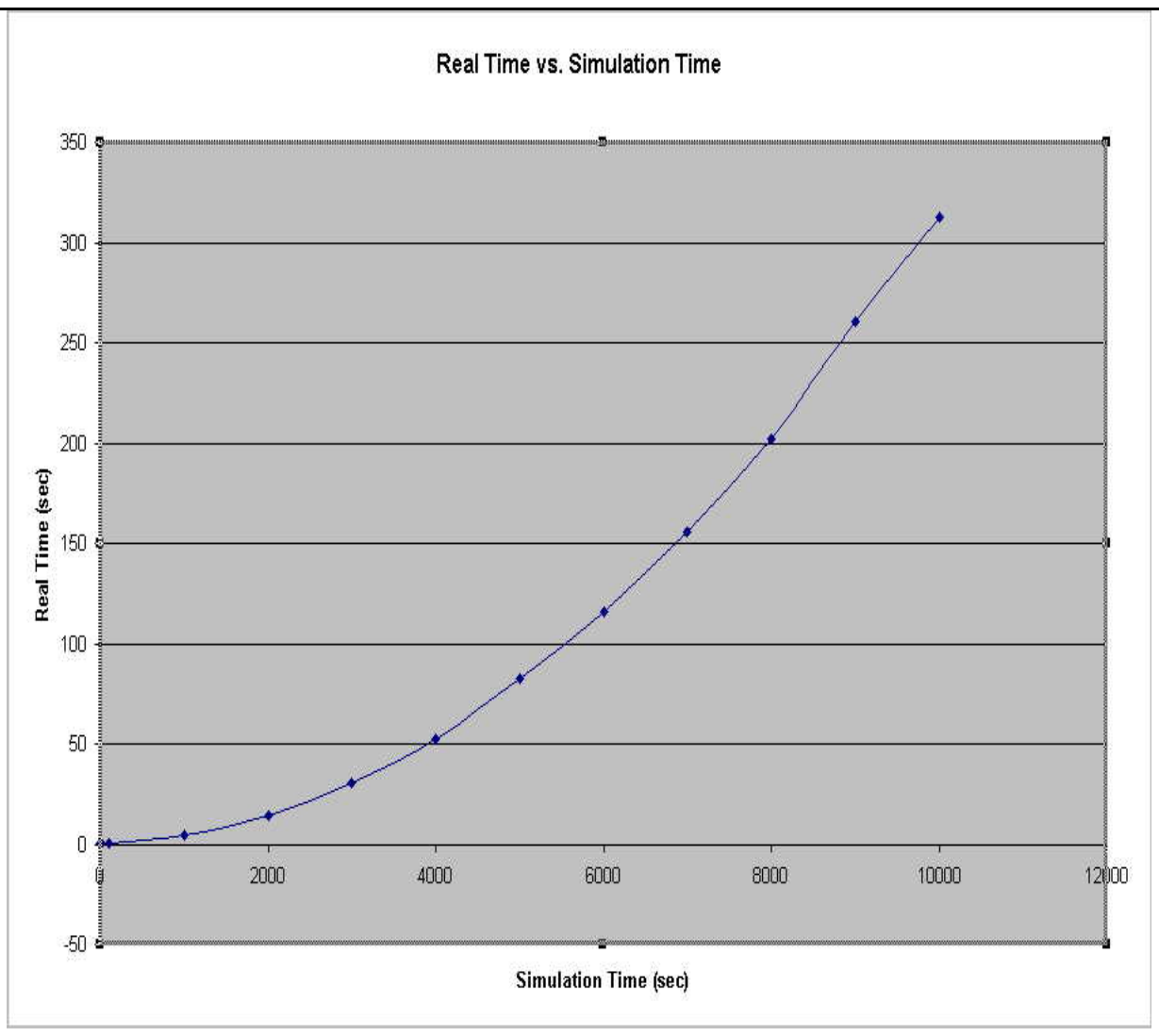

Figure 3.6: Polynomial relationship between simulation and real execution time

The analytical solutions for the validation cases presented in this subsection are based on the rigid attitude equation (2.9). Considering the case where the spacecraft reference frame is aligned with the spacecraft principal axes (i.e. the products of inertia are zero), then (2.9) can be written as the following three scalar, nonlinear and coupled equations, known as the Euler's equations of motion:

$$\begin{aligned}
\tau_x &= \dot{\omega}_x I_x + \omega_y \omega_z \left( I_z - I_y \right) \\
\tau_y &= \dot{\omega}_y I_y + \omega_x \omega_z \left( I_x - I_z \right) \\
\tau_z &= \dot{\omega}_z I_z + \omega_x \omega_y \left( I_y - I_x \right)
\end{aligned} \tag{3.1}$$

where *x*, *y*, and *z* represent the principal axes of inertia.

Although the general attitude motion of a rigid body may be modeled by the Euler's equations, they have no explicit solution unless we specify the components of $\vec{\tau}$. As was mentioned, the equations are nonlinear and coupled where the coupling effects are most present when we are dealing with large angular rates $\vec{\omega}$. In order to validate the numerical simulation of the rigid attitude motion, one needs to come up with analytical solutions of (3.1) for special cases. This involves the judicious selection of the parameters (e.g. inertia) to eliminate the nonlinearities in (3.1) and obtain a set of linear differential equations which can be more easily solved analytically and then used to verify the correctness of the results obtained from the simulator. This is done below through the following nine (9) validation case studies. Table 3.4 summarizes the specifications corresponding to these cases:

Table 3.4: Validation case study specifications for rigid spacecraft

| Case No. | Inertia Terms $\underline{I}_t = \begin{bmatrix} I_x & 0 & 0 \\ 0 & I_y & 0 \\ 0 & 0 & I_z \end{bmatrix}$ | Initial Angular Rate $\vec{\omega}_o = \begin{bmatrix} (\omega_o)_x \\ (\omega_o)_y \\ (\omega_o)_z \end{bmatrix}$ | Torque $\vec{\tau} = \begin{bmatrix} \tau_x \\ \tau_y \\ \tau_z \end{bmatrix}$ |
|---|---|---|---|
| 1 | $I_x = I_y = I_z = 0$ | $(\omega_o)_x = (\omega_o)_y = (\omega_o)_z = 0$ | $\tau_x = \tau_y = \tau_z = 0$ |
| 2 | $I_x = I_y = I_z > 0$ | $(\omega_o)_x, (\omega_o)_y, (\omega_o)_z \neq 0$ | $\tau_x = \tau_y = \tau_z = 0$ |
| 3 | $I_x \neq (I_y = I_z) > 0$ | $(\omega_o)_x, (\omega_o)_y, (\omega_o)_z \neq 0$ | $\tau_x = \tau_y = \tau_z = 0$ |
| 4 | $I_y \neq (I_x = I_z) > 0$ | $(\omega_o)_x, (\omega_o)_y, (\omega_o)_z \neq 0$ | $\tau_y = 0; \tau_x, \tau_z \neq 0$ |
| 5 | $I_y \neq (I_x = I_z) > 0$ | $(\omega_o)_x, (\omega_o)_y, (\omega_o)_z \neq 0$ | $\tau_y = 0; \tau_x = \cos(\gamma t); \tau_z = \sin(\gamma t); \gamma \neq 0$ |
| 6 | $I_x \neq (I_y = I_z) > 0$ | $(\omega_o)_x, (\omega_o)_y, (\omega_o)_z \neq 0$ | $\tau_x = \tau_z = 0; \tau_y = m[u(t-t_1) - u(t-t_2)]$<br>$m \neq 0$ , $t_2 > t_1 > 0$<br>$u(t-t_i)$ is the unit step function at time $t = t_i$ |
| 7 | $I_y \neq (I_x = I_z) > 0$ | $(\omega_o)_x, (\omega_o)_y, (\omega_o)_z \neq 0$ | $\tau_x = \tau_z = 0; \tau_y \neq 0$ |
| 8 | $I_y \neq (I_x = I_z) > 0$ | $(\omega_o)_x, (\omega_o)_y, (\omega_o)_z \neq 0$ | $\tau_x = \tau_z = 0; \tau_y = m\cos(\gamma t) \qquad m, \gamma \neq 0$ |
| 9 | $I_y \neq (I_x = I_z) > 0$ | $(\omega_o)_x, (\omega_o)_y, (\omega_o)_z \neq 0$ | $\tau_x = \tau_z = 0; \tau_y = m[u(t-t_1) - u(t-t_2)]$;<br>$m \neq 0$ , $t_2 > t_1 > 0$, |

Test case 1 is the all zero case, that is, it verifies that the code is well behaved (e.g. no division by zero). Test cases 2 to 6 correspond to Linear Time Invariant (LTI) models, whereas test cases 7 to 9 correspond to Linear Time Varying (LTV) models, as detailed in Table 3.4. It is important to note that any real values chosen for the parameters in Table 3.4 should be reasonably small to be physically meaningful. For example, an initial $y$-axis angular rate, $(\omega_o)_y$ , of 1000 rad/s, is mathematically possible but physically highly unlikely and hence if selected, would require a very small time step and relative tolerance for a convergent numerical integration (making the simulation runtime extremely long).

<u>Validation Case 1</u>

With zero torque, zero inertia and zero initial angular velocity, the angular body rates of the rigid spacecraft are expected to remain constant at zero, and this is precisely what was observed from the simulation results. This is the all zero test case, which insures that no mistakes have been made in coding the rigid dynamics equations (e.g. no division by zeros are present).

<u>Validation Case 2</u>

With zero torque, inertia matrix equal to a multiple of the identity matrix, and any initial angular rate, the angular body rates of the rigid spacecraft are expected to remain constant at their initial values, and again, this is precisely what was observed from the simulation results.

Validation Case 3

Following the specifications outlined in Table 3.4, let us set $\tau_x = \tau_y = \tau_z = 0$ and $I_y = I_z$, now (3.1) becomes

$$\dot{\omega}_x I_x = 0 \tag{3.2}$$

$$\dot{\omega}_y I_y + \omega_x \omega_z \left(I_x - I_y\right) = 0 \tag{3.3}$$

$$\dot{\omega}_z I_y + \omega_x \omega_y \left(I_y - I_x\right) = 0 \tag{3.4}$$

From (3.2), we can solve for $\omega_x$ as follows:

$$\dot{\omega}_x = 0 \quad \Rightarrow \quad \omega_x = (\omega_o)_x \tag{3.5}$$

Equations (3.3) and (3.4) can be re-written as

$$\dot{\omega}_y = -\frac{I_x - I_y}{I_y}(\omega_o)_x \omega_z \tag{3.6}$$

$$\dot{\omega}_z = -\frac{I_y - I_x}{I_y}(\omega_o)_x \omega_y \tag{3.7}$$

or in a compact matrix form as

$$\begin{bmatrix} \dot{\omega}_y \\ \dot{\omega}_z \end{bmatrix} = \begin{bmatrix} 0 & -a \\ a & 0 \end{bmatrix} \begin{bmatrix} \omega_y \\ \omega_z \end{bmatrix} \tag{3.8}$$

where $a = \frac{I_x - I_y}{I_y}(\omega_o)_x$ .

Equation (3.8) is a linear time invariant homogeneous system of the form

$$\dot{\vec{x}}(t) = \underline{A}\vec{x}(t) \tag{3.9}$$

where $\underline{A} = \begin{bmatrix} 0 & -a \\ a & 0 \end{bmatrix}$ and $\vec{x}(t) = \begin{bmatrix} \omega_y(t) & \omega_z(t) \end{bmatrix}^T$.

The solution to the homogeneous system (3.9) is given by

$$\vec{x}(t) = e^{\underline{A}t}\vec{x}(t_0) \tag{3.10}$$

and for our particular case, we obtain:

$$\begin{bmatrix} \omega_y(t) \\ \omega_z(t) \end{bmatrix} = e^{\underline{A}t} \begin{bmatrix} (\omega_o)_y \\ (\omega_o)_z \end{bmatrix} = \begin{bmatrix} \cos(at) & -\sin(at) \\ \sin(at) & \cos(at) \end{bmatrix} \begin{bmatrix} (\omega_o)_y \\ (\omega_o)_z \end{bmatrix} \tag{3.11}$$

Hence, the following equations provide the analytical solution for case 3:

$$\boxed{\begin{aligned} &\omega_x(t) = (\omega_o)_x \\ &\omega_y(t) = (\omega_o)_y \cos(at) - (\omega_o)_z \sin(at) \\ &\omega_z(t) = (\omega_o)_y \sin(at) + (\omega_o)_z \cos(at) \end{aligned}} \tag{3.12}$$

For a numerical example, we select inertia terms $I_y = I_z = 10 \text{ kg m}^2$ and $I_x = 5 \text{ kg m}^2$, and an initial angular velocity $\vec{\omega}_o^T = \begin{bmatrix} 0.3 & -0.4 & 0.5 \end{bmatrix}$. After running the simulation case, the results were analyzed and plotted. As expected, the output results for $\omega_x$ (not shown) is constant and equal to $(\omega_o)_x$. For $\omega_y$ and $\omega_z$, the results are similar, and as an example, Figure 3.7 shows the time evolution of the simulated and expected *y*-axis angular rate, $\omega_y$, which depicts visually that the results match perfectly for at least the first 1000 sec. However, as expected with any numerical simulation, the simulated results do diverge slowly from the expected analytical solution as can be easily noticed from Figure 3.8 where the simulation is allowed to continue running up to 10,000 seconds. The absolute errors are sinusoidal functions (because the solution of the angular rates, $\omega_y$ and $\omega_z$, are sinusoidal) and their worst case amplitude increases steadily at about $2.1 \times 10^{-9}$ rad/sec per second after being bounded for over 2000 seconds. This error is mainly due to numerical round-off and truncation errors which are accumulative and increase as time evolves. By conducting several simulation cases with different initial conditions, it was concluded that the error is heavily dependent on the magnitude of the initial conditions. As the initial conditions get larger, so does the rate of increase of the amplitude of the sinusoidal error, while keeping the integration time step the same. The expected and simulated numerical results along with absolute and relative errors, for 4 specific times (i.e. $t = 10,100,1000$ and $10000$ ) are presented in Table 3.5.

Table 3.5: Validation case 3 - Expected and simulated angular velocity values

| Sim. Time (sec) | $\omega_y$ (rad/sec) | | | | $\omega_z$ (rad/sec) | | | |
|---|---|---|---|---|---|---|---|---|
| | Expected | Simulated | Abs. Error | Rel. Error (%) | Expected | Simulated | Abs. Error | Rel. Error (%) |
| 10 | 0. 47045261263495 | 0. 47045261287925 | -2.44E-10 | -5.19E-08 | 0. 43436659547547 | 0. 43436659521079 | 2.65E-10 | 6.09E-08 |
| 100 | 0.62901908522209 | 0.62901908515495 | 6.71E-11 | 1.07E-08 | -0.11972882036656 | -0.11972882071903 | 3.52E-10 | -2.94E-07 |
| 1000 | -0.63713853740593 | -0.63713853736284 | -4.31E-11 | 6.76E-09 | 0.06367483138752 | 0.06367483181896 | -4.31E-10 | -6.78E-07 |
| 10000 | -0.45284401744784 | -0.45284400603377 | -1.14E-08 | 2.52E-06 | -0.45269448401953 | -0.45269449543857 | 1.14E-08 | -2.52E-06 |

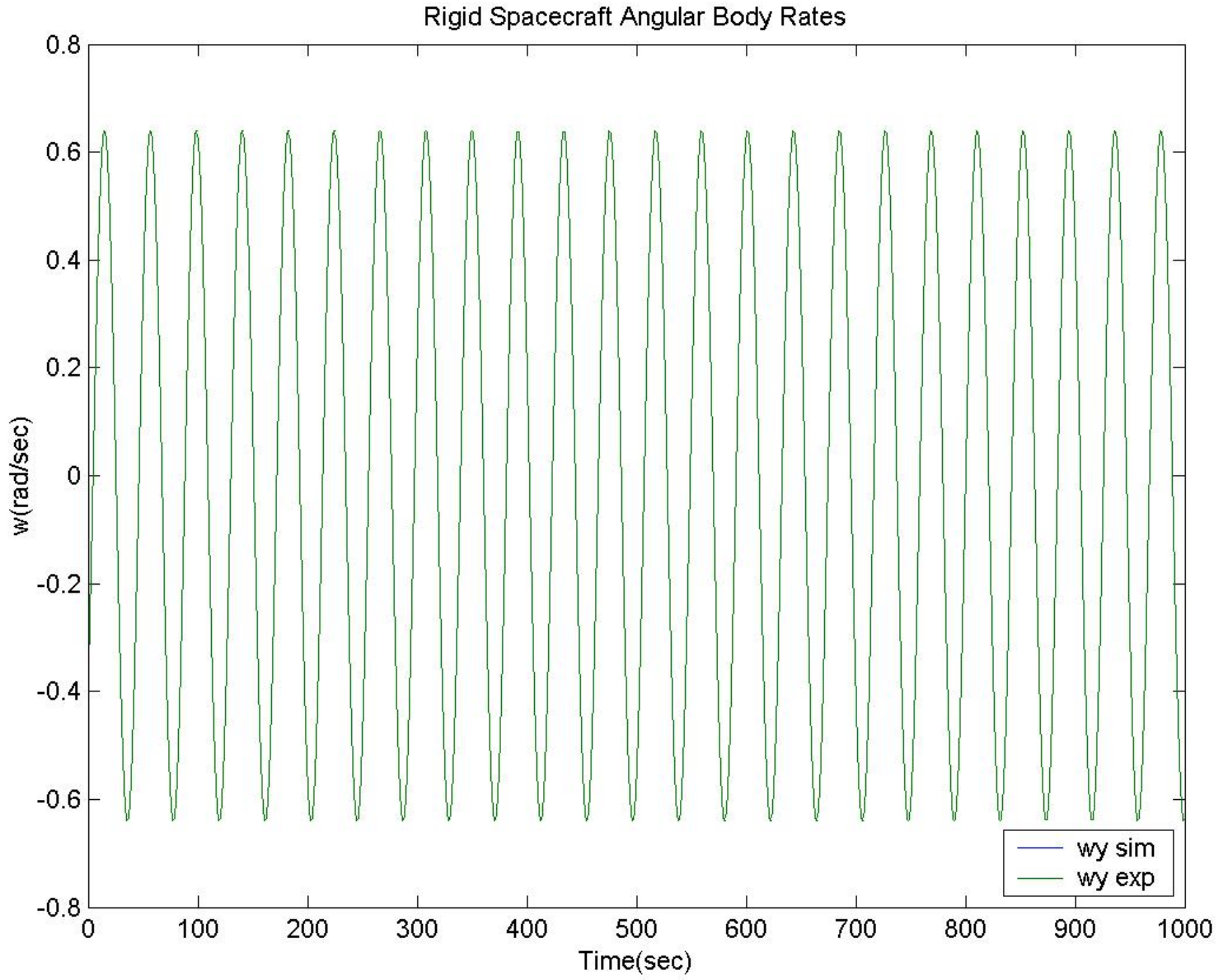

Figure 3.7: Validation case 3 – Expected and simulated *y*-axis angular body rate

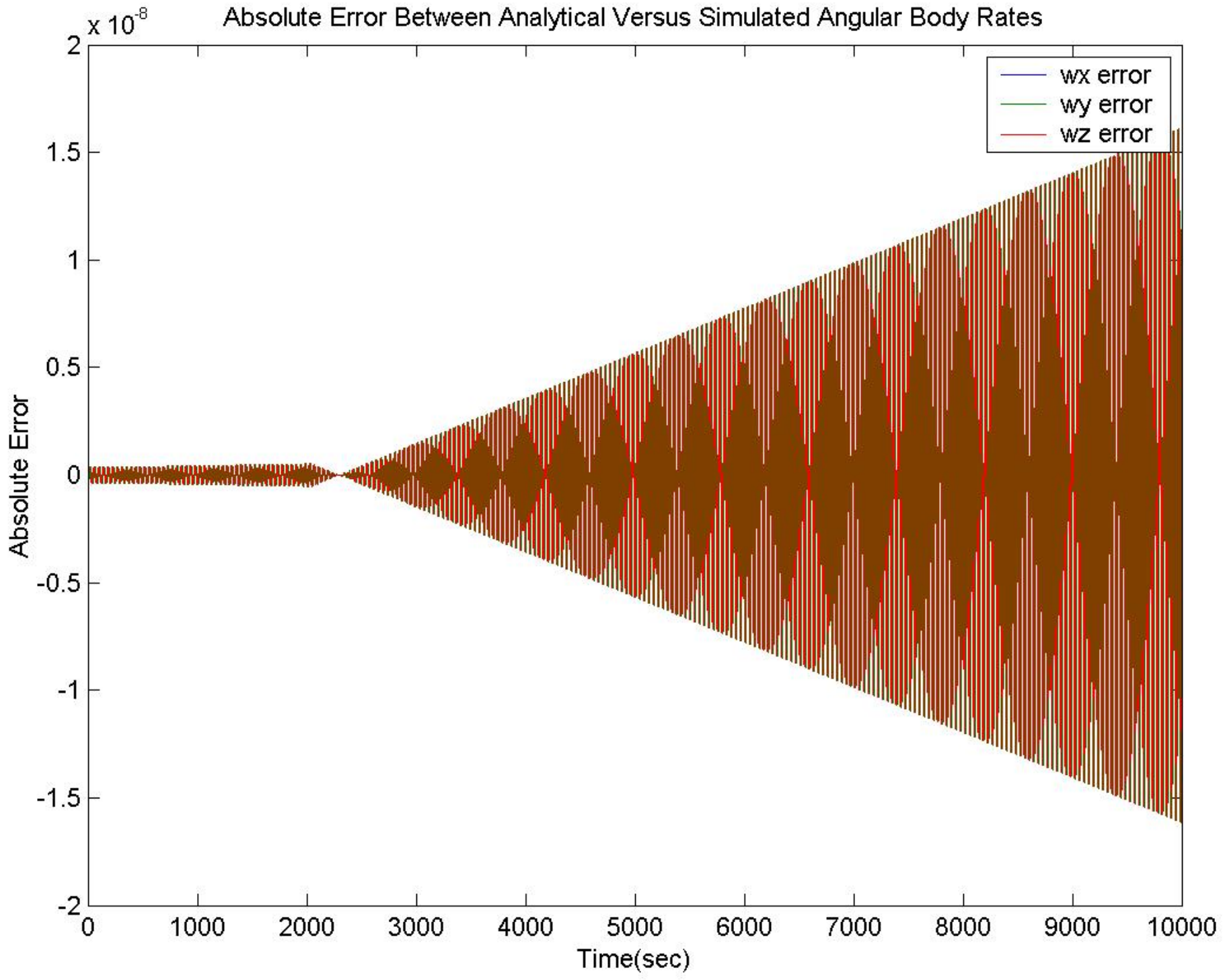

Figure 3.8: Validation case 3 - Absolute error between expected and simulated angular body rates

Validation Case 4

As per Table 3.4, let's set $\tau_x = c_x, \tau_z = c_z$ and $I_x = I_z$, and (3.1) can now be simplified to

$$\dot{\omega}_x I_x + \omega_y \omega_z \left(I_x - I_y\right) = c_x \tag{3.13}$$

$$\dot{\omega}_y I_y = 0 \tag{3.14}$$

$$\dot{\omega}_z I_z + \omega_x \omega_y \left(I_y - I_x\right) = c_z \tag{3.15}$$

From (3.14), we can solve for $\omega_y$ as follows:

$$\dot{\omega}_y = 0 \quad \Rightarrow \quad \omega_y = (\omega_o)_y \tag{3.16}$$

Equations (3.13) and (3.15) can be re-written as

$$\dot{\omega}_x = -\frac{I_x - I_y}{I_x}(\omega_o)_y \omega_z + \frac{c_x}{I_x}$$
$$\dot{\omega}_z = -\frac{I_y - I_x}{I_x}(\omega_o)_y \omega_x + \frac{c_z}{I_x} \tag{3.17}$$

or in a compact matrix form as

$$\begin{bmatrix} \dot{\omega}_x \\ \dot{\omega}_z \end{bmatrix} = \begin{bmatrix} 0 & -a \\ a & 0 \end{bmatrix} \begin{bmatrix} \omega_x \\ \omega_z \end{bmatrix} + \frac{1}{I_x} \begin{bmatrix} c_x \\ c_z \end{bmatrix} \tag{3.18}$$

where $a = \dfrac{I_x - I_y}{I_x}(\omega_o)_y$ .

Equation (3.18) is a non-homogeneous system of the form

$$\dot{\vec{x}}(t) = \underline{A}\vec{x}(t) + \vec{f}(t) \tag{3.19}$$

where $\underline{A} = \begin{bmatrix} 0 & -a \\ a & 0 \end{bmatrix}$, $\vec{f}(t) = \dfrac{1}{I_x}\begin{bmatrix} c_x \\ c_z \end{bmatrix}$ and $\vec{x}(t) = \left[\omega_x(t) \quad \omega_z(t)\right]^T$ .

The solution to the non-homogeneous system (3.19) is given by

$$\vec{x}(t) = e^{\underline{A}t}\vec{x}(t_0) + \int_0^t e^{\underline{A}(t-T)} \vec{f}(T)\mathrm{d}T \tag{3.20}$$

which for our system results in the following expression

$$\vec{x}(t) = e^{\underline{A}t}\vec{x}(t_0) + \int_0^t e^{\underline{A}(t-T)}\vec{f}(T)\mathrm{d}T$$

$$= \begin{bmatrix} \cos(at)(\omega_o)_x - \sin(at)(\omega_o)_z + \dfrac{-c_z + c_x \sin(at) + c_z \cos(at)}{I_x a} \\ \sin(at)(\omega_o)_x + \cos(at)(\omega_o)_z - \dfrac{-c_x + c_x \cos(at) - c_z \sin(at)}{I_x a} \end{bmatrix} \quad (3.21)$$

The equations below form the basis for our analytical solution in case 4:

$$\boxed{\begin{aligned} \omega_x(t) &= \cos(at)(\omega_o)_x - \sin(at)(\omega_o)_z + \frac{-c_z + c_x \sin(at) + c_z \cos(at)}{I_x a} \\ \omega_y(t) &= (\omega_o)_y \\ \omega_z(t) &= \sin(at)(\omega_o)_x + \cos(at)(\omega_o)_z + \frac{c_x - c_x \cos(at) + c_z \sin(at)}{I_x a} \end{aligned}} \quad (3.22)$$

For a specific example, we set $c_x = 0.1$ Nm and $c_z = -0.1$ Nm for the external torque, inertia terms are selected to be $I_x = I_z = 10\ \mathrm{kg\,m^2}$ and $I_y = 5\ \mathrm{kg\,m^2}$, and an initial angular velocity $\vec{\omega}_o^T = [0.3 \quad -0.4 \quad 0.5]$ rad/sec is chosen. As expected, the simulated result for $\omega_y$ is a constant matching $(\omega_o)_y$. For $\omega_x$ and $\omega_z$, the simulated results match the analytical solution very well, however as expected and explained in case 3, the simulated results do diverge ever so slowly from the expected analytical solution as shown in Figure 3.9. The expected and simulated numerical results along with absolute and relative errors, for the 4 specific times (i.e. $t = 10, 100, 1000$ and $10000$) are presented in Table 3.6.

Table 3.6: Validation case 4 - Expected and simulated angular velocity values

| Sim. Time (sec) | $\omega_x$ (rad/sec) | | | | $\omega_z$ (rad/sec) | | | |
|---|---|---|---|---|---|---|---|---|
| | Analytical | Simulated | Abs. Error | Rel. Error (%) | Analytical | Simulated | Abs. Error | Rel. Error (%) |
| 10 | 0. 30446219196263 | 0. 30446219123299 | 7.3E-10 | 2.40E-07 | -0. 59713485948992 | -0. 59713485996239 | 4.72E-10 | -7.91E-08 |
| 100 | 0.59494860953488 | 0.59494860940820 | 1.27E-10 | 2.13E-08 | -0.14508570375730 | -0.14508570461538 | 8.58E-10 | -5.91E-07 |
| 1000 | -0.35979782721525 | -0.35979782636251 | -8.5E-10 | 2.37E-07 | 0.52360727527875 | 0.52360727573912 | -4.6E-10 | -8.79E-08 |
| 10000 | 0.33291088524358 | 0.33291090267604 | -1.7E-08 | -5.24E-06 | -0.57761657855111 | -0.57761656589978 | -1.3E-08 | 2.19E-06 |

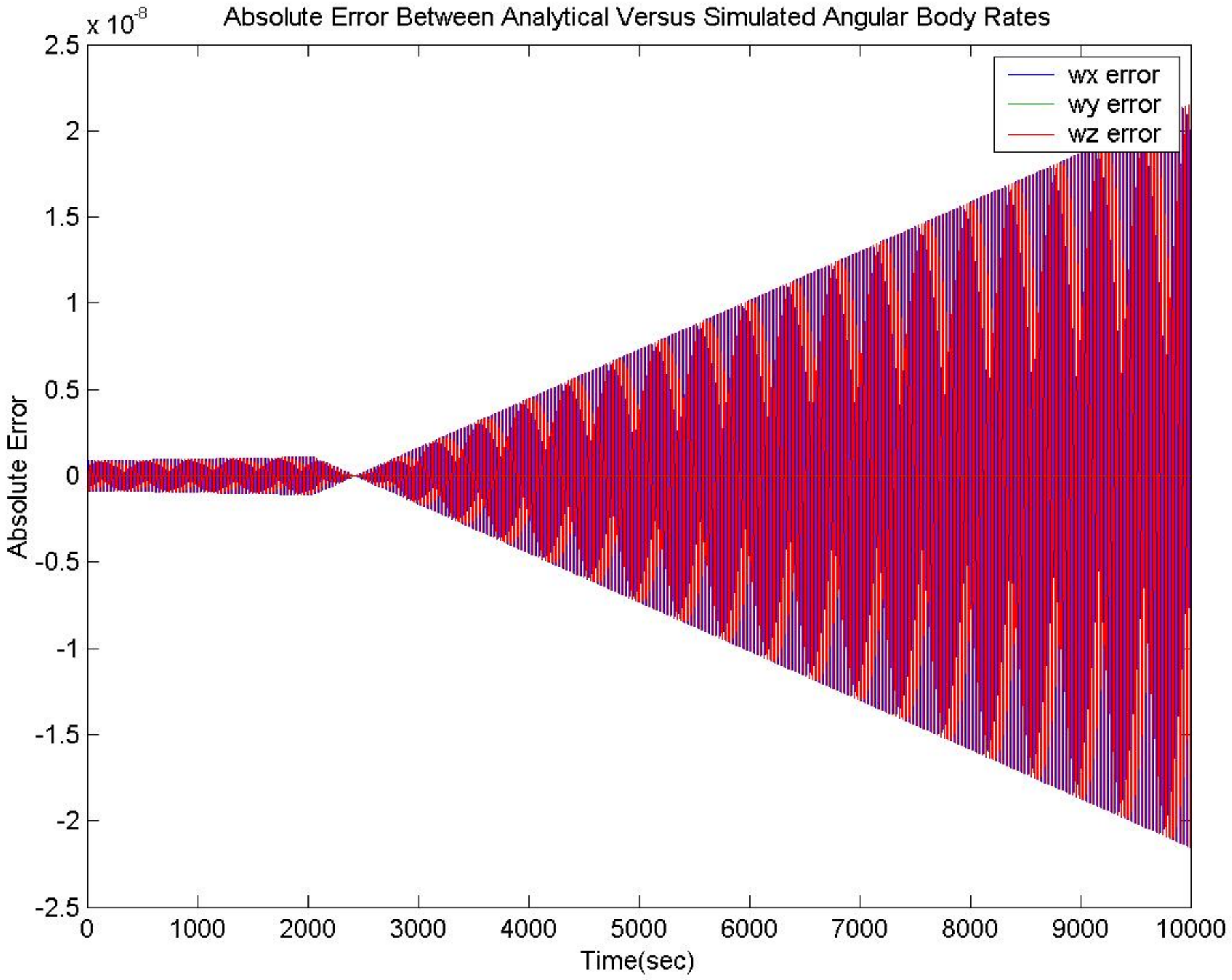

Figure 3.9: Validation case 4 - Absolute error between expected and simulated angular body rates

Validation Case 5

As per Table 3.4, by setting $\tau_x = \cos(\gamma t), \tau_z = \sin(\gamma t)$ and $I_x = I_z$, the Euler's equations (3.1) can be simplified to get

$$\dot{\omega}_x I_x + \omega_y \omega_z \left( I_x - I_y \right) = \cos(\gamma t) \tag{3.23}$$

$$\dot{\omega}_y I_y = 0 \tag{3.24}$$

$$\dot{\omega}_z I_z + \omega_x \omega_y \left( I_y - I_x \right) = \sin(\gamma t) \tag{3.25}$$

From (3.24), we can solve for $\omega_y = (\omega_o)_y$ and equations (3.23) and (3.25) can be re-written as

$$\dot{\omega}_x = -\frac{I_x - I_y}{I_x} (\omega_o)_y \omega_z + \frac{\cos(\gamma t)}{I_x} \tag{3.26}$$

$$\dot{\omega}_z = -\frac{I_y - I_x}{I_x}(\omega_o)_y \omega_x + \frac{\sin(\gamma t)}{I_x} \tag{3.27}$$

or in a compact matrix form as

$$\begin{bmatrix} \dot{\omega}_x \\ \dot{\omega}_z \end{bmatrix} = \begin{bmatrix} 0 & -a \\ a & 0 \end{bmatrix} \begin{bmatrix} \omega_x \\ \omega_z \end{bmatrix} + \frac{1}{I_x} \begin{bmatrix} \cos(\gamma t) \\ \sin(\gamma t) \end{bmatrix} \tag{3.28}$$

where $a = \frac{I_x - I_y}{I_x}(\omega_o)_y$.

Equation (3.28) is a non-homogeneous system of the form (3.19) where $\underline{A} = \begin{bmatrix} 0 & -a \\ a & 0 \end{bmatrix}$,

$\vec{f}(t) = \frac{1}{I_x}\begin{bmatrix} \cos(\gamma t) \\ \sin(\gamma t) \end{bmatrix}$ and $\vec{x}(t) = \begin{bmatrix} \omega_x(t) & \omega_z(t) \end{bmatrix}^T$.

Using (3.20), we can solve for $\vec{x}(t)$:

$$\vec{x}(t) = e^{\underline{A}t}\vec{x}(t_0) + \int_0^t e^{\underline{A}(t-T)}\vec{f}(T)\mathrm{d}T = \begin{bmatrix} \cos(at)(\omega_o)_x - \sin(at)(\omega_o)_z + \dfrac{\sin(\gamma t) - \sin(at)}{(-a+\gamma)I_x} \\ \sin(at)(\omega_o)_x + \cos(at)(\omega_o)_z - \dfrac{-\cos(\gamma t) + \cos(at)}{(-a+\gamma)I_x} \end{bmatrix}$$

The equations below form the basis for our analytical solution in case 5:

$$\boxed{\begin{aligned} \omega_x(t) &= (\cos at)(\omega_o)_x - (\sin at)(\omega_o)_z + \frac{\sin\gamma t - \sin at}{(\gamma - a)I_x} \\ \omega_y(t) &= (\omega_o)_y \\ \omega_z(t) &= (\sin at)(\omega_o)_x + (\cos at)(\omega_o)_z - \frac{\cos\gamma t - \cos at}{(\gamma - a)I_x} \end{aligned}} \tag{3.29}$$

For a specific example, we set $\gamma = \frac{2\pi}{5400}$ representing a typical frequency of a sinusoidal disturbance torque in low Earth orbit where the orbital period is about 90 minutes or 5400 sec., inertia terms are selected to be $I_x = I_z = 10\ \text{kg}\,\text{m}^2$ and $I_y = 5\ \text{kg}\,\text{m}^2$, and an initial angular velocity $\vec{\omega}_o^T = \begin{bmatrix} 0.3 & -0.4 & 0.5 \end{bmatrix}$ rad/sec is chosen. As expected, the simulated result for $\omega_y$ is a constant matching $(\omega_o)_y$ and for $\omega_x$ and $\omega_z$, the absolute errors are shown in Figure 3.10 and the error analysis is the same as for cases 3 and 4, discussion of which will not be repeated here. The expected and simulated numerical results along with absolute and relative errors, for 4 specific times, are presented in Table 3.7.

Table 3.7: Validation case 5 - Expected and simulated angular velocity values

| Sim. Time (sec) | $\omega_x$ (rad/sec) | | | | $\omega_z$ (rad/sec) | | | |
|---|---|---|---|---|---|---|---|---|
| | Analytical | Simulated | Abs. Error | Rel. Error (%) | Analytical | Simulated | Abs. Error | Rel. Error (%) |
| 1 | 0.49269315238782 | 0.49269315361116 | -1.223E-9 | -2.48E-07 | 0.42052376369752 | 0.42052376304112 | 6.564E-10 | 1.56E-07 |
| 100 | 1.09044028953033 | 1.09044028970703 | -1.767E-10 | -1.62E-08 | -0.36072838704151 | -0.36072838841553 | 1.374E-09 | -3.81E-07 |
| 1000 | -0.26816284901991 | -0.26816284790882 | -1.111E-09 | 4.14E-07 | 0.55087349190557 | 0.55087349298244 | -1.077E-09 | -1.95E-07 |
| 10000 | 0.41837010207166 | 0.41837012340459 | -2.133E-08 | -5.10E-06 | -0.94226096584060 | -0.94226093883959 | -2.700E-08 | 2.87E-06 |

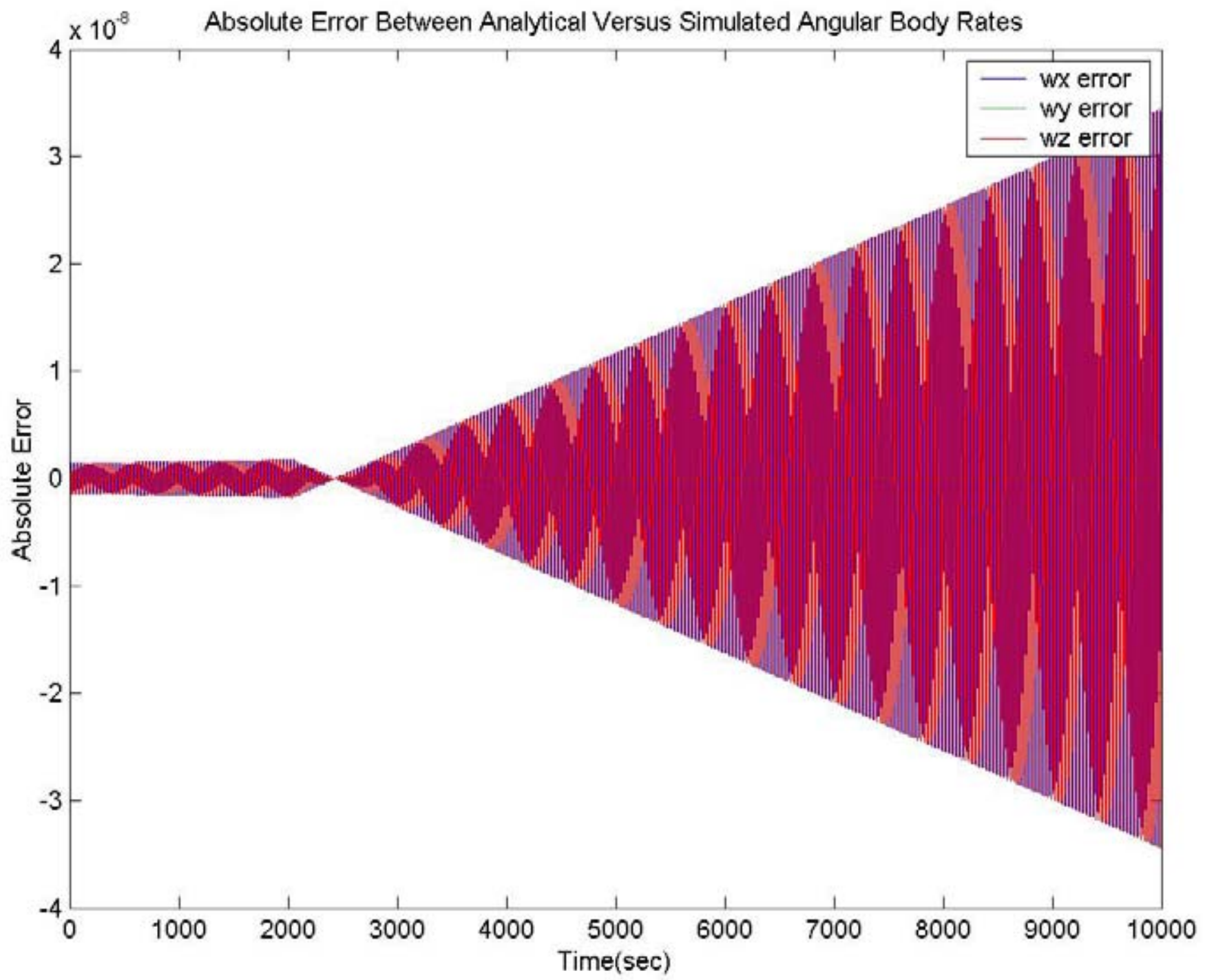

Figure 3.10: Validation case 5 - Absolute error between expected and simulated angular body rates

Validation Case 6

As per Table 3.4, let us set $\tau_y = m[u(t-t_1)-u(t-t_2)]$ which is a step function of magnitude $m$ that lasts $t_2 - t_1$ seconds and select $I_y = I_z$. Equation (3.1) can now be simplified to

$$\dot{\omega}_x I_x = 0 \tag{3.30}$$

$$\dot{\omega}_y I_y + \omega_z \omega_x \left(I_x - I_y\right) = \tau_y \tag{3.31}$$

$$\dot{\omega}_z I_y + \omega_x \omega_y \left(I_y - I_x\right) = 0 \tag{3.32}$$

From (3.30), we can solve for $\omega_x = (\omega_o)_x$ and equations (3.31) and (3.32) can be re-written as

$$\dot{\omega}_y = -\frac{I_x - I_y}{I_y}(\omega_o)_x \omega_z + \frac{\tau_y}{I_y} \tag{3.33}$$

$$\dot{\omega}_z = -\frac{I_y - I_x}{I_y}(\omega_o)_x \omega_y \tag{3.34}$$

or in a compact matrix form as

$$\begin{bmatrix} \dot{\omega}_y \\ \dot{\omega}_z \end{bmatrix} = \begin{bmatrix} 0 & -a \\ a & 0 \end{bmatrix} \begin{bmatrix} \omega_y \\ \omega_z \end{bmatrix} + \frac{1}{I_y} \begin{bmatrix} \tau_y \\ 0 \end{bmatrix} \tag{3.35}$$

where $a = \dfrac{I_x - I_y}{I_y}(\omega_o)_x$.

Equation (3.35) is a non-homogeneous system of the form (3.19) where $\underline{A} = \begin{bmatrix} 0 & -a \\ a & 0 \end{bmatrix}$,

$\vec{f}(t) = \dfrac{1}{I_y}\begin{bmatrix} m[u(t-t_1)-u(t-t_2)] \\ 0 \end{bmatrix}$ and $\vec{x}(t) = \begin{bmatrix} \omega_y(t) & \omega_z(t) \end{bmatrix}^T$.

Using (3.20), we can solve for $\vec{x}(t)$:

$$\begin{bmatrix} \omega_y(t) \\ \omega_z(t) \end{bmatrix} = \begin{bmatrix} \cos(at) & -\sin(at) \\ \sin(at) & \cos(at) \end{bmatrix} \begin{bmatrix} (\omega_o)_y \\ (\omega_o)_z \end{bmatrix} + \frac{1}{I_y} \begin{bmatrix} \cos(at) & -\sin(at) \\ \sin(at) & \cos(at) \end{bmatrix} \int_0^t \begin{bmatrix} \cos(aT) & -\sin(aT) \\ \sin(aT) & \cos(aT) \end{bmatrix} \begin{bmatrix} m[u(T-t_1)-u(T-t_2)] \\ 0 \end{bmatrix} \mathrm{d}T \tag{3.36}$$

Equation (3.36) can be re-written as

$$\begin{bmatrix} \omega_y(t) \\ \omega_z(t) \end{bmatrix} = \begin{bmatrix} \cos(at) & -\sin(at) \\ \sin(at) & \cos(at) \end{bmatrix} \begin{bmatrix} (\omega_o)_y \\ (\omega_o)_z \end{bmatrix} + \begin{bmatrix} \cos(at) & -\sin(at) \\ \sin(at) & \cos(at) \end{bmatrix} \begin{bmatrix} U \\ V \end{bmatrix} \tag{3.37}$$

where

$$U = \frac{m}{aI_y} \begin{cases} 0 & t \le t_1 \\ \sin(at) - \sin(at_1) & t_1 \le t \le t_2 \\ \sin(at_2) - \sin(at_1) & t \ge t_2 \end{cases}$$

$$V = \frac{m}{aI_y} \begin{cases} 0 & t \le t_1 \\ \cos(at) - \cos(at_1) & t_1 \le t \le t_2 \\ \cos(at_2) - \cos(at_1) & t \ge t_2 \end{cases}$$

Simplifying further, we obtain the equations below which form the basis for our analytical solution in case 6:

$$\begin{aligned} &\omega_x(t) = (\omega_o)_x \\ &\omega_y(t) = (\omega_o)_y \cos(at) - (\omega_o)_z \sin(at) + \frac{m}{aI_y} \begin{cases} 0 & t \le t_1 \\ \sin\left(a(t-t_1)\right) & t_1 \le t \le t_2 \\ \sin\left(a(t-t_1)\right) - \sin\left(a(t-t_2)\right) & t \ge t_2 \end{cases} \\ &\omega_z(t) = (\omega_o)_y \sin(at) + (\omega_o)_z \cos(at) + \frac{m}{aI_y} \begin{cases} 0 & t \le t_1 \\ 1 - \cos\left(a(t-t_1)\right) & t_1 \le t \le t_2 \\ \cos\left(a(t-t_2)\right) - \cos\left(a(t-t_1)\right) & t \ge t_2 \end{cases} \end{aligned} \tag{3.38}$$

For a numerical example, we pick a large step of magnitude $m = 10\,N$ which lasts 1 second starting at $t_1 = 1$ to $t_2 = 2$ seconds, inertia terms are selected to be $I_y = I_z = 10\,\mathrm{kg\,m^2}$ and $I_x = 5\,\mathrm{kg\,m^2}$, and an initial angular velocity $\vec{\omega}_o^T = \begin{bmatrix} 0.3 & -0.4 & 0.5 \end{bmatrix}$ rad/sec is chosen. The absolute errors for $\omega_x$, $\omega_y$ and $\omega_z$ are shown in Figure 3.11 for 10000 seconds of simulation time. The expected and simulated numerical results along with absolute and relative errors, for several times are presented in Table 3.8.

Table 3.8: Validation case 6 - Expected and simulated angular velocity values

| Sim. Time (sec) | $\omega_y$ (rad/sec) | | | | $\omega_z$ (rad/sec) | | | |
|---|---|---|---|---|---|---|---|---|
| | Analytical | Simulated | Abs. Error | Rel. Err (%) | Analytical | Simulated | Abs. Error | Rel. Err (%) |
| 0.1 | -0.39245528209058 | -0.39245528180595 | -2.846E-10 | 7.25E-08 | 0.50594352605721 | 0.50594352627792 | -2.207E-10 | -4.36E-08 |
| 1 | -0.32078936493762 | -0.32078936462587 | -3.118E-10 | 9.72E-08 | 0.55416079195746 | 0.55416079213786 | -1.804E-10 | -3.26E-08 |
| 1.5 | 0.22116692005658 | 0.22116692139654 | -1.340E-09 | -6.06E-07 | 0.55789838580189 | 0.55789838591493 | -1.130E-10 | -2.03E-08 |
| 2 | 0.76187972417109 | 0.76187972551576 | -1.345E-09 | -1.76E-07 | 0.52101684680095 | 0.52101684681328 | -1.233E-11 | -2.37E-09 |
| 10 | 0.76168109169783 | 0.76168109202675 | -3.289E-10 | -4.32E-08 | -0.52130718708207 | -0.52130718857849 | 1.496E-09 | -2.87E-07 |
| 100 | 0.03412122187747 | 0.03412122087236 | 1.005E-09 | 2.95E-06 | -0.92236381703264 | -0.92236381818721 | 1.155E-09 | -1.25E-07 |
| 1000 | -0.11549581265334 | -0.11549581164722 | -1.006E-09 | 8.71E-07 | 0.91574013017506 | 0.91574013142755 | -1.252E-09 | -1.37E-07 |
| 10000 | -0.78176931159130 | -0.78176932462735 | 1.304E-08 | -1.67E-06 | 0.49066894359252 | 0.49066892492749 | 1.867E-08 | 3.80e-006 |

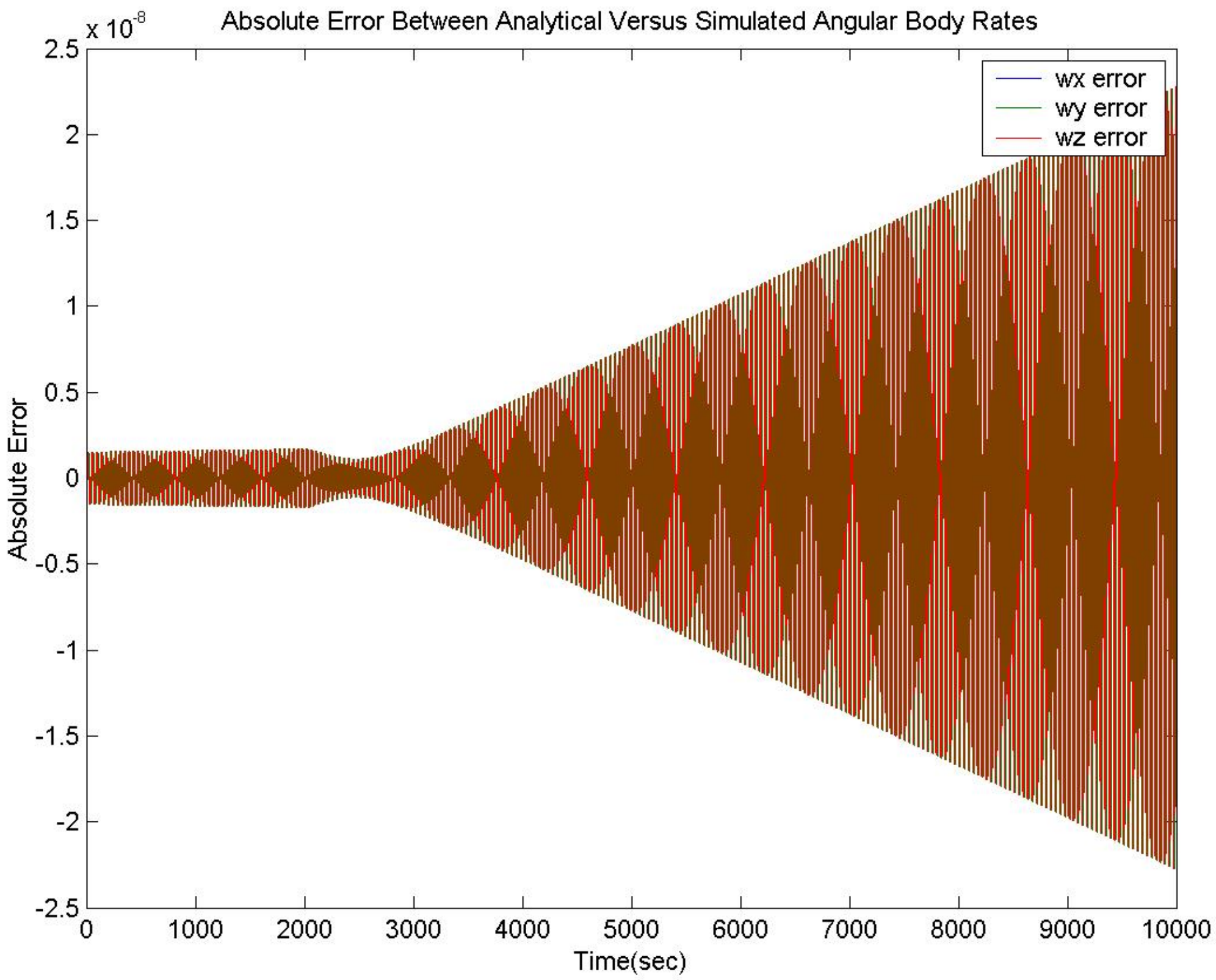

Figure 3.11: Validation case 6 - Absolute error between expected and simulated angular body rates

Validation Case 7

This is the first of the three linear time varying cases which will be presented. In all three cases, the input torque $\vec{\tau}^T = [0 \;\; \tau_y \;\; 0]$ and $I_x = I_z$ as per Table 3.4. For this case $\tau_y = c_1$ is taken to be a constant and hence equation (3.1) is simplified to

$$\dot{\omega}_x I_x + \omega_y \omega_z (I_x - I_y) = \tau_x \tag{3.39}$$

$$\dot{\omega}_y I_y = c_1 \tag{3.40}$$

$$\dot{\omega}_z I_x + \omega_x \omega_y \left(I_y - I_x\right) = \tau_z \tag{3.41}$$

From (3.40), we can solve for $\omega_y$ as follows:

$$\dot{\omega}_y = \frac{c_1}{I_y} \quad \Rightarrow \quad \omega_y = \frac{c_1}{I_y} t + (\omega_o)_y \tag{3.42}$$

Furthermore, (3.39) and (3.41) can be re-written as

$$\dot{\omega}_x = -\frac{I_x - I_y}{I_x}\left(\frac{c_1}{I_y} t + (\omega_o)_y\right)\omega_z + \frac{\tau_x}{I_x} \tag{3.43}$$

$$\dot{\omega}_z = -\frac{I_y - I_x}{I_y}\left(\frac{c_1}{I_y} t + (\omega_o)_y\right)\omega_x + \frac{\tau_z}{I_x} \tag{3.44}$$

or in a compact matrix for as

$$\begin{bmatrix} \dot{\omega}_x \\ \dot{\omega}_z \end{bmatrix} = \begin{bmatrix} 0 & -a\left(\frac{c_1}{I_y} t + (\omega_o)_y\right) \\ a\left(\frac{c_1}{I_y} t + (\omega_o)_y\right) & 0 \end{bmatrix} \begin{bmatrix} \omega_x \\ \omega_z \end{bmatrix} + \frac{1}{I_x}\begin{bmatrix} \tau_x \\ \tau_z \end{bmatrix} \tag{3.45}$$

where $a = \frac{I_x - I_y}{I_x}$ , and by defining $\alpha = a\frac{c_1}{I_y}$ and $\beta = a(\omega_o)_y$, (3.45) can be written as

$$\begin{bmatrix} \dot{\omega}_x \\ \dot{\omega}_z \end{bmatrix} = \begin{bmatrix} 0 & -(\alpha t+\beta) \\ (\alpha t+\beta) & 0 \end{bmatrix} \begin{bmatrix} \omega_x \\ \omega_z \end{bmatrix} + \frac{1}{I_x}\begin{bmatrix} \tau_x \\ \tau_z \end{bmatrix} \tag{3.46}$$

Equation (3.45) is a non-homogeneous linear time varying system of the form

$$\dot{\vec{x}}(t) = \underline{A}(t)\vec{x}(t) + \vec{f}(t) \tag{3.47}$$

where $\underline{A}(t) = \begin{bmatrix} 0 & -(\alpha t+\beta) \\ (\alpha t+\beta) & 0 \end{bmatrix} = \underline{M}_1 t + \underline{M}_2$ , $\underline{M}_1 = \begin{bmatrix} 0 & -\alpha \\ \alpha & 0 \end{bmatrix}$ and $\underline{\mathrm{M}}_2 = \begin{bmatrix} 0 & -\beta \\ \beta & 0 \end{bmatrix}$.

The solution to the non-homogeneous system (3.47) is given by

$$\vec{x}(t) = \underline{\Phi}(t)\vec{x}(t_0) + \int_0^t \underline{\Phi}(t-T)\vec{f}(T)\mathrm{d}T \tag{3.48}$$

where $\underline{\Phi}(t)$ is the state transition matrix and since $\underline{M}_1$ and $\underline{M}_2$ commute, that is $\underline{M}_1\underline{M}_2 = \underline{M}_2\underline{M}_1$, then $\underline{\Phi}(t)$ can be expressed as

$$\begin{aligned}
\underline{\Phi}(t) &= e^{\int_0^t \underline{M}_1 t\,\mathrm{d}t} e^{\int_0^t \underline{M}_2\,\mathrm{d}t} = e^{\underline{M}_1\frac{t^2}{2}} e^{\underline{M}_2 t} \\
&= \begin{bmatrix} \cos\left(\frac{1}{2}\alpha t^2\right) & -\sin\left(\frac{1}{2}\alpha t^2\right) \\ \sin\left(\frac{1}{2}\alpha t^2\right) & \cos\left(\frac{1}{2}\alpha t^2\right) \end{bmatrix} \begin{bmatrix} \cos(\beta t) & -\sin(\beta t) \\ \sin(\beta t) & \cos(\beta t) \end{bmatrix} \\
&= \begin{bmatrix} \cos\left(\frac{1}{2}\alpha t^2 + \beta t\right) & -\sin\left(\frac{1}{2}\alpha t^2 + \beta t\right) \\ \sin\left(\frac{1}{2}\alpha t^2 + \beta t\right) & \cos\left(\frac{1}{2}\alpha t^2 + \beta t\right) \end{bmatrix}
\end{aligned}$$

If we now set the input torque $\tau_x = \tau_z = 0$ then we are left with the homogeneous solution

$$\vec{x}(t) = \underline{\Phi}(t)\vec{x}(t_0) \tag{3.49}$$

That is,

$$\vec{x}(t) = \begin{bmatrix} \cos\left(\frac{1}{2}\alpha t^2 + \beta t\right) & -\sin\left(\frac{1}{2}\alpha t^2 + \beta t\right) \\ \sin\left(\frac{1}{2}\alpha t^2 + \beta t\right) & \cos\left(\frac{1}{2}\alpha t^2 + \beta t\right) \end{bmatrix} \begin{bmatrix} (\omega_o)_x \\ (\omega_o)_z \end{bmatrix} \tag{3.50}$$

Finally, the equations below form the basis for our analytical solution in case 7:

$$\boxed{\begin{aligned} \omega_x(t) &= (\omega_o)_x \cos\left(\frac{1}{2}\alpha t^2 + \beta t\right) - (\omega_o)_z \sin\left(\frac{1}{2}\alpha t^2 + \beta t\right) \\ \omega_y(t) &= \frac{c_1}{I_y} t + (\omega_o)_y \\ \omega_z(t) &= (\omega_o)_x \sin\left(\frac{1}{2}\alpha t^2 + \beta t\right) + (\omega_o)_z \cos\left(\frac{1}{2}\alpha t^2 + \beta t\right) \end{aligned}} \tag{3.51}$$

For a numerical example, we pick $c_1 = 1$ Nm, inertia terms are selected to be $I_x = I_z = 10\ \text{kg}\,\text{m}^2$ and $I_y = 5\ \text{kg}\,\text{m}^2$, and an initial angular velocity $\vec{\omega}_o^T = \begin{bmatrix} 0.3 & -0.4 & 0.5 \end{bmatrix}$ rad/sec is chosen. The absolute errors for $\omega_x$, $\omega_y$ and $\omega_z$ are shown in Figure 3.12 for 1000 sec. of simulation time. In this case, due to the large and sustained *y*-axis torque which causes a final *y*-axis speed of about 200 rad/sec, a significant increase in errors was noticed (e.g. the relative error was about 48% for $\omega_z$ at 1000 seconds using a maximum step size of 0.01 and a relative tolerance of $1\times10^{-5}$). This error is due to the fact that this minimum axis spinner has a nutation frequency of about half the spin frequency, and naturally a reduced step size and relative tolerance is required to accurately calculate the angular velocities at these high frequencies. Hence, in order to reduce this error to acceptable levels, the maximum step size and the relative tolerance, given in Table 3.2, were reduced from 0.01 sec to 0.001 sec and from $1\times10^{-5}$ to $1\times10^{-12}$, respectively. It is important to note that this (and the next) case involves a very high speed motion not normally seen in past or present spacecraft, but the idea here is to validate the simulation software under some severe and limiting cases.

The expected and simulated numerical results along with absolute and relative errors, for several times are presented in Table 3.9.

Table 3.9: Validation case 7 - Expected and simulated angular velocity values

| Sim. Time (sec) | $\omega_x$ (rad/sec) | | | |
|---|---|---|---|---|
| | Analytical | Simulated | Abs. Error | Rel. Err (%) |
| 1 | 0.37135038961761 | 0.37135038961821 | -6.0E-13 | -1.62E-10 |
| 10 | -0.36755775301007 | -0.36755775301076 | -6.9E-13 | -1.88E-10 |
| 100 | -0.54432828146037 | -0.54432828171873 | 2.584E-10 | -4.75E-08 |
| 1000 | 0.51069543501167 | 0.51069476280996 | 6.722E-07 | 1.32E-04 |

| Sim. Time (sec) | $\omega_y$ (rad/sec) | | | |
|---|---|---|---|---|
| | Analytical | Simulated | Abs. Error | Rel. Err (%) |
| 1 | -0.20000000000000 | -0.20000000000002 | 2.0E-14 | -7.63E-12 |
| 10 | 1.60000000000000 | 1.59999999999987 | 1.3E-13 | 8.41E-12 |
| 100 | 19.60000000000000 | 19.59999999995830 | 4.17E-12 | 2.13E-10 |
| 1000 | 199.5999999999990 | 199.6000000077259 | 7.726E-09 | -3.87E-09 |

| Sim. Time (sec) | $\omega_z$ (rad/sec) | | | |
|---|---|---|---|---|
| | Analytical | Simulated | Abs. Error | Rel. Err (%) |
| 1 | 0.44955409922594 | 0.44955409922545 | 5.0E-13 | 1.10E-010 |
| 10 | -0.45266024588226 | -0.45266024588169 | -5.7E-13 | 1.26E-010 |
| 100 | -0.20906152683457 | -0.20906152616162 | -6.729E-10 | 3.22E-007 |
| 1000 | 0.28140748507857 | 0.28140870110359 | -1.216E-06 | -4.32E-04 |

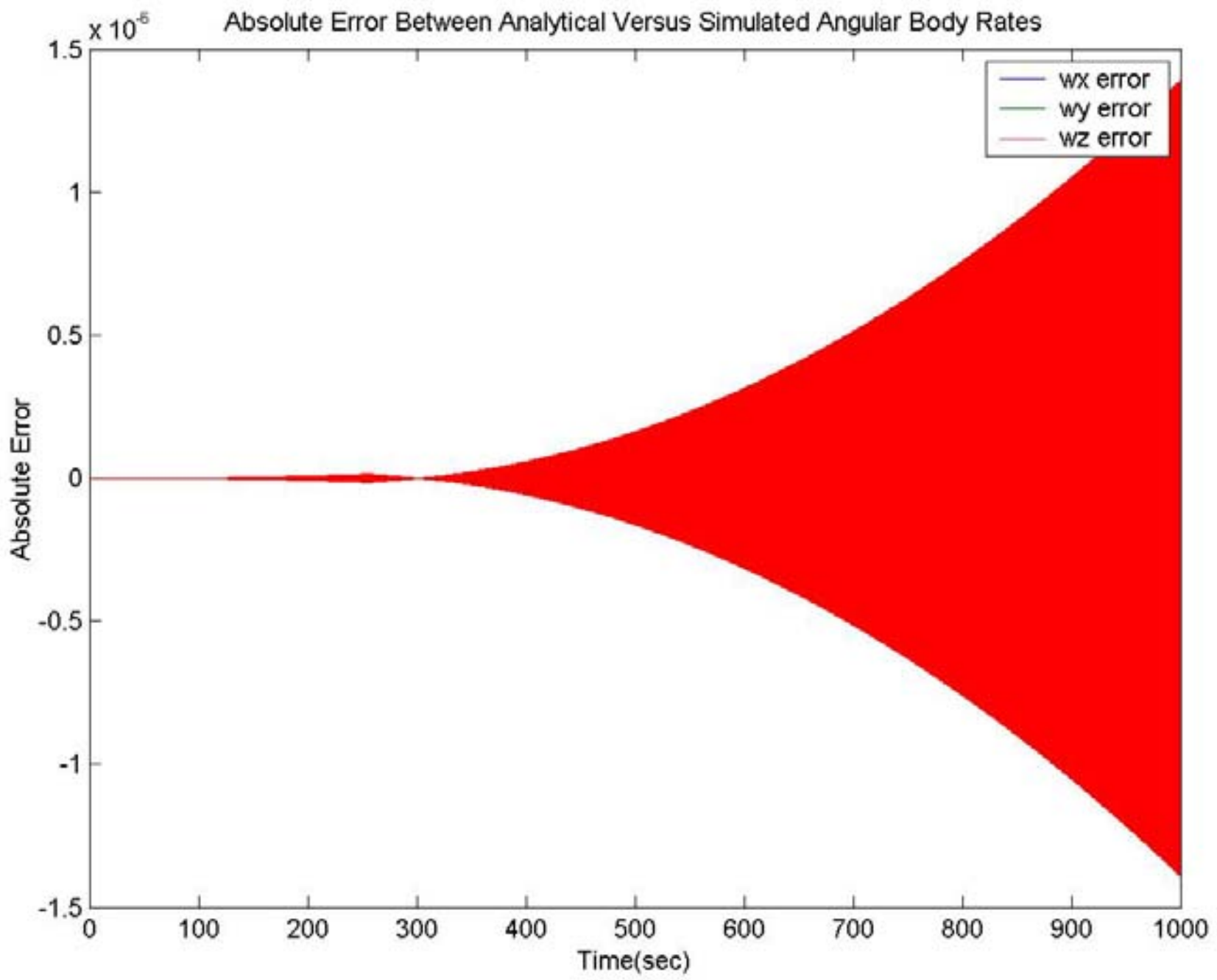

Figure 3.12: Validation case 7 - Absolute error between expected and simulated angular body rates

Validation Case 8

This is a repeat of case 7 except that for this case $\tau_y = m\cos(\gamma t)$ and hence equation (3.1) is simplified to

$$\dot{\omega}_x I_x + \omega_y \omega_z (I_x - I_y) = \tau_x \tag{3.52}$$

$$\dot{\omega}_y I_y = m\cos(\gamma t) \tag{3.53}$$

$$\dot{\omega}_z I_x + \omega_x \omega_y \left(I_y - I_x\right) = \tau_z \tag{3.54}$$

From (3.53) above, we can solve for $\omega_y$ as follows

$$\dot{\omega}_y = \frac{m}{I_y}\cos(\gamma t) \quad \Rightarrow \quad \omega_y = \frac{m}{\gamma I_y}\sin(\gamma t) + (\omega_o)_y \tag{3.55}$$

Furthermore, (3.52) and (3.54) can be re-written as

$$\dot{\omega}_x = -\frac{I_x - I_y}{I_x}\left(\frac{m}{\gamma I_y}\sin(\gamma t) + (\omega_o)_y\right)\omega_z + \frac{\tau_x}{I_x} \tag{3.56}$$

$$\dot{\omega}_z = -\frac{I_y - I_x}{I_y}\left(\frac{m}{\gamma I_y}\sin(\gamma t) + (\omega_o)_y\right)\omega_x + \frac{\tau_z}{I_x} \tag{3.57}$$

or in a compact matrix for as

$$\begin{bmatrix}\dot{\omega}_x \\ \dot{\omega}_z\end{bmatrix} = \begin{bmatrix} 0 & -a\left(\frac{m}{\gamma I_y}\sin(\gamma t) + (\omega_o)_y\right) \\ a\left(\frac{m}{\gamma I_y}\sin(\gamma t) + (\omega_o)_y\right) & 0 \end{bmatrix}\begin{bmatrix}\omega_x \\ \omega_z\end{bmatrix} + \frac{1}{I_x}\begin{bmatrix}\tau_x \\ \tau_z\end{bmatrix} \tag{3.58}$$

where $a = \frac{I_x - I_y}{I_x}$ , and by defining $\alpha = a\frac{m}{\gamma I_y}$ and $\beta = a(\omega_o)_y$, (3.58) can be written as

$$\begin{bmatrix}\dot{\omega}_x \\ \dot{\omega}_z\end{bmatrix} = \begin{bmatrix} 0 & -(\alpha\sin(\gamma t) + \beta) \\ (\alpha\sin(\gamma t) + \beta) & 0 \end{bmatrix}\begin{bmatrix}\omega_x \\ \omega_z\end{bmatrix} + \frac{1}{I_x}\begin{bmatrix}\tau_x \\ \tau_z\end{bmatrix} \tag{3.59}$$

where $\underline{A}(t) = \begin{bmatrix} 0 & -(\alpha\sin(\gamma t) + \beta) \\ (\alpha\sin(\gamma t) + \beta) & 0 \end{bmatrix} = \underline{M}_1\sin(\gamma t) + \underline{M}_2$, $\underline{M}_1 = \begin{bmatrix} 0 & -\alpha \\ \alpha & 0 \end{bmatrix}$ and $\underline{M}_2 = \begin{bmatrix} 0 & -\beta \\ \beta & 0 \end{bmatrix}$.

Since $\underline{M}_1$ and $\underline{M}_2$ commute, $\underline{\Phi}(t)$ can be expressed as

$$\underline{\Phi}(t) = e^{\int_0^t \underline{M}_1 \sin(\gamma t)\mathrm{d}t} e^{\int_0^t \underline{M}_2 \,\mathrm{d}t} = e^{-\frac{1}{\gamma}\underline{M}_1 \cos(\gamma t)} e^{\frac{1}{\gamma}\underline{M}_1} e^{\underline{M}_2 t}$$

$$= \begin{bmatrix} \cos\left(\frac{\alpha}{\gamma}\cos(\gamma t)\right) & -\sin\left(\frac{\alpha}{\gamma}\cos(\gamma t)\right) \\ \sin\left(\frac{\alpha}{\gamma}\cos(\gamma t)\right) & \cos\left(\frac{\alpha}{\gamma}\cos(\gamma t)\right) \end{bmatrix} \begin{bmatrix} \cos\left(\frac{1}{\gamma}\alpha\right) & -\sin\left(\frac{1}{\gamma}\alpha\right) \\ \sin\left(\frac{1}{\gamma}\alpha\right) & \cos\left(\frac{1}{\gamma}\alpha\right) \end{bmatrix} \begin{bmatrix} \cos(\beta t) & -\sin(\beta t) \\ \sin(\beta t) & \cos(\beta t) \end{bmatrix}$$

$$= \begin{bmatrix} \cos\left(\alpha\frac{\cos(\gamma t)-1}{\gamma} - \beta t\right) & -\sin\left(\alpha\frac{\cos(\gamma t)-1}{\gamma} - \beta t\right) \\ \sin\left(\alpha\frac{\cos(\gamma t)-1}{\gamma} - \beta t\right) & \cos\left(\alpha\frac{\cos(\gamma t)-1}{\gamma} - \beta t\right) \end{bmatrix}$$

If we now set the input torque $\tau_x = \tau_z = 0$, we are then left with the homogeneous solution (3.49) which for this case becomes

$$\vec{x}(t) = \begin{bmatrix} \cos\left(\alpha\frac{\cos(\gamma t)-1}{\gamma} - \beta t\right) & -\sin\left(\alpha\frac{\cos(\gamma t)-1}{\gamma} - \beta t\right) \\ \sin\left(\alpha\frac{\cos(\gamma t)-1}{\gamma} - \beta t\right) & \cos\left(\alpha\frac{\cos(\gamma t)-1}{\gamma} - \beta t\right) \end{bmatrix} \begin{bmatrix} (\omega_o)_x \\ (\omega_o)_z \end{bmatrix} \tag{3.60}$$

And finally, the equations below form the basis for our analytical solution in case 8:

$$\boxed{\begin{aligned} \omega_x(t) &= (\omega_o)_x \cos\left(\alpha\frac{\cos(\gamma t)-1}{\gamma} - \beta t\right) - (\omega_o)_z \sin\left(\alpha\frac{\cos(\gamma t)-1}{\gamma} - \beta t\right) \\ \omega_y(t) &= \frac{m}{\gamma I_y}\sin(\gamma t) + (\omega_o)_y \\ \omega_z(t) &= (\omega_o)_x \sin\left(\alpha\frac{\cos(\gamma t)-1}{\gamma} - \beta t\right) + (\omega_o)_z \cos\left(\alpha\frac{\cos(\gamma t)-1}{\gamma} - \beta t\right) \end{aligned}} \tag{3.61}$$

For a numerical example, we pick $m = 1$ Nm and $\gamma = \frac{2\pi}{5400}$, inertia terms are selected to be $I_x = I_z = 10\ \mathrm{kg\,m^2}$ and $I_y = 5\ \mathrm{kg\,m^2}$, and an initial angular velocity $\vec{\omega}_o^T = \begin{bmatrix} 0.3 & -0.4 & 0.5 \end{bmatrix}$ rad/sec is chosen. The absolute errors for $\omega_x$, $\omega_y$ and $\omega_z$ are shown in Figure 3.13 for 1000 sec. of simulation time. In this case again, as was explained in case 7, due to the high speed spin motion, a significant increase in errors was noticed (e.g. the relative error was about 38% for $\omega_x$ at 1000 seconds using a maximum step size of 0.01 and a relative tolerance of $1\times10^{-5}$). In order to reduce this error to acceptable levels, the maximum step size and the relative tolerance used, which are given in Table 3.2 for all cases expect cases 7 and 8, were reduced to 0.001 sec and to $1\times10^{-12}$, respectively. The expected and simulated numerical results along with absolute and relative errors, for several times are presented in Table 3.10.

Table 3.10: Validation case 8 - Expected and simulated angular velocity values

| Sim. Time (sec) | $\omega_x$ (rad/sec) | | | |
|---|---|---|---|---|
| | Analytical | Simulated | Abs. Error | Rel. Err (%) |
| 1 | 0.37135039215324 | 0.37135039215417 | -9.3E-13 | -2.51E-10 |
| 10 | -0.36758328716307 | -0.36758328716172 | -1.35E-12 | 3.67E-10 |
| 100 | -0.57179982492277 | -0.57179982479658 | -1.262E-10 | 2.21E-08 |
| 1000 | 0.27644889520778 | 0.27644962629059 | -7.311E-07 | -2.65E-04 |

| Sim. Time (sec) | $\omega_y$ (rad/sec) | | | |
|---|---|---|---|---|
| | Analytical | Simulated | Abs. Error | Rel. Err (%) |
| 1 | -0.20000004512850 | 0.2000000451285 | 0.0 | -4.44E-13 |
| 10 | 1.59995487179885 | 1.59995487179887 | -2.0E-14 | -1.15E-12 |
| 100 | 19.55490203225110 | 19.5549020322284 | 2.27E-11 | 1.16E-10 |
| 1000 | 157.4297228155190 | 157.429722817886 | -2.4E-09 | -1.50E-9 |

| Sim. Time (sec) | $\omega_z$ (rad/sec) | | | |
|---|---|---|---|---|
| | Analytical | Simulated | Abs. Error | Rel. Err (%) |
| 1 | 0.44955409713141 | 0.44955409713064 | 7.7E-13 | 1.71E-10 |
| 10 | -0.45263951108845 | -0.45263951108955 | 1.09E-12 | -2.42E-10 |
| 100 | 0.11421453593254 | 0.11421453656374 | -6.3E-10 | -5.53E-07 |
| 1000 | -0.51339654102691 | -0.51339614733452 | -3.9E-07 | 7.67E-05 |

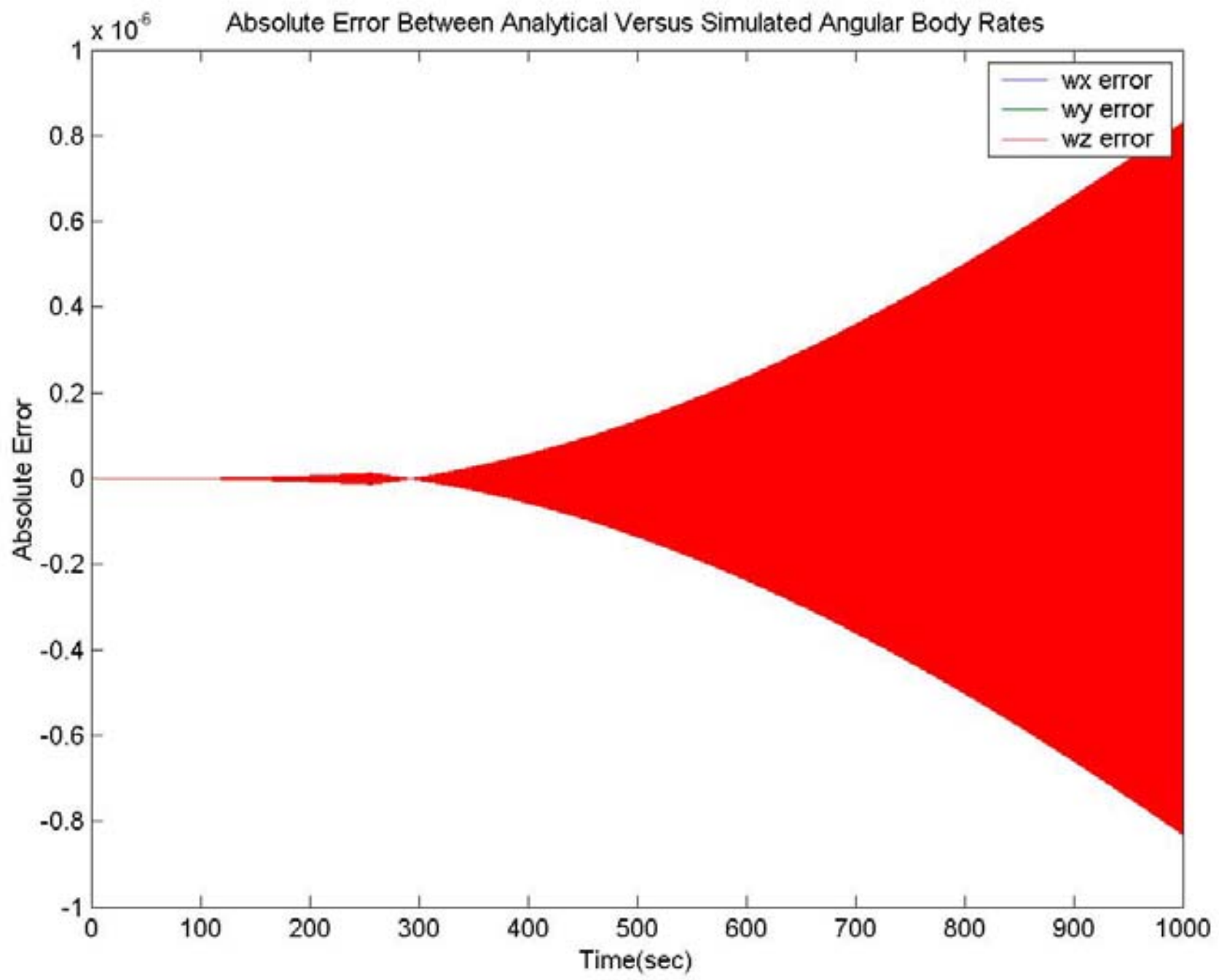

Figure 3.13: Validation case 8 - Absolute error between expected and simulated angular body rates

Validation Case 9

This is a repeat of case 7 except that for this case $\tau_y = m[u(t-t_1)-u(t-t_2)]$ and hence equation (3.1) is simplified to

$$\dot{\omega}_x I_x + \omega_y \omega_z \left(I_x - I_y\right) = \tau_x \tag{3.62}$$

$$\dot{\omega}_y I_y = m[u(t-t_1)-u(t-t_2)] \tag{3.63}$$

$$\dot{\omega}_z I_y + \omega_x \omega_y \left(I_y - I_x\right) = \tau_z \tag{3.64}$$

From (3.63) above, we can solve for $\omega_y$ as follows:

$$\dot{\omega}_y = \frac{m}{I_y}[u(t-t_1) - u(t-t_2)]$$

$$\begin{aligned}
t \le t_1: &\quad \omega_y = \int_0^t 0\,\mathrm{d}t \quad \Rightarrow \quad \omega_y = (\omega_o)_y \\
t_1 \le t \le t_2: &\quad \omega_y = \frac{m}{I_y}\int_{t_1}^t \mathrm{d}t \quad \Rightarrow \quad \omega_y = \frac{m}{I_y}(t-t_1) + (\omega_o)_y \\
t \ge t_2: &\quad \omega_y = \int_{t_2}^t 0\,\mathrm{d}t \quad \Rightarrow \quad \omega_y = \frac{m}{I_y}(t_2-t_1) + (\omega_o)_y
\end{aligned} \tag{3.65}$$

Furthermore, (3.62) and (3.64) can be re-written as

$$\dot{\omega}_x = -\frac{(I_x - I_y)}{I_x}\omega_y\omega_z + \frac{\tau_x}{I_x} \tag{3.66}$$

$$\dot{\omega}_z = -\frac{(I_y - I_x)}{I_x}\omega_y\omega_x + \frac{\tau_z}{I_x} \tag{3.67}$$

or in a compact matrix form as

$$\begin{bmatrix} \dot{\omega}_x \\ \dot{\omega}_z \end{bmatrix} = \begin{bmatrix} 0 & -a\omega_y \\ a\omega_y & 0 \end{bmatrix} \begin{bmatrix} \omega_x \\ \omega_z \end{bmatrix} + \frac{1}{I_y}\begin{bmatrix} \tau_x \\ \tau_z \end{bmatrix} \tag{3.68}$$

where $a = \dfrac{I_x - I_y}{I_x}$.

If we now set the input torque $\tau_x = \tau_z = 0$, the solution is divided into three parts:

***Case*** $t \le t_1$***:***

Using (3.65), system (3.68) can be expressed as:

$$\begin{bmatrix} \dot{\omega}_x \\ \dot{\omega}_z \end{bmatrix} = \begin{bmatrix} 0 & -a(\omega_o)_y \\ a(\omega_o)_y & 0 \end{bmatrix} \begin{bmatrix} \omega_x \\ \omega_z \end{bmatrix} \tag{3.69}$$

which is a linear time invariant homogeneous system of the form (3.9) where $\underline{A} = \begin{bmatrix} 0 & -\alpha \\ \alpha & 0 \end{bmatrix}$, $\alpha = a(\omega_o)_y$ and $\vec{x}(t) = \begin{bmatrix} \omega_x(t) & \omega_z(t) \end{bmatrix}^T$.

Using (3.10), the solution for $t \le t_1$ is

$$\begin{aligned} \omega_x &= (\omega_o)_x \cos(\alpha t) - (\omega_o)_z \sin(\alpha t) \\ \omega_y &= (\omega_o)_y \\ \omega_z &= (\omega_o)_x \sin(\alpha t) + (\omega_o)_z \cos(\alpha t) \end{aligned} \tag{3.70}$$

***Case*** $t_1 \le t \le t_2$**:**

Using (3.65), system (3.68) can be expressed as:

$$\begin{bmatrix} \dot{\omega}_x \\ \dot{\omega}_z \end{bmatrix} = \begin{bmatrix} 0 & -a\left(\dfrac{m}{I_y}(t-t_1) + (\omega_o)_y\right) \\ a\left(\dfrac{m}{I_y}(t-t_1) + (\omega_o)_y\right) & 0 \end{bmatrix} \begin{bmatrix} \omega_x \\ \omega_z \end{bmatrix} \tag{3.71}$$

Let $\alpha = a\dfrac{m}{I_y}$, $\beta = a\left(-\dfrac{m}{I_y}t_1 + (\omega_o)_y\right)$ and $A(t) = \begin{bmatrix} 0 & -(\alpha t + \beta) \\ (\alpha t + \beta) & 0 \end{bmatrix} = \underline{M}_1 t + \underline{M}_2$

where $\underline{M}_1 = \begin{bmatrix} 0 & -\alpha \\ \alpha & 0 \end{bmatrix}$ and $\underline{M}_2 = \begin{bmatrix} 0 & -\beta \\ \beta & 0 \end{bmatrix}$ and since $\underline{M}_1$ and $\underline{M}_2$ commute, $\underline{\Phi}(t)$ can be expressed as

$$\underline{\Phi}(t) = e^{\int_{t_1}^{t} \underline{M}_1 t\,\mathrm{d}t}\, e^{\int_{t_1}^{t} \underline{M}_2\,\mathrm{d}t} = e^{\underline{M}_1 \frac{t^2}{2}} e^{-\underline{M}_1 \frac{t_1^2}{2}} e^{\underline{M}_2 t} e^{-\underline{M}_2 t_1}$$

$$= \begin{bmatrix} \cos\left(\frac{1}{2}\alpha t^2\right) & -\sin\left(\frac{1}{2}\alpha t^2\right) \\ \sin\left(\frac{1}{2}\alpha t^2\right) & \cos\left(\frac{1}{2}\alpha t^2\right) \end{bmatrix} \begin{bmatrix} \cos\left(\frac{1}{2}\alpha t_1^2\right) & \sin\left(\frac{1}{2}\alpha t_1^2\right) \\ -\sin\left(\frac{1}{2}\alpha t_1^2\right) & \cos\left(\frac{1}{2}\alpha t_1^2\right) \end{bmatrix} \begin{bmatrix} \cos(\beta t) & -\sin(\beta t) \\ \sin(\beta t) & \cos(\beta t) \end{bmatrix} \begin{bmatrix} \cos(\beta t_1) & \sin(\beta t_1) \\ -\sin(\beta t_1) & \cos(\beta t_1) \end{bmatrix}$$

$$= \begin{bmatrix} \cos\left(-\frac{1}{2}\alpha t^2 + \frac{1}{2}\alpha t_1{}^2 - \beta t + \beta t_1\right) & \sin\left(-\frac{1}{2}\alpha t^2 + \frac{1}{2}\alpha t_1{}^2 - \beta t + \beta t_1\right) \\ -\sin\left(-\frac{1}{2}\alpha t^2 + \frac{1}{2}\alpha t_1{}^2 - \beta t + \beta t_1\right) & \cos\left(-\frac{1}{2}\alpha t^2 + \frac{1}{2}\alpha t_1{}^2 - \beta t + \beta t_1\right) \end{bmatrix}$$

and using (3.49), the solution for $t_1 \le t \le t_2$ is:

$$\begin{aligned} \omega_x &= (\omega_{t_1})_x \cos\left(-\frac{1}{2}\alpha t^2 + \frac{1}{2}\alpha t_1{}^2 - \beta t + \beta t_1\right) + (\omega_{t_1})_z \sin\left(-\frac{1}{2}\alpha t^2 + \frac{1}{2}\alpha t_1{}^2 - \beta t + \beta t_1\right) \\ \omega_y &= \frac{m}{I_y}(t - t_1) + (\omega_o)_y \\ \omega_z &= -(\omega_{t_1})_x \sin\left(-\frac{1}{2}\alpha t^2 + \frac{1}{2}\alpha t_1{}^2 - \beta t + \beta t_1\right) + (\omega_{t_1})_z \cos\left(-\frac{1}{2}\alpha t^2 + \frac{1}{2}\alpha t_1{}^2 - \beta t + \beta t_1\right) \end{aligned} \tag{3.72}$$

***Case*** $t \ge t_2$***:***

Using (3.65), system (3.68) can be expressed as:

$$\begin{bmatrix} \dot{\omega}_x \\ \dot{\omega}_z \end{bmatrix} = \begin{bmatrix} 0 & -a\left(\frac{m}{I_y}(t_2 - t_1) + (\omega_o)_y\right) \\ a\left(\frac{m}{I_y}(t_2 - t_1) + (\omega_o)_y\right) & 0 \end{bmatrix} \begin{bmatrix} \omega_x \\ \omega_z \end{bmatrix} \tag{3.73}$$

which is a linear time invariant homogeneous system of the form (3.9) where $\underline{A} = \begin{bmatrix} 0 & -\alpha \\ \alpha & 0 \end{bmatrix}$, $\alpha = a\left(\frac{m}{I_y}(t_2 - t_1) + (\omega_o)_y\right)$ and $\vec{x}(t) = \left[\omega_x(t) \quad \omega_z(t)\right]^T$ .

Using (3.10), the solution for $t \geq t_2$ is

$$\begin{aligned}
\omega_x &= (\omega_{t_2})_x \cos\left(\alpha(t-t_2)\right) - (\omega_{t_2})_z \sin\left(\alpha(t-t_2)\right) \\
\omega_y &= a\left(\frac{m}{I_y}(t_2 - t_1) + (\omega_o)_y\right) \\
\omega_z &= (\omega_{t_2})_x \sin\left(\alpha(t-t_2)\right) + (\omega_{t_2})_z \cos\left(\alpha(t-t_2)\right)
\end{aligned} \tag{3.74}$$

And finally, the equations below form the basis for our analytical solution in case 9:

| | |
|---|---|
| $t \leq t_1$ | $\omega_x(t) = (\omega_o)_x \cos(\alpha t) - (\omega_o)_z \sin(\alpha t)$<br>$\omega_y(t) = (\omega_o)_y$ where $\alpha = a(\omega_o)_y$<br>$\omega_z(t) = (\omega_o)_x \sin\left(\alpha t\right) + (\omega_o)_z \cos(\alpha t)$ |
| $t_1 \leq t \leq t_2$ | $\omega_x(t) = (\omega_{t_1})_x \cos\left(-\frac{1}{2}\alpha t^2 + \frac{1}{2}\alpha t_1{}^2 - \beta t + \beta t_1\right) + (\omega_{t_1})_z \sin\left(-\frac{1}{2}\alpha t^2 + \frac{1}{2}\alpha t_1{}^2 - \beta t + \beta t_1\right)$<br>$\omega_y(t) = \frac{m}{I_y}(t - t_1) + (\omega_o)_y$ where $\alpha = a\frac{m}{I_y}$<br>$\omega_z(t) = -(\omega_{t_1})_x \sin\left(-\frac{1}{2}\alpha t^2 + \frac{1}{2}\alpha t_1{}^2 - \beta t + \beta t_1\right) + (\omega_{t_1})_z \cos\left(-\frac{1}{2}\alpha t^2 + \frac{1}{2}\alpha t_1{}^2 - \beta t + \beta t_1\right)$ |
| $t \geq t_2$ | $\omega_x(t) = (\omega_{t_2})_x \cos\left(\alpha(t-t_2)\right) - (\omega_{t_2})_z \sin\left(\alpha(t-t_2)\right)$<br>$\omega_y(t) = a\left(\frac{m}{I_y}(t_2 - t_1) + (\omega_o)_y\right)$ where $\alpha = a\left(\frac{m}{I_y}(t_2 - t_1) + (\omega_o)_y\right)$<br>$\omega_z(t) = (\omega_{t_2})_x \sin\left(\alpha(t-t_2)\right) + (\omega_{t_2})_z \cos\left(\alpha(t-t_2)\right)$ |

For a numerical example, we pick $m = 10$ Nm, inertia terms are selected to be $I_x = I_z = 10\ \text{kg}\,\text{m}^2$ and $I_y = 5\ \text{kg}\,\text{m}^2$, and an initial angular velocity $\vec{\omega}_o^T = \begin{bmatrix} 0.3 & -0.4 & 0.5 \end{bmatrix}$ rad/sec is chosen. For this last rigid spacecraft validation case, the absolute errors for $\omega_x$, $\omega_y$ and $\omega_z$ are shown in Figure 3.14 for both 1000 and 10000 sec. of simulation times. The expected and simulated numerical results along with absolute and relative errors, for several times are presented in Table 3.11.

Table 3.11: Validation case 9 - Expected and simulated angular velocity values

| Sim. Time (sec) | $\omega_x$ (rad/sec) | | | |
|---|---|---|---|---|
| | Analytical | Simulated | Abs. Error | Rel. Err (%) |
| 0.1 | 0.30993933534664 | 0.30993933600509 | -6.58E-10 | -2.12E-07 |
| 1 | 0.39335463874990 | 0.39335463932370 | -5.74E-10 | -1.46E-07 |
| 1.5 | 0.38247203046828 | 0.38247202968844 | 7.80E-10 | 2.04E-07 |
| 2 | 0.24858454125999 | 0.24858454103434 | 2.26E-10 | 9.08E-08 |
| 10 | 0.18541629367450 | 0.18541629178236 | 1.89E-09 | 1.02E-06 |
| 100 | -0.31966514036139 | -0.31966513815886 | -2.20E-09 | 6.89E-07 |
| 1000 | 0.00288835725161 | 0.00288835556425 | 1.69E-09 | 5.84E-05 |
| 10000 | 0.29744412333327 | 0.29744419011824 | -6.68E-08 | -2.24E-05 |

| Sim. Time (sec) | $\omega_y$ (rad/sec) | | | |
|---|---|---|---|---|
| | Analytical | Simulated | Abs. Error | Rel. Err (%) |
| 0.1 | -0.40000000000000 | -0.40000000000000 | 0.00E+00 | 0.00E+00 |
| 1 | 0.60000000000000 | 0.60000000000000 | 0.00E+00 | 0.00E+00 |
| 1.5 | 1.60000000000000 | 1.60000000000000 | 0.00E+00 | 0.00E+00 |
| 2 | 1.60000000000000 | 1.60000000000000 | 0.00E+00 | 0.00E+00 |
| 10 | 1.60000000000000 | 1.60000000000000 | 0.00E+00 | 0.00E+00 |
| 100 | 1.60000000000000 | 1.60000000000000 | 0.00E+00 | 0.00E+00 |
| 1000 | 1.60000000000000 | 1.60000000000000 | 0.00E+00 | 0.00E+00 |
| 10000 | 1.60000000000000 | 1.60000000000000 | 0.00E+00 | 0.00E+00 |

| Sim. Time (sec) | $\omega_z$ (rad/sec) | | | |
|---|---|---|---|---|
| | Analytical | Simulated | Abs. Error | Rel. Err (%) |
| 0.1 | 0.49390040332529 | 0.49390040291190 | 4.13E-10 | 8.37E-08 |
| 1 | 0.43043248968210 | 0.43043248915751 | 5.25E-10 | 1.22E-07 |
| 1.5 | 0.44013082817439 | 0.44013082624144 | 1.93E-09 | 4.39E-07 |
| 2 | 0.52745210763306 | 0.52745210556099 | 2.07E-09 | 3.93E-07 |
| 10 | 0.55282980929036 | 0.55282980780140 | 1.49E-09 | 2.69E-07 |
| 100 | -0.48766197108010 | -0.48766197011649 | -9.64E-10 | 1.98E-07 |
| 1000 | 0.58308803571364 | 0.58308803370840 | 2.01E-09 | 3.44E-07 |
| 10000 | 0.50152466887931 | 0.50152462692731 | 4.20E-08 | 8.37E-06 |

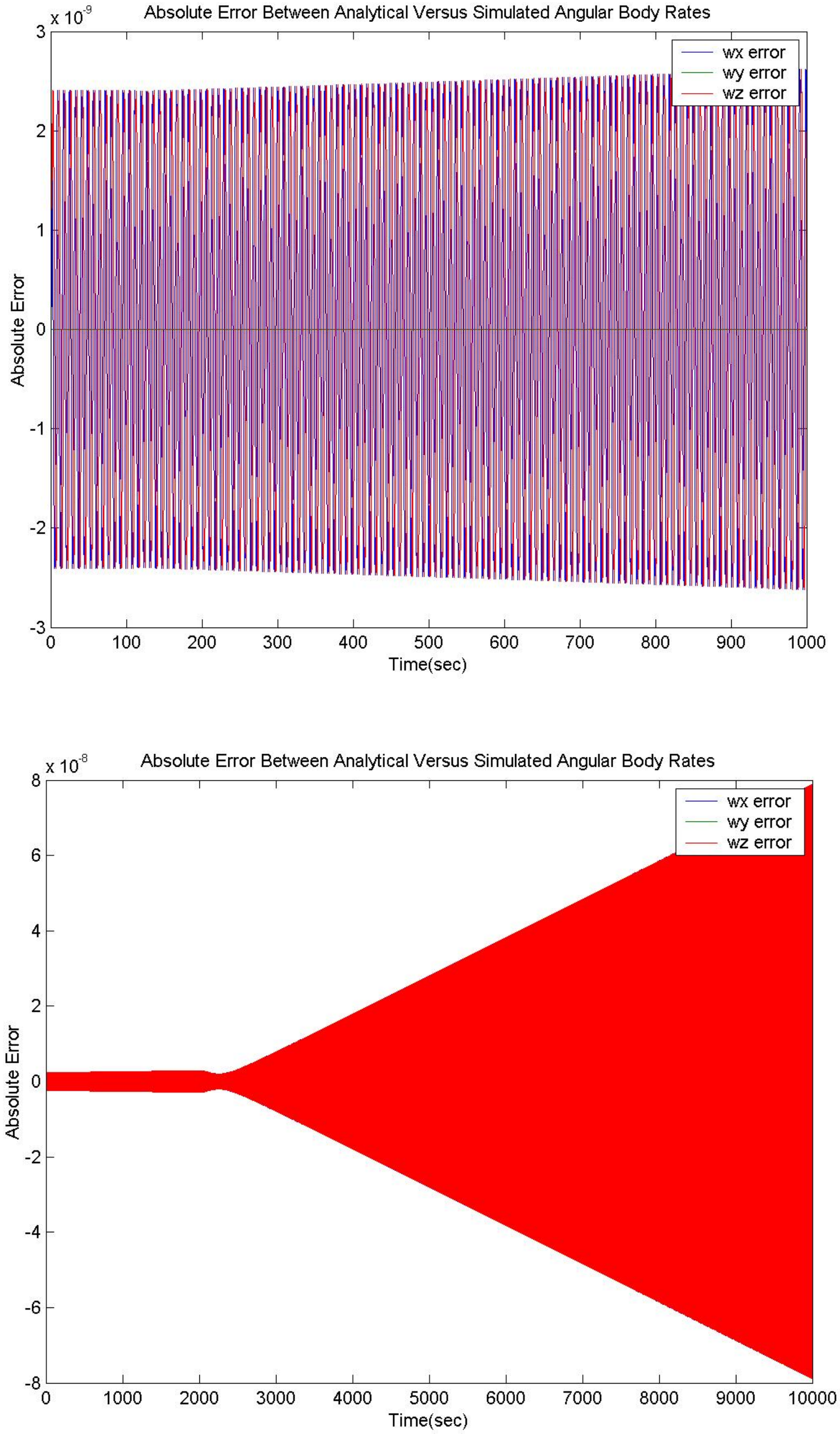

Figure 3.14: Validation case 9 - Absolute error between expected and simulated angular body rates

### 3.4.2 Flexible Spacecraft Validation

All test cases for the flexible spacecraft have been validated using both the RK and ABM integration techniques, however, only the results obtained from the RK solver (with 0.01 sec. time step) are presented here. For the flexible spacecraft, validation cases with an analytical solution are not always possible, and we have to hence rely on more fundamental concepts (i.e. conservation of energy and angular momentum) to validate the flexible motion. After careful study of the flexible spacecraft equations of motion (2.57), three validation cases were devised where the solution is analytically known and where the flexibility of the system is not excited using a judicious choice of system parameters and initial conditions. However for more comprehensive testing, including flexible motion, one can resort to conservation of energy and momentum principles which state that when the there is no torque applied to the spacecraft, the kinetic energy, *T*, and the magnitude of the angular momentum vector, $\vec{l}$, of the spacecraft must remain constant, assuming that there is no damping in the appendages. The kinetic and potential energies of the flexible spacecraft are given by equations (2.25) and (2.44), respectively. The angular momentum vector of the flexible spacecraft is given by

$$\vec{l} = \underline{I}_t \vec{\omega} + \sum_{i=1}^{n} \vec{\kappa}_i \qquad (3.75)$$

This section comprises seven (7) validation case studies. Table 3.12 summarizes the specifications corresponding to these cases, and Figure 2.6 shows the layout of the flexible spacecraft:

Table 3.12: Validation case study specifications for flexible spacecraft

| Case No. | Total Inertia, $\underline{I}_t$ | Initial Rate, $\vec{\omega}_o$ | Torque, $\vec{\tau}$ | Appendage Specifications |
|---|---|---|---|---|
| 10 | $\underline{I}_t = \underline{0}$ | $(\omega_o)_x = 0$<br>$(\omega_o)_y = 0$<br>$(\omega_o)_z = 0$ | $\tau_x = 0$<br>$\tau_y = 0$<br>$\tau_z = 0$ | No. of appendages: $n = 0$<br>No. of modes: $p = q = 0$<br>Appendage location: $\vec{d}^T = [0\ 0\ 0]$<br>Modulus of Elasticity: $E = 0$<br>Poisson ratio: $\gamma = 0$<br>Width: $a = 0$; Length: $b = 0$<br>Thickness: $h = 0$<br>Area density: $\rho = 0$<br>Initial deflection: $\vec{\chi}_o = \vec{0}$<br>Initial rate of deflection: $\dot{\vec{\chi}}_o = \vec{0}$ |
| 11 | Any inertia matrix | $(\omega_o)_x = 0$<br>$(\omega_o)_y = 0$<br>$(\omega_o)_z = 0$ | $\tau_x = 0$<br>$\tau_y = 0$<br>$\tau_z = 0$ | Arbitrary number of appendages located at any rigid bus attachment point and in any direction with arbitrary but physically meaningful properties<br>Initial appendage deflection:<br>$\vec{\chi}_o = \vec{0}$<br>Initial appendage rate of def.:<br>$\dot{\vec{\chi}}_o = \vec{0}$ |
| 12 | $\underline{I}_t = \begin{bmatrix} I_x & 0 & 0 \\ 0 & I_y & 0 \\ 0 & 0 & I_z \end{bmatrix}$<br>with $I_x, I_y, I_z > 0$ | $(\omega_o)_x \neq 0$<br>$(\omega_o)_y = 0$<br>$(\omega_o)_z = 0$ | $\tau_x = 0$<br>$\tau_y = 0$<br>$\tau_z = 0$ | Same as case 11 except for appendage locations, which must have $d_z = 0$ and must result in the total inertia being a positive diagonal matrix |
| 13 | Same as case 12 | $(\omega_o)_x = 0$<br>$(\omega_o)_y \neq 0$<br>$(\omega_o)_z = 0$ | $\tau_x = 0$<br>$\tau_y = 0$<br>$\tau_z = 0$ | Same as case 12 |
| 14 | Same as case 12 | $(\omega_o)_x = 0$<br>$(\omega_o)_y = 0$<br>$(\omega_o)_z \neq 0$ | $\tau_x = 0$<br>$\tau_y = 0$<br>$\tau_z = 0$ | Same as case 12 except there is no requirement for $d_z$ to equal zero as long as the total inertia is a positive diagonal matrix |
| 15 | Any inertia matrix | Any initial angular rate less then the critical angular rate+ | $\tau_x = 0$<br>$\tau_y = 0$<br>$\tau_z = 0$ | Same as case 11 |
| 16 | Any inertia matrix | Any initial angular rate less then the critical angular rate+ | $\tau_i = m_i [u(t-t_1) - u(t-t_2)]$<br>$m_i \neq 0$<br>$t_2 > t_1 > 0$ | Same as case 11 |

+ The critical angular rate is the rate above which the numerical solution diverges. The physical meaning of this situation is that structural damage (e.g. buckling) occurs in the appendages since the elastic restoring force in the appendages is weaker than the centrifugal inertial force, created by the large angular rate.

Validation Case 10

With all system parameters and initial conditions set to zero (e.g. zero torque, zero inertia, zero modulus of elasticity, etc.), the angular body rates and appendage deflections of the flexible spacecraft are expected to remain constant at zero, and this is precisely what was observed from the simulation results. This is the all zero test case, which insures that no mistakes have been made in coding the flexible dynamics equations (e.g. no division by zero is present).

Validation Case 11

With zero torque, zero initial angular velocity, zero appendage deflection and rate of deflections, and any inertia matrix and physically meaningful appendage specifications, the angular body rates and the deflections of the flexible spacecraft are expected to remain constant at zero. This is precisely what was observed from the simulation results shown in Figure 3.15 and Figure 3.16.

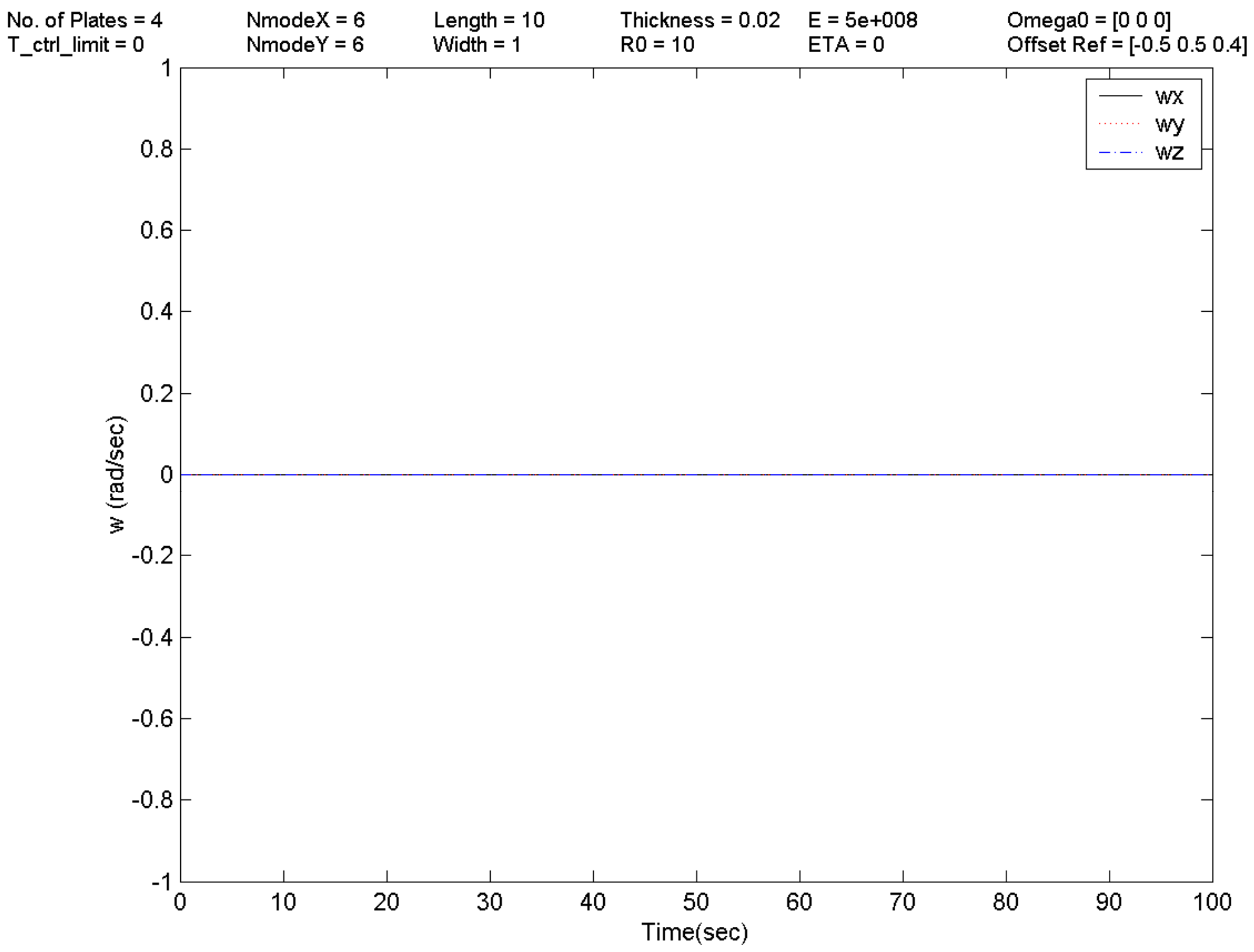

Figure 3.15 Validation case 11 - Spacecraft body angular rates ( $\vec{\omega}$ )

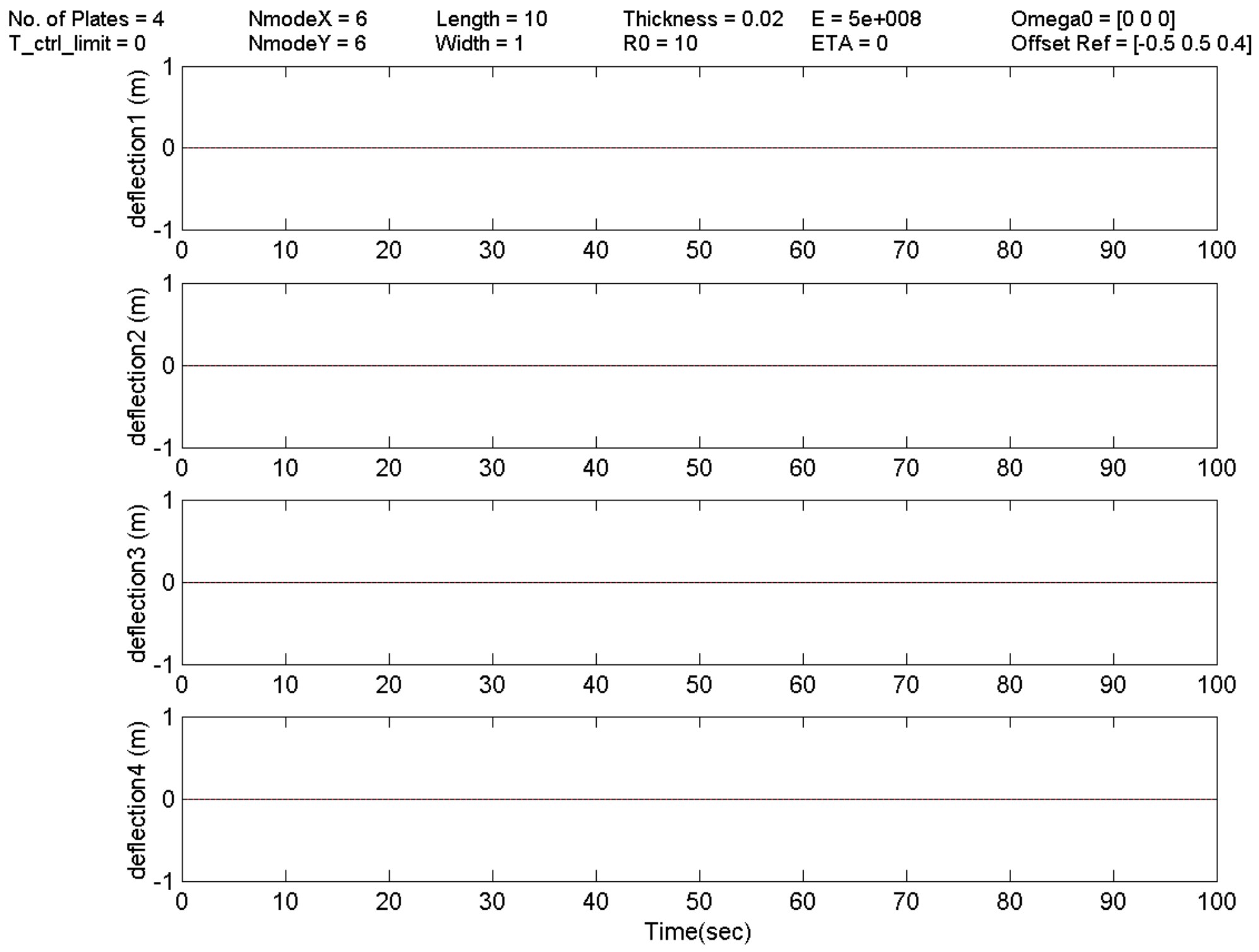

Figure 3.16 Validation case 11 - Deflection ( $w$ ) at tip $x = .5; y = 10$ and midway $x = .5; y = 5$

Validation Case 12

As per Table 3.12, by setting $(\omega_o)_y = (\omega_o)_z = d_z = 0$, $\vec{\chi}_o = \dot{\vec{\chi}}_o = \vec{0}$ and $\vec{\tau} = 0$, many of the terms in the spacecraft attitude and flexible dynamics equation (2.52) drop to zero and we obtain the following simple equations:

$$\begin{aligned} \underline{I}_t \dot{\vec{\omega}} &= -\vec{\omega} \times (\underline{I}_t \vec{\omega}) \\ \underline{M}_{\phi\psi} \ddot{\vec{\chi}} &= \vec{0} \end{aligned} \tag{3.76}$$

From (3.76) and noting that $\underline{M}_{\phi\psi}$ is simply a positive multiple of the identity matrix, we can solve for $\vec{\chi}$ and $\dot{\vec{\chi}}$ by integration:

$$\ddot{\vec{\chi}} = \vec{0} \quad \Rightarrow \quad \dot{\vec{\chi}}(t) = \dot{\vec{\chi}}_o = \vec{0} \quad \Rightarrow \quad \vec{\chi}(t) = \vec{\chi}_o = \vec{0} \tag{3.77}$$

Also from (3.76), we note that the attitude of the flexible spacecraft behaves exactly like the rigid spacecraft as long as $\vec{\tau} = 0$ and hence, we recover the Euler's equations of motion (3.1), by placing the appendages such that the total inertia matrix, $\underline{I}_t$, is diagonal. Then, from (3.1) and knowing that $(\omega_o)_y = (\omega_o)_z = 0$, we simply have the following solutions:

$$\begin{aligned} \omega_x(t) &= (\omega_o)_x \\ \omega_y(t) &= (\omega_o)_y = 0 \\ \omega_z(t) &= (\omega_o)_z = 0 \end{aligned} \tag{3.78}$$

For a numerical example, we pick our typical spacecraft (Figure 2.6) with a rigid bus which is a $1\text{m} \times 1\text{m} \times 1\text{m}$ cube and 2 appendages attached at mid-way along two opposing sides of the rigid bus. This configuration will result in a diagonal inertia matrix and the attachment vectors are $\vec{d}_1^T = [-0.5 \;\; 0.5 \;\; 0.0]\text{m}$ and $\vec{d}_2^T = [0.5 \;\; -0.5 \;\; 0.0]\text{m}$. A summary of these and other system parameters can be found in Table 3.13.

Table 3.13: Specifications for the flexible spacecraft used for validation

| |
|---|
| Rigid bus mass: $m_c = 2000\text{kg}$ |
| Rigid bus dimension: $1\text{m} \times 1\text{m} \times 1\text{m}$ cube |
| Rigid bus inertia: $\underline{I}_c = 333.\overline{3}\,\underline{i}\ \ \text{kg}\,\text{m}^2$; $\underline{i}$ is the identity matrix |
| Number of appendages: $n = 2$ |
| Appendage mass: $m_i = 100\text{kg}$ |
| Number of modes: $p_i = q_i = 6$ |
| Appendages locations: $\vec{d}_1^T = [-0.5\ \ 0.5\ \ 0.0]m$ and $\vec{d}_2^T = [0.5\ \ -0.5\ \ 0.0]m$ |
| Modulus of elasticity: $E_i = 5.0 \times 10^8\ \text{N/m}^2$ |
| Poisson ratio: $\gamma_i = 0.3$ |
| Width: $a_i = 1\text{m}$ |
| Length: $b_i = 10\text{m}$ |
| Thickness: $h_i = 0.02\text{m}$ |
| Area density: $\rho_i = 10\text{kg/m}^2$ |
| Initial angular rate: $\vec{\omega}^T = [0.1\ \ 0.0\ \ 0.0]\text{rad/sec}$ |
| Initial deflection: $\vec{\chi}_o = \vec{0}$ |
| Initial rate of deflection: $\dot{\vec{\chi}}_o = \vec{0}$ |

Note: Both appendages have the same specifications (i.e. $i = 1, 2$). The total system inertia matrix, $\underline{I}_t$, is diagonal with entries $I_x = 8050\ \text{kg}\,\text{m}^2$, $I_y = 350\ \text{kg}\,\text{m}^2$ and $I_z = 8067\ \text{kg}\,\text{m}^2$.

With the above setting, the simulation was run for 200 sec. and $\omega_y$ and $\omega_z$ were observed to remain at zero and $\omega_x$ remained at its initial value of 0.1 rad/sec as expected. The time evolution of the Euler angles are shown in Figure 3.17 which clearly indicates that the spacecraft is rotating about its *x*-axis with a constant rate of $\approx 6\,\text{deg/sec}$ (i.e. slope of each saw tooth in Fig. 3.17). Figure 3.17 looks discontinuous, however in reality, the angle is continuously and smoothly oscillating between $\pm 180^o$, where attitude angle $+180^o$ is of course the same as $-180^o$. There are no appendage vibrations in this case, as can be deduced from the zero potential energy plot in Figure 3.18.

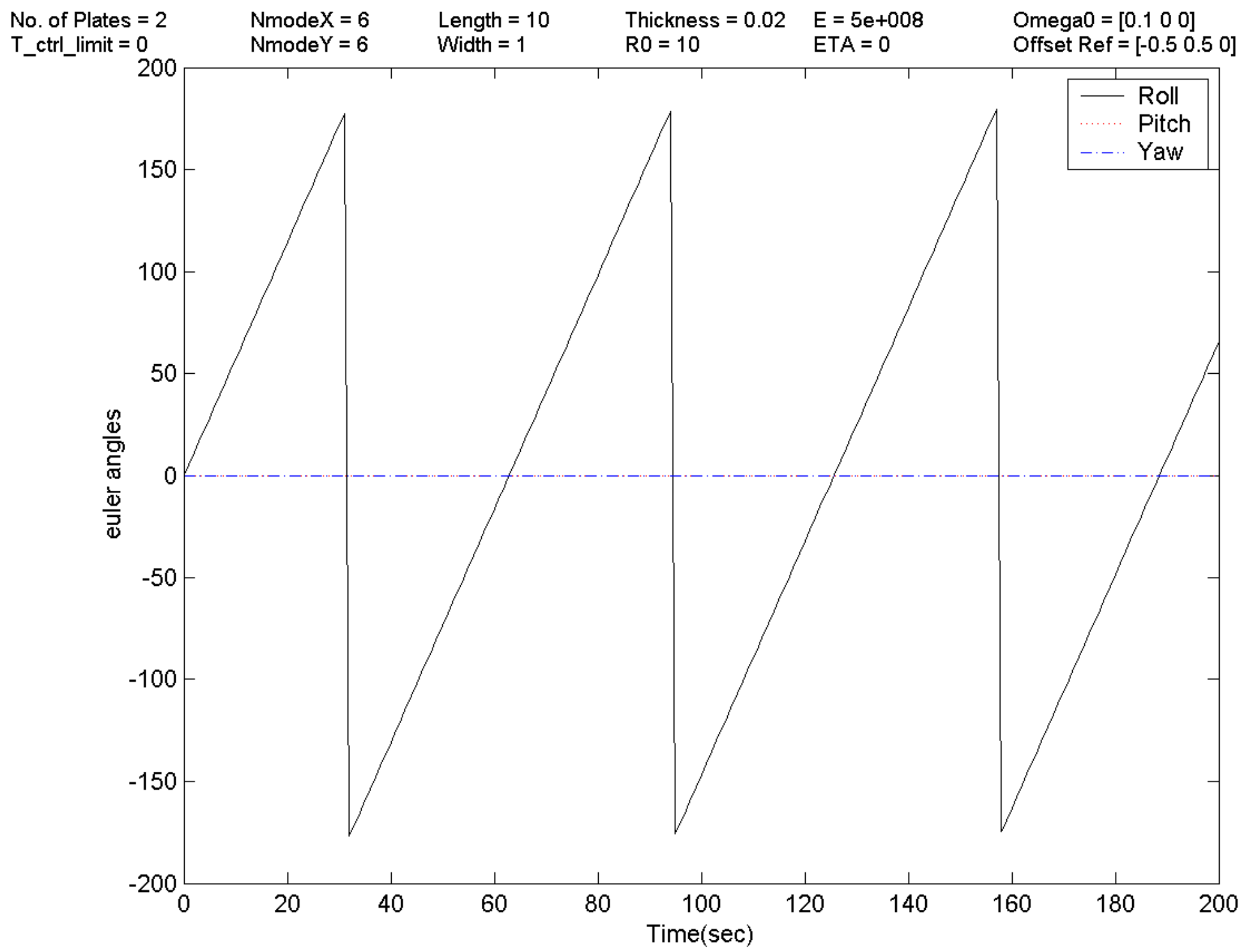

Figure 3.17: Validation case 12 - Spacecraft attitude (Euler angles)

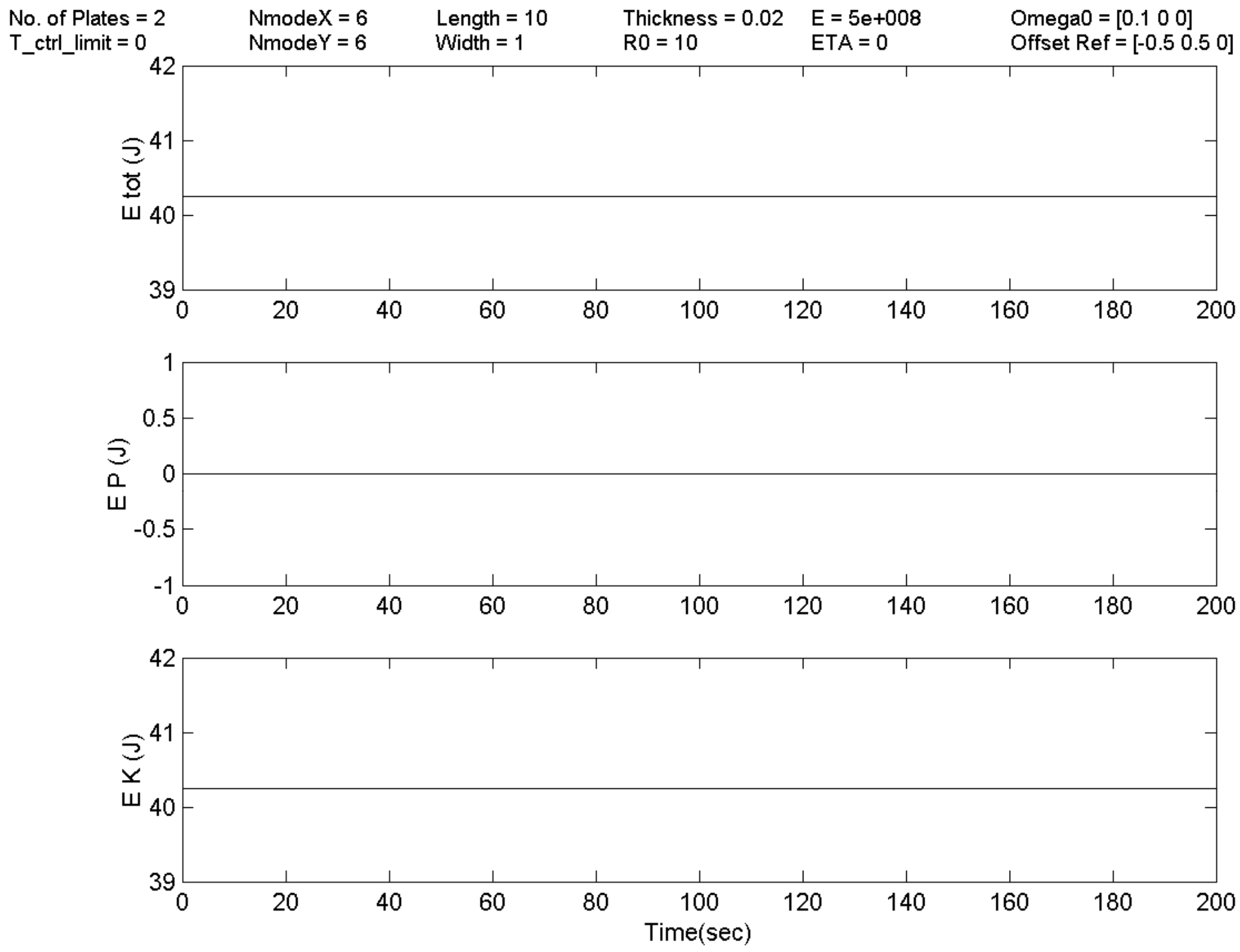

Figure 3.18: Validation case 12 - Spacecraft total, potential, and kinetic energies

Validation Case 13

This is a repeat of case 12 except that for this case $(\omega_o)_x = (\omega_o)_z = 0$ and $(\omega_o)_y = 0.1\,\text{rad/sec}$, hence all the analysis done in case 12 apply here. Using the settings in Table 3.13 (except for the initial angular rate), the simulation was run for 200 seconds and $\omega_x$ and $\omega_z$ were observed to remain at zero and $\omega_y$ remained at its initial value of 0.1 rad/sec, as expected. This time the spacecraft starts to rotate about its *y*-axis with the constant rate of $\approx 6\,\text{deg/sec}$ as seen in Figure 3.19. Note that unlike all the other cases presented in this thesis where the (3,2,1) Euler rotation sequence has been used to obtain the Euler angles from the quaternions, for the present case, the (3,1,2) Euler rotation sequence was used in order to make the pure pitch spin detectable. Again, since there are no acceleration or disturbing torques of any kind, no appendage vibrations are induced and the potential energy of the system, which is due entirely to the elastic deformation of the appendages, remains zero. Figure 3.20 shows that the system energy which is due to the rotational kinetic energy of the spacecraft ($E = 1.75\,\text{J}$) is much less than its counterpart value in case 12 ($E = 40.25\,\text{J}$). This is expected as the *y*-axis inertia term ($I_y = 350\,\text{kgm}^2$) is smaller than the *x*-axis inertia term ($I_x = 8050\,\text{kgm}^2$). In fact this case represents a rotation about the spacecraft's minor axis.

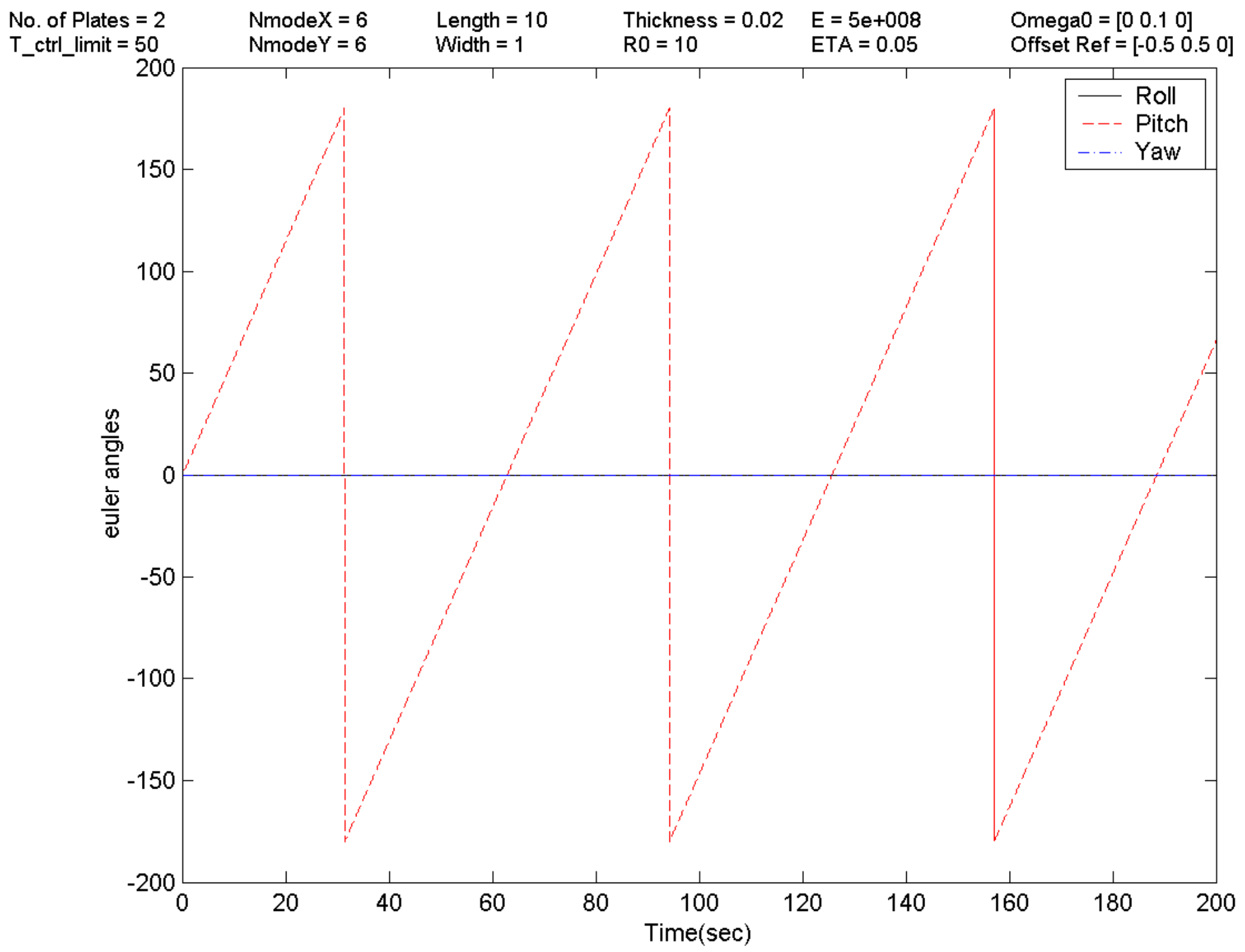

Figure 3.19: Validation case 13 - Spacecraft attitude (Euler angles)

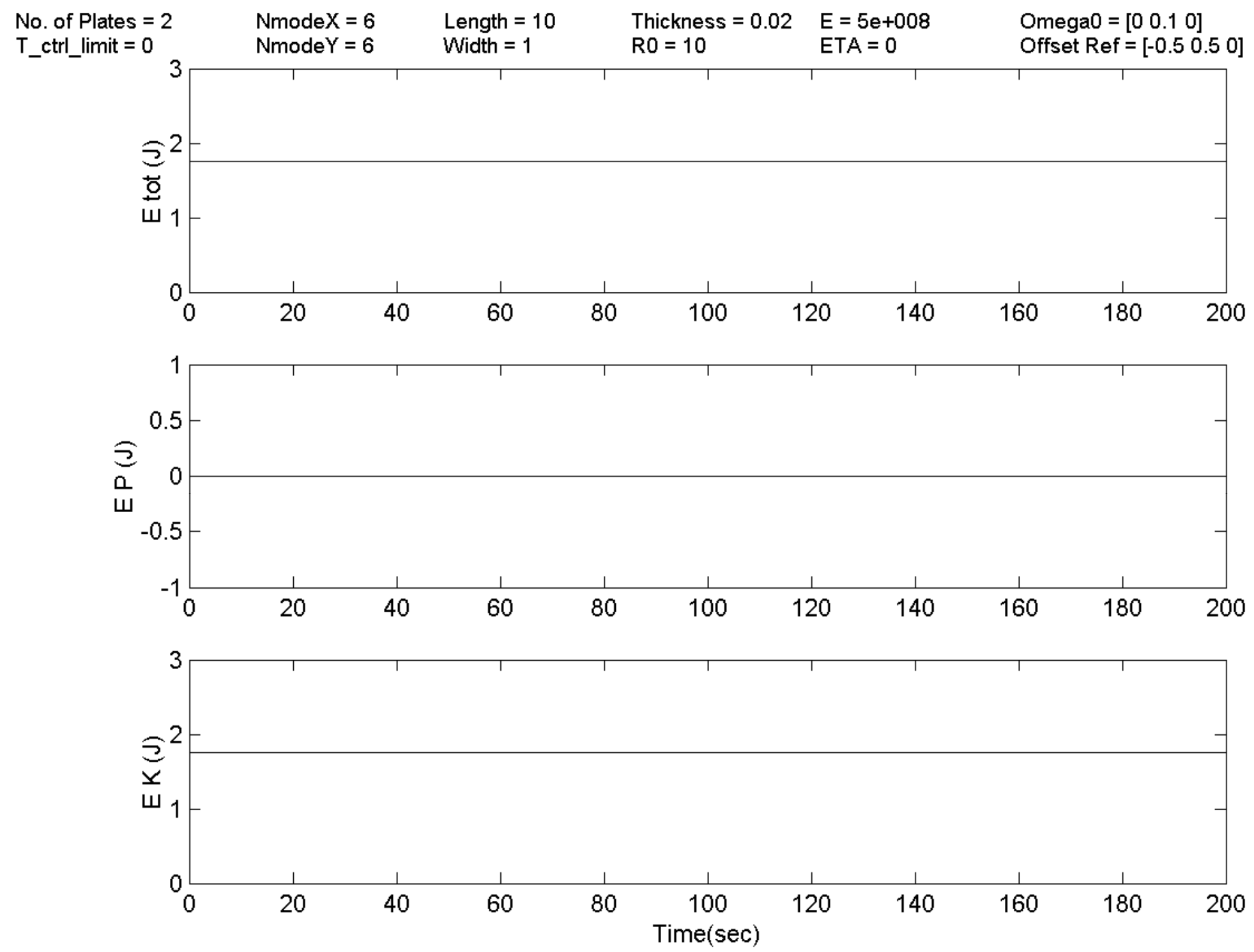

Figure 3.20: Validation case 13 - Spacecraft total, potential, and kinetic energies

Validation Case 14

This is a repeat of case 12 except that for this case $(\omega_o)_x = (\omega_o)_y = 0$ and $(\omega_o)_z = -0.2$ rad/sec. Using the settings in Table 3.13 (except for the initial angular rate), the simulation was run for 200 sec. and $\omega_x$ and $\omega_y$ were observed to remain at zero and $\omega_z$ remained at its initial value of -0.2 rad/sec, as expected. This time the spacecraft starts to rotate about its z-axis with the constant rate of $\approx -12\,\mathrm{deg/sec}$ as seen in Figure 3.21. Compared with Figure 3.19 in case 13, the Euler angle starts in the negative direction in this case and the saw teeth have steeper slopes, in fact the magnitude of slope in Figure 3.21 in double the one in Figure 3.19, as it is anticipated. In this case as was in cases 12 and 13, because there are no acceleration or disturbing torques, the appendages are not deflected and the potential energy of the system remains zero.

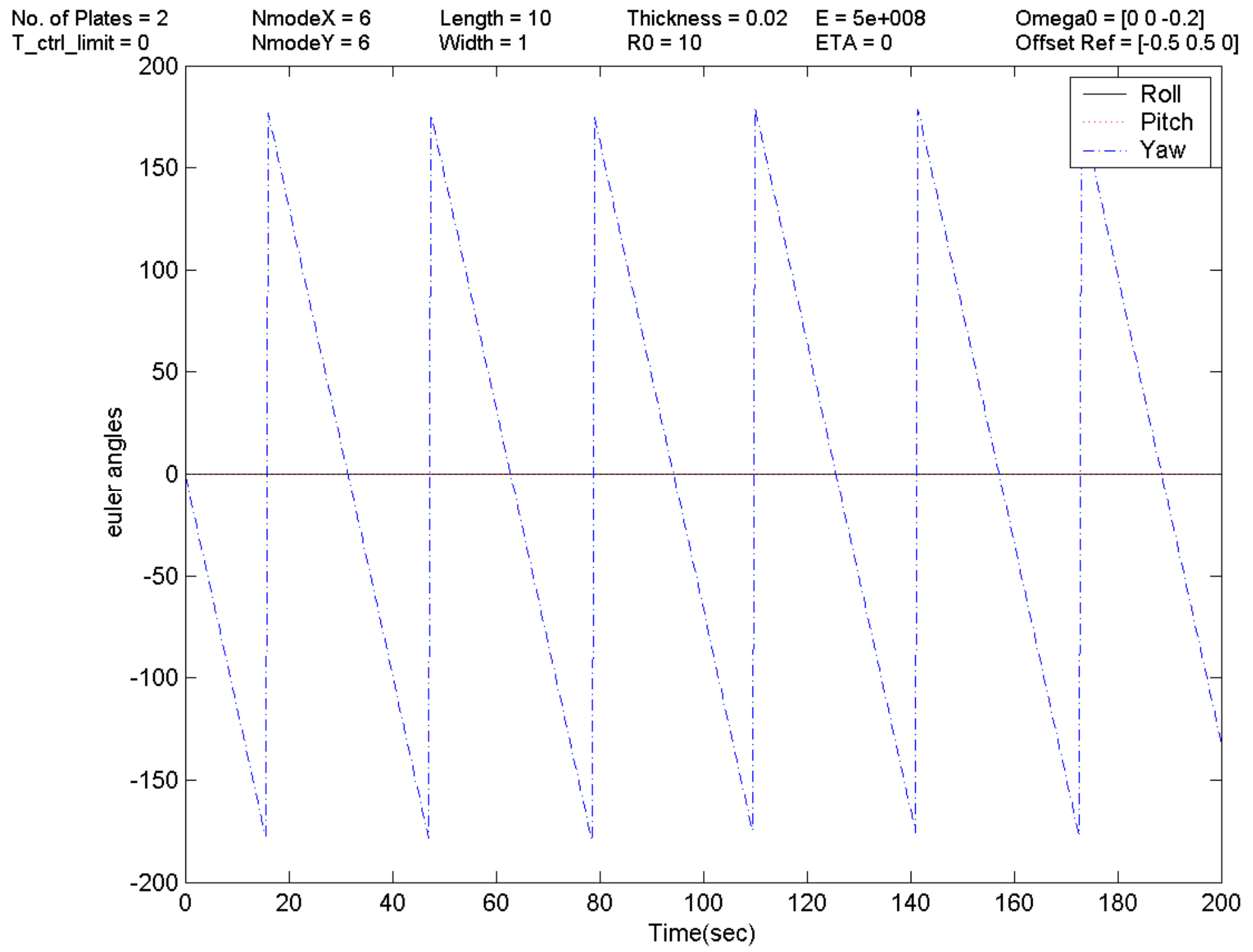

Figure 3.21: Validation case 14 - Spacecraft attitude (Euler angles)

Validation Case 15

The next two cases (cases 15 and 16) are the most general validation cases for the flexible spacecraft, as they include accelerating motion and hence appendage vibrations. The simulator was set up as per specifications in Table 3.13, except that in this case we have added two extra appendages for a total of four (i.e. $n$=4), all of which have now 3 modes in each $x$ and $y$ direction (i.e. $p_i = q_i = 3$) bringing the total system dimension to $m = 3 + 2(4)(3)(3) = 75$. The two new appendages are attached at the following locations: $\vec{d}_3^T = [-0.5 \ -0.5 \ 0.0]\text{m}$ and $\vec{d}_4^T = [0.5 \ 0.5 \ 0.0]\text{m}$. The initial angular body rate is selected to be $\vec{\omega}_o = [0.1 \ \ -0.1 \ \ 0.1]^T$ rad/sec which could represent an unknown malfunction or simply the launcher's tip-off rates. Also, the modal damping of the appendages has been set to zero in order to show the correctness of the simulation results. As expected in this case, the total energy (Figure 3.22) and angular momentum of the system (not shown), are conserved (i.e. constant). The spacecraft angular body rate and the appendages deflections are shown in Figures 3.23 and 3.24, respectively.

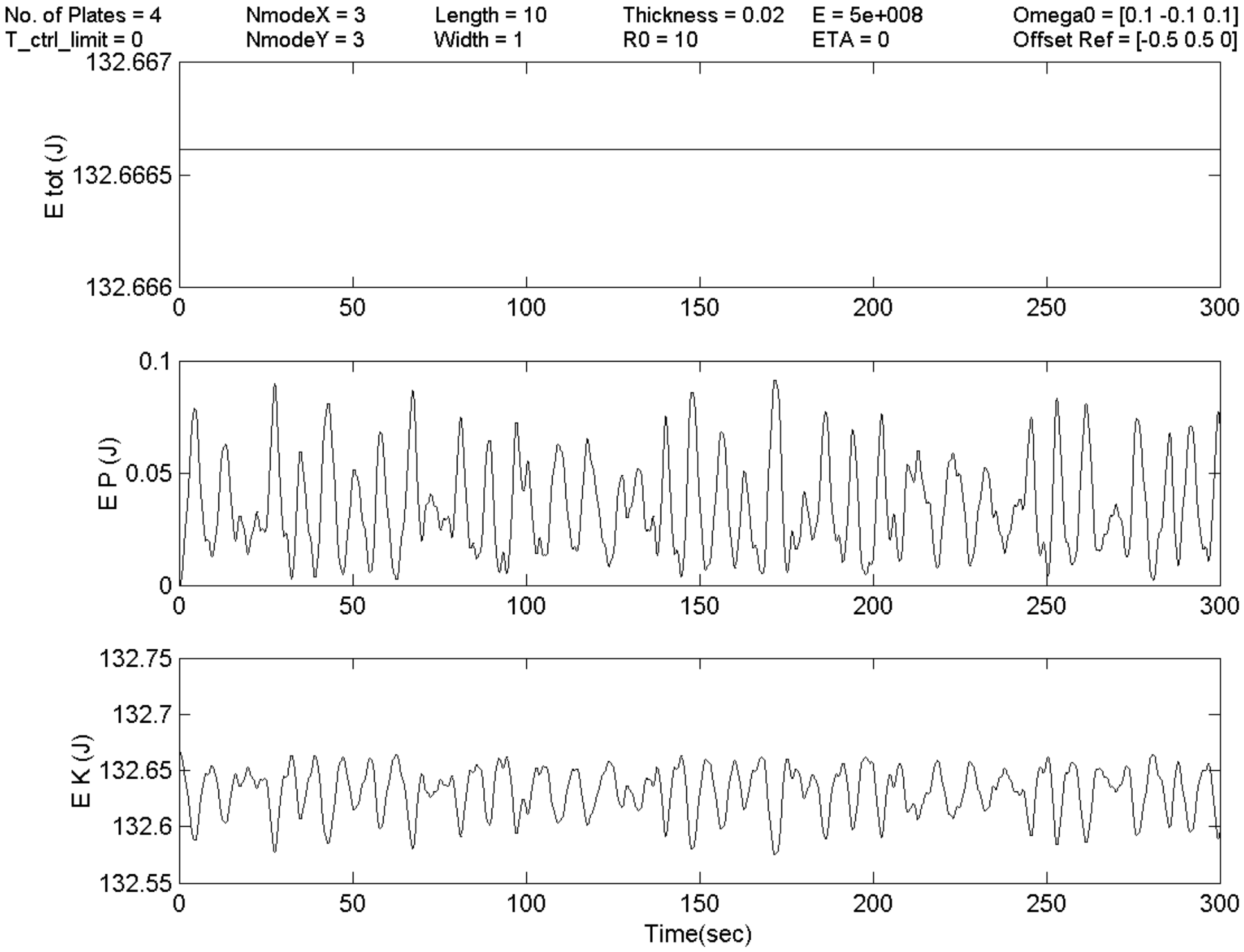

Figure 3.22 Validation case 15 - Spacecraft total, potential, and kinetic energies

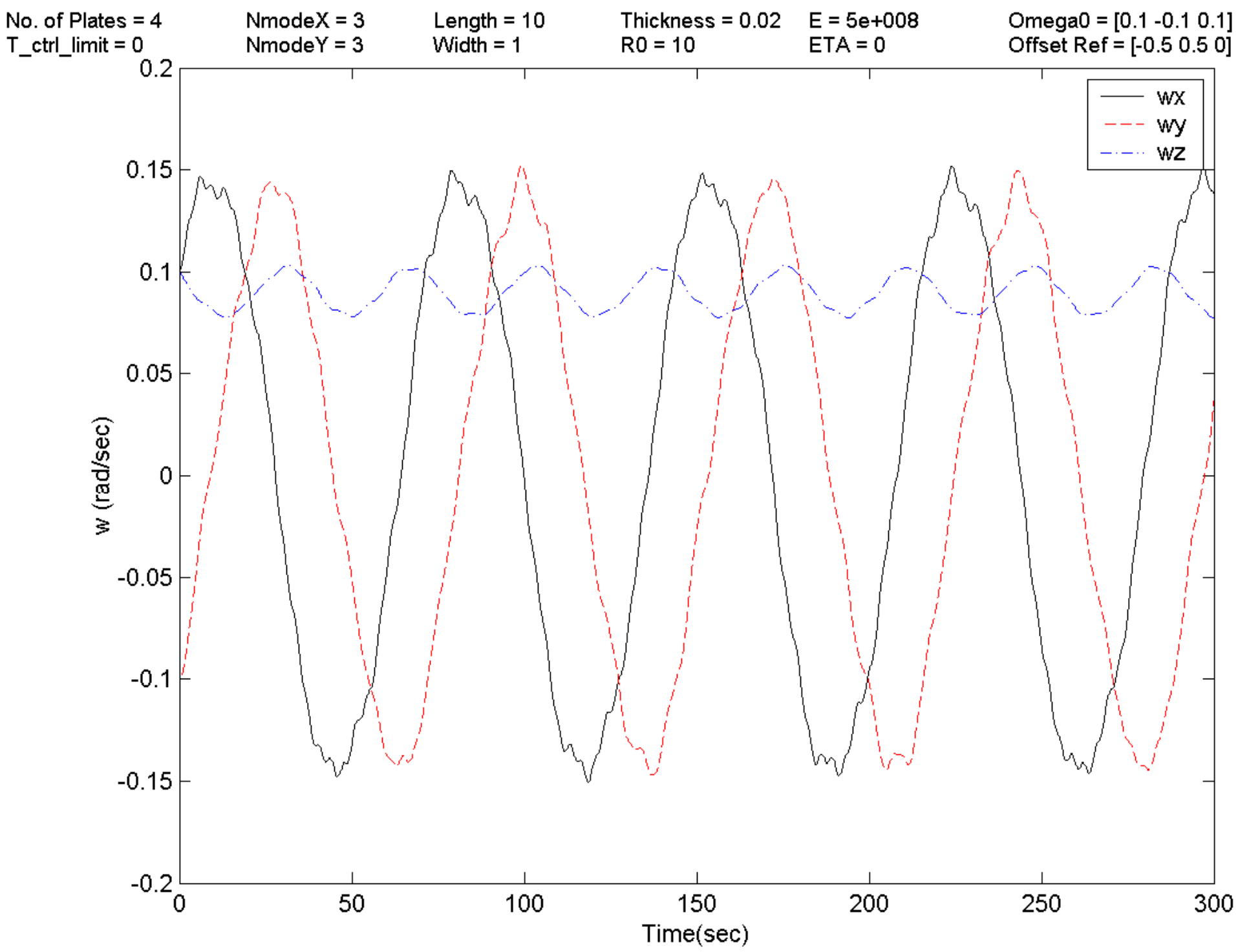

Figure. 3.23 Validation case 15 - Spacecraft body angular rates ( $\vec{\omega}$ )

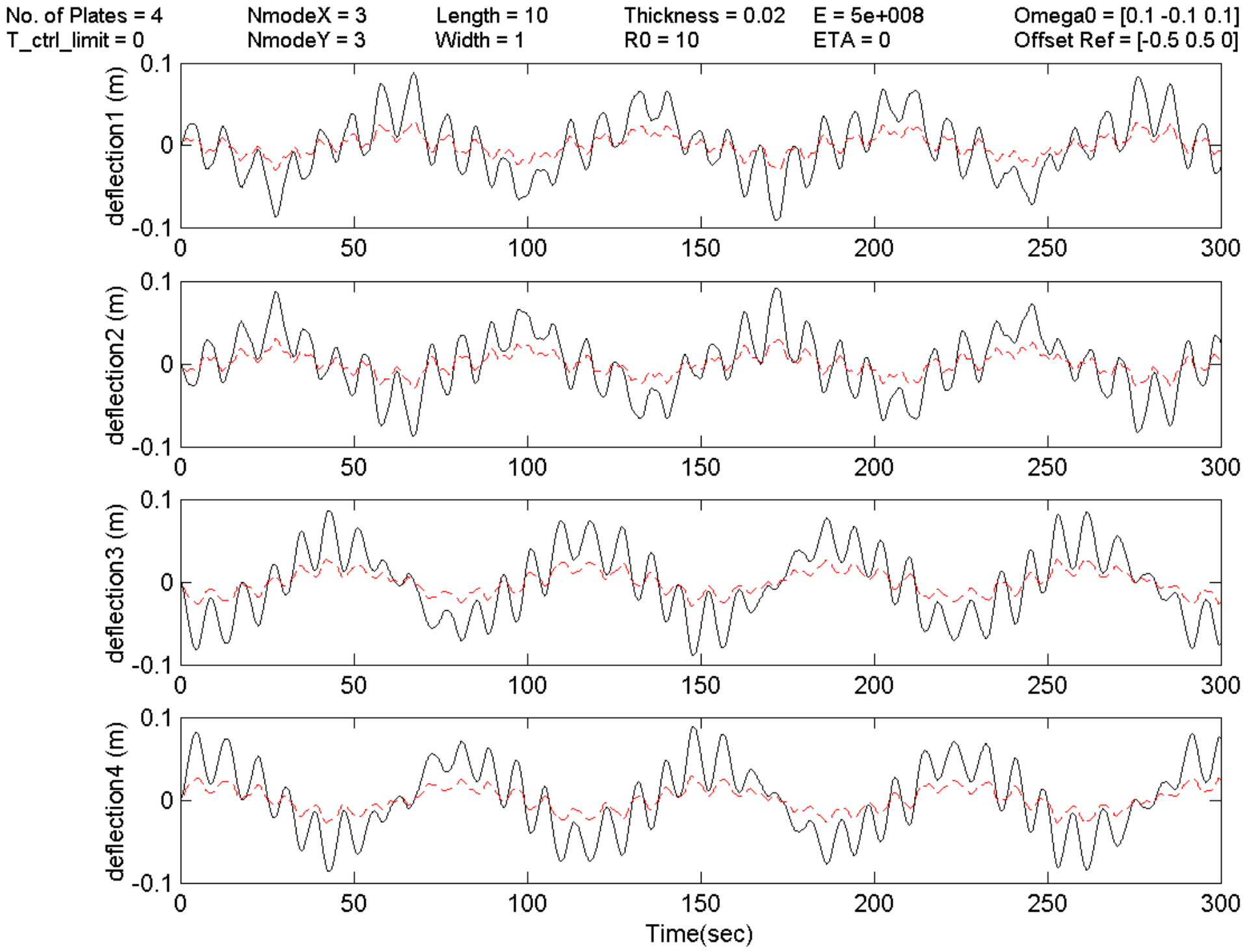

Figure 3.24 Validation case 15 - Deflection ( $w$ ) at tip $x = .5; y = 10$ and midway $x = .5; y = 5$

Validation Case 16

This is a repeat of case 15 except that for this case $(\omega_o)_x = (\omega_o)_y = (\omega_o)_z = 0$ and we subject the spacecraft to an actual large thruster misfire of $\vec{\tau} = [100 \quad -100 \quad 100]^T$ Nm which starts at the 5 seconds mark and lasts for 2 seconds. In this case, a single appendage is included which is clamped mid-way along the right side of the rigid bus and has the same properties as described in Table 3.13 with 2 modes in each $x$ and $y$ directions of the appendage (i.e. $p = q = 2$), bringing the total system dimension to $m = 3 + 2(1)(2)(2) = 11$. Again, the modal damping of the appendage has been set to zero in order to show the correctness of the simulation results. As expected in this case also, the total energy (Figure 3.25) and angular momentum of the system (not shown), are conserved, after the initial energy injection due to the thrusters' disturbance torque. The spacecraft angular body rate and appendage deflections are also shown in Figures 3.26 and 3.27, respectively.

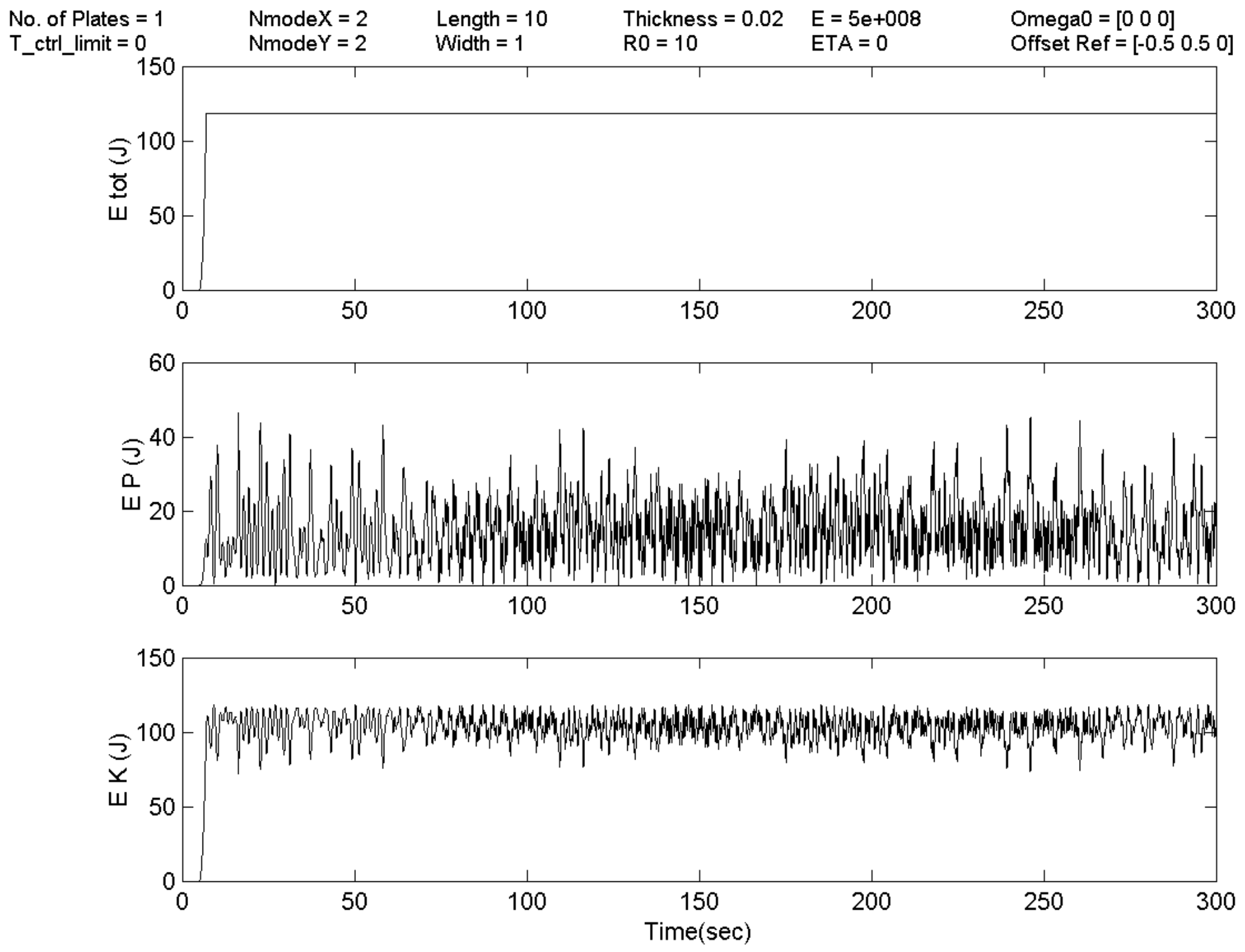

Figure 3.25 Validation case 16 - Spacecraft total, potential, and kinetic energies

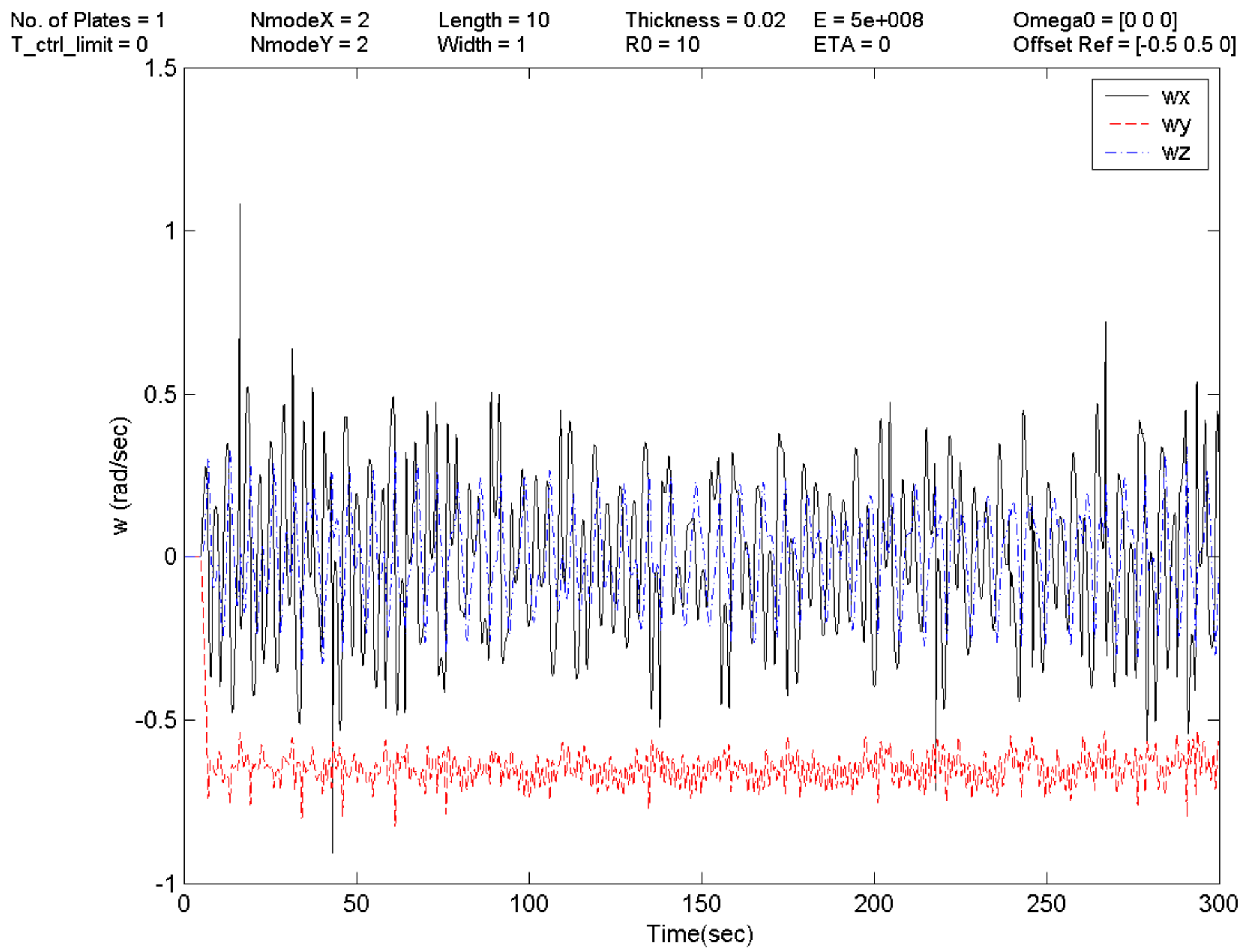

Figure. 3.26 Validation case 16 - Spacecraft body angular rates ( $\vec{\omega}$ )

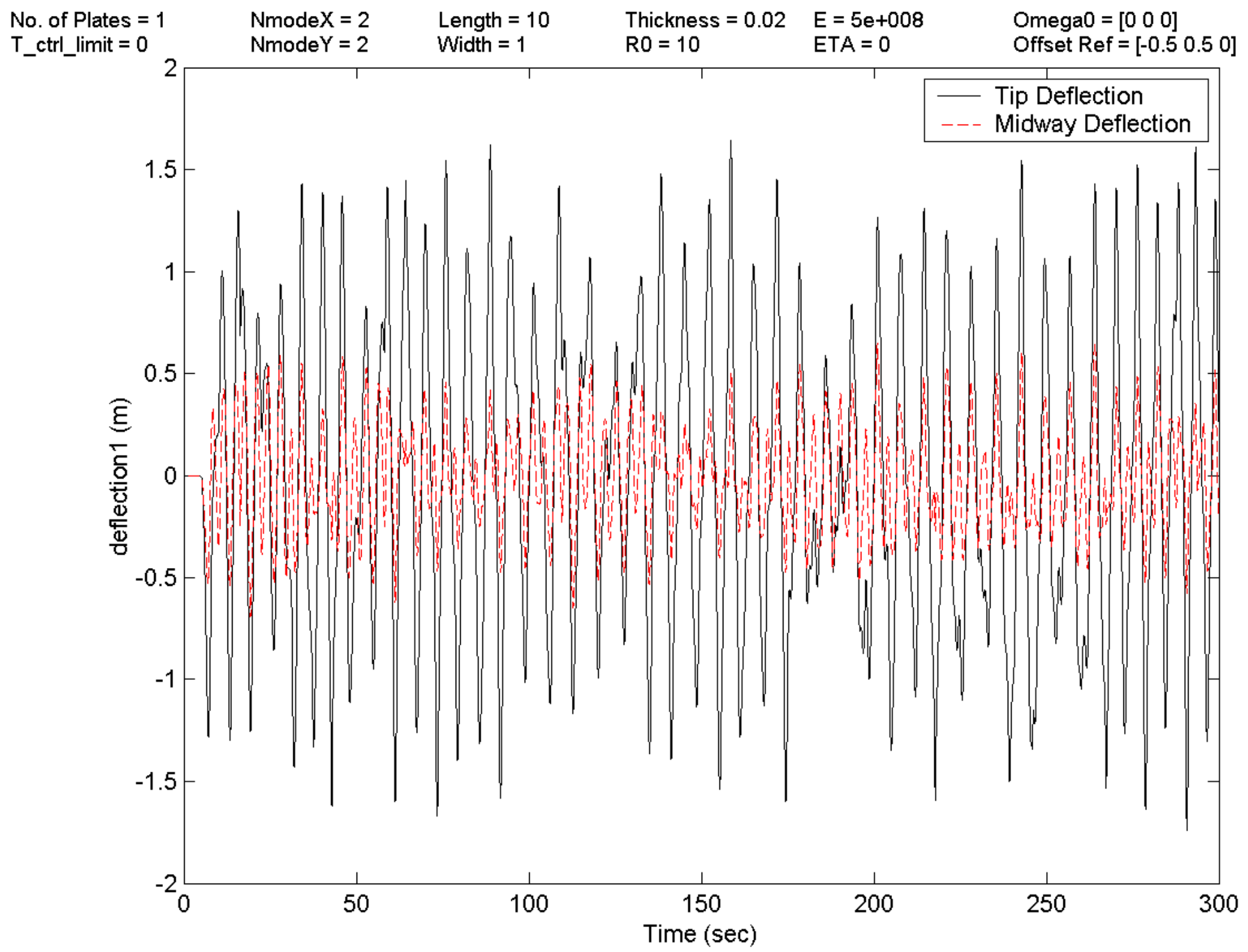

Figure 3.27 Validation case 16 - Deflection ( $w$ ) at tip $x = .5; y = 10$ and midway $x = .5; y = 5$

## 3.5 Concluding Remarks

A survey of simulators, which have been mentioned in the literature and include both rigid and flexible dynamics, was undertaken and resulted in the identification of several possible candidates for use in this thesis. However, after careful study of these simulators, it was found that one or several of the following drawbacks were present:

(i) they do not include all the required functionalities;
(ii) they are provided without source code;
(iii) they are expensive for an academic setting;
(iv) they are not user-friendly.

Therefore for the purposes of this thesis, a high-fidelity, user-friendly modeling and control software environment was developed. The dynamics engine of this simulation environment is based on the generic formulation presented in Chapter 2. In this chapter, we covered the architecture and environment of the simulator and provided detailed study of the important topic of the simulation validation where closed form solutions of some rigid and flexible dynamic cases and conservation principles (e.g. conservation of energy and angular momentum) were utilized to achieve this validation goal. It is envisaged that the software simulator will be further validated through comparative simulation studies with one of the commercially or otherwise available simulators, and eventually using actual flight data obtained from a flexible spacecraft.

In the next two chapters, the simulator is the key tool used for the investigation of the attitude recovery of both rigid and flexible spacecrafts using the input-output feedback linearization technique.

# CHAPTER 4

## RIGID SPACECRAFT ATTITUDE RECOVERY SYSTEM

The attitude recovery for rigid spacecraft is investigated in this chapter. Given that there are three outputs which are the same as the system states (e.g. 3 of the 4 quaternion parameters) and that need to be controlled using three inputs (e.g. 3 torques), full state linearization can easily be achieved. The output error dynamics equation which is used to determine stable controller gains will be derived in Section 4.1. Sections 4.2 and 4.3 present results of the attitude recovery simulations performed using the proposed controller for a rigid spacecraft without or with parameter and model uncertainties, respectively. The simulation cases are for two different failures severely affecting the attitude motion of the spacecraft with some cases including parameter uncertainties to study the robustness of the controller. The source of the very large disturbance torque is not really important and has nothing to do with the main thrust of the thesis. The large disturbance is being used only to show that even under exceptionally severe conditions, the proposed controller would complete a maneuver and remain stable.

## 4.1 Controller based on Input-Output Feedback Linearization

The problem of rigid attitude recovery can be stated as the regulating (or zeroing) of the Euler angle errors or angular rates given a specified reference and starting from an unknown initial condition. Otherwise stated mathematically there are two different scenarios:

(i) Zeroing of angular body rates:

$$\lim_{t \to \infty} \vec{\omega} = 0 \tag{4.1}$$

(ii) Regulating Euler angles (or quaternions):

$$\lim_{t \to \infty} \vec{q} = \vec{q}_{\mathrm{ref}}, \quad \lim_{t \to \infty} \dot{\vec{q}} = 0 \tag{4.2}$$

Case (ii) is the more general case, as it also implicitly requires that $\lim_{t \to \infty} \vec{\omega} = 0$. Case (i) is the less restrictive, as it does not require the attitude recovery maneuver to point the spacecraft in any particular direction, it only requires that the body's angular motion is brought to zero.

We make the assumption that attitude knowledge of the spacecraft is available to the controller through the use of gyros, with the possible addition of a magnetometer. In fact, after some careful thought about our attitude determination problem, it was found that many standard attitude sensors, such as Earth, Sun and star sensors, are not ideal for the attitude recovery system proposed, specially with the fast motions sometimes involved. These is mainly due to the fact that these sensors are made to work on a fairly stable spacecraft with slow motions. In reality, any sensors relying on the environment (e.g. Earth, Sun, Moon, stars) would not be appropriate and it seems that the only viable sensor in case of rapid attitude recovery maneuvers would be a set of mechanical or laser gyros. The magnetometer has also been used as an attitude sensor in the past (e.g. Radarsat) and in our case, its 3-axis magnetic field measurements could be used to complement the gyro data (as a separate source of measurement). Of course, there would be a need for processing of these raw attitude data obtained from the sensors to calculate the quaternions used by the controller, and this processing is performed by the attitude determination algorithms which is a major field of investigation on its own right [6].

As for actuation, reaction wheels can not muster up enough torque authority and the two real viable choices with today's existing actuators are thrusters (e.g. bi-propellants) or control moment gyros (capable of producing very large control torques). In this thesis, we have assumed the availability of control moment gyros, providing continuous, smooth and high authority control in all three spacecraft axis. However, the dynamics of a control moment gyro has not been modeled, nor has any internal angular momentum been included in the simulations presented. Having made the above assumptions, it is understood that for a real implementation of an attitude recovery system on an actual spacecraft, considerable effort would be needed to study the effects of actuator non-linearities, sensor time delays, noise, etc. on the control system, and many of these issues are identified for future work in Chapter 7.

In order to design attitude controllers, we could linearize the nonlinear attitude dynamics equations, about an operating point and then use linear control theory to design a simple stable controller. However, this linear controller will not be suitable for large re-orientation or attitude recovery maneuvers which will take the satellite away from the operating/equilibrium point and invoke the non-linearities inherent in the attitude dynamics. In cases where we need large maneuvers, it is best to directly use the nonlinear dynamic equations and modern control theory (e.g. feedback linearization) to design more capable control algorithm.

In order to implement the Feedback Linearization Controller (FLC), we start with the attitude dynamics equation (2.20) which is in a form suitable for feedback linearization with the state variable, $\vec{q}$. To proceed, let us set the feedback linearization control torque to be:

$$\vec{\tau} = \underline{G}_{\vec{q}}^{-1}[-\vec{f}_{\vec{q}} + \vec{v}] \tag{4.3}$$

By substituting (4.3) into (2.20), we will obtain a linear relationship between the output/states and the new input $\vec{v}$ :

$$\ddot{\vec{q}} = \vec{v} \tag{4.4}$$

We will now propose and define a control error signal based on addition of the vector parts of the actual and reference quaternions:

$$\vec{q}_e = \vec{q}_{\text{ref}} - \vec{q} \tag{4.5}$$

In is important to note that the error signal (4.5) is not the standard quaternion error vector found in literature for control purposes, which is calculated using multiplication as opposed to addition. For example, [143]-[144] define the quaternion error vector as

$$\vec{q}_{\text{error}} = \begin{bmatrix} -q_1 & q_0 & q_3 & -q_2 \\ -q_2 & -q_3 & q_0 & q_1 \\ -q_3 & q_2 & -q_1 & q_0 \end{bmatrix} \tilde{q}_{\text{ref}} \qquad \text{where} \quad \tilde{q}_{\text{ref}} = \begin{bmatrix} q_{0\text{ref}} \\ \vec{q}_{\text{ref}} \end{bmatrix} \tag{4.6}$$

or equivalently expressed as $\vec{q}_{\text{error}} = q_0 \vec{q}_{\text{ref}} - q_{0\text{ref}} \vec{q} - \vec{q}^{\times} \vec{q}_{\text{ref}}$.

The quaternion error vector (4.6) gives the actual rotational error between the two reference frames expressed by a quaternion where in our case, the error term constructed by the addition of quaternions does not represent a rotation. A few words are then in order regarding the geometrical interpretation of a quaternion addition given by $\tilde{q}_e = \tilde{q}_{\text{ref}} - \tilde{q}$ where $\tilde{q}_{\text{ref}}$ is the reference/target attitude and $\tilde{q}$ is the present attitude, both expressed with respect to the same reference frame. Considering the quaternion as a four dimensional (4D) vector, then $\tilde{q}_e$ represents the error vector between the target and the present attitudes in the 4D vector space. As mentioned before however, $\tilde{q}_e$ does not represent a rotation and the quaternion unitary condition no longer holds for this special definition of error vector. Now, the quaternion unitary condition must hold for both $\tilde{q}$ and $\tilde{q}_{ref}$ separately and hence imposes a restriction on the location where these quaternions can be in our 4D vector space. The set of all the allowable points for the quaternions representing 3D attitude motion in the 4D space is the 4D unit hypersphere given by the unitary condition. The attitude motion of the spacecraft is then represented by the trajectory traversed on the surface of the hypersphere starting from the present to the reference point and the error term $\vec{q}_e$ in (4.5) can be thought of as the projection of $\tilde{q}_e$ onto the 3D space of $\vec{q}$.

Indeed, quaternion addition has only been used recently for control purposes, by Lin and Lin [107]-[109]. However, their approach has a major problem of non-uniqueness which is addressed in this thesis. This non-uniqueness refers to the existence of the two solutions (i.e. equilibrium points where $q_0 = \pm\sqrt{1-\vec{q}^T\vec{q}}$) if the control signal is only controlling the vector part of the quaternion, $\vec{q}$. Physically what happens is that as the control signal based on the error vector defined in (4.5) is driven to zero, the spacecraft can converge to two different attitudes which at worst case can be 180 degrees apart! (i.e. when $\vec{q}^T\vec{q} = 1/2$ such that $q_0 = \pm\sqrt{1-\vec{q}^T\vec{q}} = \pm\sqrt{2}/2$ and hence the Euler angle can be $\pm 90^{\circ}$). This problem can be solved by taking the vector part of the reference quaternion to be $\vec{q}_{\text{ref}} = [0\,0\,0]^T$. In this case the $\vec{q}^T\vec{q} = 0$ such that $q_0 = \pm\sqrt{1-0} = \pm 1$ and hence the Euler angle can be 0 or 360 degrees which represents the same spacecraft attitude. In order to achieve this, one would implement in the attitude determination algorithm the appropriate transformation so that the attitude of the spacecraft body is given with respect to the control or recovery reference frame (e.g. orbital reference frame, celestial reference frame or any other appropriate reference frame pre-selected for the attitude recovery maneuver). Hence by regulating or zeroing $\vec{q}_e$, as defined in (4.5), the spacecraft reference frame will converge towards and eventually coincide with the recovery reference frame (e.g. for simulations performed in this thesis, the celestial inertial frame was chosen to be the target attitude recovery frame). Of course, depending on $q_0$ converging to +1 or –1, the spacecraft attitude motion in 3D space (and the attitude path on the 4D hypersphere) will be different, but this is not of concern in an attitude recovery maneuver, as long as the reference/target attitude is finally achieved.

It is important to note that *infinitely* many other solutions, besides $\vec{q}_{\text{ref}} = [0\ 0\ 0]^T$, exist as a valid $\vec{q}_{\text{ref}}$ command such that a unique attitude is obtained. These solutions all lie on the surface of the 3D unit sphere defined by $\vec{q}^T\vec{q} = q_1^2 + q_2^2 + q_3^2 = 1$ such that, $q_0 = \pm\sqrt{1-1} = 0$ and hence, $q_0$ is always unique. In fact, $q_0 = 0$ gives the Euler angle $\pm 180^o$ which results in the same spacecraft attitude. Without loss of generality, we have selected $\vec{q}_{\text{ref}} = [0\ 0\ 0]^T$ in all the simulation cases presented in this and the next chapter.

We assume a PD control law for the new input torque $\vec{v}$ in (4.4) which is given by:

$$\vec{v} = \underline{K}_{Pq}\vec{q}_e + \underline{K}_{Dq}\dot{\vec{q}}_e + \ddot{\vec{q}}_{\text{ref}} \tag{4.7}$$

Substituting (4.3) and (4.7) into (2.20) gives:

$$\ddot{\vec{q}} = \vec{f}_{\bar{q}} + \underline{G}_{\bar{q}}\underline{G}_{\bar{q}}^{-1}[-\vec{f}_{\bar{q}} + \underline{K}_{Pq}\vec{q}_e + \underline{K}_{Dq}\dot{\vec{q}}_e + \ddot{\vec{q}}_{\text{ref}}] \tag{4.8}$$

Simplifying (4.8), we get:

$$\ddot{\vec{q}} = \underline{K}_{Pq}\vec{q}_e + \underline{K}_{Dq}\dot{\vec{q}}_e + \ddot{\vec{q}}_{\text{ref}} \tag{4.9}$$

and hence the quaternion error equation becomes:

$$\ddot{\vec{q}}_e + \underline{K}_{Dq}\dot{\vec{q}}_e + \underline{K}_{Pq}\vec{q}_e = 0 \tag{4.10}$$

The stability of (4.10) can be guaranteed by selecting diagonal positive definite gain matrices $\underline{K}_{Dq}$ and $\underline{K}_{Pq}$ so that we obtain three decoupled, linear time-invariant, second order systems which are asymptotically stable. Gains for each second order system can be selected to satisfy control design specifications (e.g. settling time, overshoot, etc.).

The simulation results for the controller are presented next. Note that in all simulation cases presented in this chapter, the Roll, Pitch and Yaw refer to Euler angles defining the orientation between the spacecraft and the celestial reference frame, since an inertial pointing maneuver is being considered. In all cases, the (3,2,1) Euler rotation sequence was used to obtain the Euler angles from the quaternions. Also, it is noteworthy to underline the fact that the spacecraft used for simulation purposes does not include models of the ACS sensors, actuators and attitude determination algorithms.

## 4.2 Simulation Results for the Rigid Spacecraft

The feedback linearization controller was used to recover a rigid spacecraft from 2 different malfunctions which throw the spacecraft into a tumble. The controller must proceed to detumble the spacecraft towards a safe hold mode of zero angular velocity and some inertial attitude pointing. Without loss of generality, we will set the reference attitude pointing to be the vernal equinox direction which means that the controller must align the spacecraft reference frame with the celestial reference frame. In doing so, the controller will be reducing the imparted kinetic energy back to zero. The spacecraft has a configuration as shown in Fig. 2.6 with a cube rigid bus of dimensions $1\text{m}\times 1\text{m}\times 1\text{m}$ and mass 200 kg and two *rigid* plate-type appendages with area density of $\rho = 10\ \text{kg/m}^2$, width $a = 1\ \text{m}$, different lengths $b_1 = 2\ \text{m}$ and $b_2 = 4\ \text{m}$ and thickness $h = 0.02\ \text{m}$. The first appendage is clamped at the bottom of the right side panel, $\vec{d}_1^T = \left[-0.5\ 0.5\ 0.5\right]$ m and the second appendage is clamped at the top of the left side panel, $\vec{d}_2^T = \left[-0.5\ -0.5\ -0.5\right]$ m. With this configuration, the spacecraft's inertia matrix with its appendages fully deployed is:

$$\underline{I}_t = \begin{bmatrix} 143.3 & 60 & 30 \\ 60 & 193.3 & -35 \\ 30 & -35 & 273.3 \end{bmatrix} \text{kgm}^2$$

Finally, controller gains $\underline{K}_{Pq} = 0.08\underline{i}$ and $\underline{K}_{Dq} = 0.57\underline{i}$ were used in (4.7) where $\underline{i}$ is the $3\times 3$ identity matrix.

*Simulation Case 1*: As the first Attitude Recovery Maneuver (ARM) simulation case for the rigid spacecraft (Figs. 4.1-4.4), the closed loop response of the system to an initial disturbance is studied. The disturbance is due to some malfunction which causes an initial large disturbance torque mainly about the *x*-axis given by $\vec{\tau} = \begin{bmatrix} 100 & -10 & -10 \end{bmatrix}^T$ Nm. The disturbance is assumed to start at the 5-seconds mark (e.g. $t = 5$) and lasts for 5 seconds, with constrained control torques ($\pm 50Nm$), starting 10 seconds after the end of the disturbance torque (i.e. $t = 20$) to allow for extra deviation of the spacecraft's states from the equilibrium point.

It is noted that this is a severe malfunction which throws the spacecraft into spin about all three axis such that the attitude angles go through several $360^o$ rotation (Fig. 4.2) within the first 30 seconds of the maneuver before staring to stabilize towards zero around $t = 40$ seconds. In fact, we see that the *y*-axis angular rate ($\omega_y$) fluctuates about the –3 rad/sec mark which is a very fast rotation rate for a spacecraft (i.e. about a half-rotation every second). The control commands are saturated between $t = 20$ to $t = 35$ seconds (Fig. 4.3) when the main effort of the controller is taking place as can be also seen from the kinetic energy plot of Fig. 4.4. In this case, as will be in all subsequent attitude recovery cases in the present and the next chapter, the energy plot is a good indication of the overall performance of the system. In the present case, the initial disturbance injects just over 500 Joules into the system from $t = 5$ to $t = 10$ seconds after which the energy remains constant until $t = 20$ seconds when the attitude recovery controller becomes active and slowly dissipates most of the system energy during the next 20 seconds, however another 20 seconds are needed to completely null the angular rates and attitude angles. Hence, it takes about 40 seconds to recover the spacecraft from the initial large disturbance.

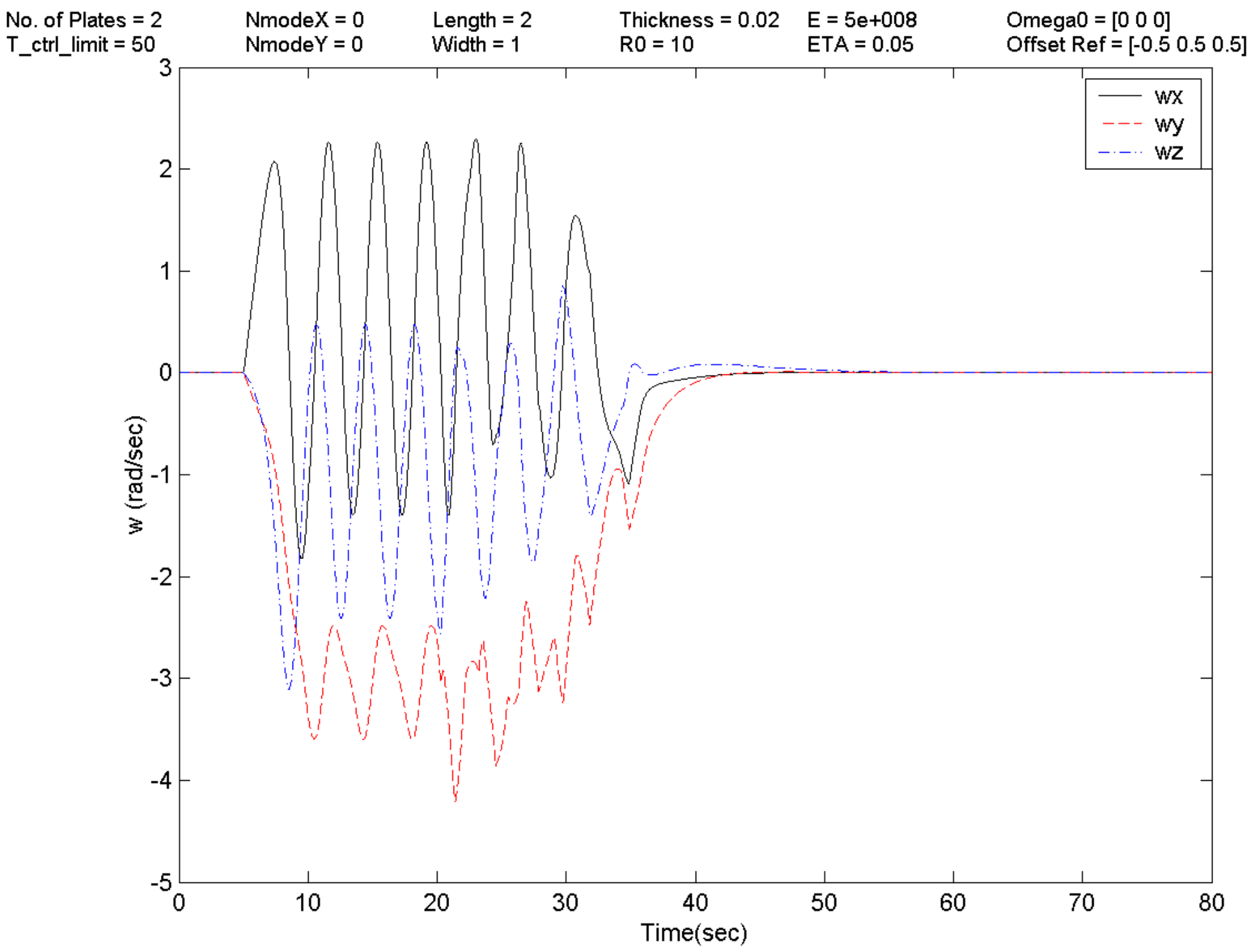

Figure 4.1: ARM case 1 - Spacecraft body angular rates ( $\vec{\omega}$ )

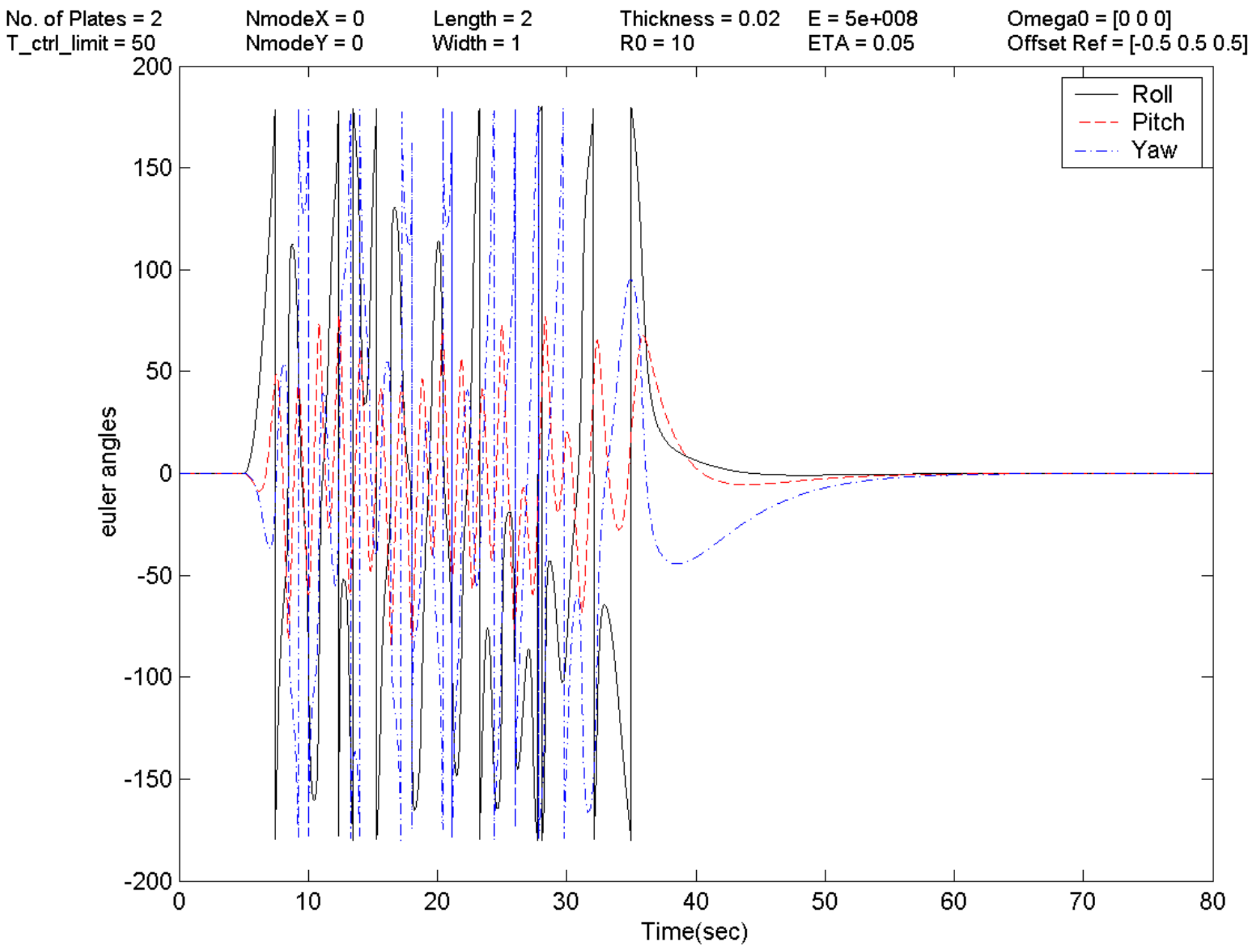

Figure 4.2: ARM case 1 - Spacecraft attitude (Euler angles)

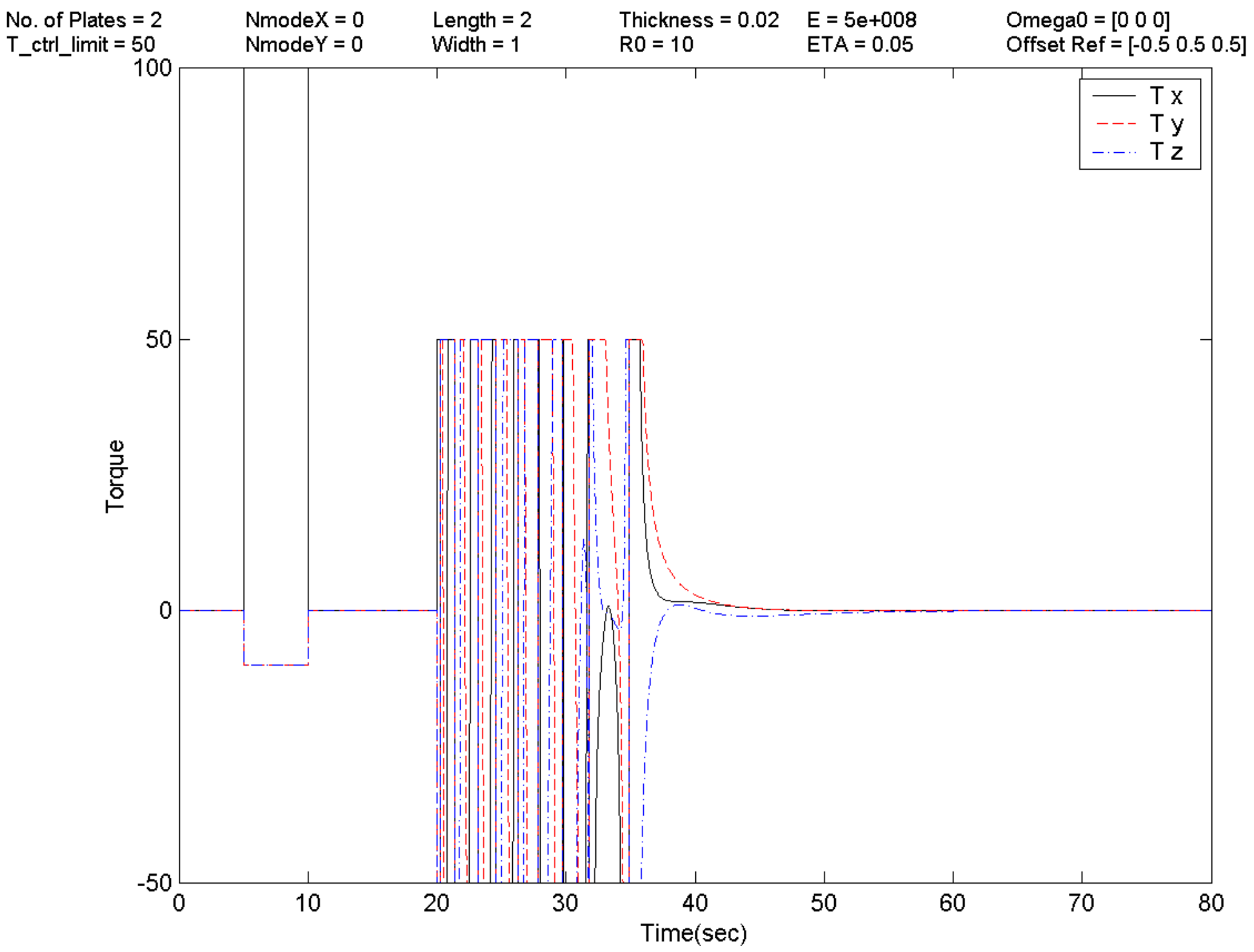

Figure 4.3: ARM case 1 - Initial disturbance and attitude recovery torques ( $\vec{\tau}$ )

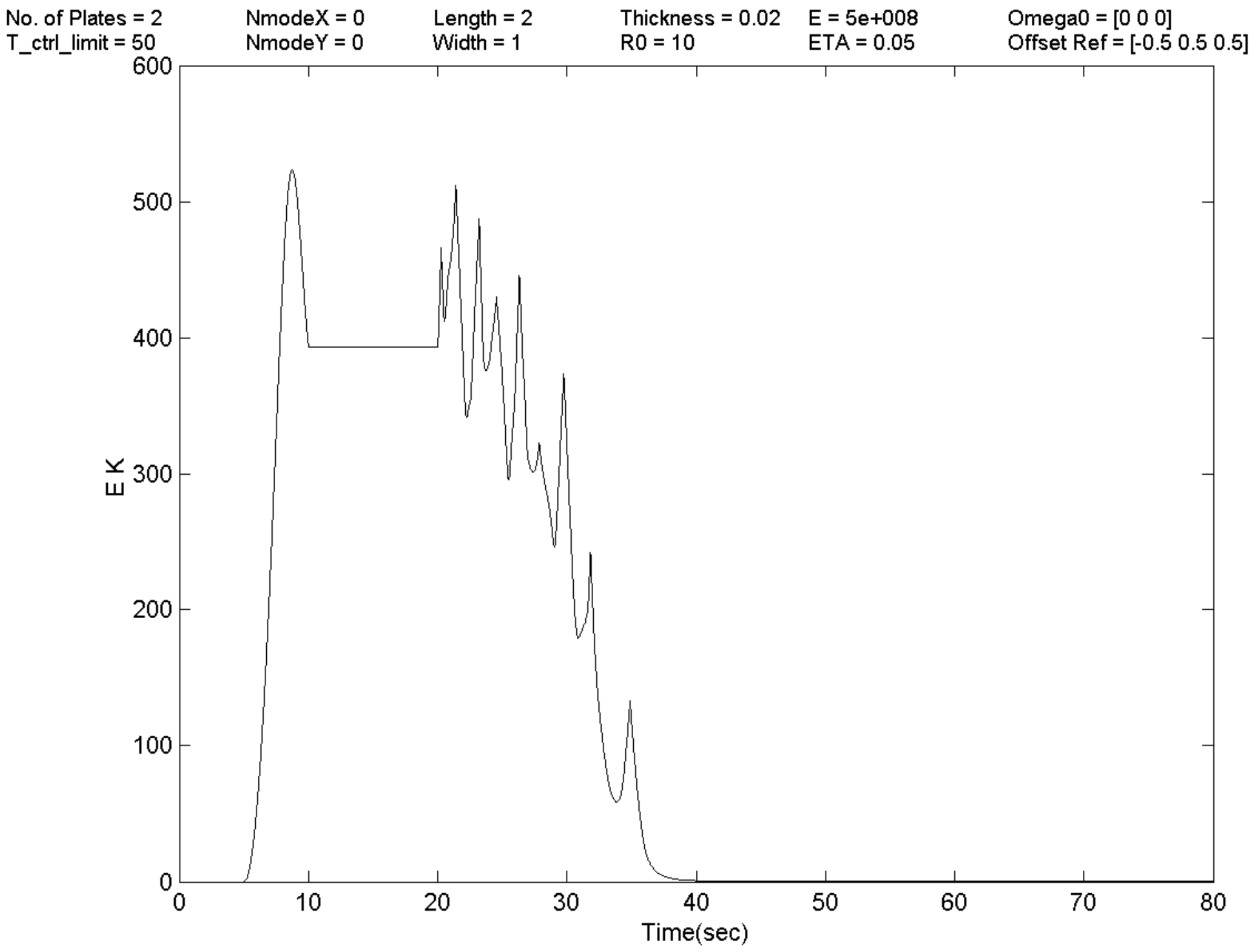

Figure 4.4: ARM case 1 - Spacecraft kinetic energy

*Simulation Case 2*: Figures 4.5 to 4.8 show the closed loop response of the rigid spacecraft to an initial torque disturbance of $\vec{\tau} = \begin{bmatrix} 100 & -100 & 100 \end{bmatrix}^T$ Nm, starting at $t = 5$ seconds and ending after 4 seconds, with constrained control torques ($\pm 50\ \text{Nm}$), starting 1.0 second after the end of the disturbance torque (i.e. $t = 10$) to allow for extra deviation of the spacecraft's states from the equilibrium point. All other parameters were kept the same as in case 1. Simply considering the magnitude of the disturbance torque, it is obvious that this is a more severe malfunction than case 1 which throws the spacecraft into a fast spin about all three axis. Compared to case 1, we note more frequent full $360^o$ Euler angle rotations (Fig. 4.2) before the attitude angles start to stabilize towards zero around $t = 45$ seconds. This time, the *y*-axis angular rate ($\omega_y$) reaches below the –6 rad/sec mark which is an extremely fast rotation rate for a spacecraft (i.e. a full rotation every second). The control commands are saturated at $\pm 50\ \text{Nm}$ for just over 25 seconds of the total ARM time (Fig. 4.11) and the kinetic energy plot (Fig. 4.12) depicts an initial energy injections of over 1200 Joules which is mostly dissipated by the controller over 30 seconds. However another 20 seconds are needed to completely regulate the angular rates and attitude angles, taking the total attitude recovery maneuver time up to about 50 seconds, namely 10 seconds more than case 1.

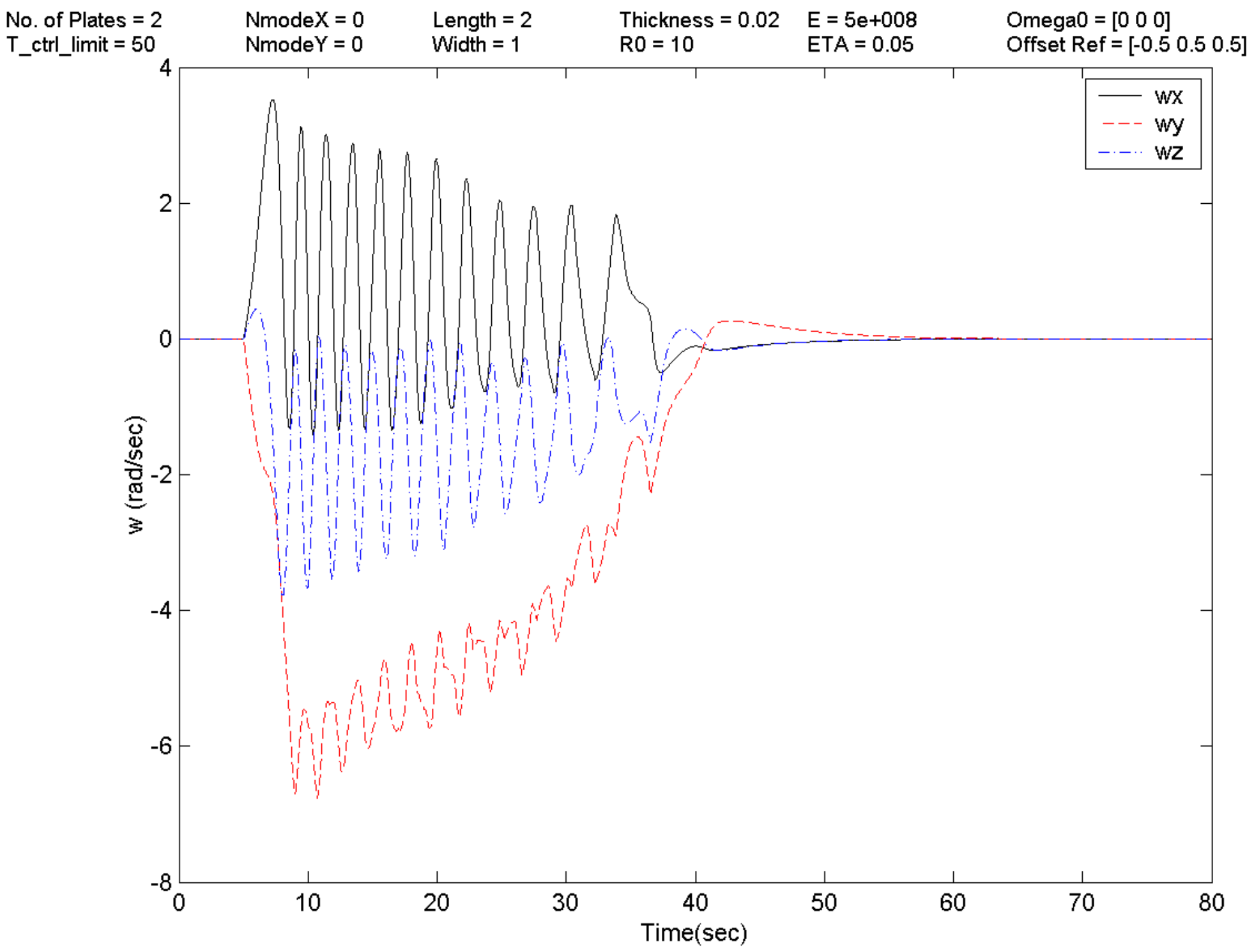

Figure 4.5: ARM case 2 - Spacecraft body angular rates ( $\vec{\omega}$ )

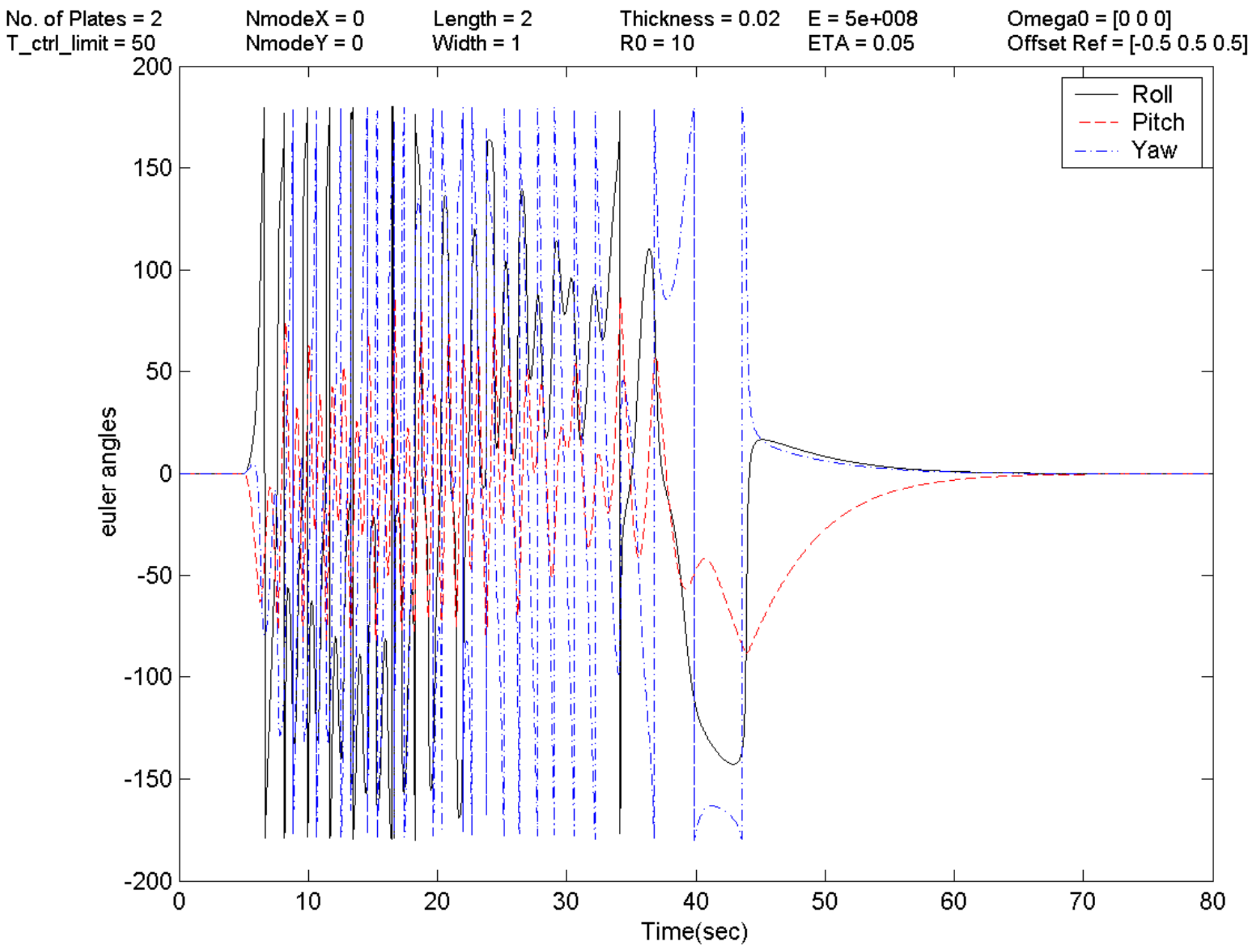

Figure 4.6: ARM case 2 - Spacecraft attitude (Euler angles)

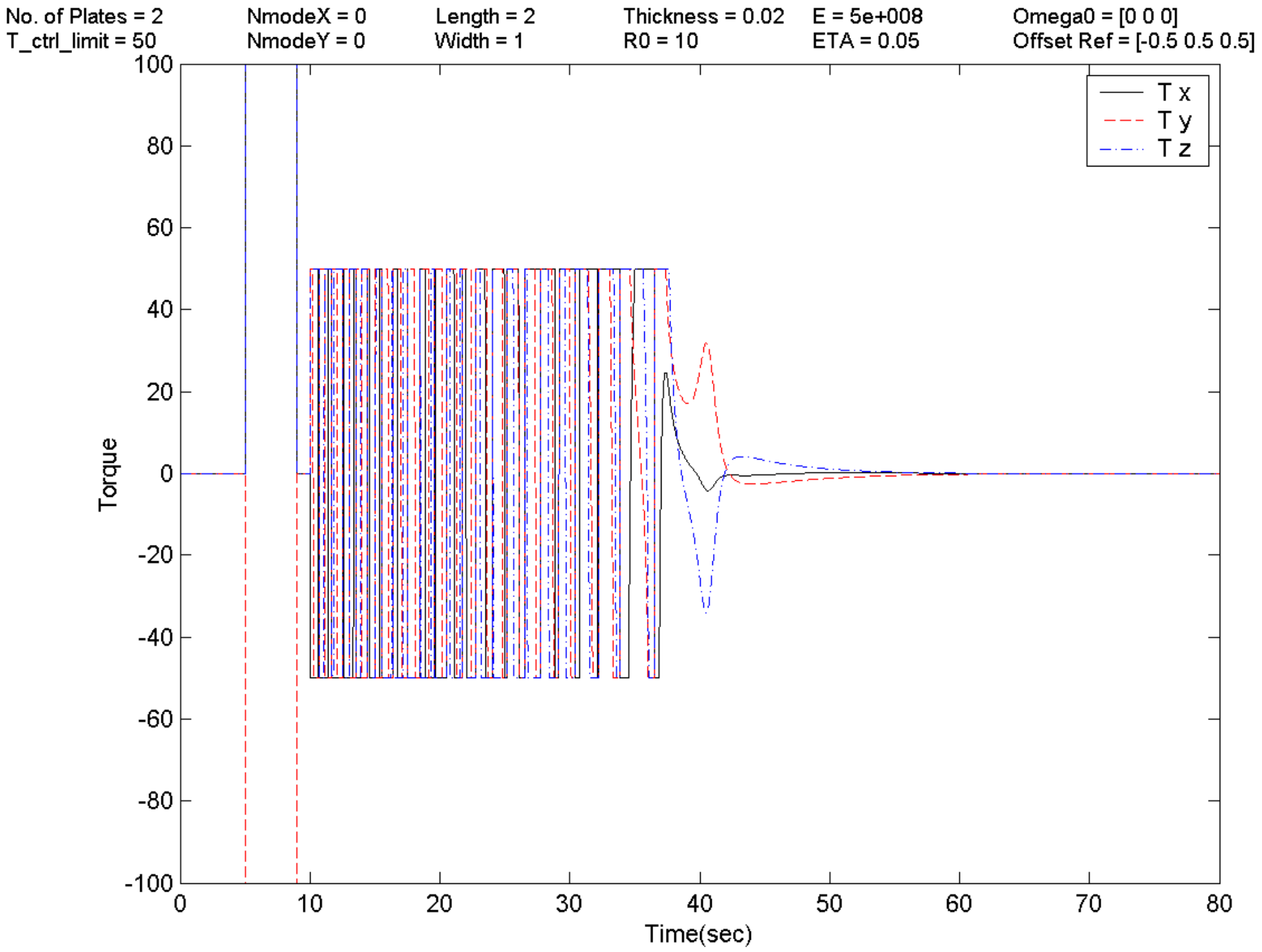

Figure 4.7: ARM case 2 - Initial disturbance and attitude recovery torques ( $\vec{\tau}$ )

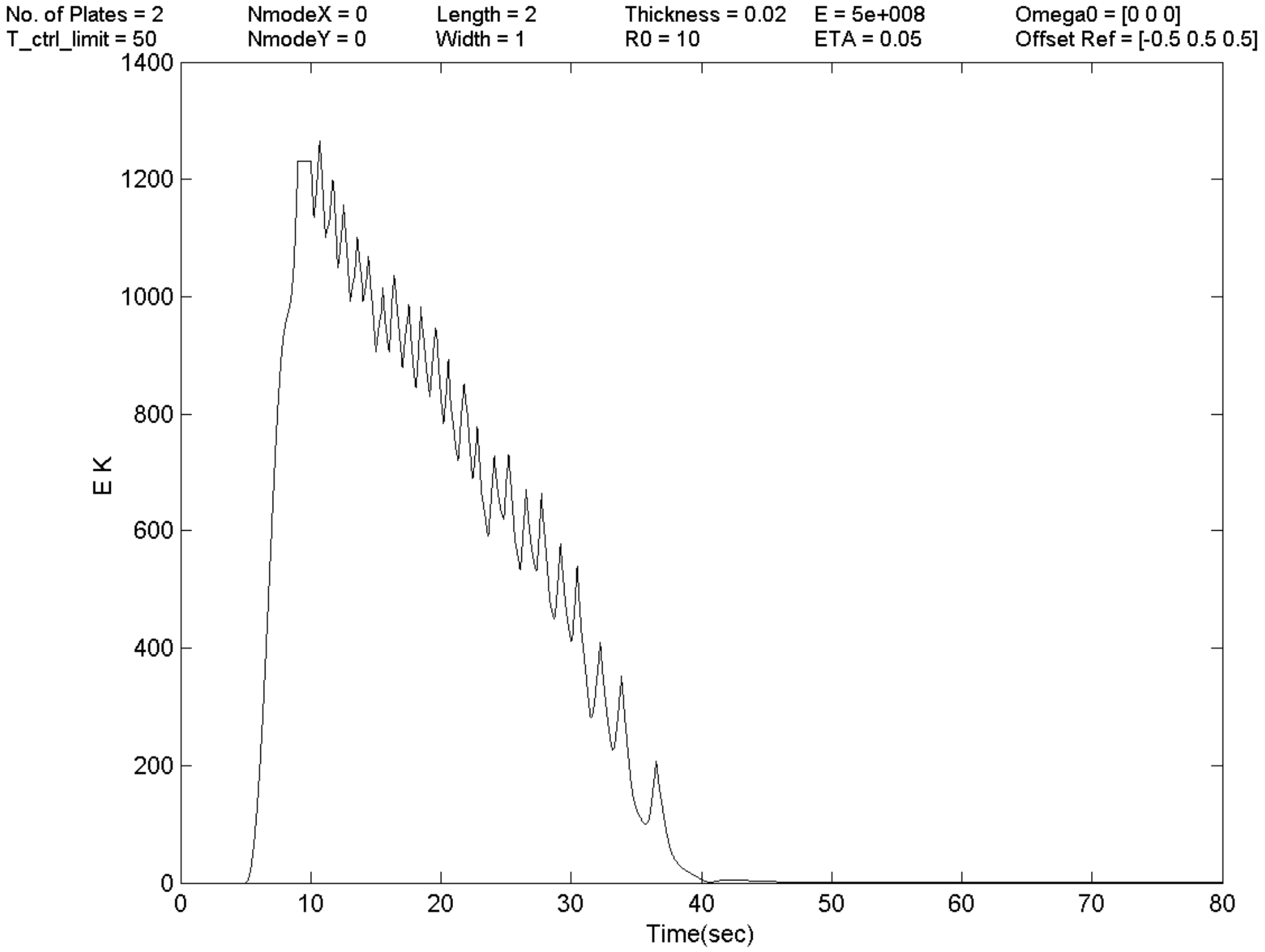

Figure 4.8: ARM case 2 - Spacecraft kinetic energy

## 4.3 Robustness Considerations

In order to validate and investigate the capability of the controller and its robustness against parameters and model uncertainties, which will be undoubtedly present in any complex system model, a series of simulations cases were carried out. For the rigid spacecraft, the main source of parameter uncertainties is the inertia matrix and model uncertainties could arise from the fact that no spacecraft is really entirely rigid and there are always some flexibilities at play. As for inertia parameters, these are quite difficult to obtain with precision and are usually off from the actual values. Inertia values may also vary unpredictably once the spacecraft is in orbit due to deploying appendages, fuel slosh, astronaut motion, etc. The following two simulation cases are repeats of the simulation cases of the last subsection and are devoted to examining the controller's robustness characteristics.

*Simulation Case 3*: This is a repeat of simulation case 1 with 50% increase in the values of $\vec{f}_{\vec{q}}$ and $\underline{G}_{\vec{q}}$ in (4.3) which would represent the model/parameter uncertainties in the system. This large increase in $\vec{f}_{\vec{q}}$ and $\underline{G}_{\vec{q}}$ values is chosen in order to magnify the effect it has on the overall system response. Figures 4.9 to 4.12 show the closed loop response which, by comparison to case 1, depicts a more sluggish response of about 10 seconds. The associated specific deteriorations in angular rate (Fig. 4.9), attitude angles (Fig. 4.10), control effort (Figs. 4.11-4.12) are noted in comparing every corresponding figure from this case to case 1. From Fig. 4.9, we notice more angular rate oscillations which lead to a more dynamic attitude motion (Fig. 4.10) with many more full $360^o$ angular rotations. Also, the control torques (Fig. 4.11) are saturated for approximately 10 seconds longer than case 1 and the energy plot indicates a corresponding sluggishness to move towards the zero energy level. One way to assess this sluggishness from the energy plot, is to visualize a line through the many energy oscillations (after $t = 20$) and note that the slope of this line (e.g. average rate of change of the kinetic energy) is less steep than the similar line for case 1 (Fig. 4.4).

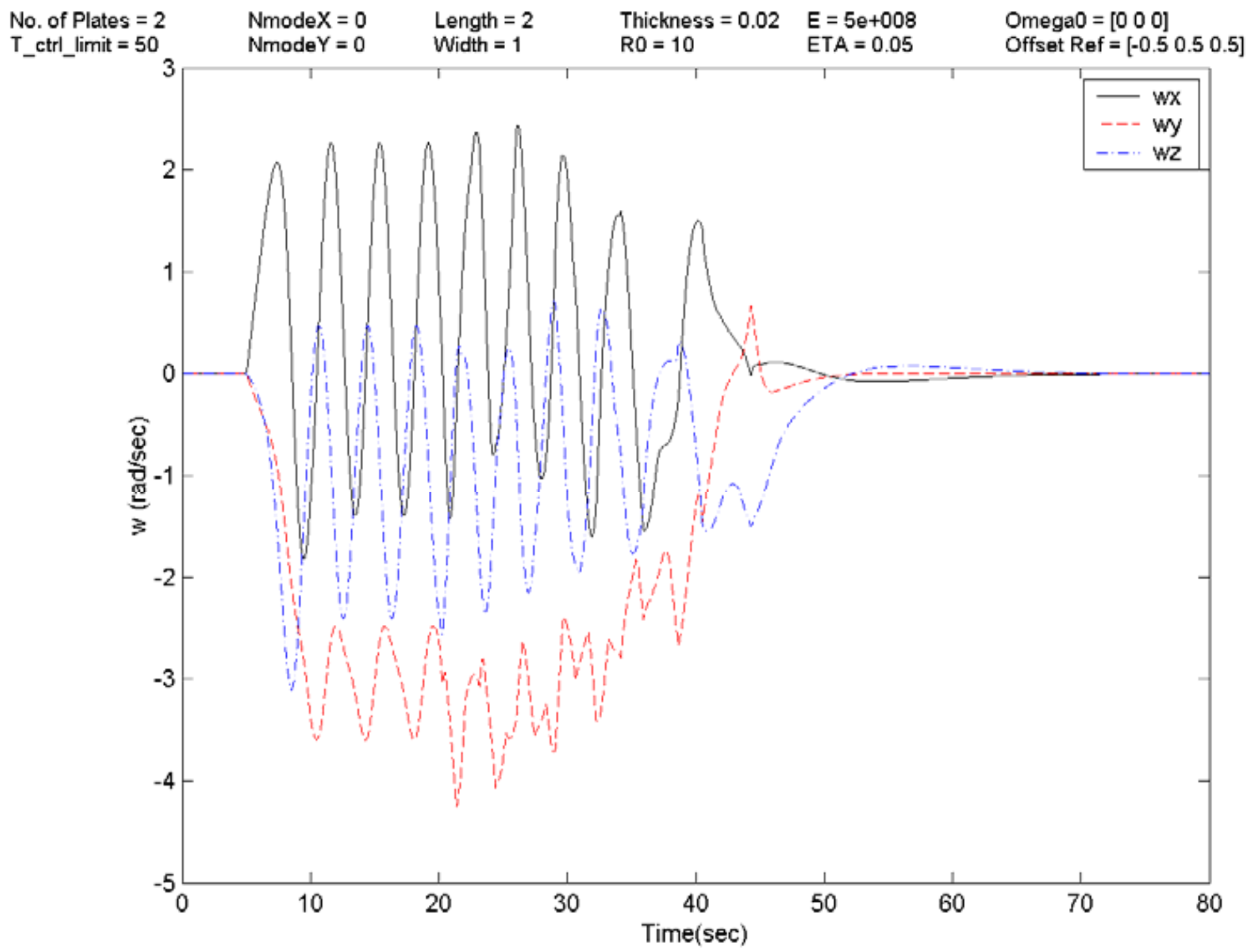

Figure 4.9: ARM case 3 - Spacecraft body angular rates ( $\vec{\omega}$ )

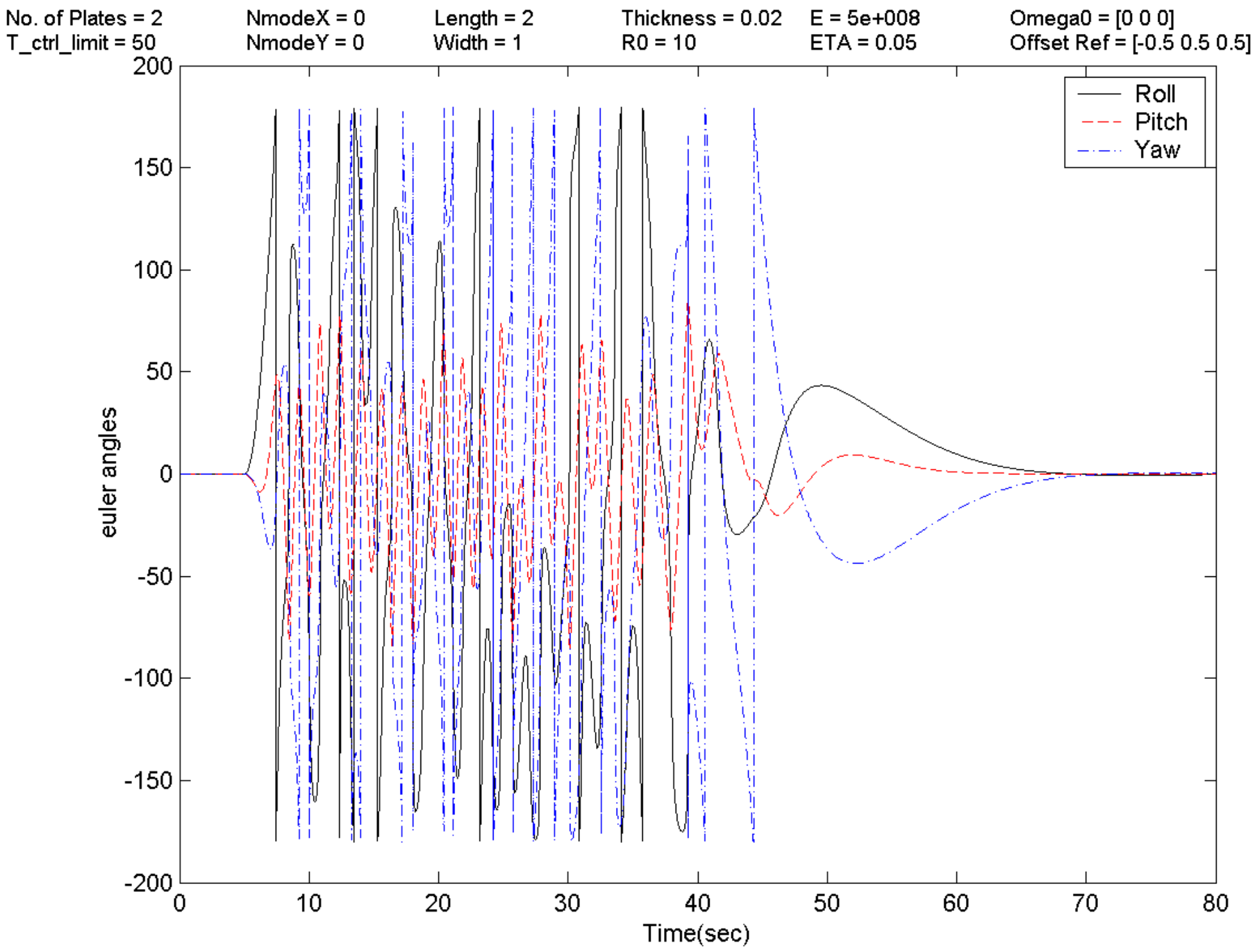

Figure 4.10: ARM case 3 - Spacecraft attitude (Euler angles)

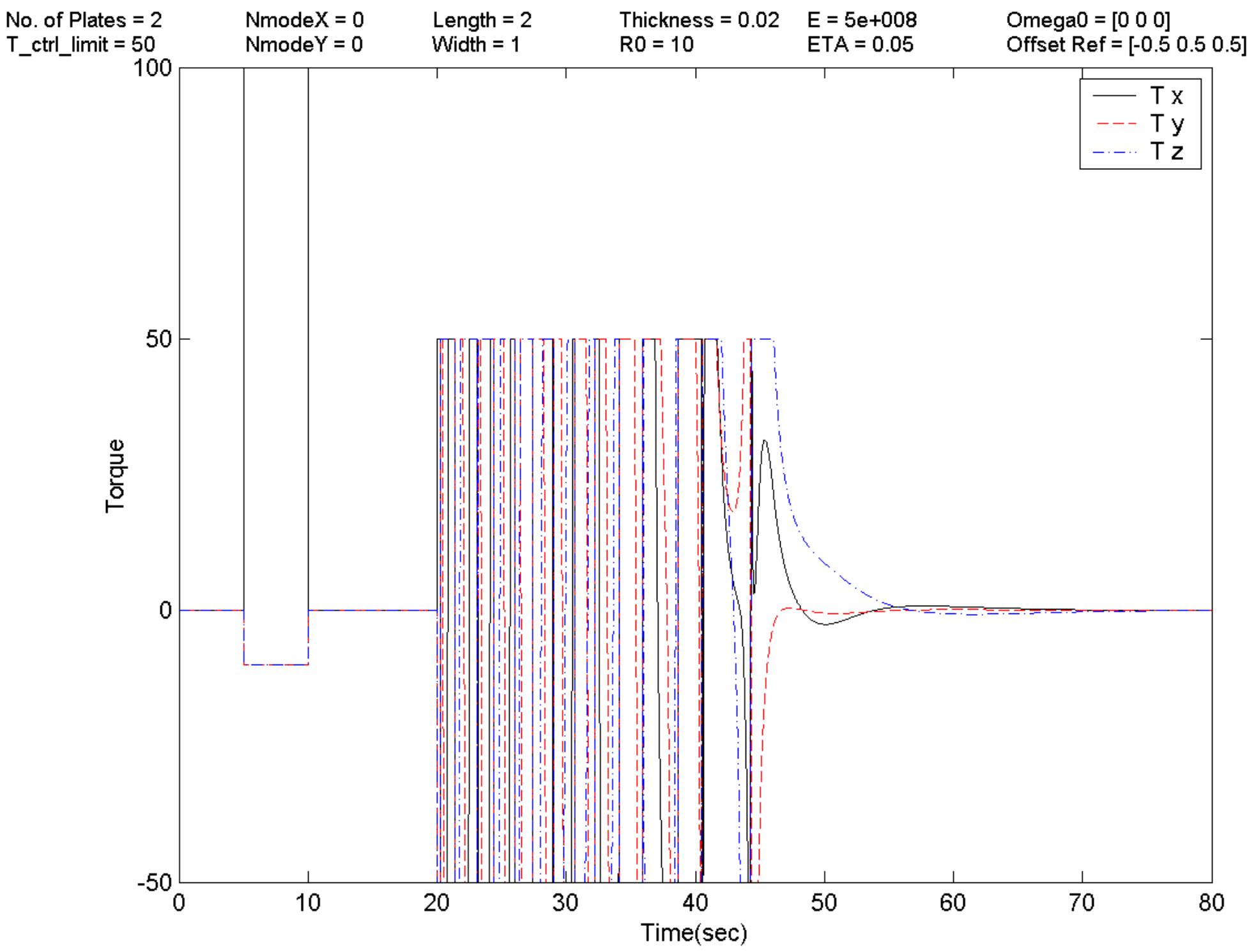

Figure 4.11: ARM case 3 - Initial disturbance and attitude recovery torques ( $\vec{\tau}$ )

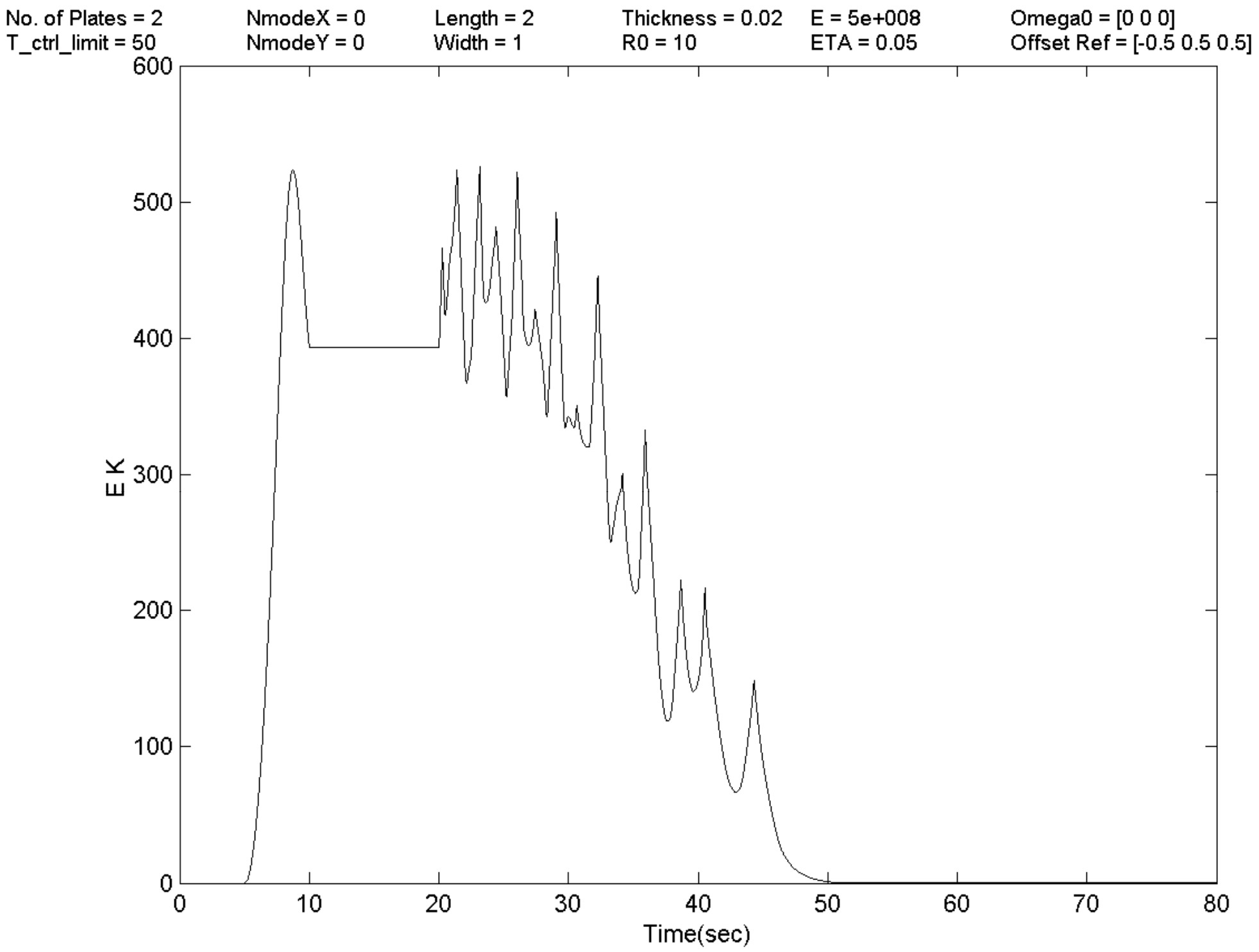

Figure 4.12: ARM case 3 - Spacecraft kinetic energy

*Simulation Case 4*: Finally, we present a repeat of simulation case 2 with 50% increase in the values of $\vec{f}_{\bar{q}}$ and $\underline{G}_{\bar{q}}$ in (4.3). Figures 4.13 to 4.16 show the closed loop response in this simulation case which again is slower than when compared to case 2. The increase in the attitude recovery maneuver time is once again about 10 seconds and all the deteriorations discussed in the previous case are also present here and can be noted by comparing Figs. 4.13-4.16 to Figs. 4.5-4.8, respectively.

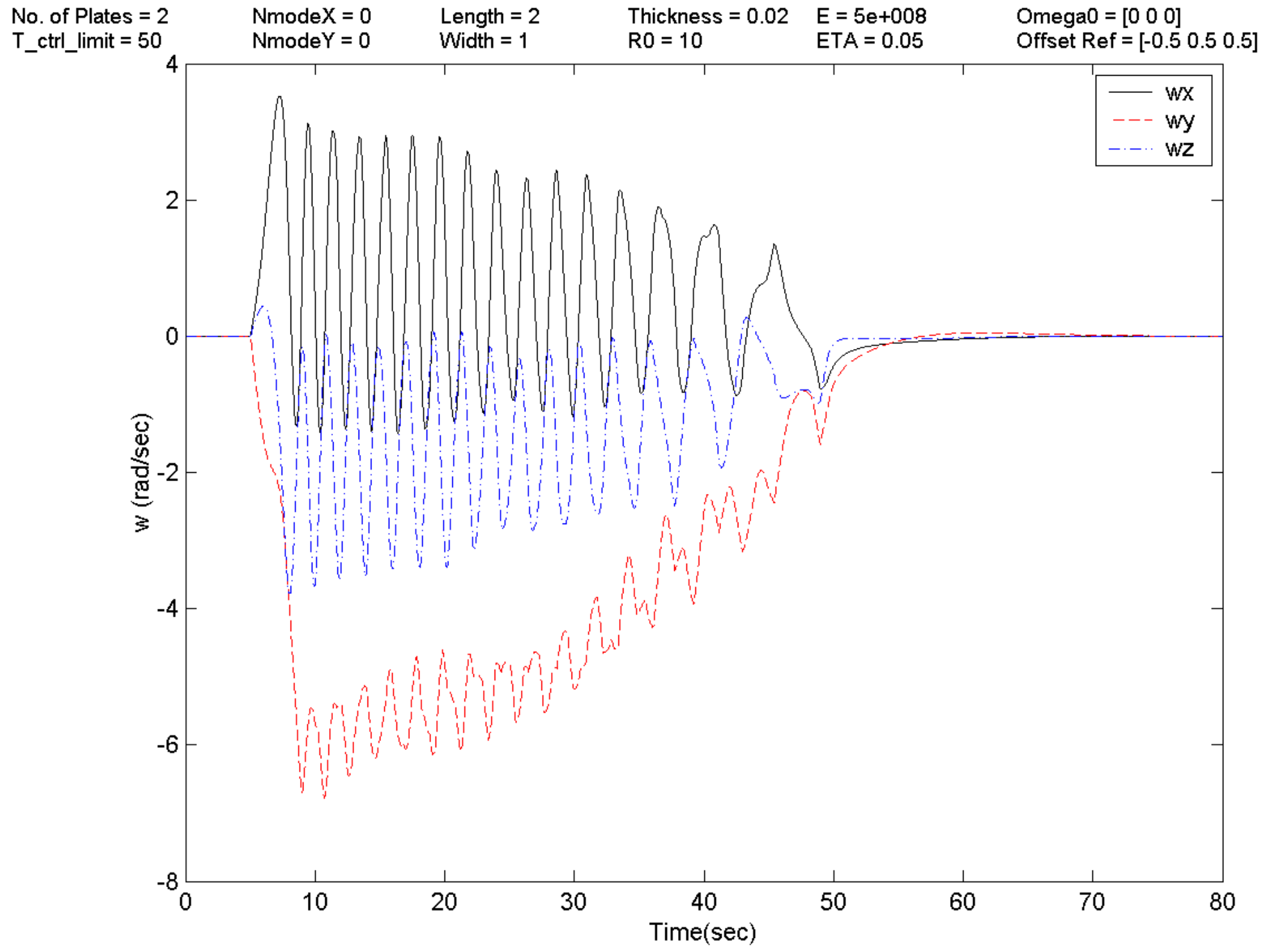

Figure 4.13: ARM Case 4 - Spacecraft body angular rates ( $\vec{\omega}$ )

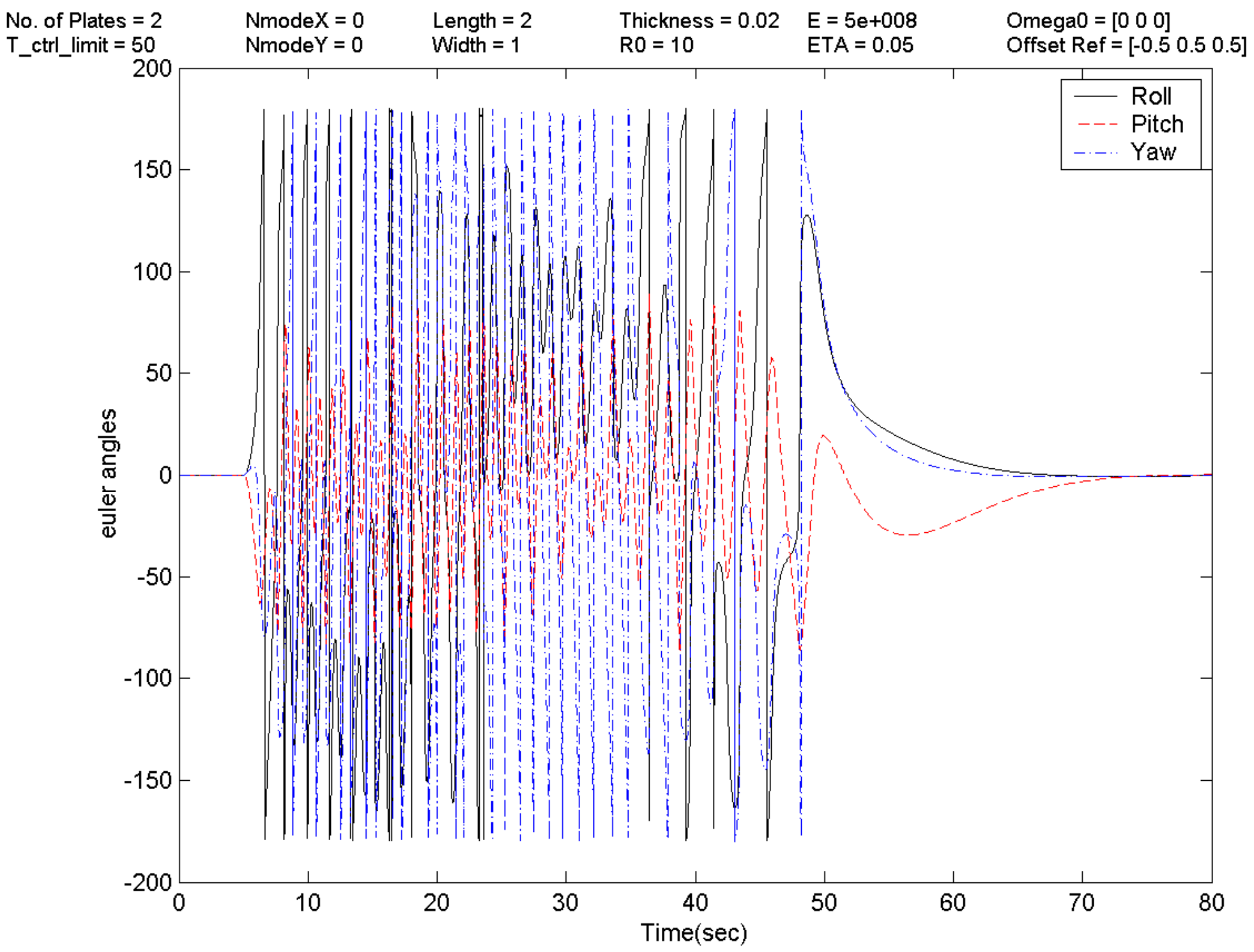

Figure 4.14: ARM Case 4 - Spacecraft attitude (Euler angles)

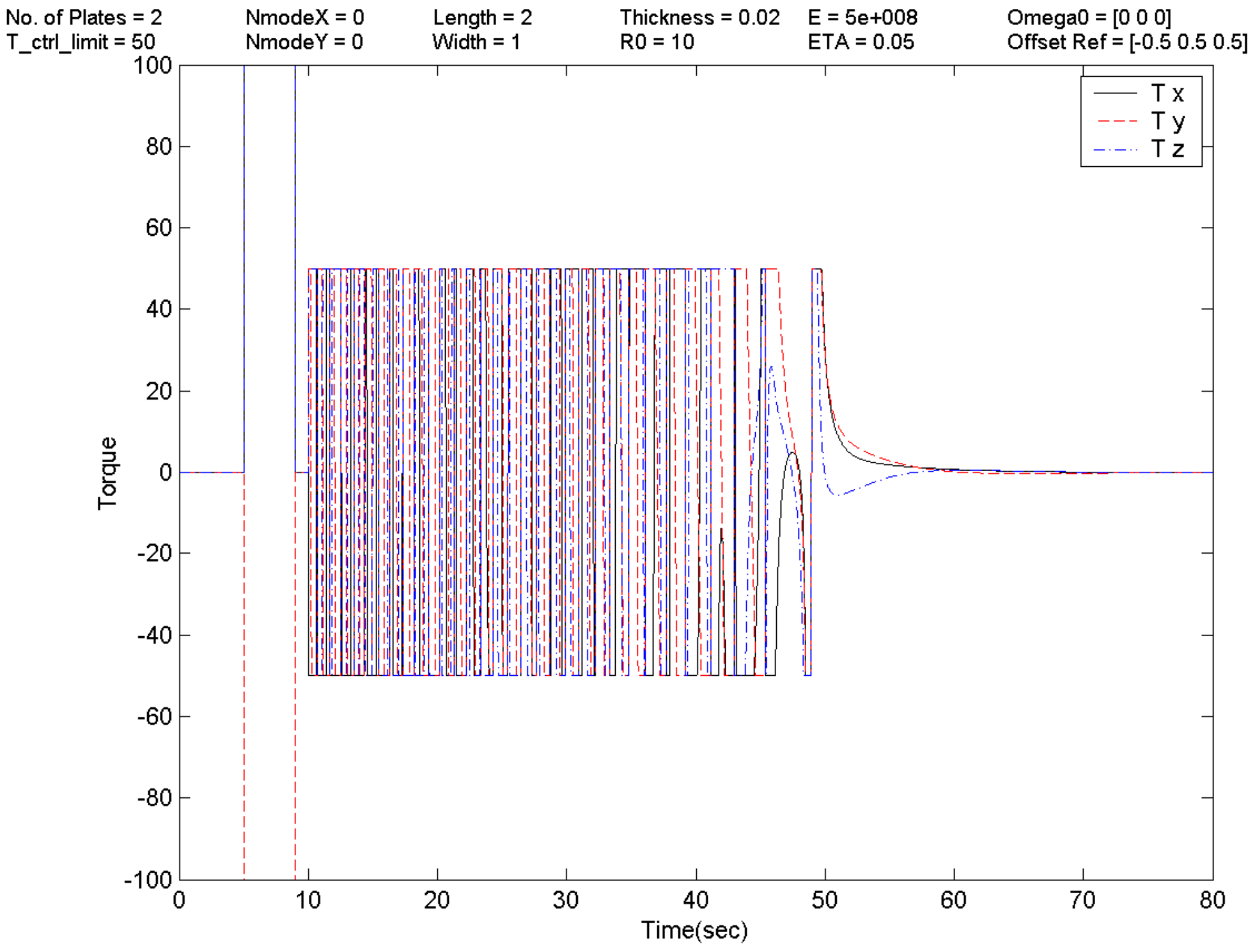

Figure 4.15: ARM Case 4 - Initial disturbance and attitude recovery torques ( $\vec{\tau}$ )

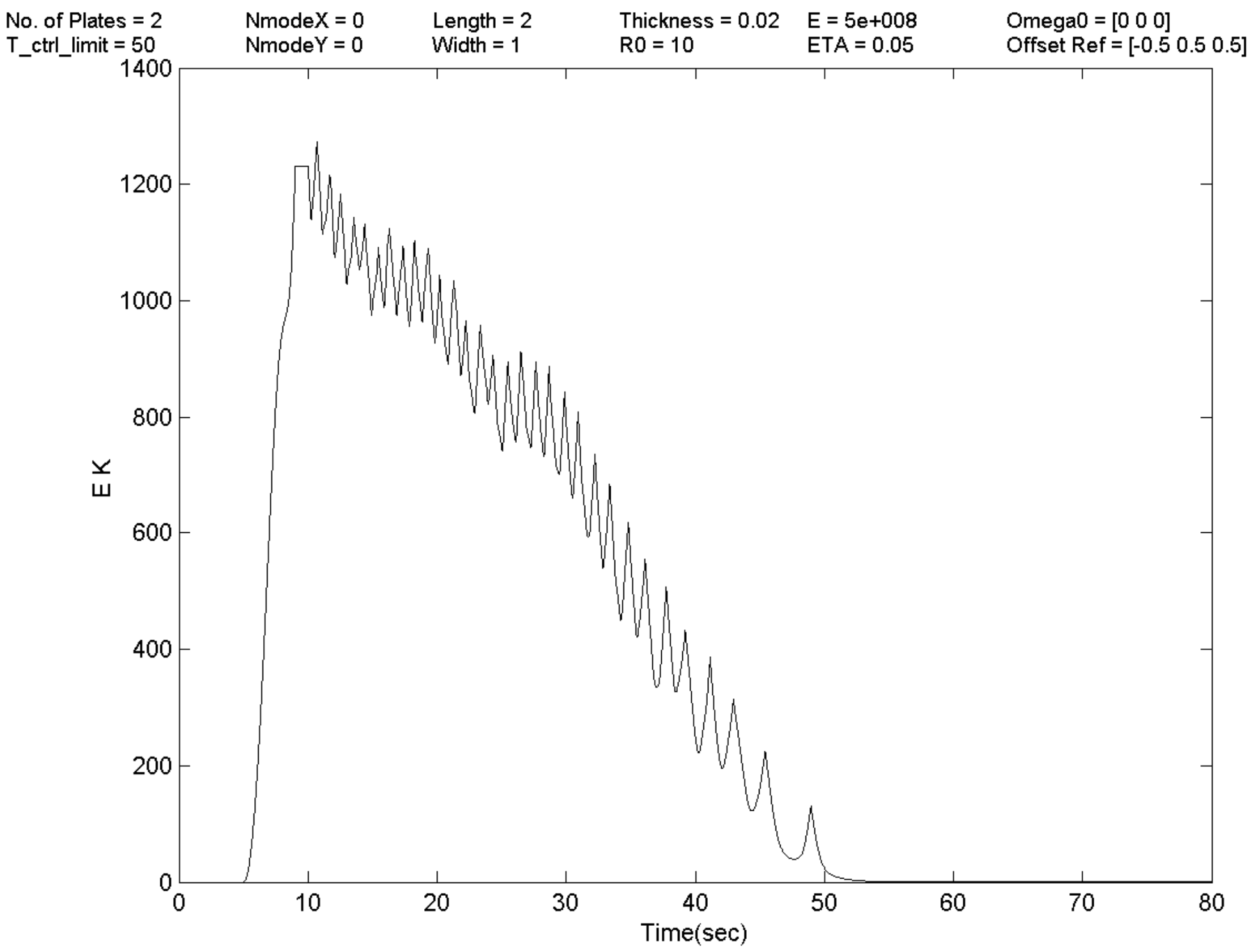

Figure 4.16: ARM Case 4 - Spacecraft kinetic energy

## 4.4 Concluding Remarks

Six different simulation cases were presented in this chapter based on two different malfunctions, namely, (i) a failure imparting a large disturbance torque mainly about the *x*-axis; and (ii) a severe malfunction affecting all three attitude axes. For all the simulation cases presented here, the system performance requirements of Section 1.3 were met and of course, the controller gains can be further tuned to decrease the attitude recovery maneuver time, if needed. It is also noted that the main effect of parameter/model uncertainty is the deterioration of the close-loop system performance in its response time which becomes slow and sluggish.

# CHAPTER 5

## FLEXIBLE SPACECRAFT ATTITUDE RECOVERY SYSTEM

In this chapter, the attitude recovery for flexible spacecraft is investigated. Since the flexible spacecraft is under-actuated, input-state feedback linearization can not be achieved and we turn to a controller based on input-output feedback linearization. General theoretical results and methodology for input-output feedback linearization exist in the literature, however, the application of these general results to our complex system has not been undertaken, and is the subject of Section 5.1. Sections 5.2 and 5.3 present results of the attitude recovery simulations performed using the proposed controller. Similar to Chapter 4, the simulation cases here are for two different failures severely affecting the attitude of the spacecraft, and some cases also include model/parameter uncertainties to study the robustness of the controller. As was mentioned before, the source of the disturbance torque is not really important and the large torque is only being used to show that even under exceptionally severe conditions, the proposed controller can perform a stable maneuver for a typical flexible spacecraft.

## 5.1 Controller based on Input-Output Feedback Linearization

The problem of flexible attitude recovery can be stated as the regulating (or zeroing) of the Euler angle errors or angular rates given a specified reference and starting from an unknown initial condition as was the case with the rigid spacecraft. However, in the case of a flexible spacecraft, we must as well include the requirement of vibration suppression of the appendages. Otherwise stated mathematically, there are two different scenarios:

(i) Zeroing of angular body rates and vibration suppression

$$\lim_{t\to\infty}\vec{\omega} = 0, \quad \lim_{t\to\infty}\vec{\chi} = 0, \quad \lim_{t\to\infty}\dot{\vec{\chi}} = 0 \tag{5.1}$$

(ii) Regulating Euler angles (or quaternions) and vibration suppression

$$\lim_{t\to\infty}\vec{q} = \vec{q}_{\text{ref}}, \quad \lim_{t\to\infty}\dot{\vec{q}} = 0, \quad \lim_{t\to\infty}\vec{\chi} = 0, \quad \lim_{t\to\infty}\dot{\vec{\chi}} = 0 \tag{5.2}$$

Case (ii) above is the more general case, as it implicitly requires that $\lim_{t\to\infty}\vec{\omega} = 0$. Case (i) is the less restrictive one, as it does not require the recovery maneuver to point the spacecraft in any particular direction, it only requires that all system motions are brought to zero.

We assume that the full state measurement of the system is available through attitude (e.g. gyros) and structural (e.g. strain gauges) sensors. Strain gauges have been successfully used on flexible robotic links to measure the displacements from which generalized displacements ($\vec{\chi}$) are derived and and then a nonlinear observer can be constructed to also obtain the associated rates ($\dot{\vec{\chi}}$) [145]-[147]. Note that an observer is necessary since simple numerical differentiation of the $\vec{\chi}$ states to get $\dot{\vec{\chi}}$ states will introduce too much noise and is not a viable option in a real implementation. It is also important to note that there would be no simple way of estimating the $\dot{\vec{\chi}}$ states without the *measurements* of $\vec{\chi}$ states. The control actuators will be the same as for the rigid spacecraft case (i.e. control moment gyros) and we do not assume the use of piezoelectric actuators on the appendages, as these are not commonly found on today's spacecrafts. However, it is envisaged that some time in the future, spacecraft manufacturers will consider using these actuation mechanisms to damp out vibrations of flexible parts of the spacecraft. Hence, without the use of any direct actuation mechanisms for the appendages, our system is in fact under-actuated. For this specific reason, we can not use the straight forward input-state feedback linearization approach and must turn to the input-output feedback linearization techniques.

Input-output linearization of our multi-input multi-output system is obtained by differentiating the outputs $y_i$ until the inputs $\tau_i$ appear. Using the notation of differential geometry, the process of repeated differentiation means that

$$\dot{y}_i = \nabla h_i \dot{\vec{x}} = \nabla h_i (\vec{f} + \underline{G}\vec{\tau}) = L_{\vec{f}} h_i + \sum_{j=1}^{3} (L_{\vec{g}_j} h_i)\, \tau_j \qquad i = 1, 2, 3 \tag{5.3}$$

where the Lie derivative $L_{\vec{f}} h$ of the scalar function $h$ along the vector field $\vec{f}$ is defined as

$$L_{\vec{f}} h = \nabla h \; \vec{f} = \frac{\partial h}{\partial \vec{x}} \vec{f} = \begin{bmatrix} \frac{\partial h}{\partial x_1} \cdots \frac{\partial h}{\partial x_m} \end{bmatrix} \begin{bmatrix} f_1 \\ \vdots \\ f_m \end{bmatrix} = \frac{\partial h}{\partial x_1} f_1 + \cdots + \frac{\partial h}{\partial x_m} f_m$$

If $L_{\vec{g}_j} h_i = 0$ for $\forall j$ in a neighborhood $\Omega_i$ of the point $\vec{x}_o$, the differentiation process needs to be continued. Assuming that $r_i$ is the smallest integer such that at least one of the inputs, $\tau_j$, appears in $y_i^{(r_i)}$, then

$$y_i^{(r_i)} = L_{\vec{f}}^{r_i} h_i + \sum_{j=1}^{3} (L_{\vec{g}_j} L_{\vec{f}}^{r_i - 1} h_i)\, \tau_j \tag{5.4}$$

with $L_{\vec{g}_j} L_{\vec{f}}^{r_i - 1} h_i \neq 0$ for at least one *j*, in a neighborhood $\Omega_i$ of the point $\vec{x}_o$. Note that the repeated Lie derivatives can be defined recursively as

$$L_{\vec{f}}^{0} h = h$$
$$L_{\vec{f}}^{r} h = L_{\vec{f}}\left(L_{\vec{f}}^{r-1} h\right) = \nabla\left(L_{\vec{f}}^{r-1} h\right)\vec{f}$$

and similarly, if $\vec{g}$ is another vector field, then the scalar function $L_{\vec{g}} L_{\vec{f}} h = \nabla\left(L_{\vec{f}} h\right)\vec{g}$. The differentiation process above is done on all outputs which yields the following

$$\begin{bmatrix} y_1^{(r_1)} \\ y_2^{(r_2)} \\ y_3^{(r_3)} \end{bmatrix} = \begin{bmatrix} L_{\vec{f}}^{r_1} h_1 \\ L_{\vec{f}}^{r_2} h_2 \\ L_{\vec{f}}^{r_3} h_3 \end{bmatrix} + E(\vec{x})\,\vec{\tau} \tag{5.5}$$

where the $3 \times 3$ matrix E(x) is defined as

$$E(\vec{x}) = \begin{bmatrix} L_{\vec{g}_1} L_{\vec{f}}^{r_1-1} h_1 & L_{\vec{g}_2} L_{\vec{f}}^{r_1-1} h_1 & L_{\vec{g}_3} L_{\vec{f}}^{r_1-1} h_1 \\ L_{\vec{g}_1} L_{\vec{f}}^{r_2-1} h_2 & L_{\vec{g}_2} L_{\vec{f}}^{r_2-1} h_2 & L_{\vec{g}_3} L_{\vec{f}}^{r_2-1} h_2 \\ L_{\vec{g}_1} L_{\vec{f}}^{r_3-1} h_3 & L_{\vec{g}_2} L_{\vec{f}}^{r_3-1} h_3 & L_{\vec{g}_3} L_{\vec{f}}^{r_3-1} h_3 \end{bmatrix} \tag{5.6}$$

$E(\vec{x})$ is the decoupling matrix and is invertible in the region $\Omega$, so that input-output linearization can be achieved. Note that $\Omega$ is the intersection of the $\Omega_i$ and for well-defined *relative degrees*, $r_i$, $\Omega$ is itself a finite neighborhood of $\vec{x}_o$.

Now the input-output linearizing control vector is chosen as

$$\vec{\tau} = E^{-1} \begin{bmatrix} v_1 - L_{\vec{f}}^{r_1} h_1 \\ v_2 - L_{\vec{f}}^{r_2} h_2 \\ v_3 - L_{\vec{f}}^{r_3} h_3 \end{bmatrix} \tag{5.7}$$

where $\vec{v}$ is a $3\times 1$ vector, representing the new control input torque. Substituting (5.7) in the output equations in (5.5), we obtain the three input-output decoupled and linearized ordinary differential equations as

$$\begin{bmatrix} y_1^{(r_1)} \\ y_2^{(r_2)} \\ y_3^{(r_3)} \end{bmatrix} = \begin{bmatrix} v_1 \\ v_2 \\ v_3 \end{bmatrix} \tag{5.8}$$

Since the input $v_i$ only affects the output $y_i$, we have a decoupling control law where, as mentioned before, $E(\vec{x})$ plays the role of the decoupling matrix of the system. The system is then said to have relative degrees, $r_i$, for $i = 1, 2, 3$ at $\vec{x}_o$. The total relative degree of the system may be defined as $r = \sum_{i=1}^{3} r_i$ at $\vec{x}_o$.

In our case the relative degree $r_i = 1\, or\, 2$ for every output $y_i$, as it easy to see that 1 or 2 differentiation steps are needed for the input to appear in systems (2.57)-(2.58) or (2.59)-(2.60), respectively. The total relative degree of the system is $r = (3\, or\, 6) < m$ and hence there are $m-3$ or $m-6$ states that belong to the internal dynamics of systems (2.57)-(2.58) or (2.59)-(2.60), respectively.

We will now concentrate on the more general attitude recovery problem stated by requirements (5.2) applied to the system (2.59)-(2.60). In order to obtain the feedback linearization control torque, the decoupling matrix, $E(\vec{x})$, and the $L_{\vec{f}}^{r_i} h_i$ ($i = 1, 2, 3$) terms in (5.7) must be derived for our specific system (2.59)-(2.60). Let's begin by finding the $L_{\vec{f}}^{r_1} h_1$ term where $r_1 = 2$ and $h_1 = x_2$ from (2.60). We know $L_{\vec{f}}^{2} h_1 = L_{\vec{f}}\left(L_{\vec{f}} h_1\right)$ and $L_{\vec{f}} h_1$ can be derived as follows

$$L_{\vec{f}} h_1 = \underbrace{\left[ \frac{\partial h_1}{\partial x_1} \frac{\partial h_1}{\partial x_2} \cdots \frac{\partial h_1}{\partial x_m} \right]}_{\nabla h_1} \underbrace{\begin{bmatrix} x_5 \\ x_6 \\ x_7 \\ x_8 \\ f_{q_0}(\tilde{q}, \dot{\tilde{q}}, \vec{\chi}, \dot{\vec{\chi}}) \\ \vec{f}_{\vec{q}}(\tilde{q}, \dot{\tilde{q}}, \vec{\chi}, \dot{\vec{\chi}}) \\ x_{npq+9} \\ \vdots \\ x_{2npq+8} \\ \vec{f}_{\vec{\chi}}(\tilde{q}, \dot{\tilde{q}}, \vec{\chi}, \dot{\vec{\chi}}) \end{bmatrix}}_{\vec{f}} = \frac{\partial h_1}{\partial x_1} x_5 + \frac{\partial h_1}{\partial x_2} x_6 + \cdots + \frac{\partial h_1}{\partial x_m} f_m \tag{5.9}$$

All the terms is (5.9) are zero except the second term which is $\frac{\partial h_1}{\partial x_2} x_6 = \frac{\partial x_2}{\partial x_2} x_6 = x_6$ and hence $L_{\vec{f}} h_1 = x_6$ and we can proceed to find $L_{\vec{f}}^2 h_1 = L_{\vec{f}} \left( L_{\vec{f}} h_1 \right) = L_{\vec{f}} x_6$ to be

$$L_{\vec{f}}^2 h_1 = L_{\vec{f}} x_6 = \underbrace{\left[ \frac{\partial x_6}{\partial x_1} \cdots \frac{\partial x_6}{\partial x_6} \cdots \frac{\partial x_6}{\partial x_m} \right]}_{\nabla x_6} \underbrace{\begin{bmatrix} x_5 \\ x_6 \\ x_7 \\ x_8 \\ f_{q_0}(\tilde{q}, \dot{\tilde{q}}, \vec{\chi}, \dot{\vec{\chi}}) \\ \vec{f}_{\vec{q}_1}(\tilde{q}, \dot{\tilde{q}}, \vec{\chi}, \dot{\vec{\chi}}) \\ \vec{f}_{\vec{q}_2}(\tilde{q}, \dot{\tilde{q}}, \vec{\chi}, \dot{\vec{\chi}}) \\ \vec{f}_{\vec{q}_3}(\tilde{q}, \dot{\tilde{q}}, \vec{\chi}, \dot{\vec{\chi}}) \\ \vdots \\ \vec{f}_{\vec{\chi}}(\tilde{q}, \dot{\tilde{q}}, \vec{\chi}, \dot{\vec{\chi}}) \end{bmatrix}}_{\vec{f}} = \frac{\partial x_6}{\partial x_6} \vec{f}_{\vec{q}_1}(\tilde{q}, \dot{\tilde{q}}, \vec{\chi}, \dot{\vec{\chi}}) = \vec{f}_{\vec{q}_1} \tag{5.10}$$

So $L_{\vec{f}}^2 h_1$ was found to be simply the sixth element of the vector field $\vec{f}$ given in (2.59), and the other two $L_{\vec{f}}^{r_i} h_i$ ($i = 2, 3$) terms and the 9 elements in the decoupling matrix $E(\vec{x})$ can also be derived similarly (not shown) to get the final results below:

$$\begin{bmatrix} L_{\vec{f}}^2 h_1 \\ L_{\vec{f}}^2 h_2 \\ L_{\vec{f}}^2 h_3 \end{bmatrix} = \begin{bmatrix} \vec{f}_{\vec{q}_1} \\ \vec{f}_{\vec{q}_2} \\ \vec{f}_{\vec{q}_3} \end{bmatrix} = \vec{f}_{\vec{q}} \qquad E(\vec{x}) = \begin{bmatrix} \underline{G}_{\vec{q}}(1,1) & \underline{G}_{\vec{q}}(1,2) & \underline{G}_{\vec{q}}(1,3) \\ \underline{G}_{\vec{q}}(2,1) & \underline{G}_{\vec{q}}(2,2) & \underline{G}_{\vec{q}}(2,3) \\ \underline{G}_{\vec{q}}(3,1) & \underline{G}_{\vec{q}}(3,3) & \underline{G}_{\vec{q}}(3,3) \end{bmatrix} = \underline{G}_{\vec{q}} \tag{5.11}$$

Finally, substituting (5.11) in (5.7), the feedback linearization control torque is obtained to be

$$\vec{\tau} = \underline{G}_{\vec{q}}^{-1}[-\vec{f}_{\vec{q}} + \vec{v}] \tag{5.12}$$

where the closed-form expression for $\vec{f}_{\vec{q}}$ and $\underline{G}_{\vec{q}}$ of system (2.59) are available.

By substituting the control torque (5.12) into the system dynamics (2.59), we will obtain (i) a simple linear relationship between the output $\vec{y}$ and the new input $\vec{v}$ which constitutes the new external and observable dynamics of the system (5.13) and (ii) an internal unobservable dynamics (5.14) (i.e. dynamics of $\vec{\chi}, \dot{\vec{\chi}}$ states):

$$\ddot{\vec{q}} = \vec{v} \tag{5.13}$$

$$\ddot{\vec{\chi}} = \vec{f}_{\vec{\chi}} - \underline{G}_{\vec{\chi}} \underline{G}_{\vec{q}}^{-1} \vec{f}_{\vec{q}} + \underline{G}_{\vec{\chi}} \underline{G}_{\vec{q}}^{-1} \vec{v} \tag{5.14}$$

We will now define the quaternion and the generalized displacement error vectors as

$$\begin{aligned} \vec{q}_e &= \vec{q}_{\text{ref}} - \vec{q} \\ \vec{\chi}_e &= \vec{\chi}_{\text{ref}} - \vec{\chi} \end{aligned} \tag{5.15}$$

where both $\vec{q}_{\text{ref}}$ and $\vec{\chi}_{\text{ref}}$ are set to zero. Note that, as it was explained in Section 4.1, we will be using quaternion addition to define the control error signal and we'll only need to consider the vector part ($\vec{q}$) of the quaternion for control and stability analysis.

Assuming a PD control law for the new input torque $\vec{v}$ in (5.13), we have

$$\vec{v} = \ddot{\vec{q}}_{\text{ref}} + \underline{K}_{Pq}\vec{q}_e + \underline{K}_{Dq}\dot{\vec{q}}_e + \underline{K}_{P\chi}\vec{\chi}_e + \underline{K}_{D\chi}\dot{\vec{\chi}}_e \tag{5.16}$$

The two terms $\underline{K}_{P\chi}\vec{\chi}_e$ and $\underline{K}_{D\chi}\dot{\vec{\chi}}_e$, in (5.16), are included to increase the rate of vibration suppression of the appendages. Without these terms, the time constant and energy spent for vibration suppression are too high and not optimal for a practical implementation. However, as will be discussed in Section 6.3, if these two terms are dropped, a larger stability region of convergence for the closed-loop system can be found. In a sense we can trade-off the performance of the controller versus a better stability proof!

Finally, substituting (5.16) in (5.13) and using (5.15), the output error dynamics for the flexible spacecraft, is governed by

$$\ddot{\vec{q}}_e + \underline{K}_{Dq}\dot{\vec{q}}_e + \underline{K}_{Pq}\vec{q}_e - \underline{K}_{P\chi}\vec{\chi} - \underline{K}_{D\chi}\dot{\vec{\chi}} = 0 \tag{5.17}$$

The next step is to select the appropriate gain matrices $\underline{K}_{Dq}, \underline{K}_{Pq}, \underline{K}_{P\chi}$ and $\underline{K}_{D\chi}$ to insure that the error dynamics is at worst, locally asymptotically stable. This stability analysis is covered in Chapter 6, and the simulation results for the controller are presented next.

## 5.2 Simulation Results for the flexible Spacecraft

The feedback linearization controller was used to recover a typical flexible spacecraft from 2 different malfunctions which throws the spacecraft into a tumble. The controller must proceed to detumble the spacecraft towards a safe hold mode of zero angular velocity and some inertial attitude pointing. Without loss of generality, we will set the reference attitude pointing to be the vernal equinox direction ($\vec{q}_{\text{ref}}^{\text{T}} = \begin{bmatrix} 0 & 0 & 0 \end{bmatrix}$) which means that the controller must align the spacecraft reference frame with the celestial reference frame. In doing so, the controller will be reducing the imparted kinetic energy back to zero. The spacecraft has a configuration as shown in Fig. 2.6 with a cube rigid bus of dimensions $1\text{m}\times1\text{m}\times1\text{m}$ and mass 2000 kg and one single appendage with area density of $\rho = 10\ \text{kg/m}^2$, width $a = 1\ \text{m}$, length $b = 10\ \text{m}$ and thickness $h = 0.02\ \text{m}$. The appendage is clamped midway along the side panel, $\vec{d}^T = [-0.5\ 0.5\ 0.0]$ m. The appendage modal damping is assigned to be 5% (i.e. $\xi = 0.05$), a low modulus of elasticity is used to increase the flexibility of the appendage, $E = 5.0 \times 10^8\ \text{N/m}^2$, and the Poisson ratio is set to $\gamma = 0.3$. With this configuration, the spacecraft's inertia matrix with its appendage fully deployed and undeflected is:

$$\underline{I}_t = \begin{bmatrix} 4192 & 0 & 0 \\ 0 & 342 & 0 \\ 0 & 0 & 4200 \end{bmatrix} \text{kgm}^2$$

The followings are the Attitude Recovery Maneuver (ARM) simulation cases for the flexible spacecraft and we resume the case numbering from where we left off in Chapter 4. Note that in all simulation cases presented in this chapter, the Roll, Pitch and Yaw refer to Euler angles defining the orientation between the spacecraft and the celestial reference frame, since an inertial pointing maneuver is being considered. Also in all cases, the (3,2,1) Euler rotation sequence was used to obtain the Euler angles from the quaternions.

*Simulation Case 5:* As the first simulation case (Figs. 5.1-5.4), the closed loop response of the system for a controller where direct vibration suppression has been turned off (i.e. $\underline{K}_{P\chi} = \underline{K}_{D\chi} = \underline{0}$) is presented. The initial disturbance is due to a malfunction imparting an initial large disturbance torque mainly about the *x*-axis given by $\vec{\tau} = [100 \quad -10 \quad -10]^T$ Nm. The disturbance is assumed to start at the 5-seconds mark ($t = 5$ seconds) and lasts for 5 seconds, with constrained control torques ($\pm 50 Nm$), starting 10 seconds after the end of the disturbance torque ($t = 20$ seconds) to allow for extra deviation of the spacecraft's states from the equilibrium point. In this case, 2 modes in each *x* and *y* directions of the appendage (i.e. $p = q = 2$) were activated, bringing the total system dimension to $m = 8 + 2(1)(2)(2) = 16$.

As expected, the performance of the controller is deteriorated (compared to other simulation cases below where $\underline{K}_{P\chi} \neq \underline{0}$; $\underline{K}_{D\chi} \neq \underline{0}$) as far as vibration suppression and control effort is concerned (Figs. 5.3 and 5.4). It takes about 1600 seconds to damp out the vibrations through the natural damping of the appendage. Incidentally, an approximation of this time constant can be obtained analytically using the zero dynamics (refer to Section 6.2) given by (6.42) which is a second order system. The settling time for a second order system is given by $t_s \approx 4/\xi\omega_n$ where $\xi$ and $\omega_n$ are the damping ratio and the natural frequency. Using the parameters for our flexible spacecraft, the settling time is obtained to be 1600 seconds which matches our simulation results. Hence, this simulation run can be used also as another validation case where predicted analytical results matches our simulation output.

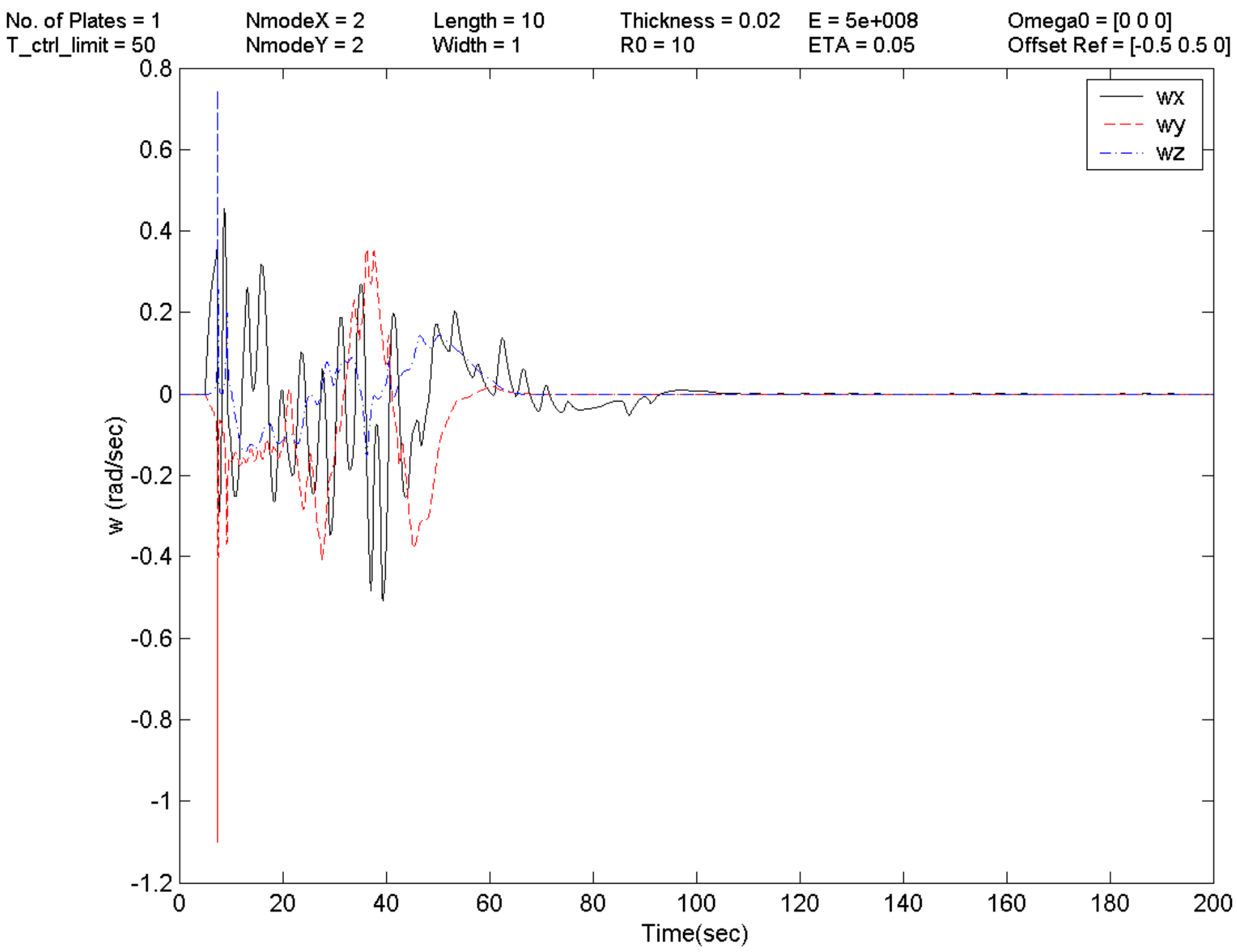

Figure 5.1: ARM case 5 - Spacecraft body angular rates ($\vec{\omega}$)

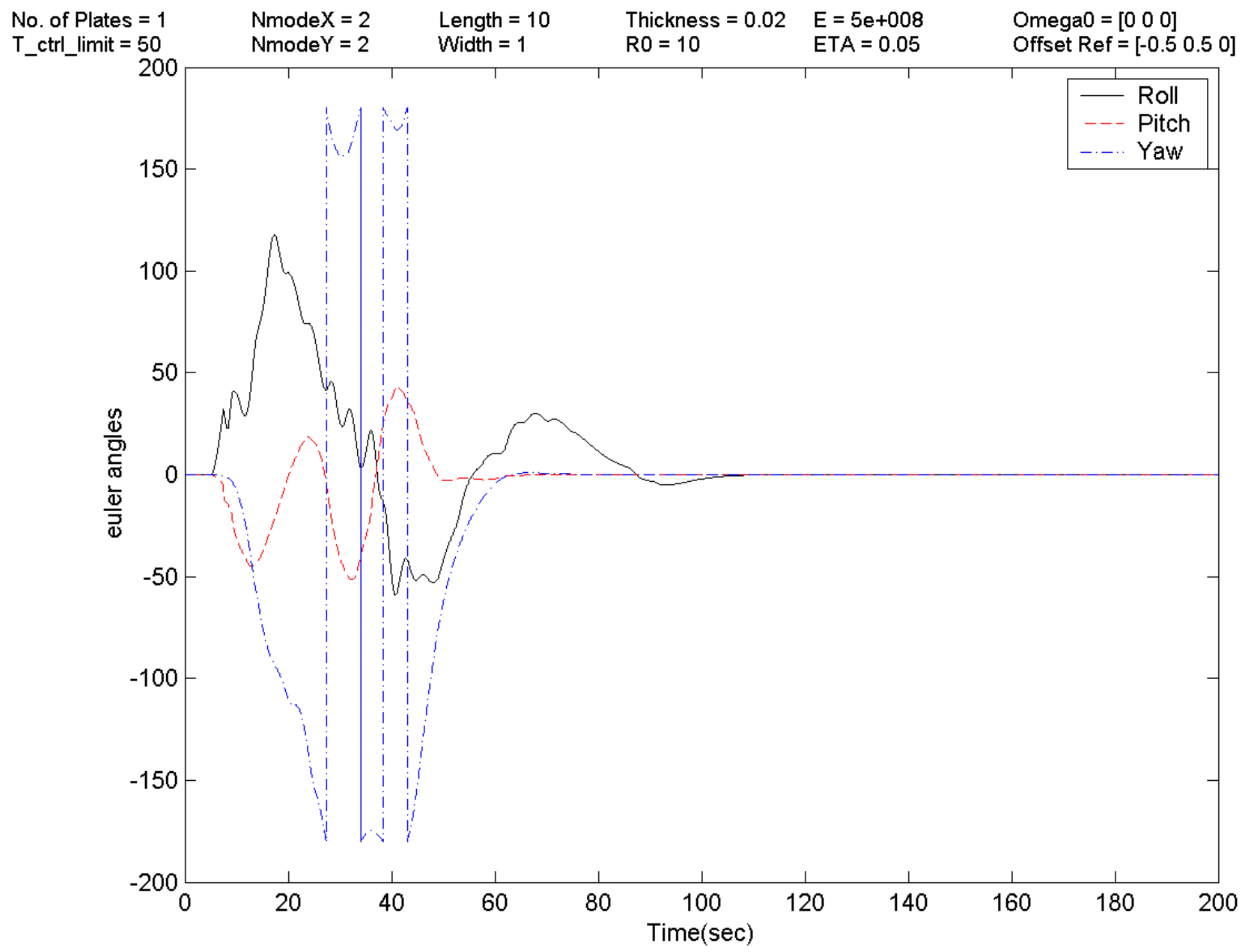

Figure 5.2: ARM case 5 - Spacecraft attitude (Euler angles)

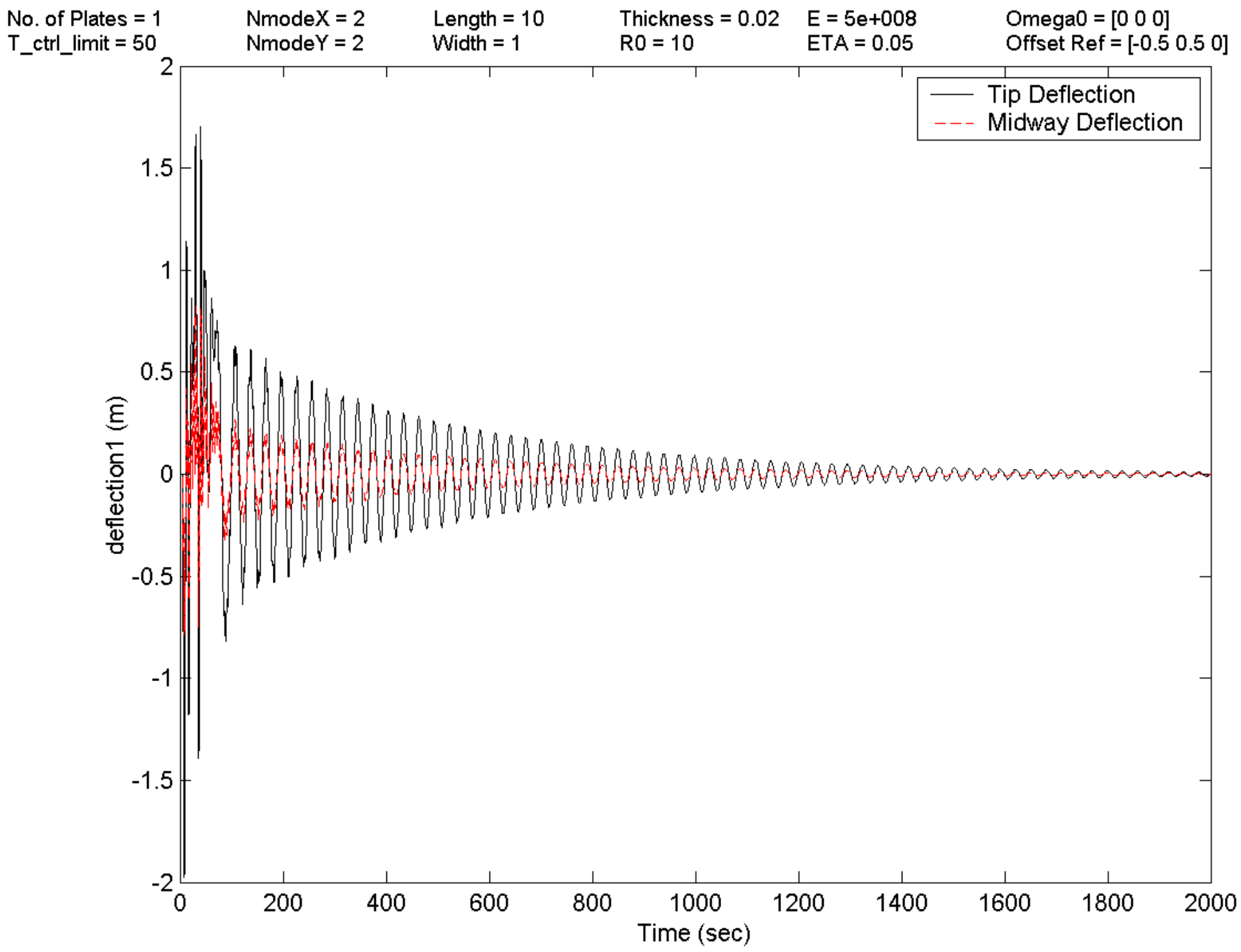

Figure 5.3: ARM case 5 - Deflection ( $w$ ) at tip $x = .5; y = 10$ and midway $x = .5; y = 5$

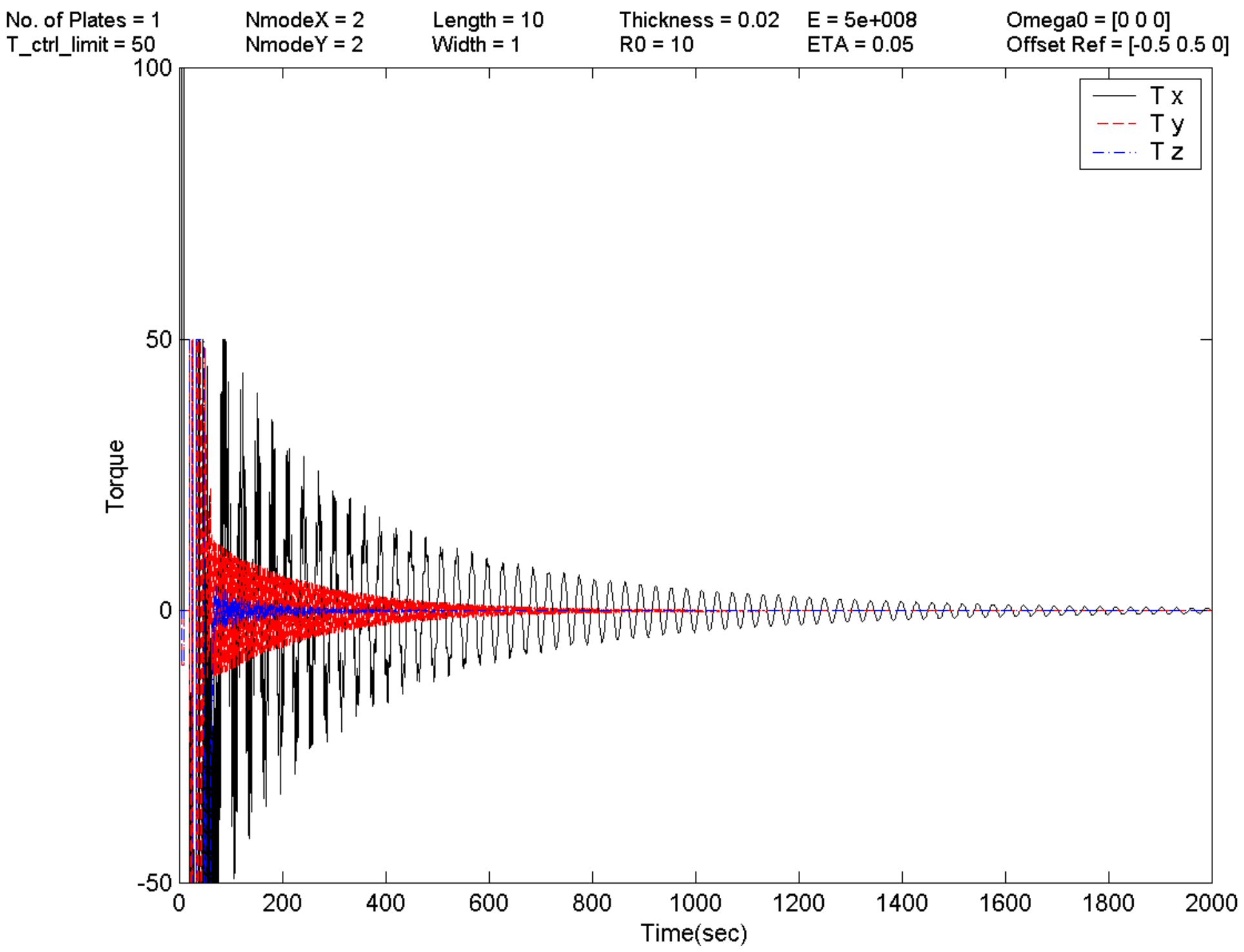

Figure 5.4: ARM case 5 - Initial disturbance and attitude recovery torques ( $\vec{\tau}$ )

*Simulation Case 6:* This simulation case represents the closed loop response of the flexible spacecraft (Figs. 5.5 to 5.9) to the same *x*-axis momentum wheel failure as in case 5. In this case again, 2 modes in each *x* and *y* directions of the appendage were activated, however, the same simulation case was run with higher numbers of model vibration modes (i.e. $p = q = 3 \; or \; 4$ ) but did not result in any noticeable difference in the dynamic response, indicating that the higher flexible modes are not excited as much as the first 2 modes. The controller gains $\underline{K}_{Pq} = 0.08\underline{i}$, $\underline{K}_{Dq} = 0.57\underline{i}$, $\underline{K}_{P\chi} = 0.01\underline{1}$ and $\underline{K}_{D\chi} = 0.001\underline{1}$ were used in (5.16) where $\underline{i}$ is the $3 \times 3$ identity matrix and $\underline{1}$ is the $3 \times 4$ matrix having 1 for all its entries.

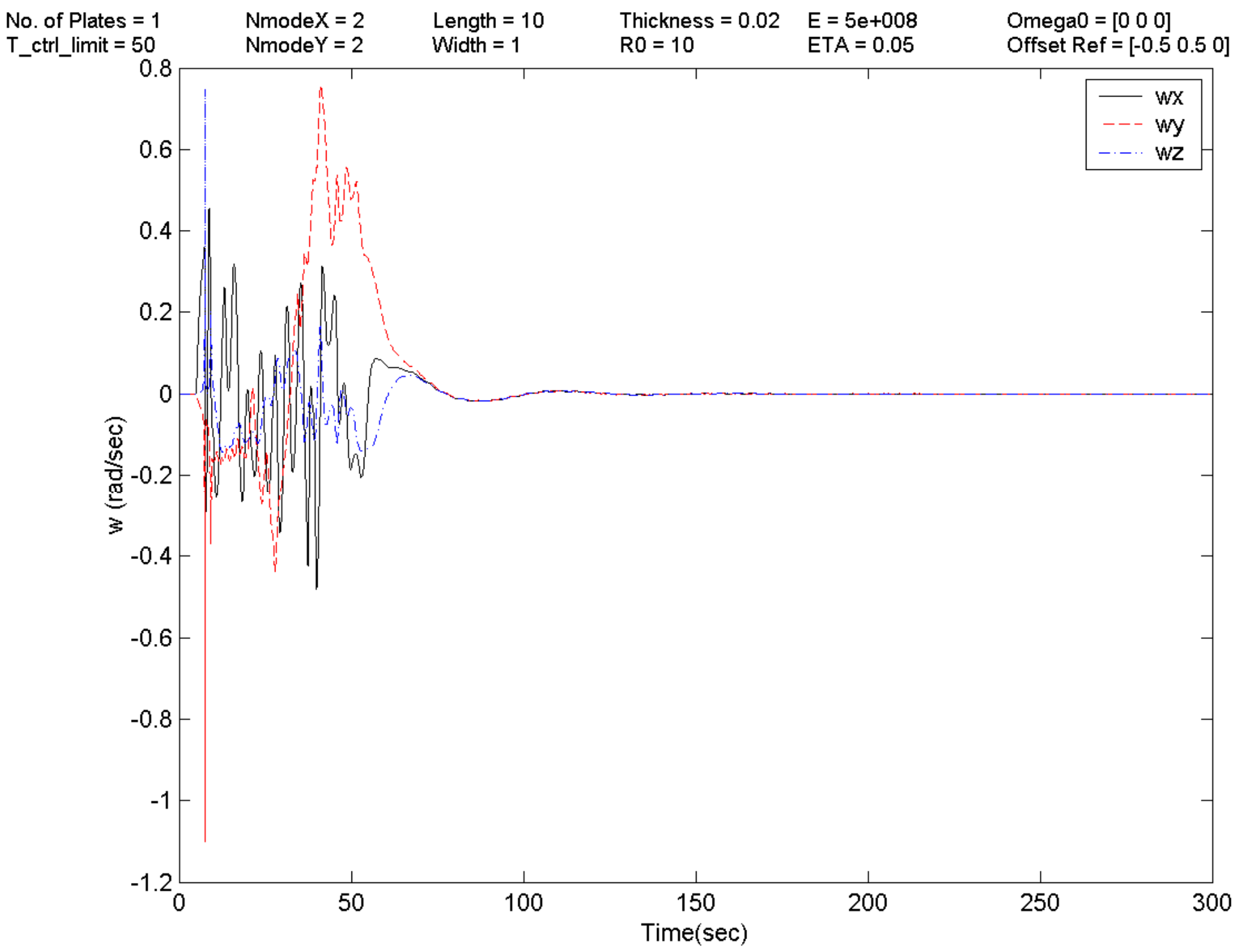

Figure 5.5: ARM case 6 - Spacecraft body angular rates ( $\vec{\omega}$ )

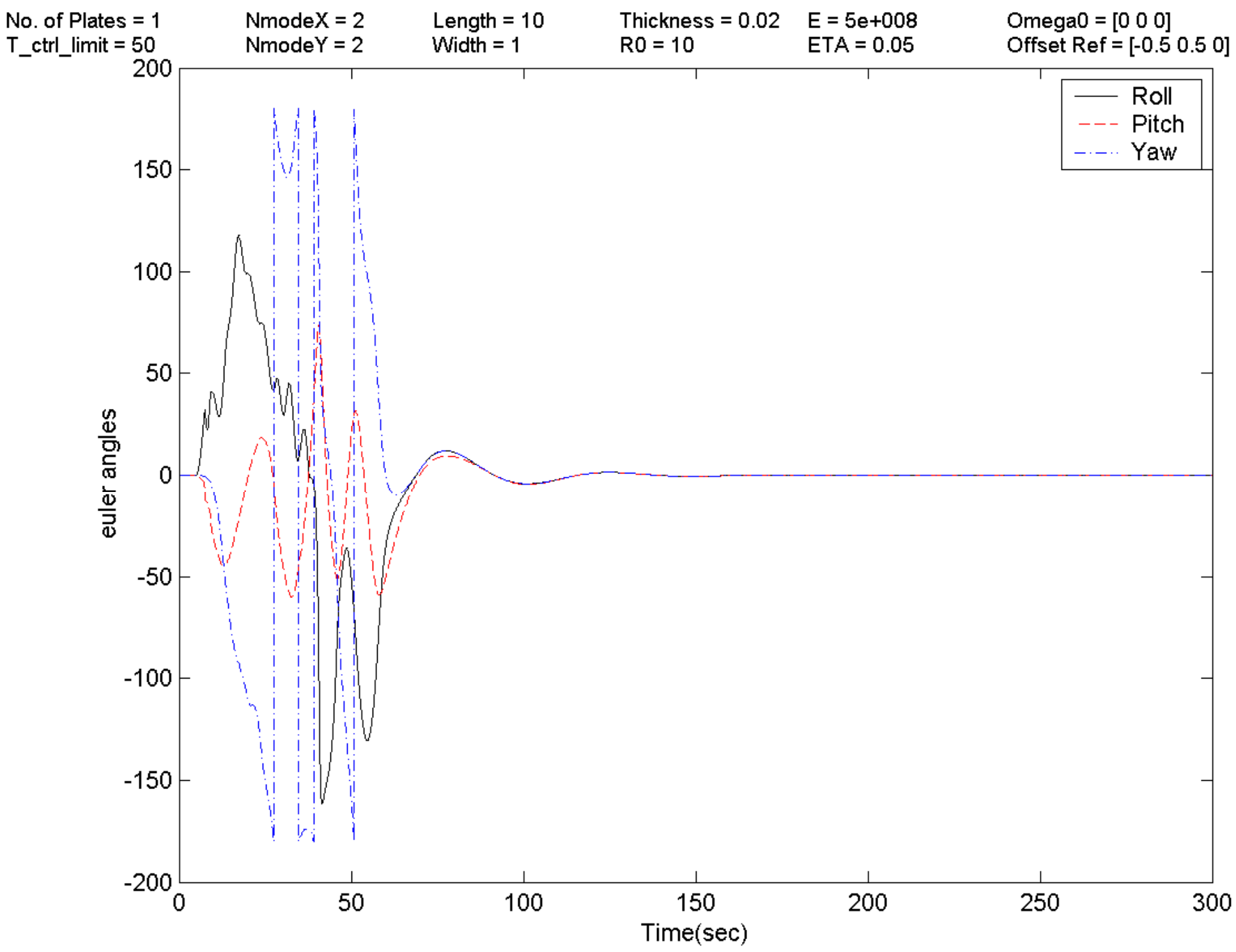

Figure 5.6: ARM case 6 - Spacecraft attitude (Euler angles)

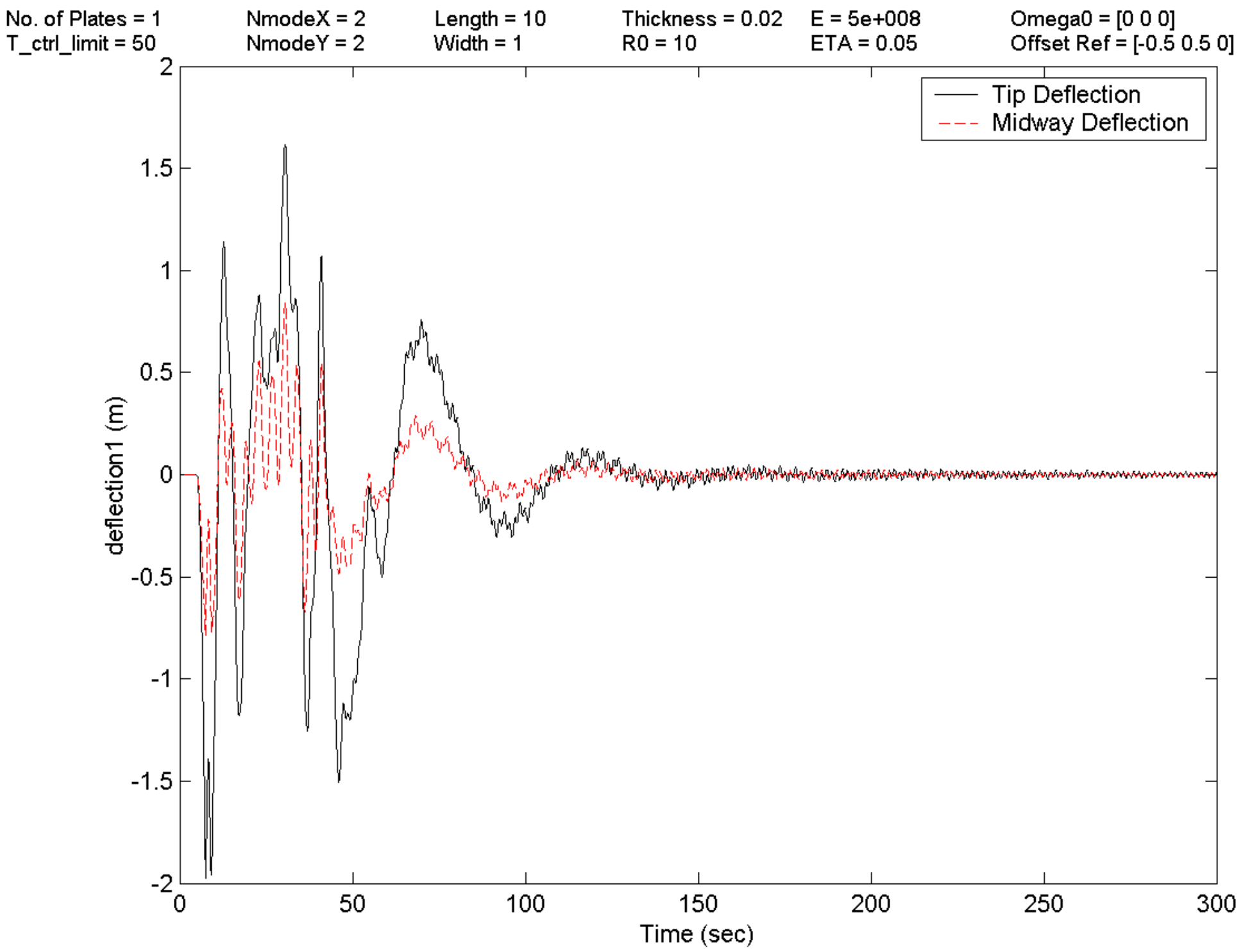

Figure 5.7: ARM case 6 - Deflection ( $w$ ) at tip $x = .5; y = 10$ and midway $x = .5; y = 5$

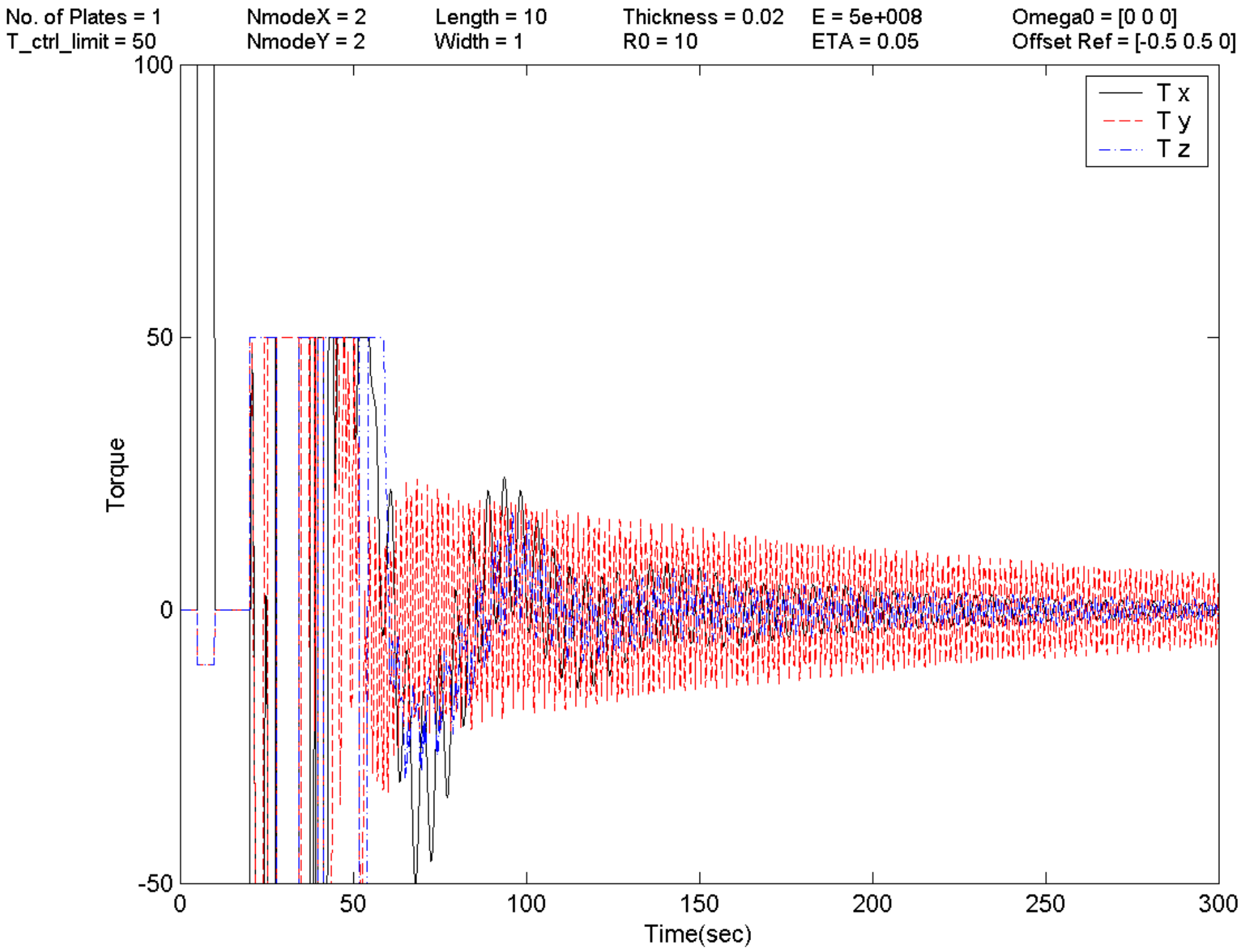

Figure 5.8: ARM case 6 - Initial disturbance and attitude recovery torques ( $\vec{\tau}$ )

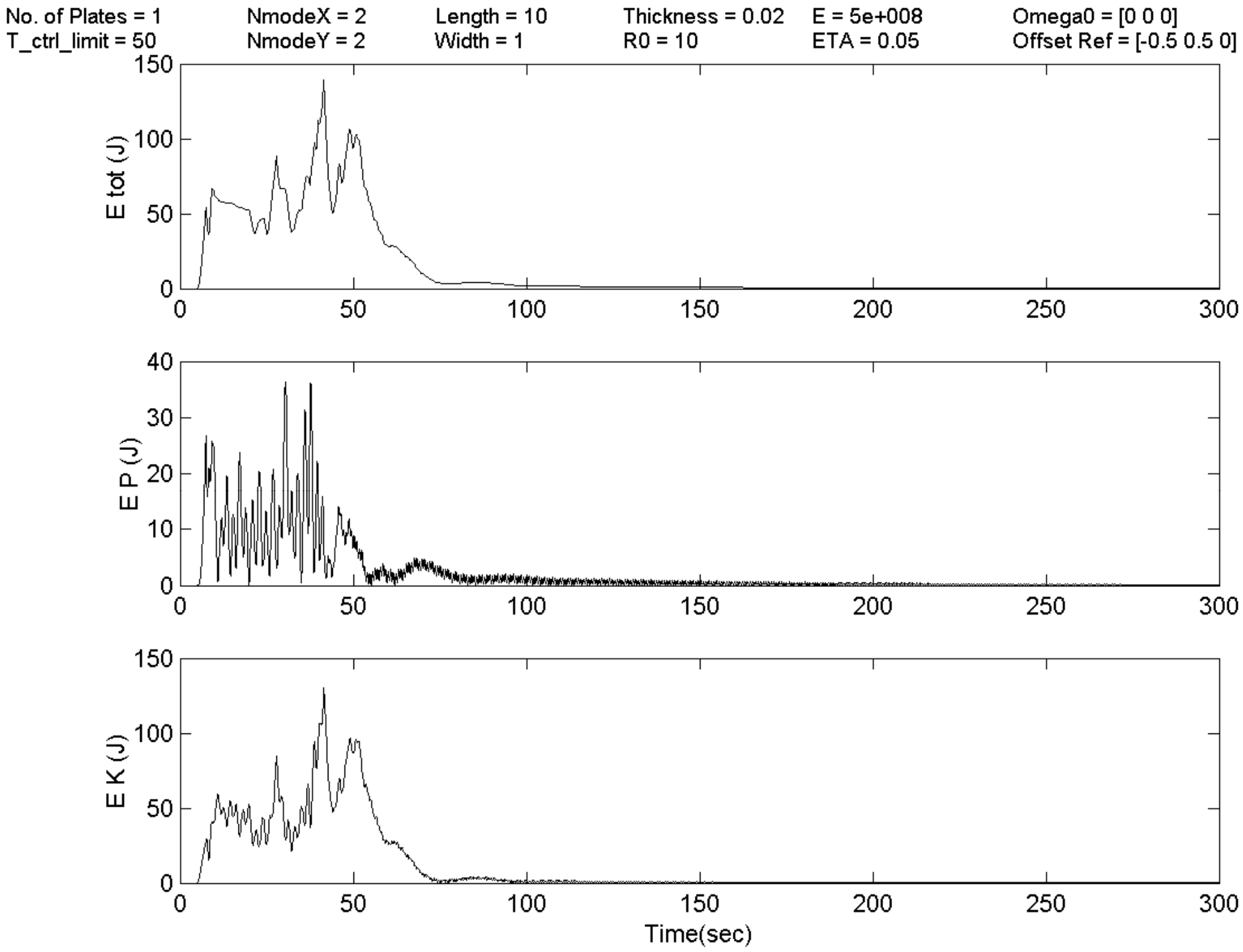

Figure 5.9: ARM case 6 - Spacecraft total, potential, and kinetic energies

We notice a max tip deflection of about 2m (Fig. 5.7) which represents 20% of the 10m appendage length at an angle of deflection of about $10^o$. Although, this is certainly not the usual small deflections found on space appendages (1-2%), we intended to allow more flexibility for the appendages in anticipation of more flexible materials to be used in the construction of future spacecraft and also to test out our controller in the worst of conditions. We believe that the 20% tip deflections is at the limit allowed by the linear theory.

*Simulation Case 7:* Figures 5.10 to 5.14 show the closed loop response of the flexible spacecraft to a severe disturbing torque of $\vec{\tau} = \begin{bmatrix} 100 & -100 & 100 \end{bmatrix}^T$ Nm, starting at the 5 seconds mark and ending after 4 seconds, with constrained control torques ($\pm 50 Nm$), starting 1.0 second after the end of the disturbance torque ($t = 10$ seconds) to allow for extra deviation of the spacecraft's states from the equilibrium point. All other parameters were kept the same as in case 6.

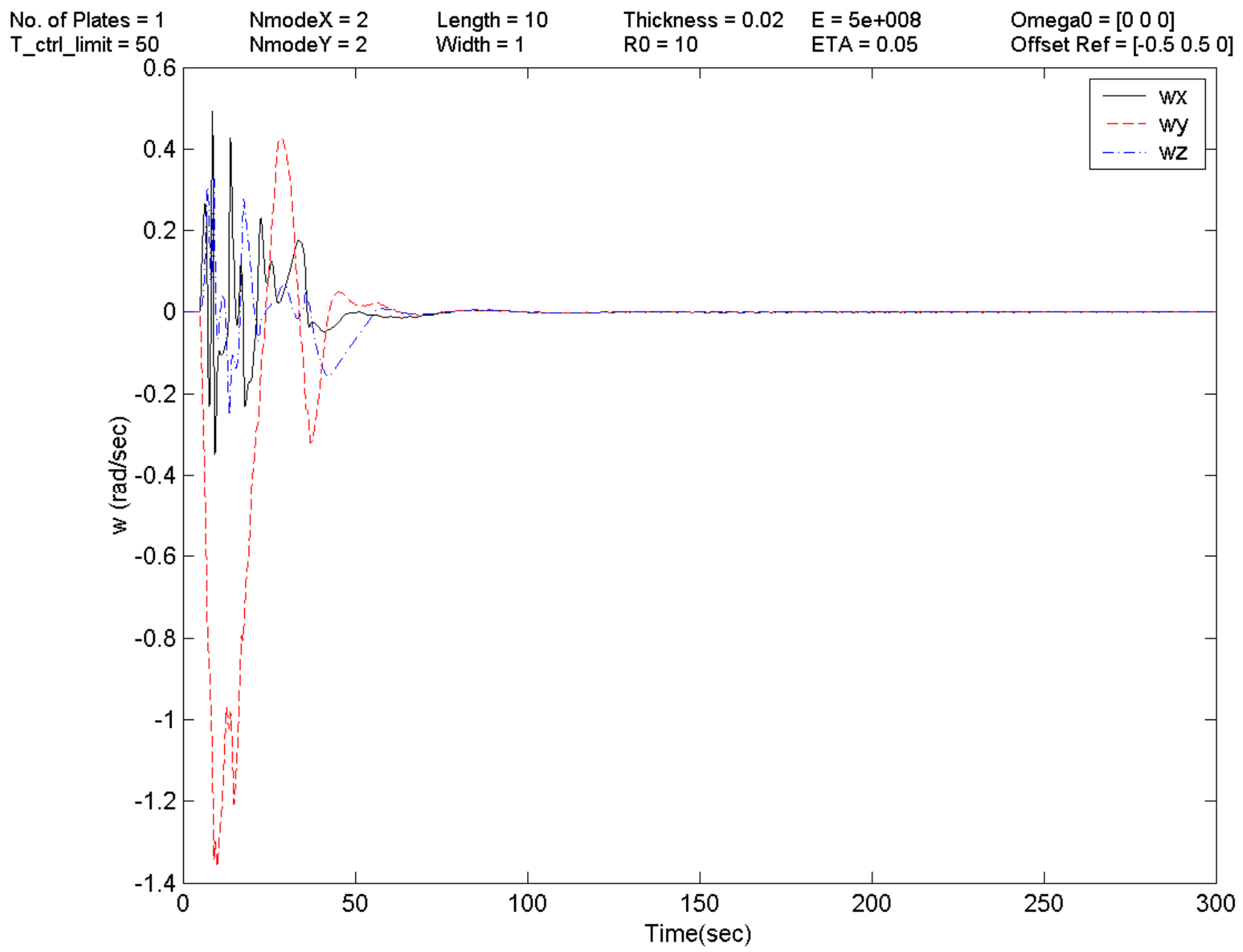

Figure 5.10: ARM case 7 - Spacecraft body angular rates ( $\vec{\omega}$ )

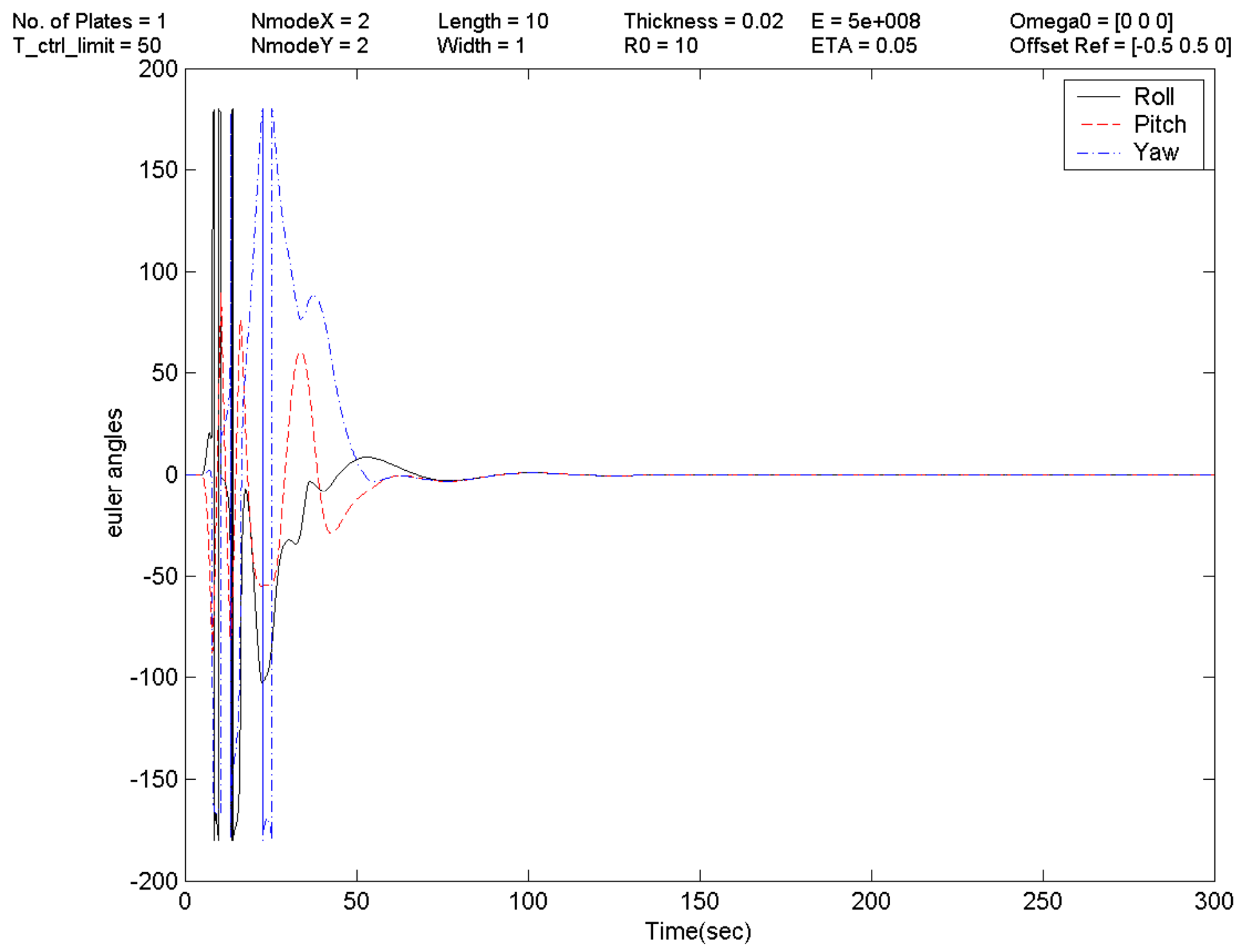

Figure 5.11: ARM case 7 - Spacecraft attitude (Euler angles)

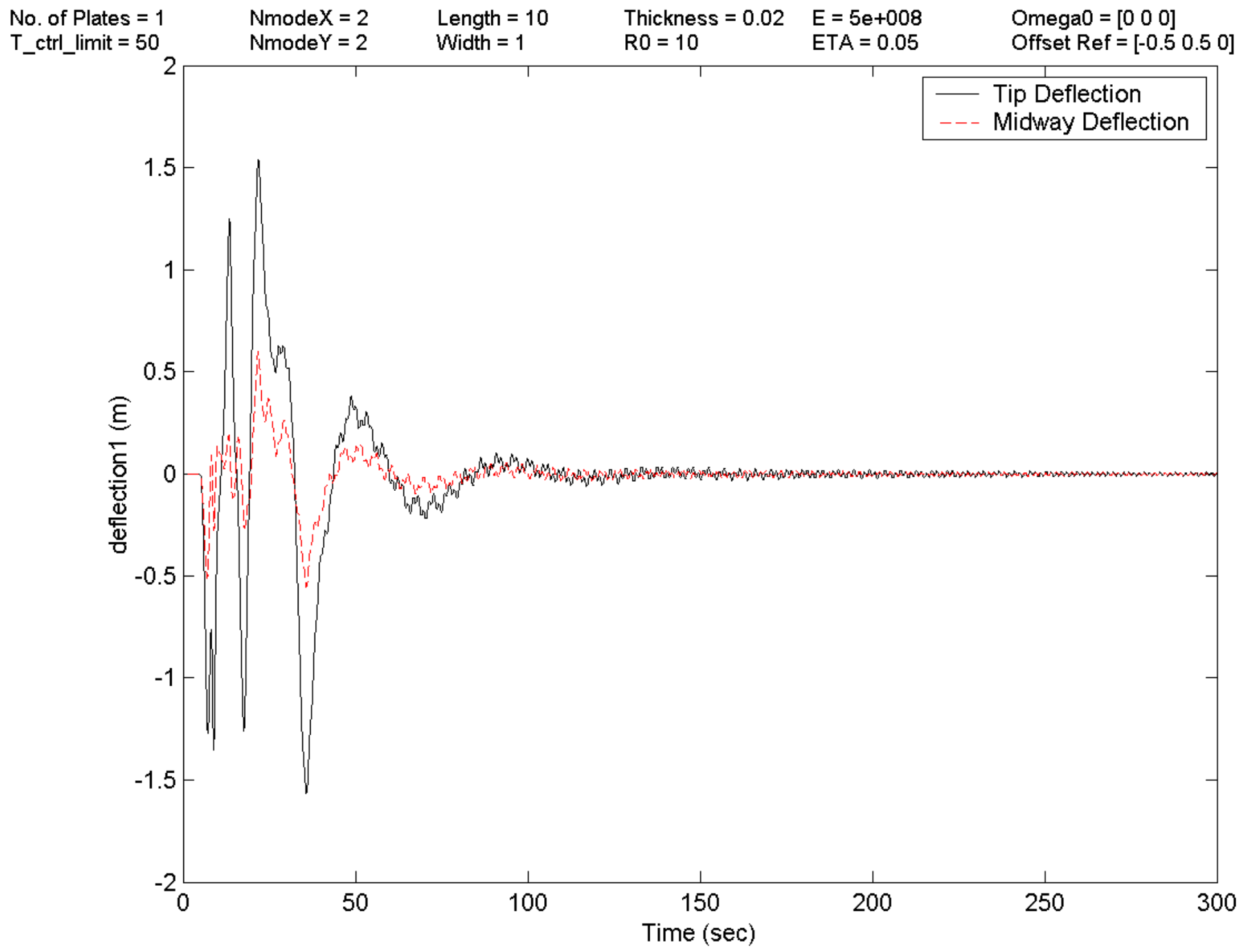

Figure 5.12: ARM case 7 - Deflection ( $w$ ) at tip $x = .5; y = 10$ and midway $x = .5; y = 5$

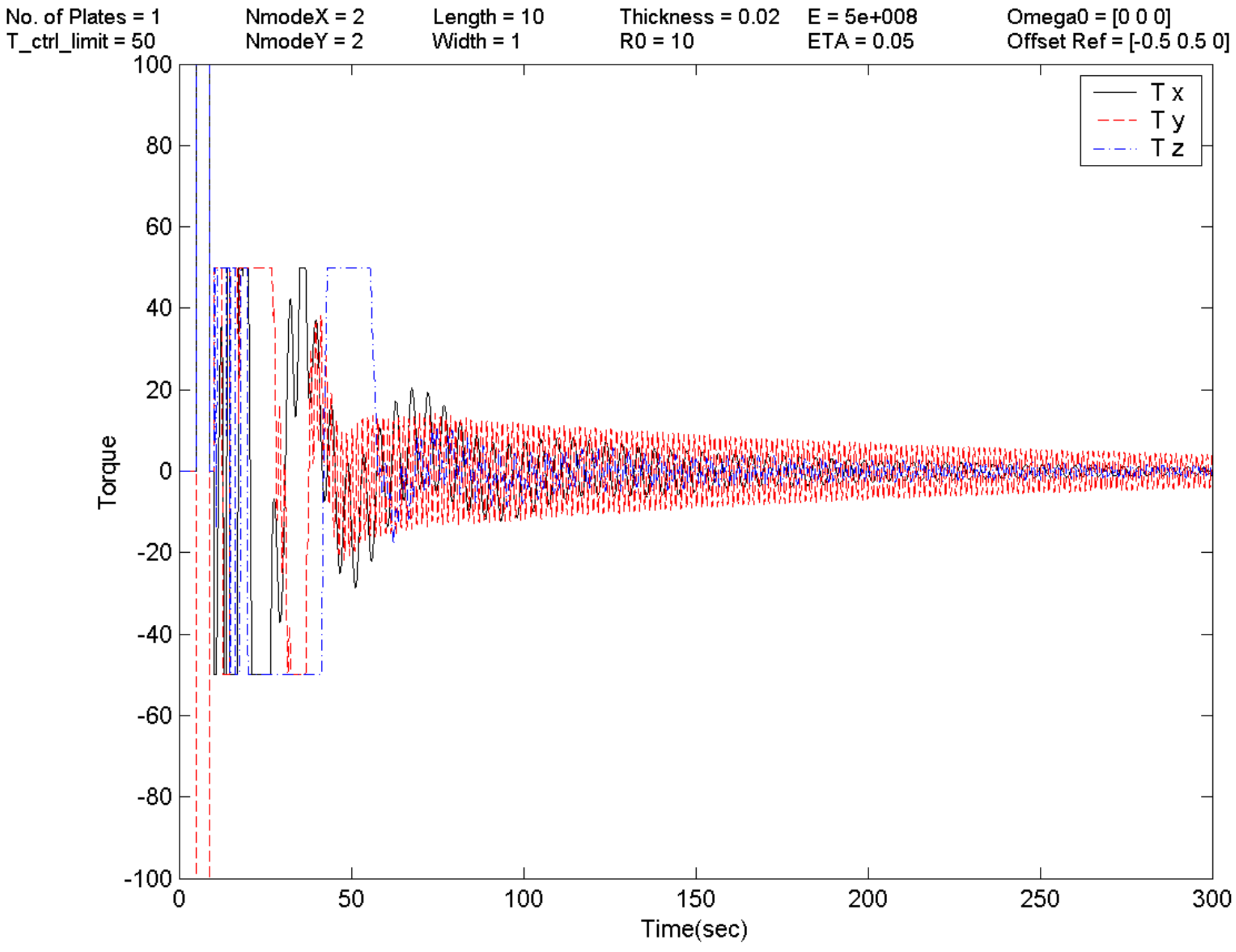

Figure 5.13: ARM case 7 - Initial disturbance and attitude recovery torques ( $\vec{\tau}$ )

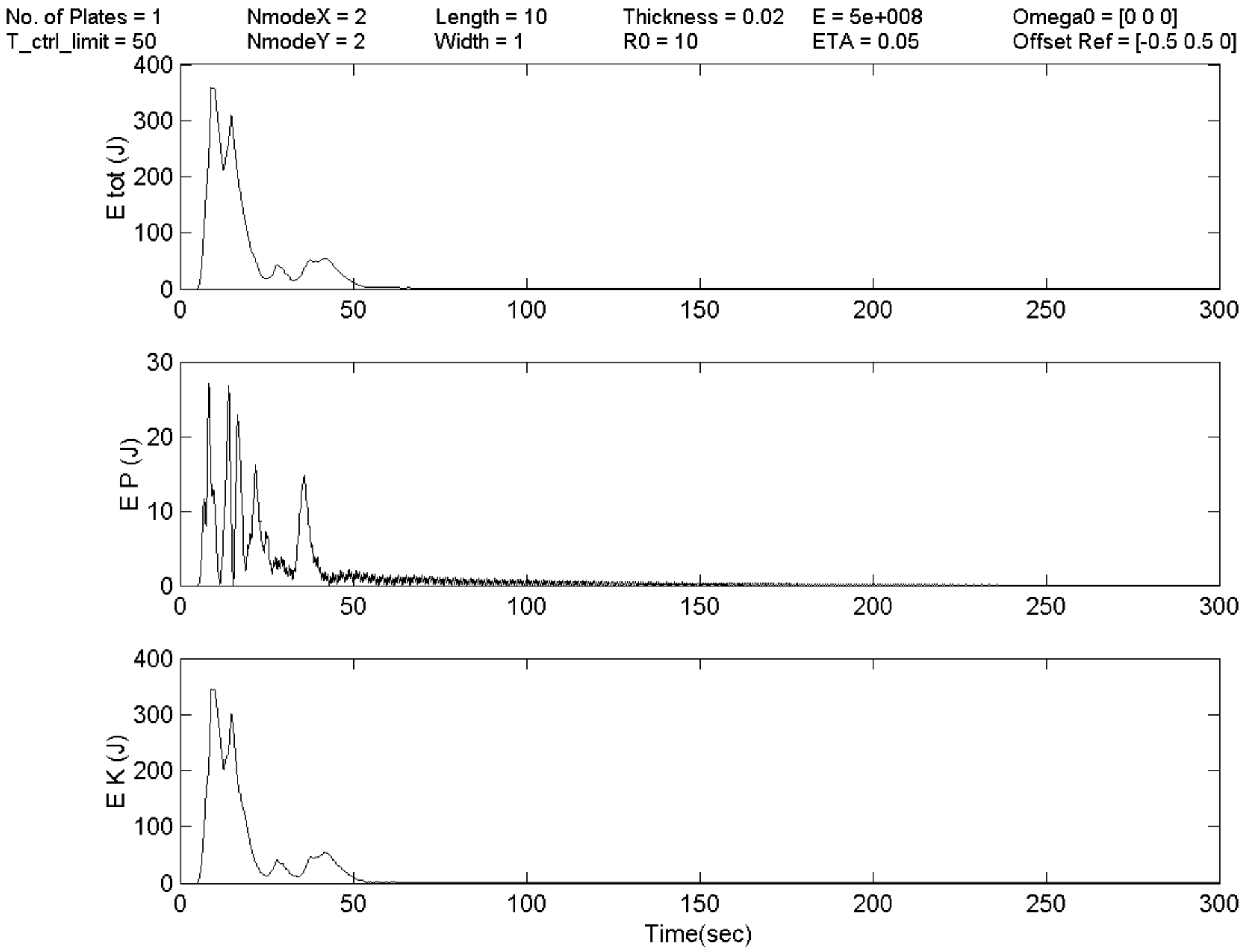

Figure 5.14: ARM case 7 - Spacecraft total, potential, and kinetic energies

## 5.3 Robustness Considerations

In order to validate and investigate the capability of the controller and its robustness against model and parameters uncertainties, which will be undoubtedly present in any complex system model, numerous simulation cases were carried out (6 of which are presented in this section). The model uncertainties are due to various factors such as

(i) Assumptions made in the formulation of the flexible dynamics might not be entirely correct in the real system (e.g. the appendages were considered to be of uniform density but could in fact have significant density variation).

(ii) The appendage mode shapes were modeled using two separate beam eigenfunctions (e.g. free-free and clamped-free in *x* and *y* direction, respectively) which most likely will not represent the actual, more complex mode shapes of a flexible plate-type appendage.

(iii) The appendages were discretized using a finite number of modes (e.g. $p = 3,\ q = 2$) where in fact the real system has infinite modes!

(iv) Some second and higher order nonlinear terms were ignored in the formulation (e.g. $\dot{v}^2$ term was dropped from equation (A18); refer to Appendix A).

The parameter uncertainties are mainly due to the fact that accurate knowledge of the spacecraft parameters (e.g. appendage density $\rho$, damping $\xi$, modulus of elasticity $E$, inertia matrix $\underline{I}$, etc.) are difficult. In fact, some parameters such as inertia could be quite difficult to obtain with precision and are usually only an estimate to the actual values.

In order to simulate model and parameter uncertainties, large increase in the values of $\vec{f}_{\bar{q}}$ and $\underline{G}_{\bar{q}}$ were introduced. In some cases, we also reduced the number of controller modes (e.g. low-order controller) as another mechanism to simulate model uncertainties. The following three subsections are devoted to simulation results examining robustness of the proposed controller.

### 5.3.1 Simulation Results for the Flexible Spacecraft with Model/Parameter Uncertainties

*Simulation Case 8:* This is a repeat of simulation case 6 with 20% increase in the values of $\vec{f}_{\bar{q}}$ and $\underline{G}_{\bar{q}}$ in (5.12) which would represents the model/parameter uncertainties in the system. Figures 5.15 to 5.19 show the closed loop response and it is noted that the system has become somewhat sluggish due to the introduction of errors in the $\vec{f}_{\bar{q}}$ and $\underline{G}_{\bar{q}}$ terms of the controller. However, the overall settling time of the system motions is still about 150 seconds which indicates that the controller must be working harder to achieve this result. For example, Fig. 5.17 shows an extra low frequency oscillation of the appendage compared to case 6, which also shows up in the potential energy plot in Fig. 5.19 as two extra peaks between the 50 and 100 seconds time interval. The maximum positive tip deflection of the appendage is larger in this case as well and can be seen to be around 1.75m in Fig. 5.17. Overall, comparing the plots in this case to the corresponding ones in case 6, we notice more activity between the 50 and 100 seconds time interval in case 8.

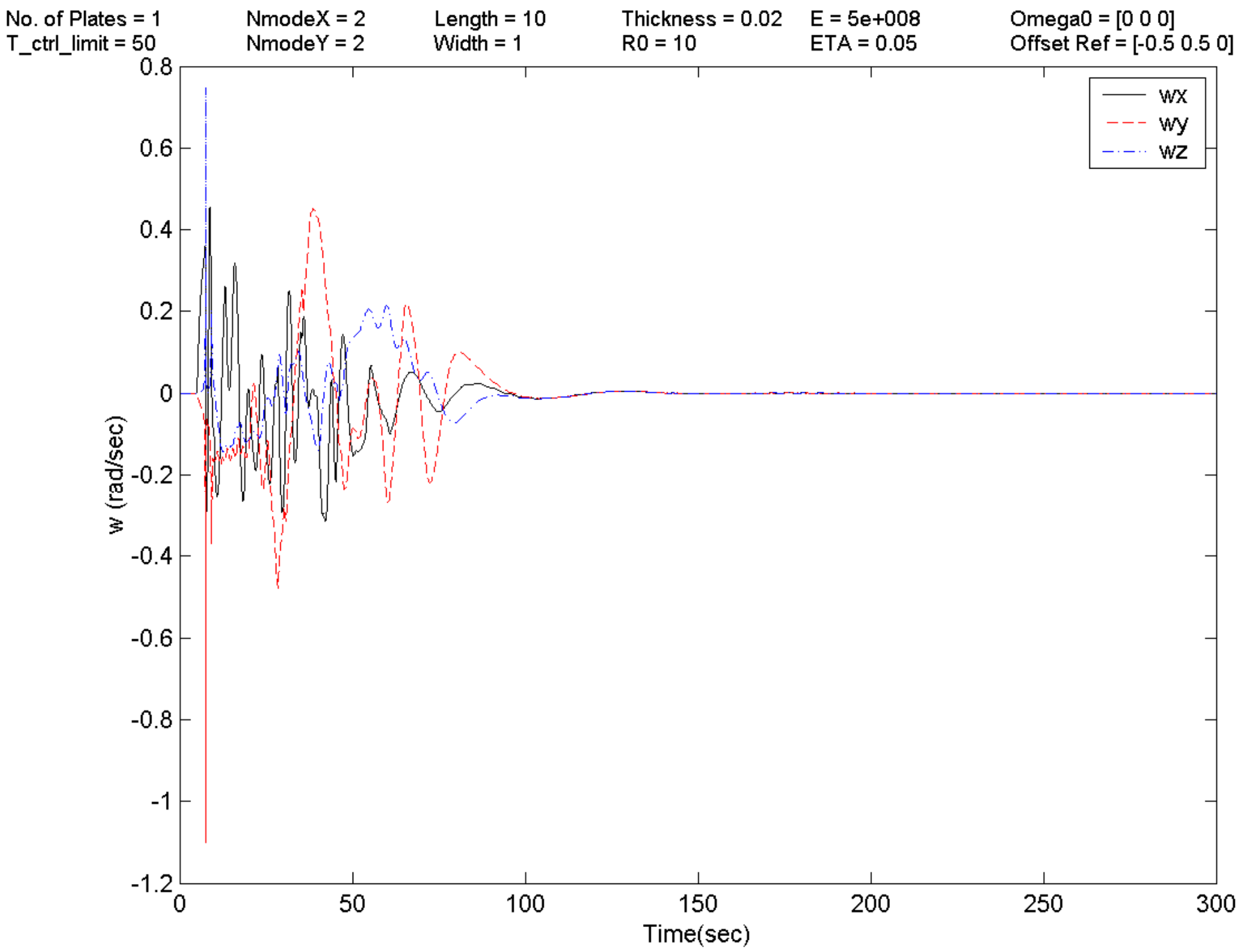

Figure 5.15: ARM case 8 - Spacecraft body angular rates ($\vec{\omega}$)

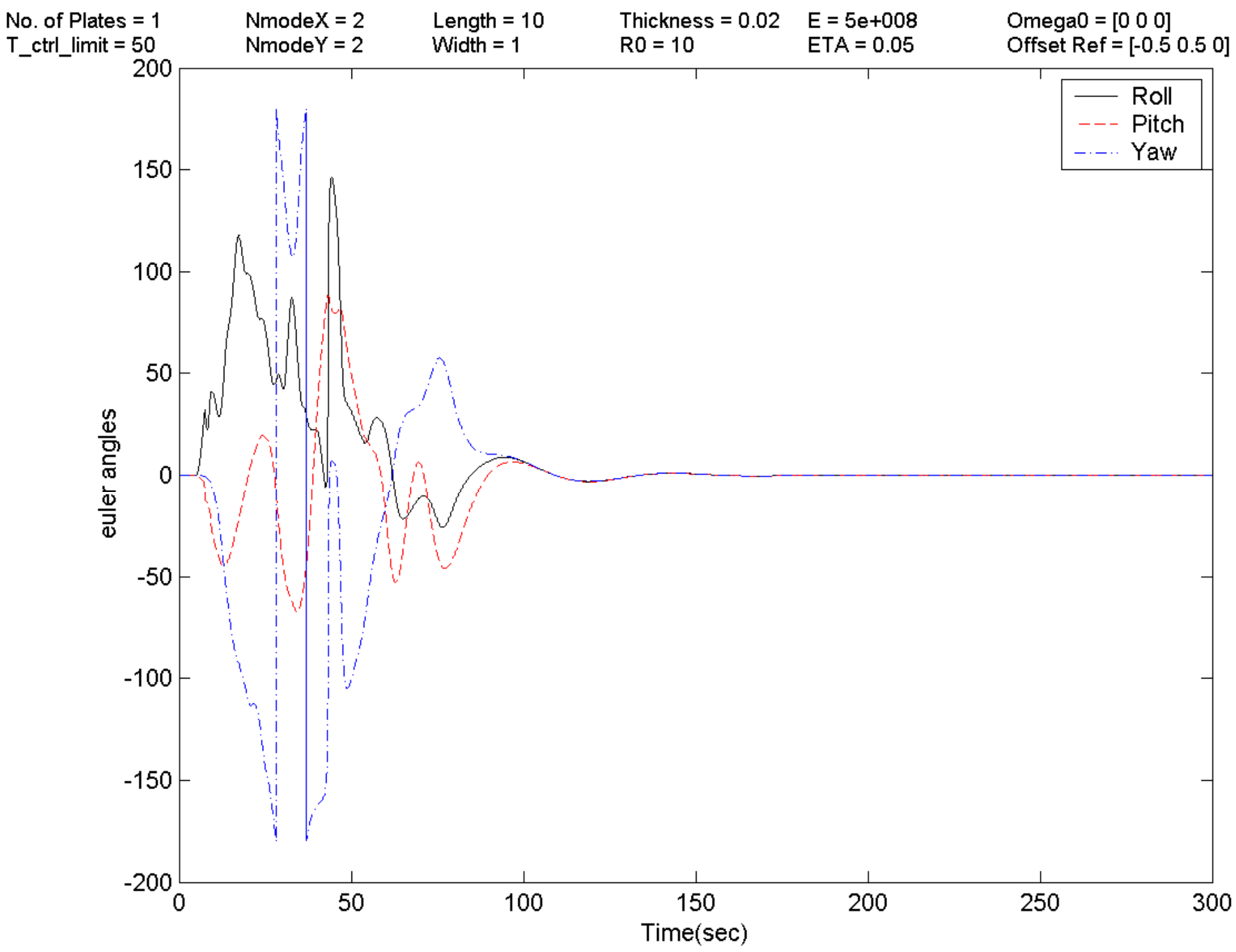

Figure 5.16: ARM case 8 - Spacecraft attitude (Euler angles)

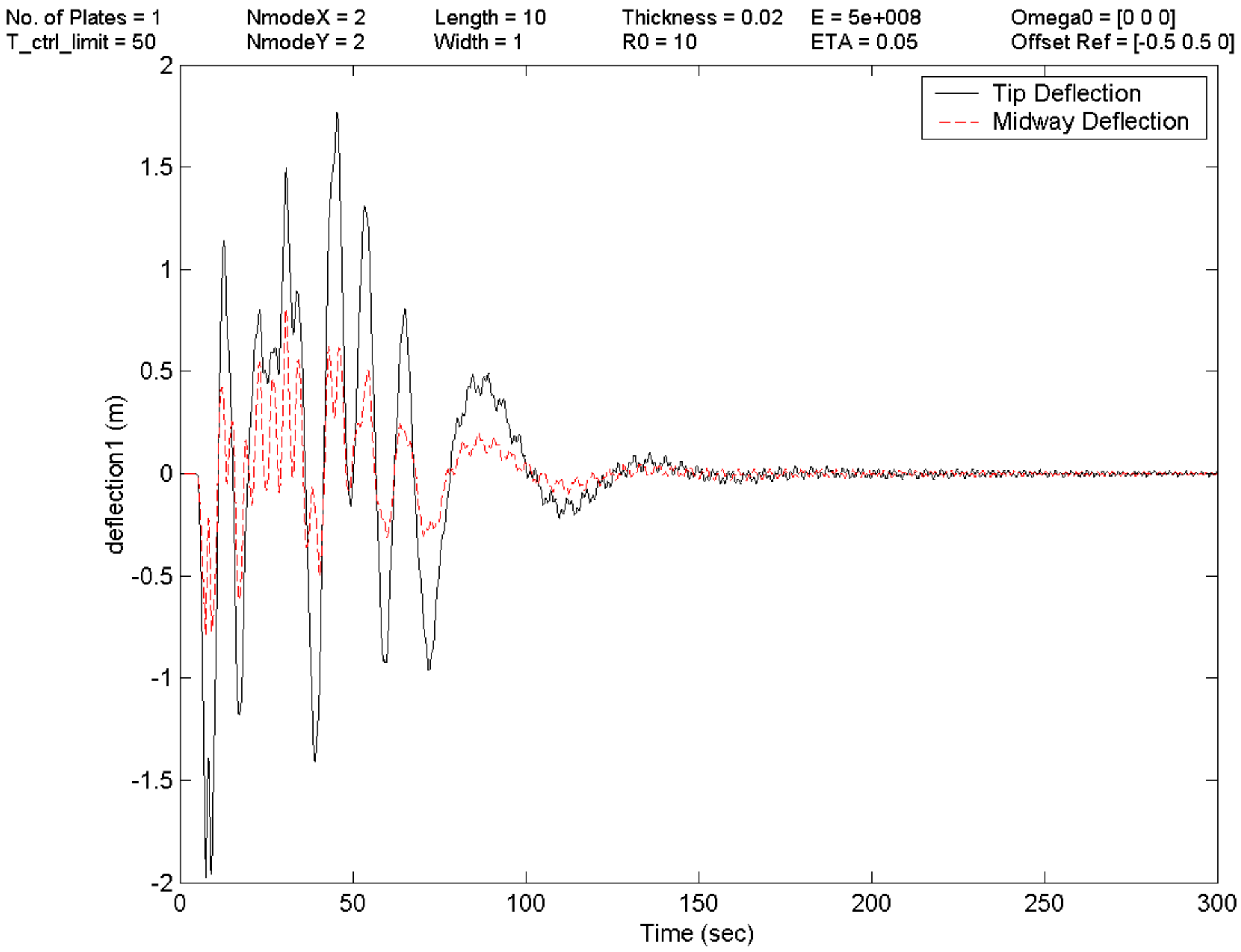

Figure 5.17: ARM case 8 - Deflection ( $w$ ) at tip $x = .5; y = 10$ and midway $x = .5; y = 5$

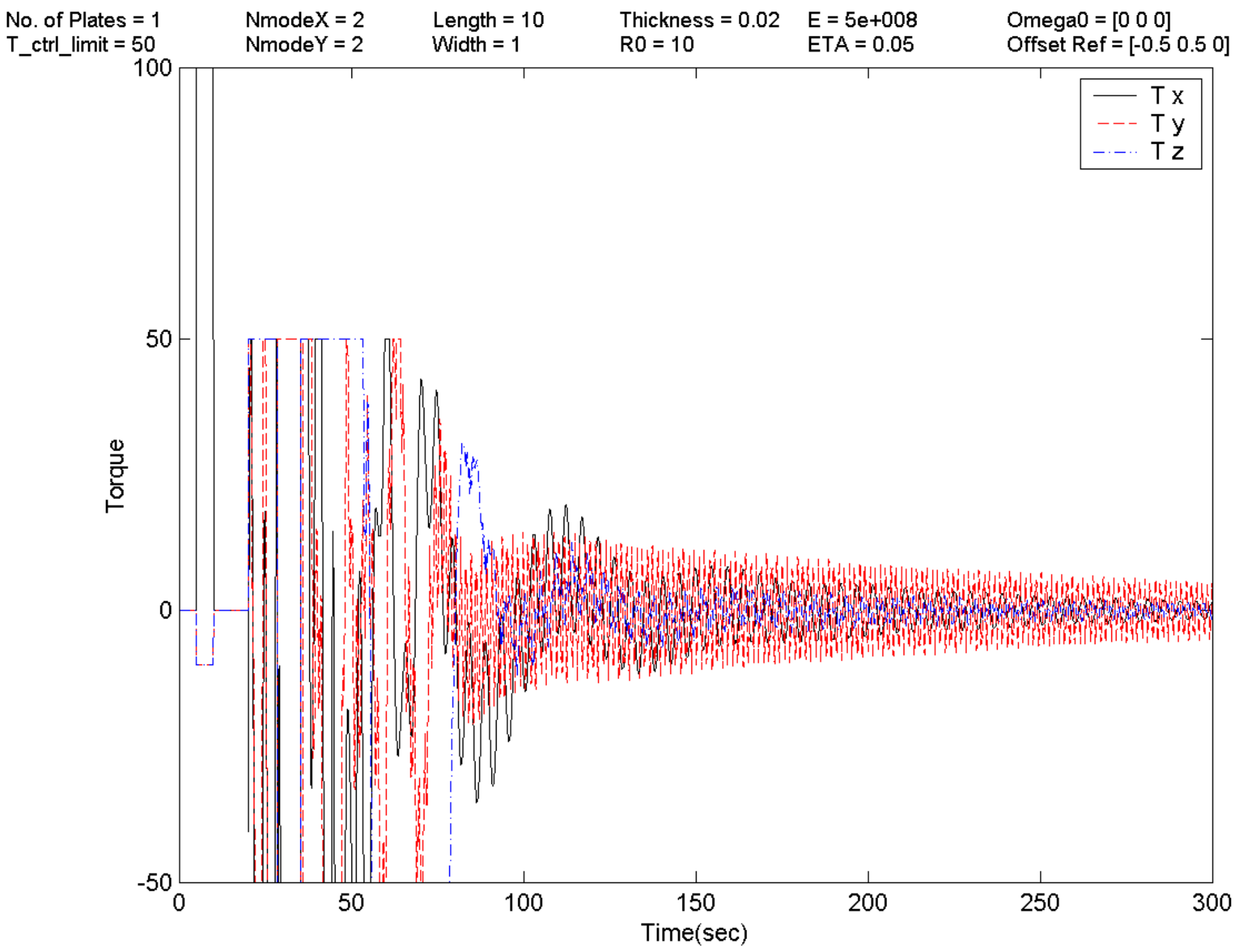

Figure 5.18: ARM case 8 - Initial disturbance and attitude recovery torques ( $\vec{\tau}$ )

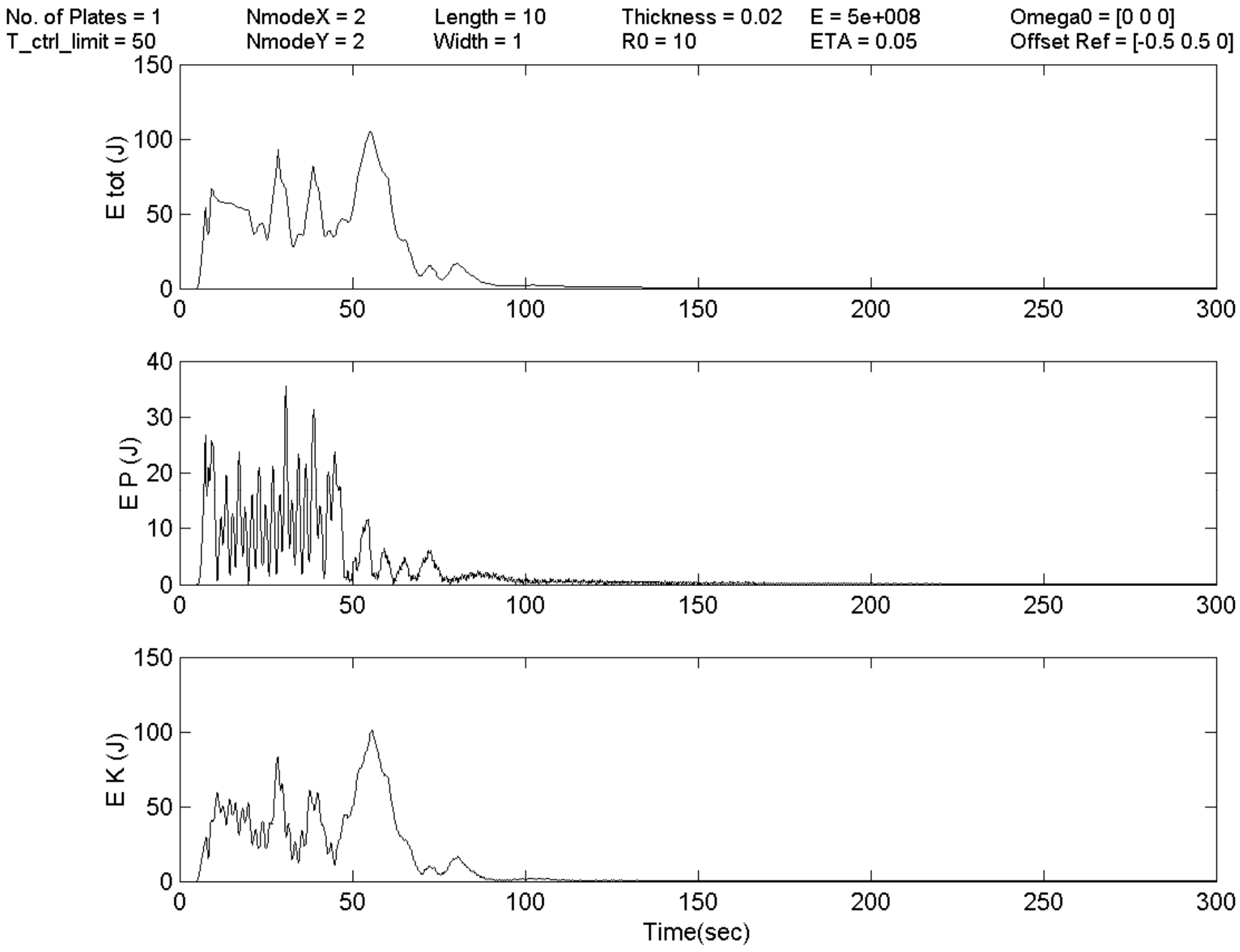

Figure 5.19: ARM case 8 - Spacecraft total, potential, and kinetic energies

*Simulation Case 9:* This is a repeat of simulation case 7 with 20% increase in the values of $\vec{f}_{\vec{q}}$ and $\underline{G}_{\vec{q}}$ in (5.12) representing the model/parameter uncertainties in the system. Figures 5.20 to 5.24 show the closed loop response and we notice again a bit of a more sluggish response than case 7 and Fig. 5.24 shows one higher potential energy peak than case 7. Otherwise, the response in case 9 is fairly similar to case 7.

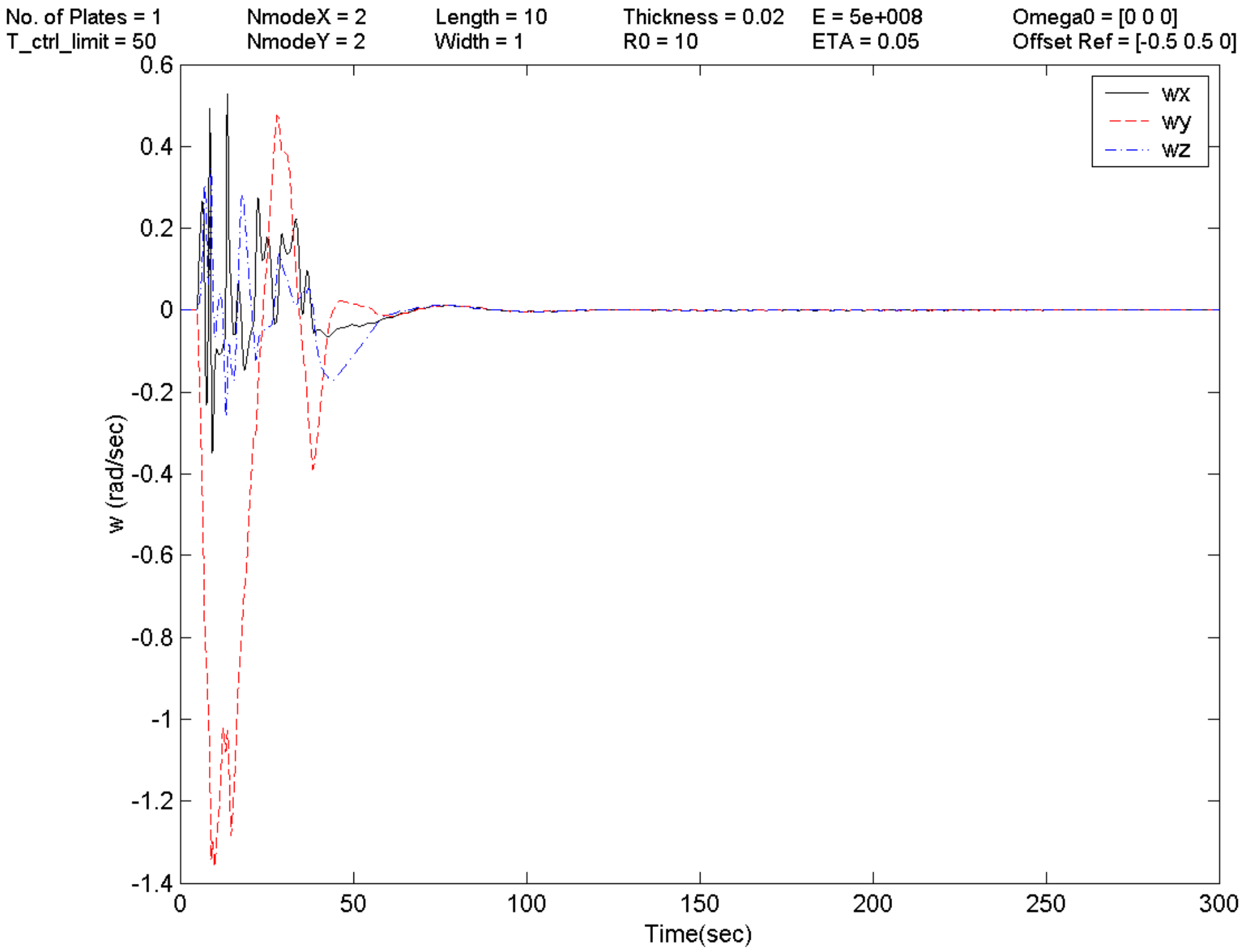

Figure 5.20: ARM case 9 - Spacecraft body angular rates ($\vec{\omega}$)

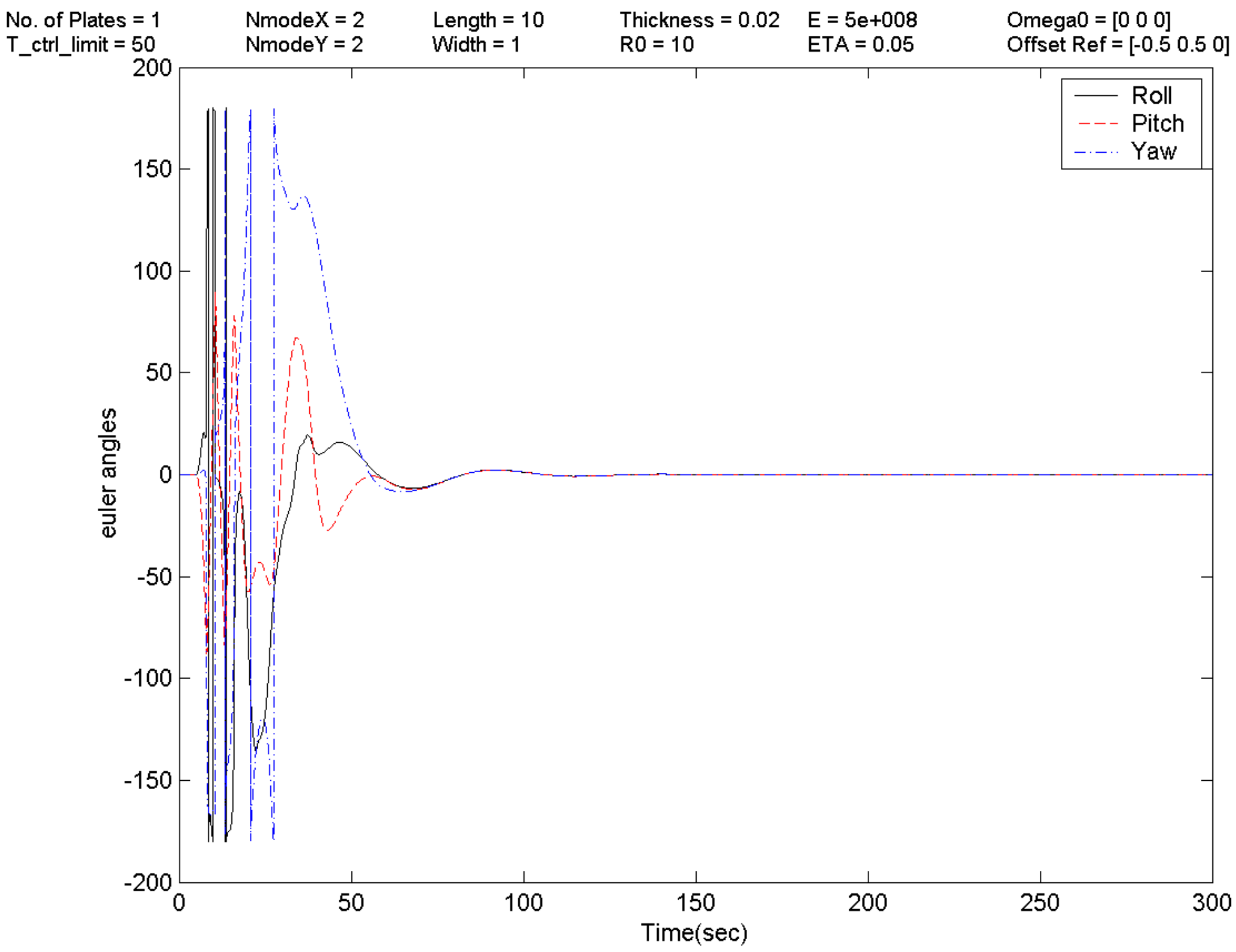

Figure 5.21: ARM case 9 - Spacecraft attitude (Euler angles)

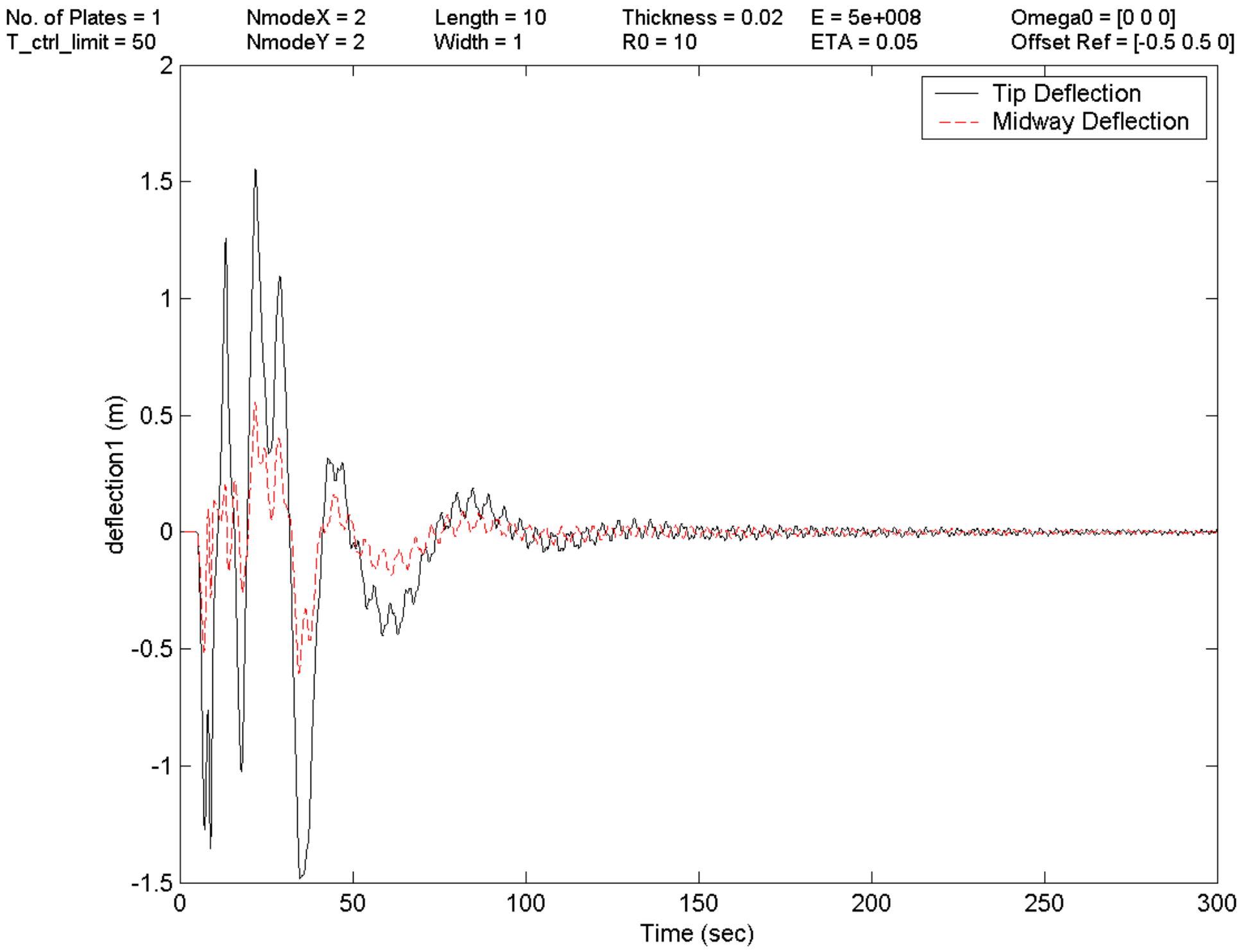

Figure 5.22: ARM case 9 - Deflection ( $w$ ) at tip $x = .5; y = 10$ and midway $x = .5; y = 5$

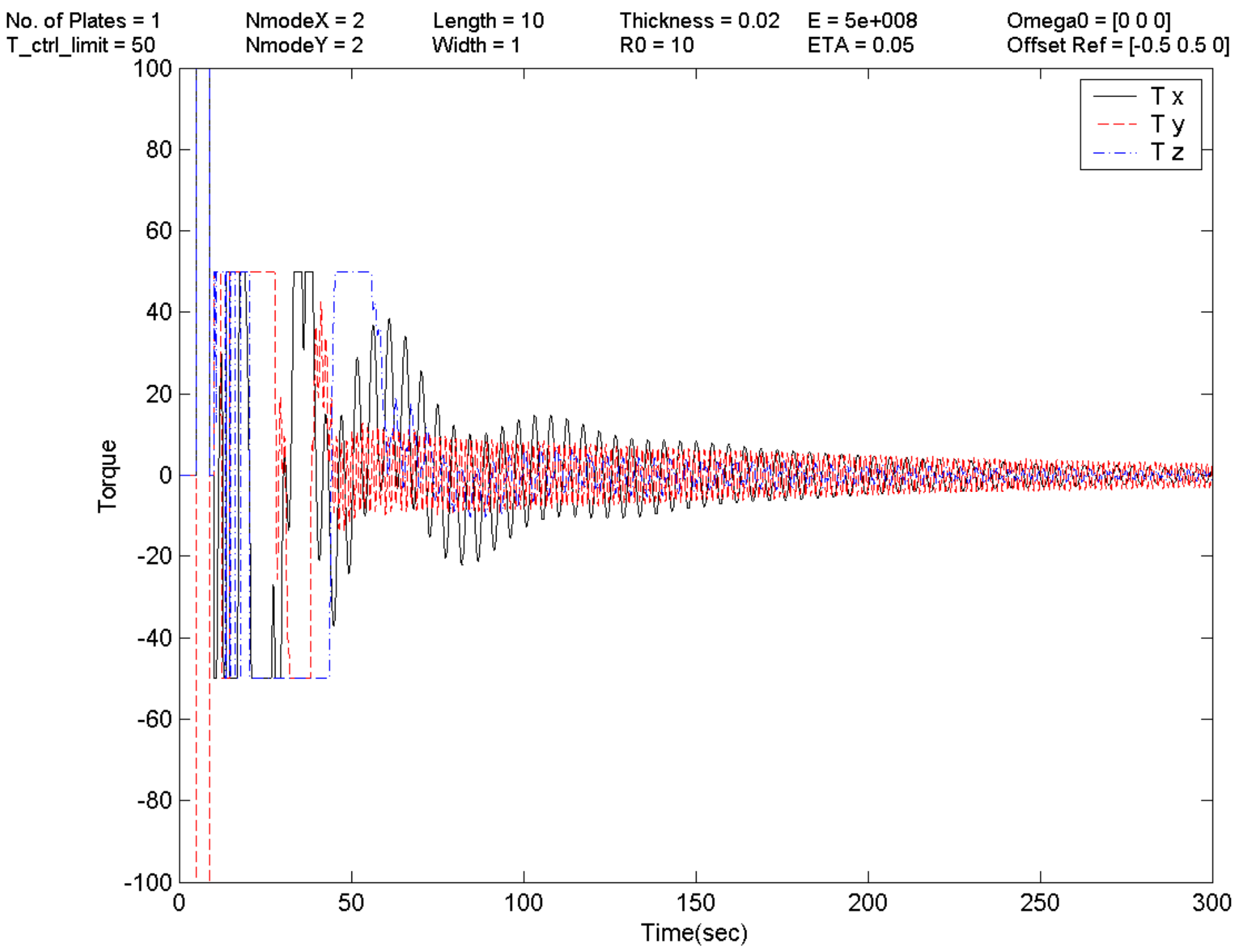

Figure 5.23: ARM case 9 - Initial disturbance and attitude recovery torques ( $\vec{\tau}$ )

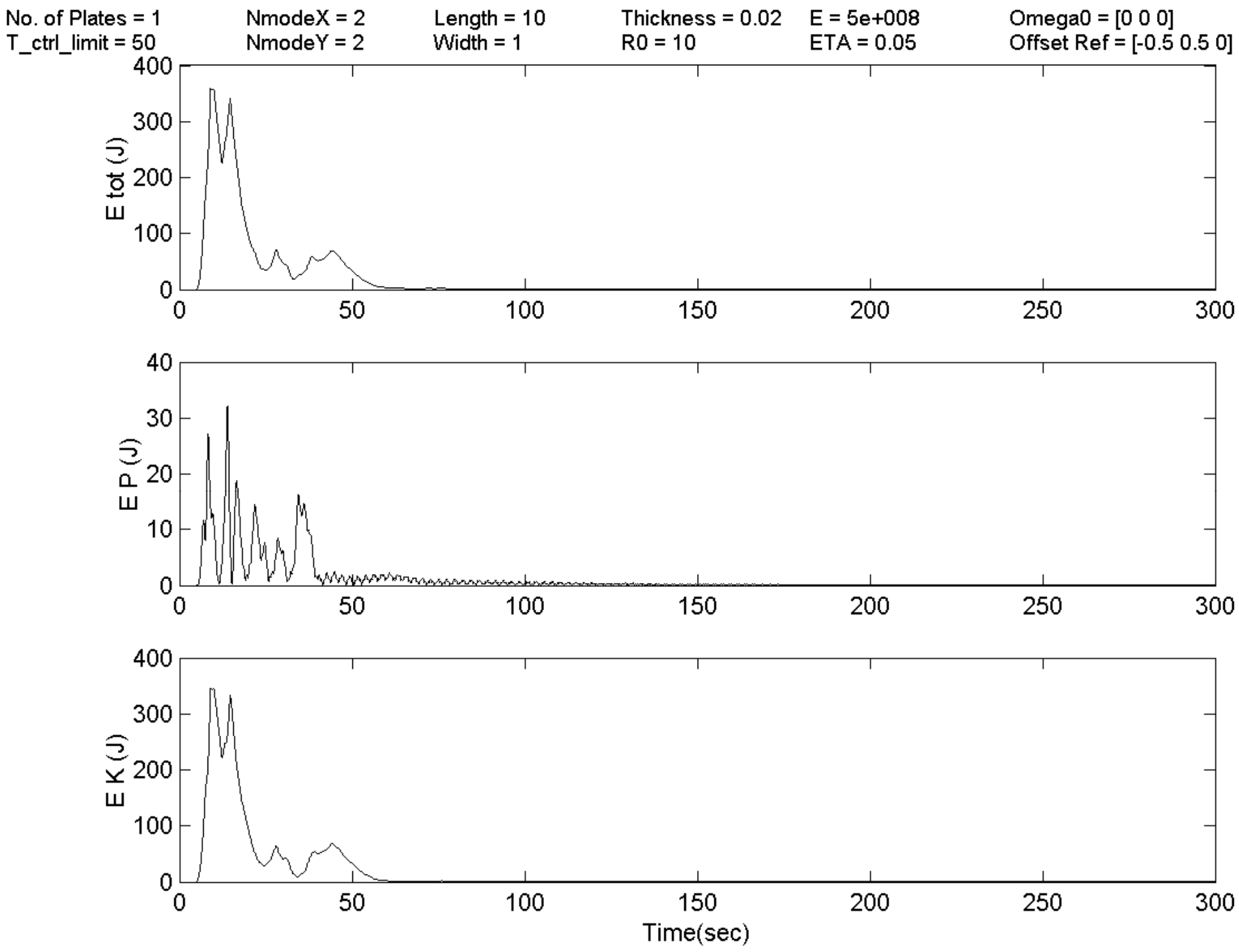

Figure 5.24: ARM case 9 - Spacecraft total, potential, and kinetic energies

Overall, for both simulation cases 8 and 9, small noticeable performance deteriorations were observed, as was discussed. However, the controller was still able to perform the attitude recovery maneuver under 3 minutes.

### 5.3.2 Simulation Results for the Flexible Spacecraft with a Low-Order Controller

*Simulation Case 10:* In order to investigate the performance of a low-order controller (i.e. $p = q = 1$) applied to a higher-order plant model (i.e. $p = q = 2$), we repeat the simulation case 6 with the appropriate reduced mode settings in the controller. Figures 5.25 to 5.29 show the closed loop response and one minor deterioration is noticed in Fig. 5.27 where the higher frequency vibrations are not damped as well as in case 6. This is expected as the controller does not have the information about the higher modes and hence must use only the indirect information embedded in the first vibration mode and the spacecraft angular rates to suppress these higher order vibrations. Corresponding to this effect, the control effort about the *x*-axis, $\tau_x$, is significantly increased compared to case 6 (compare Fig. 5.28 to Fig. 5.8) since the appendage vibrations are about the *x*-axis.

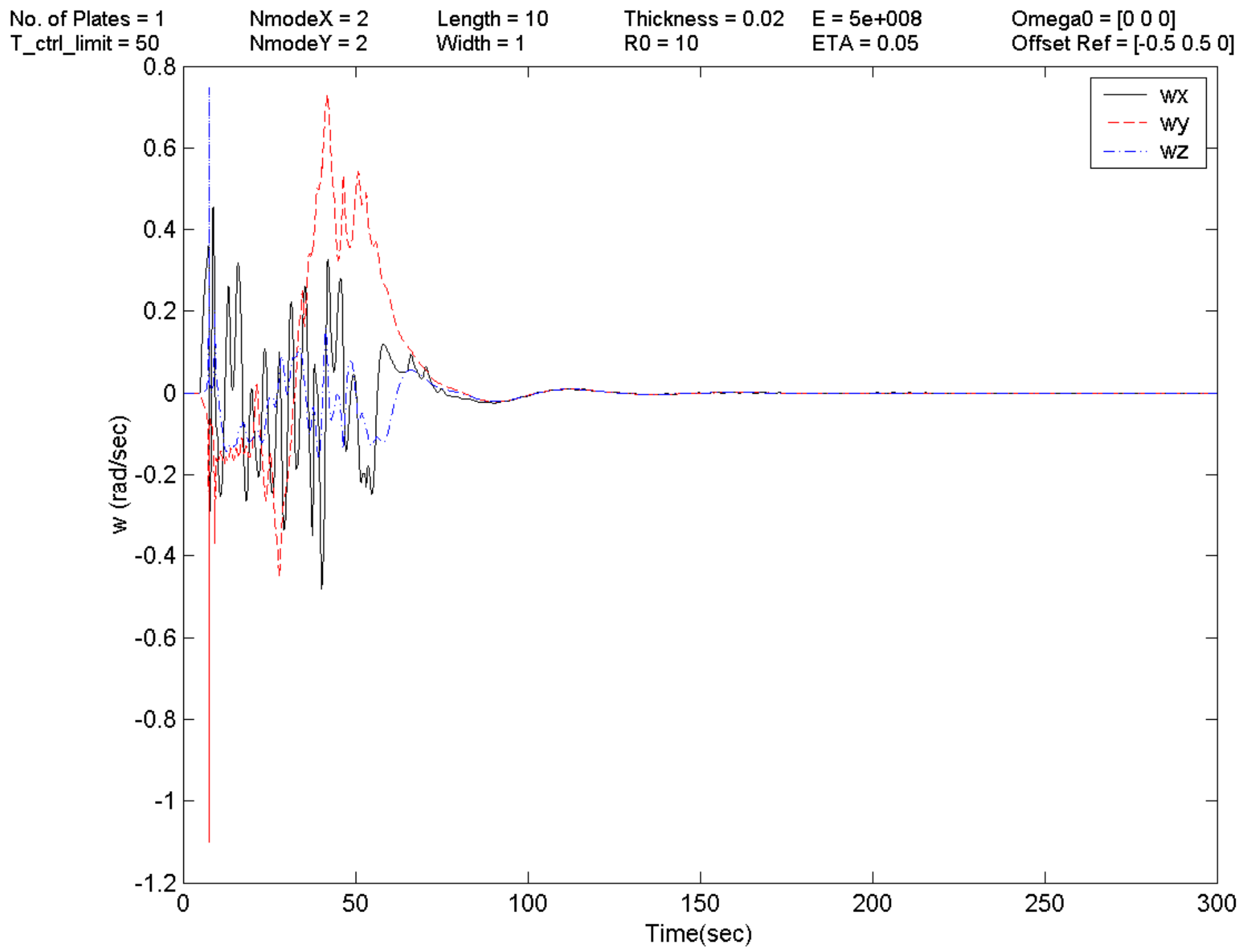

Figure 5.25: ARM case 10 - Spacecraft body angular rates ($\vec{\omega}$)

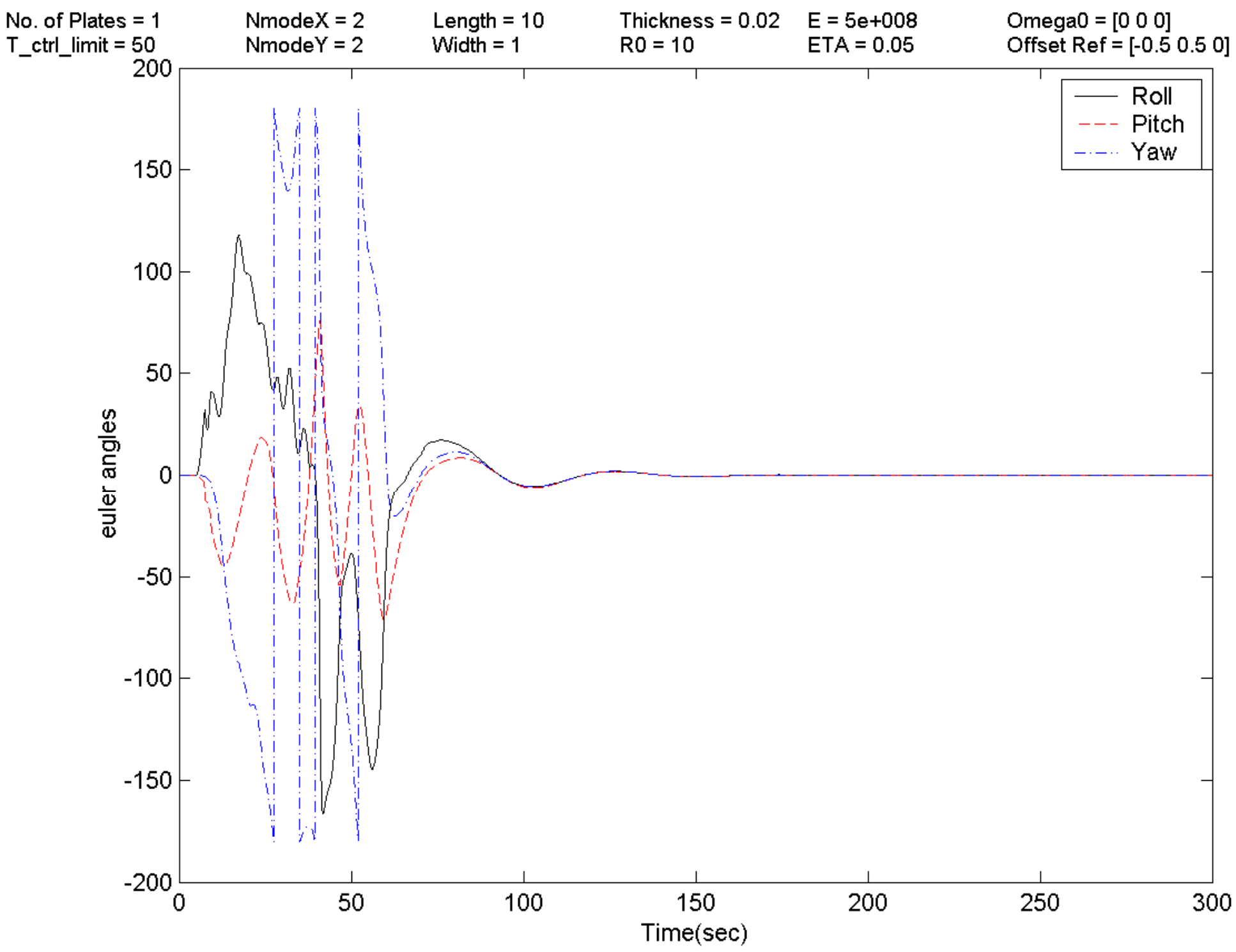

Figure 5.26: ARM case 10 - Spacecraft attitude (Euler angles)

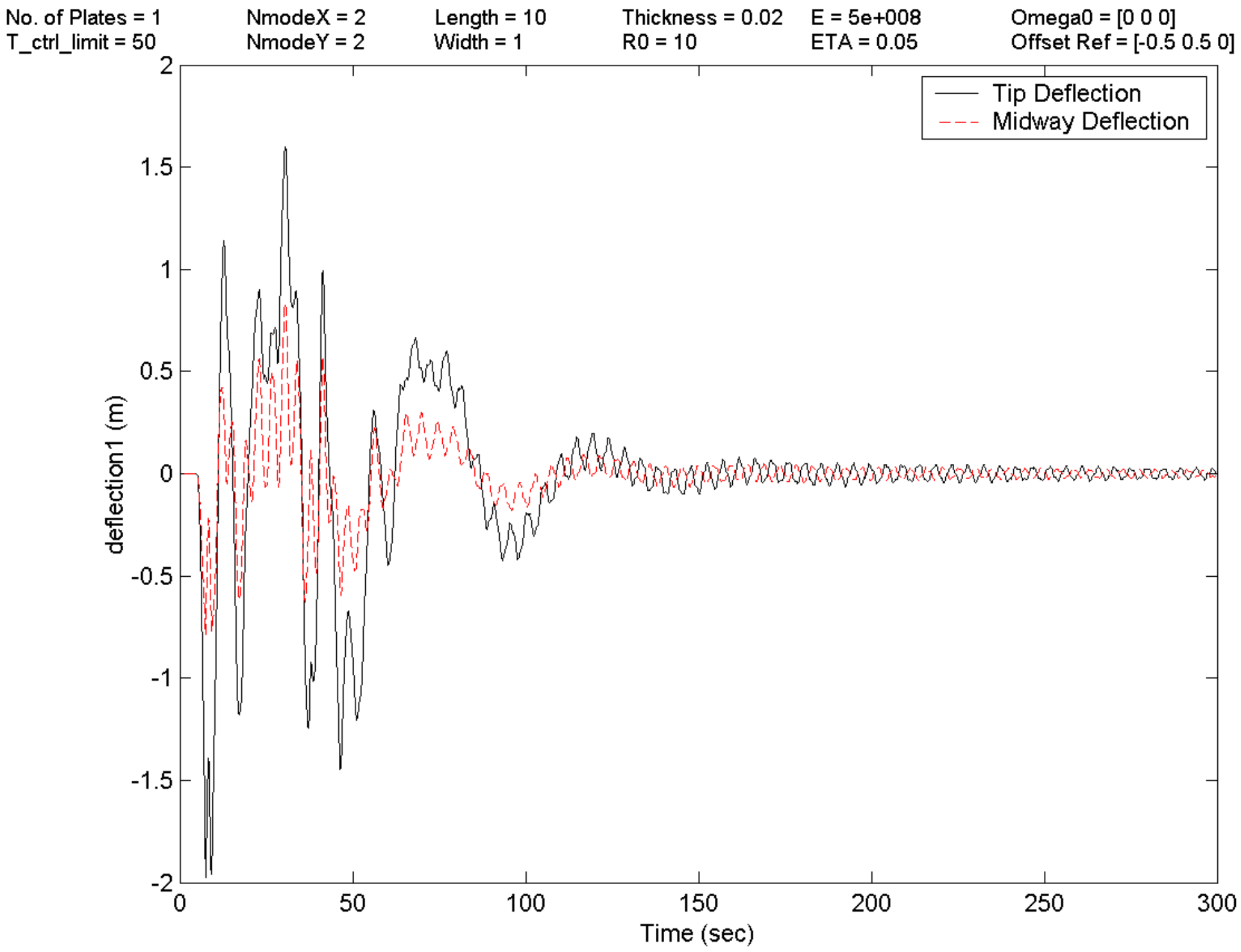

Figure 5.27: ARM case 10 - Deflection ( $w$ ) at tip $x = .5; y = 10$ and midway $x = .5; y = 5$

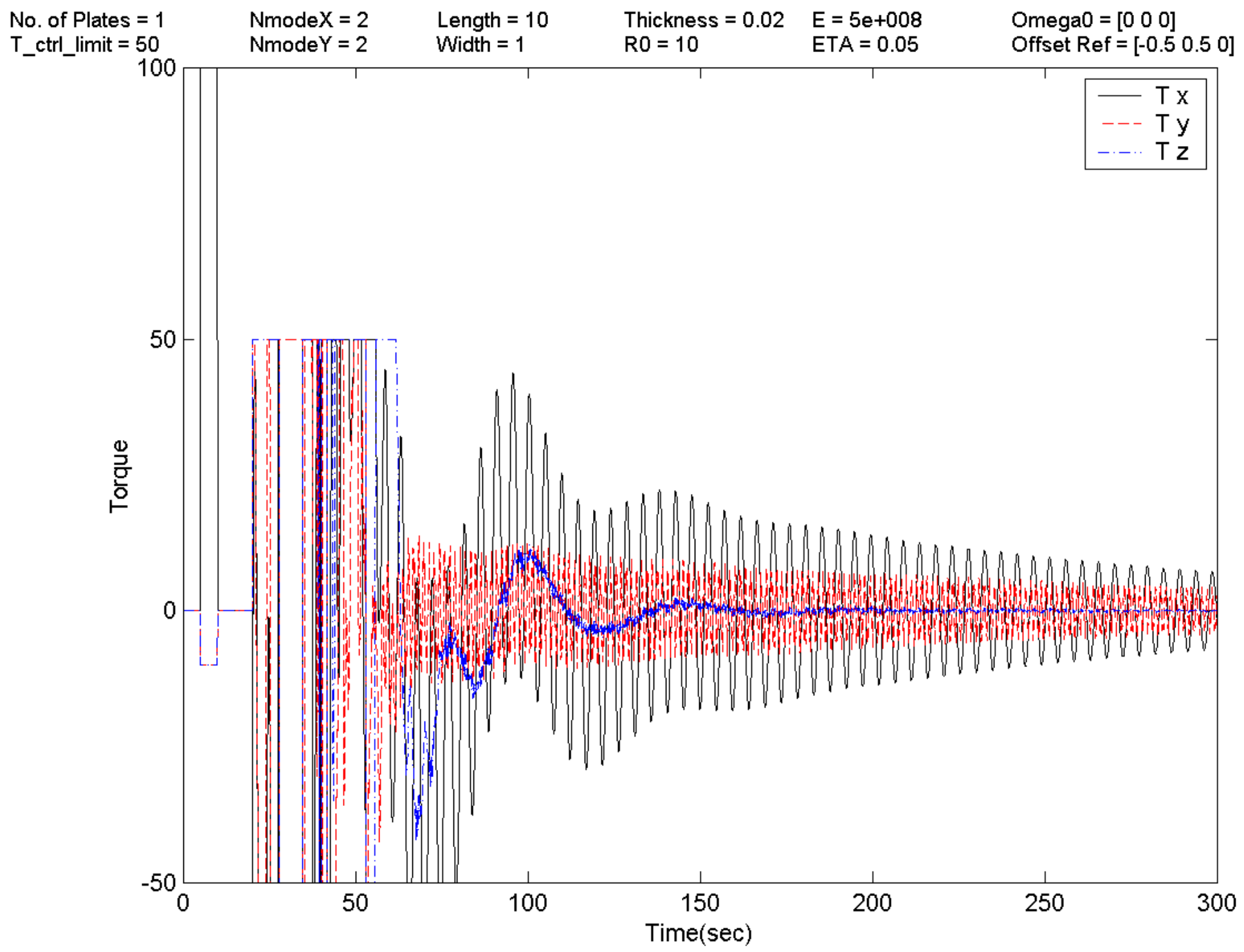

Figure 5.28: ARM case 10 – Initial disturbance and attitude recovery torques ( $\vec{\tau}$ )

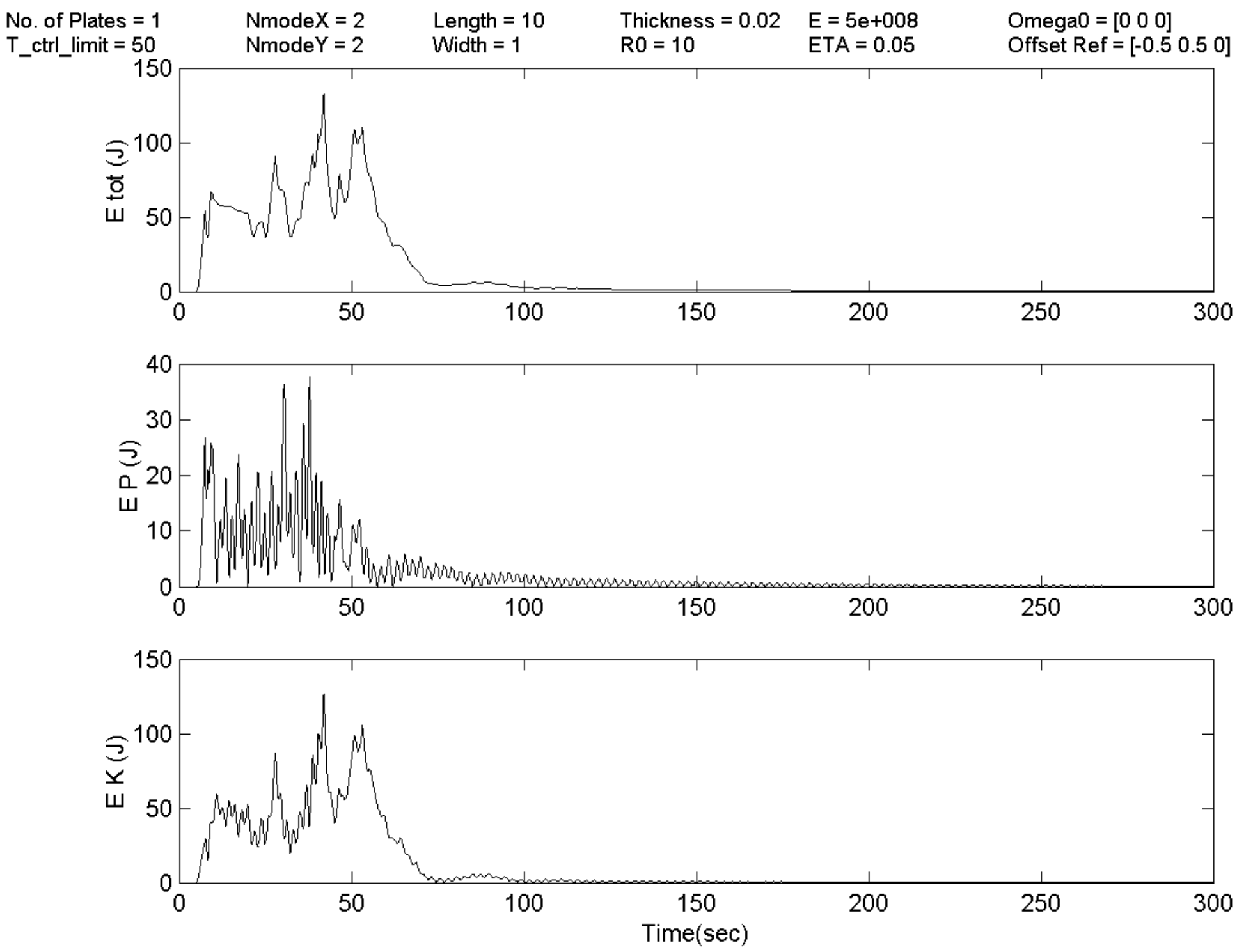

Figure 5.29: ARM case 10 - Spacecraft total, potential, and kinetic energies

*Simulation Case 11:* This is a repeat of simulation case 7 with the appropriate reduced mode settings in the controller. Figures 5.30 to 5.34 show the closed loop response and the same minor response deterioration, which was discussed in the previous case, is noticed. Figure 5.32 shows that the higher frequency vibrations are not damped as well as in case 7 and the deflection plot is rather choppy after the 50 seconds mark. As was explained previously, this also affects the *x*-axis control torque, $\tau_x$, whose level is now higher compared to case 7 (see Fig. 5.33 vs. Fig. 5.13). The controller is still able to recover the attitude of the spacecraft and suppress most of the appendage vibrations within about 5 minutes.

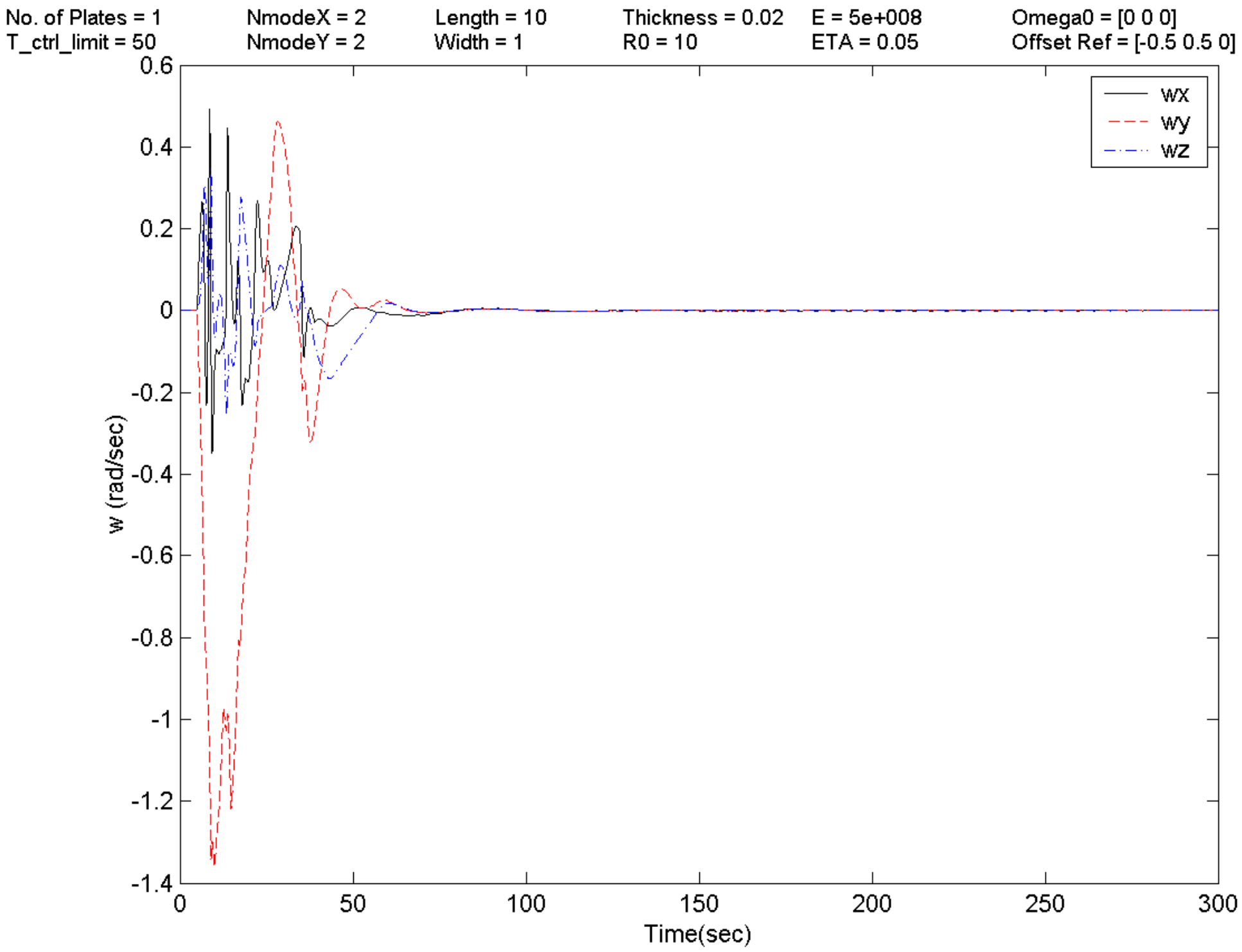

Figure 5.30: ARM case 11 - Spacecraft body angular rates ($\vec{\omega}$)

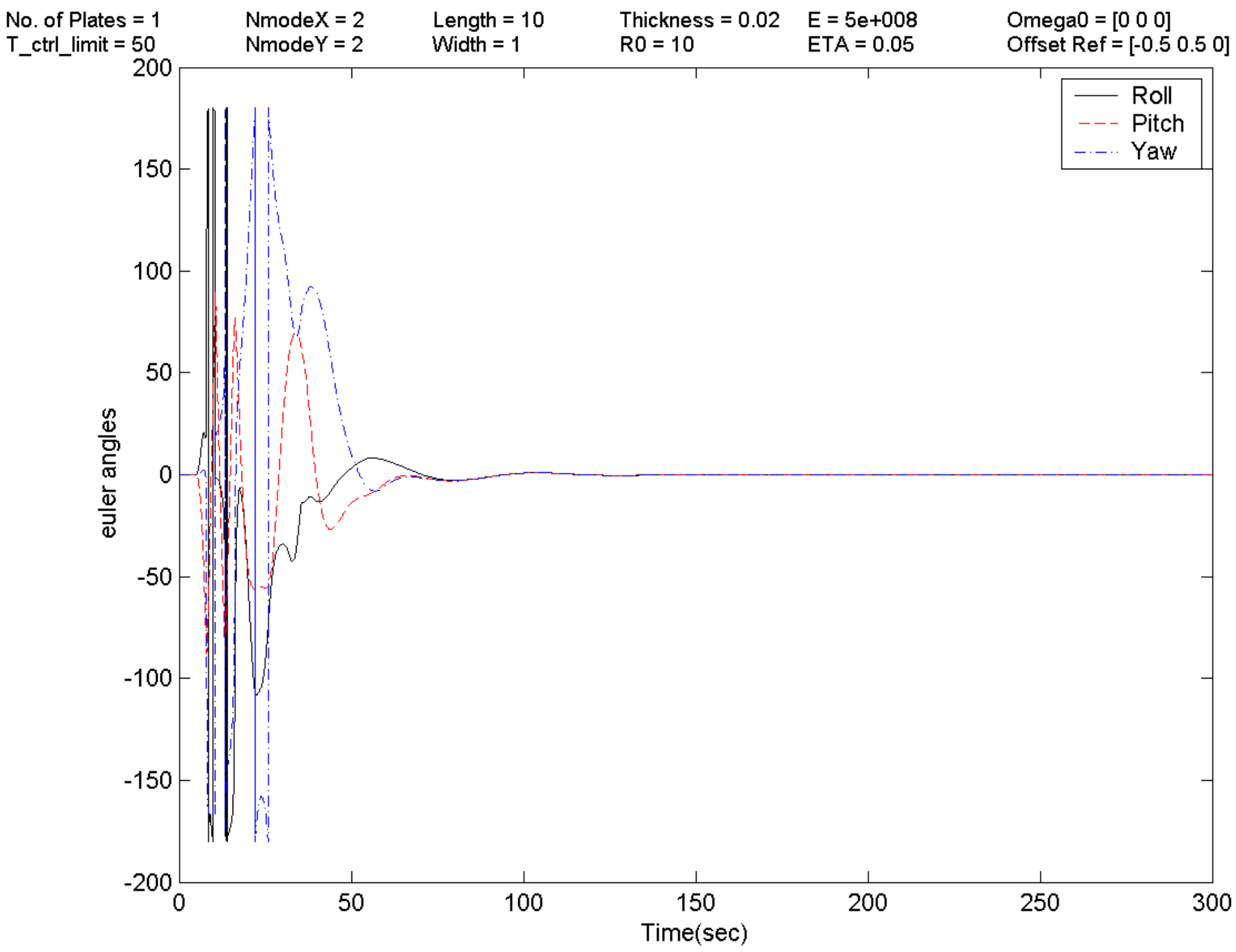

Figure 5.31: ARM case 11 - Spacecraft attitude (Euler angles)

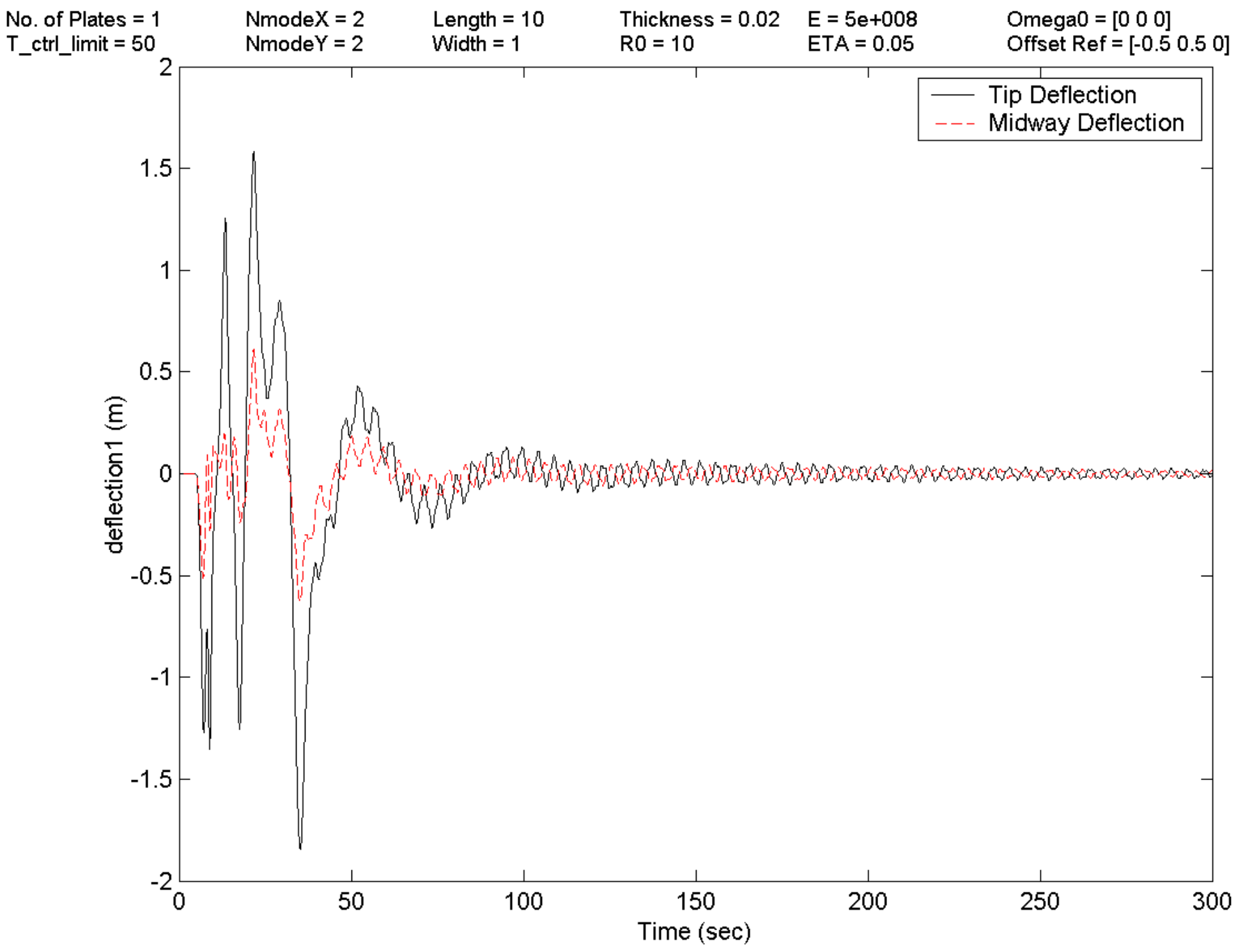

Figure 5.32: ARM case 11 - Deflection ( $w$ ) at tip $x = .5; y = 10$ and midway $x = .5; y = 5$

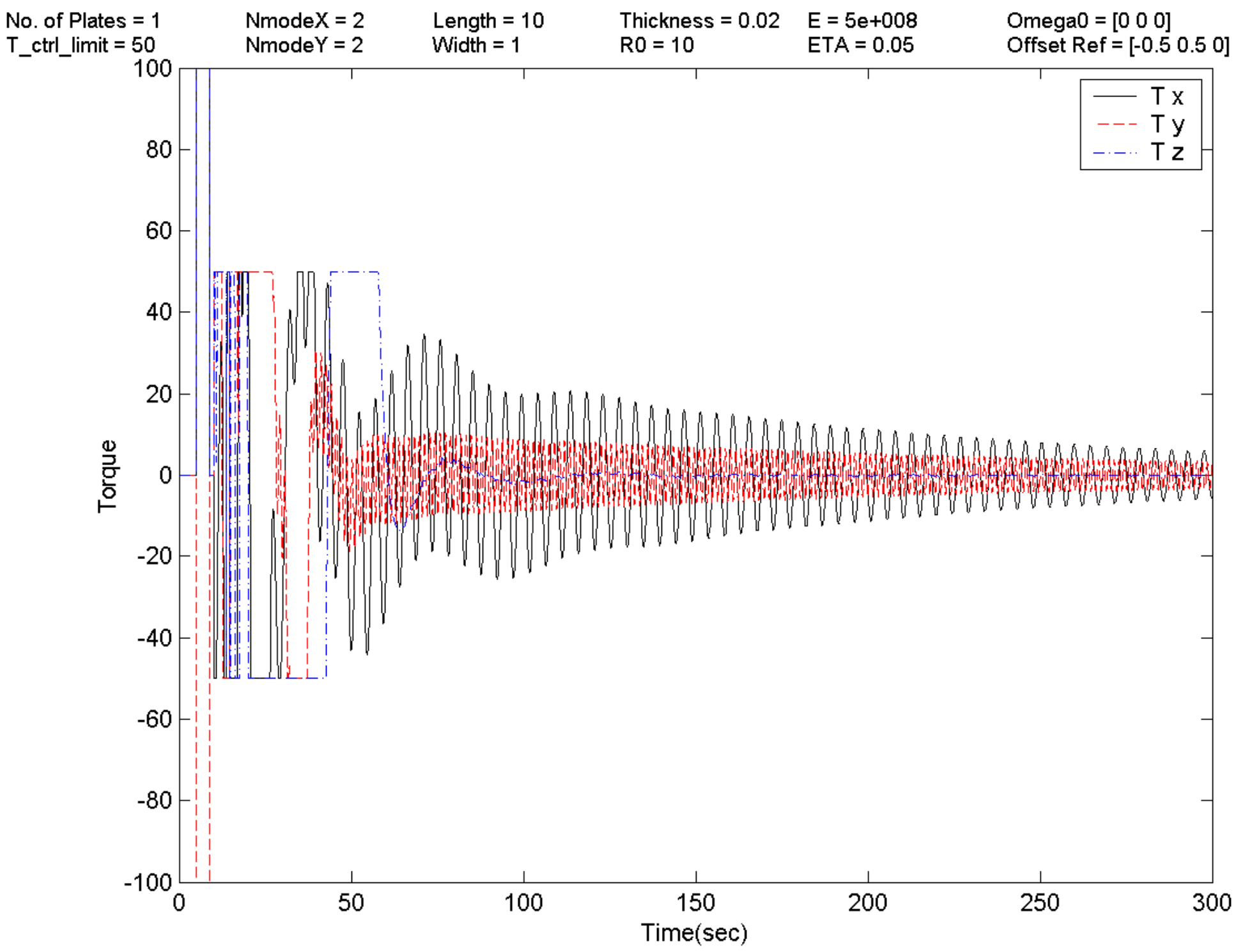

Figure 5.33: ARM case 11 - Initial disturbance and attitude recovery torques ( $\vec{\tau}$ )

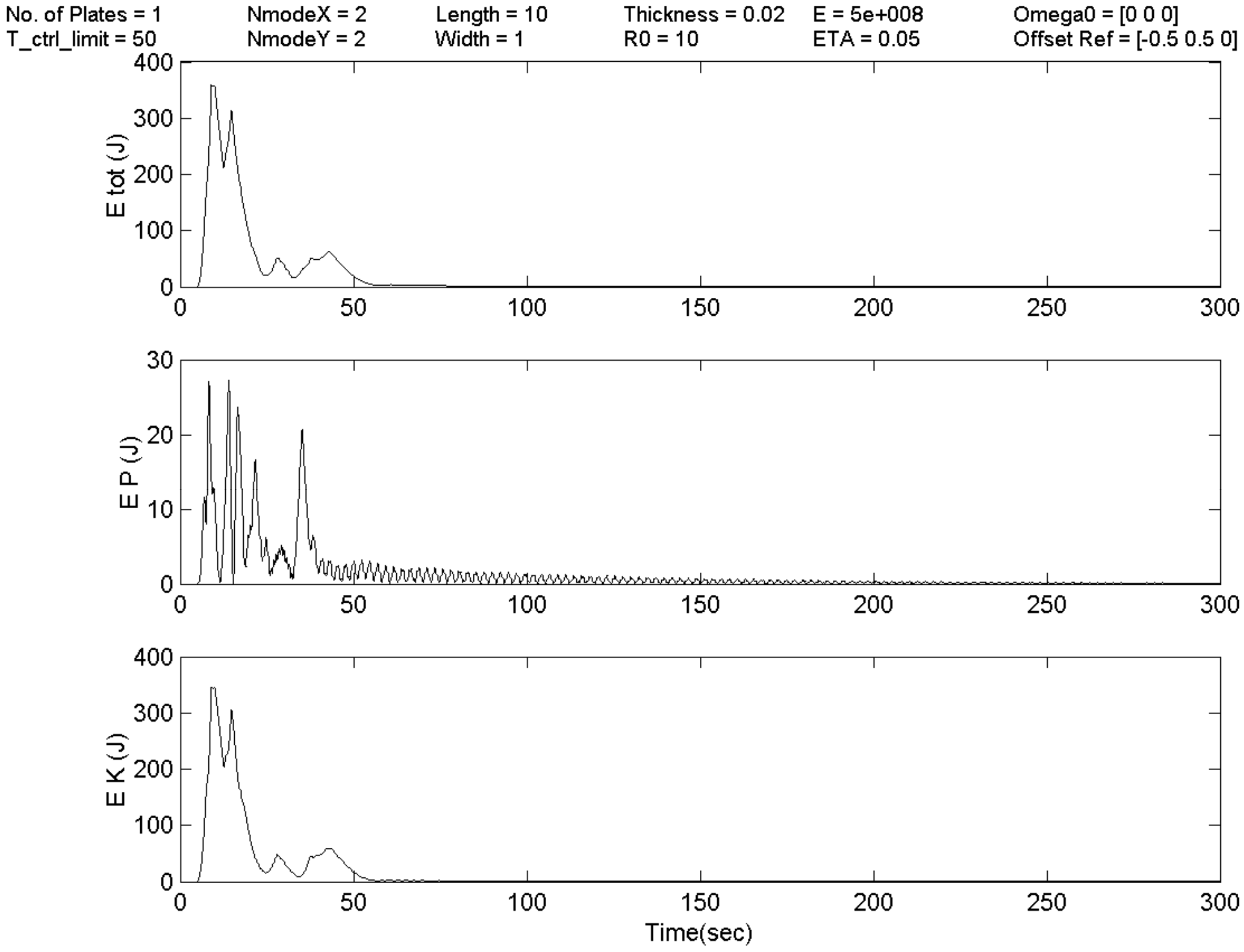

Figure 5.34: ARM case 11 - Spacecraft total, potential, and kinetic energies

### 5.3.3 Simulation Results for the Flexible Spacecraft with a Low-Order Controller and Model/Parameter Uncertainties

*Simulation Case 12:* This case is in essence a combination of cases 8 and 10 where we use a low-order controller with the additional introduction of the 20% increase in the values of $\vec{f}_{\vec{q}}$ and $\underline{G}_{\vec{q}}$ in (5.12) representing the model/parameter uncertainties in the system. Figures 5.35 to 5.39 show the closed loop response and we notice the following two points:

- There is an extra low frequency oscillation compared to case 6 as seen in Fig. 5.37. The same oscillation was also present in case 8, however this time, its amplitude is even larger, about 2 meters.
- The high frequency vibrations which were not suppressed as easily in case 10, are now damped quite smoothly as can be seen from Fig. 5.37 and the corresponding *x*-axis torque is no longer as large as it was in case 10, during the last 200 seconds of the simulation. Here, the only difference from case 10 is the 20% increase in the values of $\vec{f}_{\vec{q}}$ and $\underline{G}_{\vec{q}}$ which must in sort help smooth out the vibration suppression.

The controller is able to recover the attitude of the spacecraft and suppress all relevant appendage vibrations within about 200 seconds.

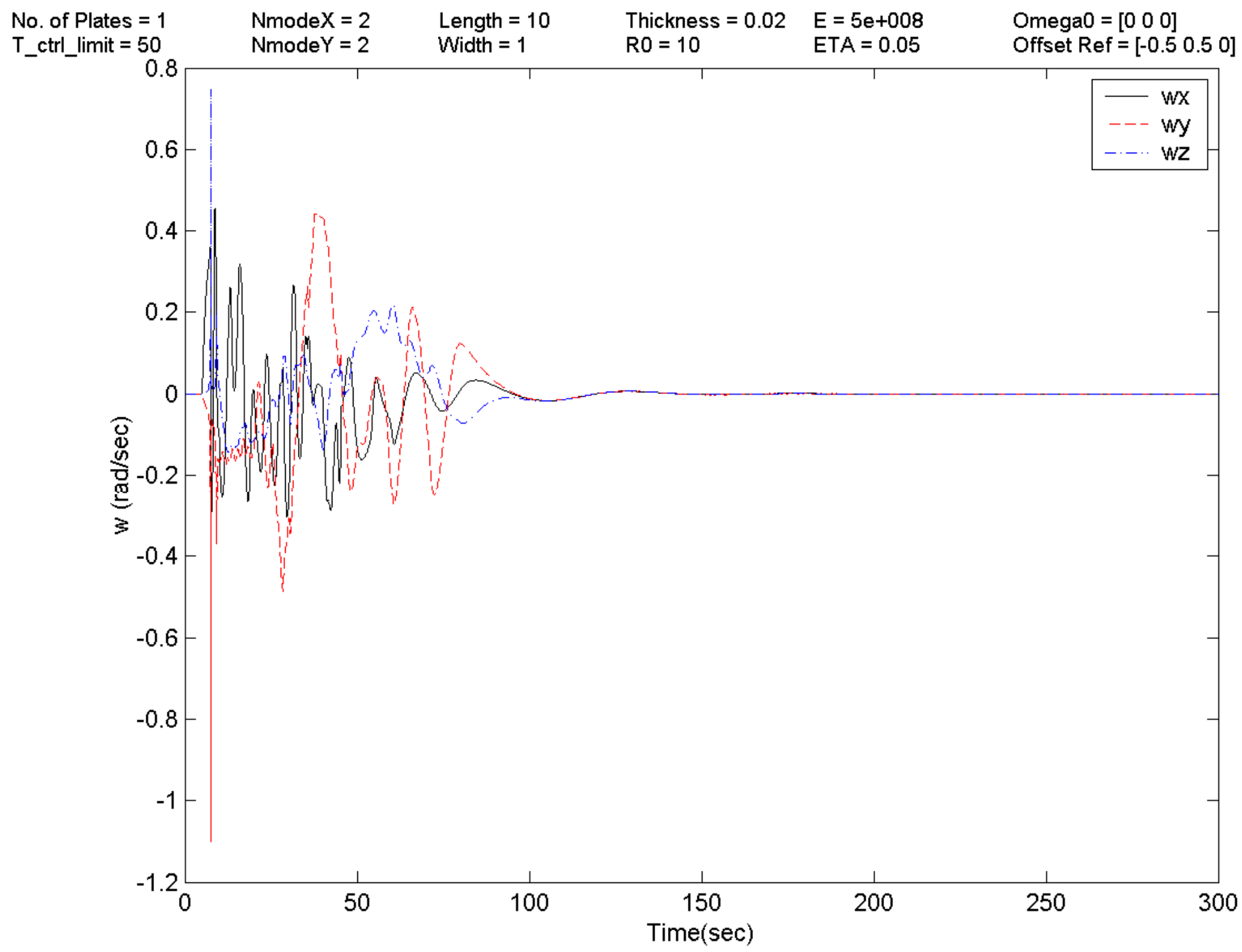

Figure 5.35: ARM case 12 - Spacecraft body angular rates ( $\vec{\omega}$ )

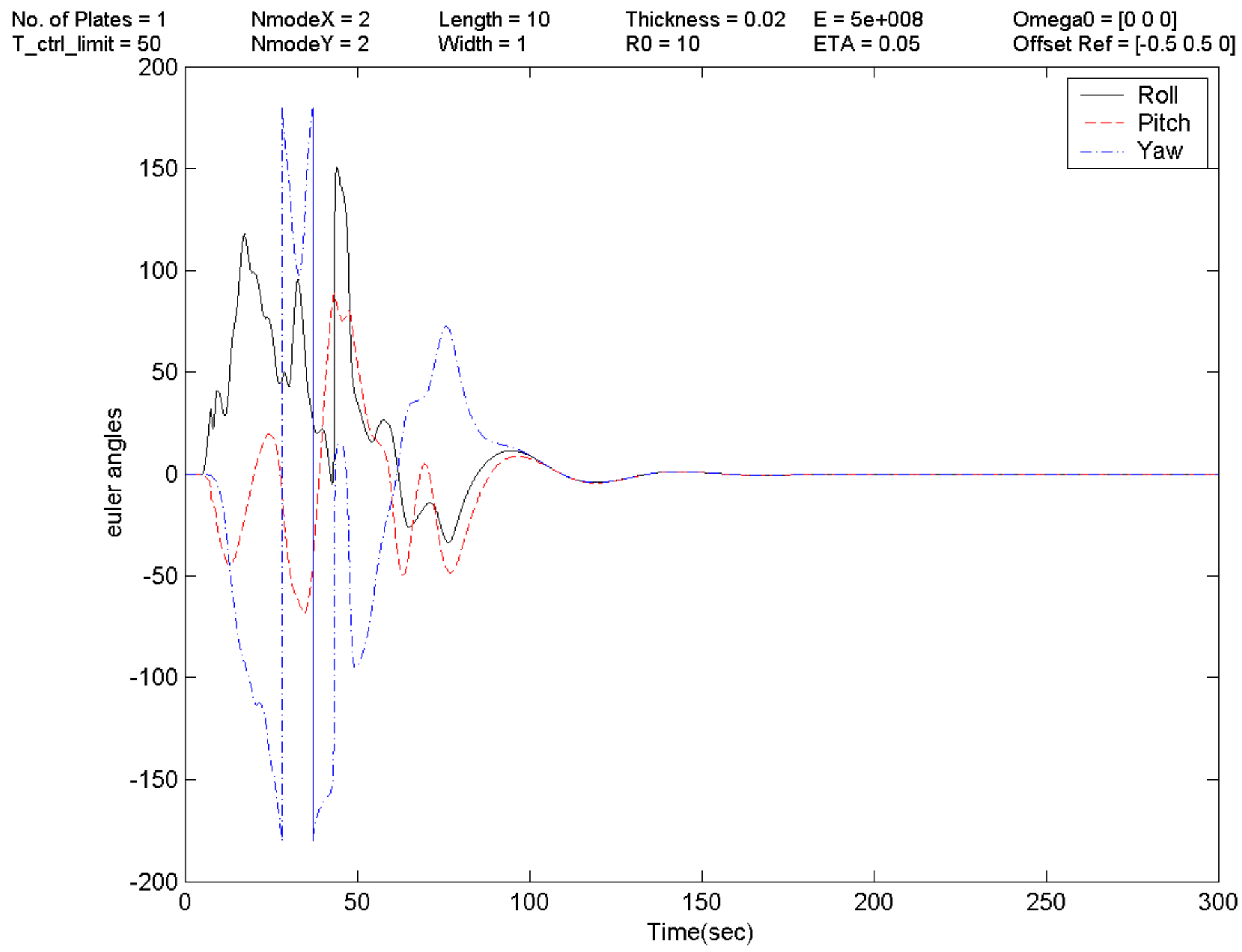

Figure 5.36: ARM case 12 - Spacecraft attitude (Euler angles)

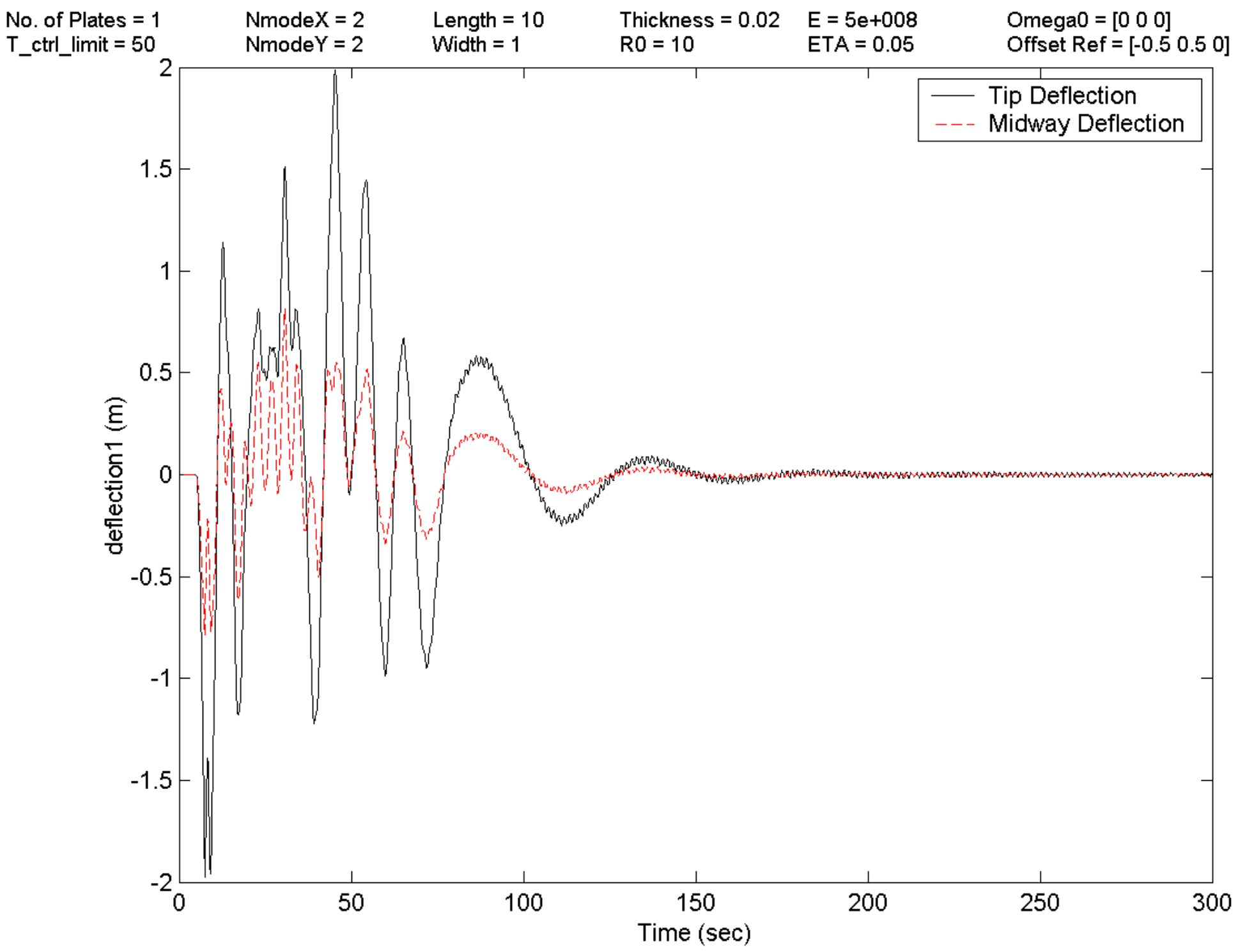

Figure 5.37: ARM case 12 - Deflection ( $w$ ) at tip $x = .5; y = 10$ and midway $x = .5; y = 5$

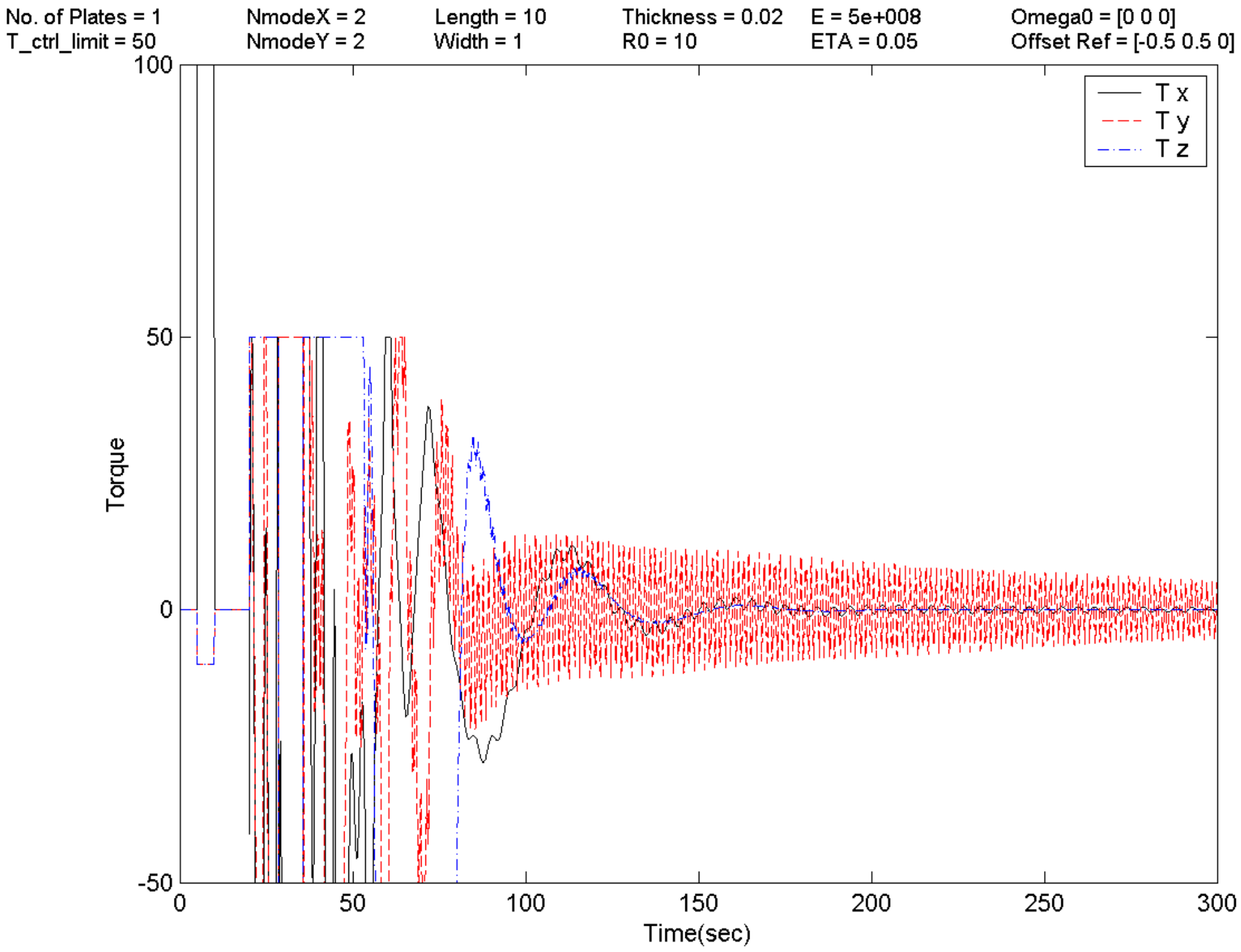

Figure 5.38: ARM case 12 - Initial disturbance and attitude recovery torques ( $\vec{\tau}$ )

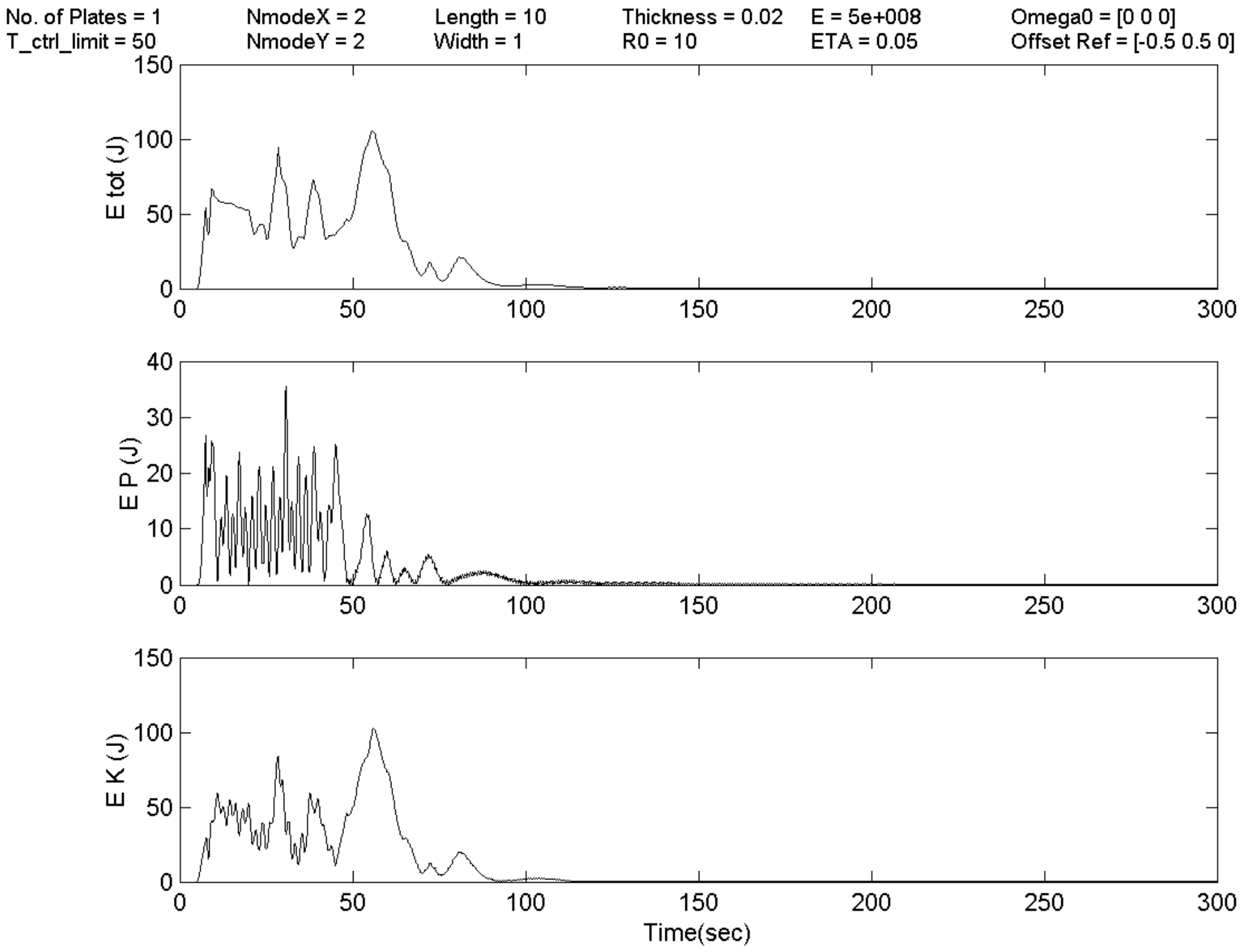

Figure 5.39: ARM case 12 - Spacecraft total, potential, and kinetic energies

*Simulation Case 13:* Finally, we present the simulation case combining cases 9 and 11 where we use a low-order controller with 20% increase in the values of $\vec{f}_{\bar{q}}$ and $\underline{G}_{\bar{q}}$ in (5.12). Figures 5.40 to 5.44 show the closed loop response and it is noted that the results are very similar to case 9 and hence the low order of the controller does not affect much this particular simulation scenario.

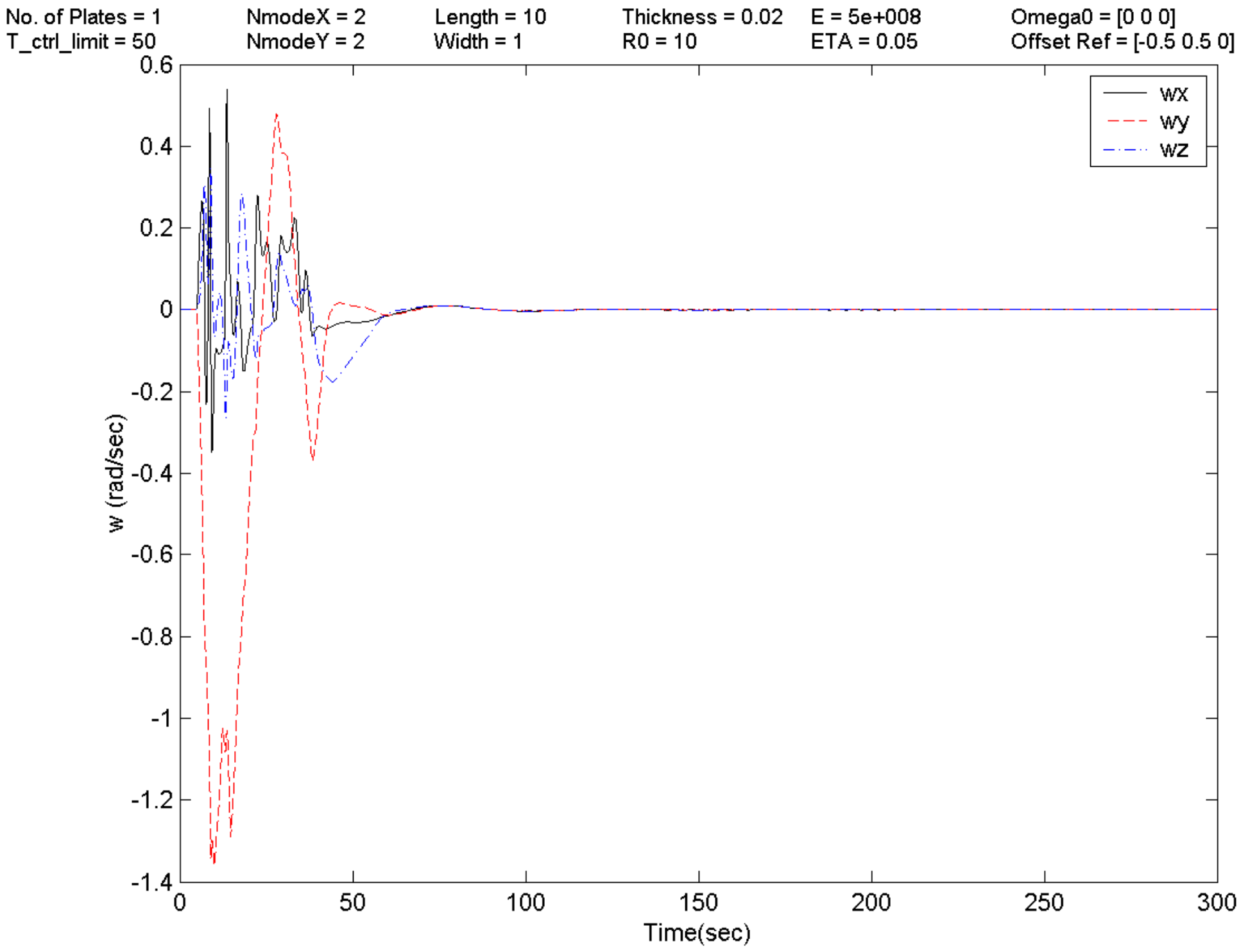

Figure 5.40: ARM case 13 - Spacecraft body angular rates ( $\vec{\omega}$ )

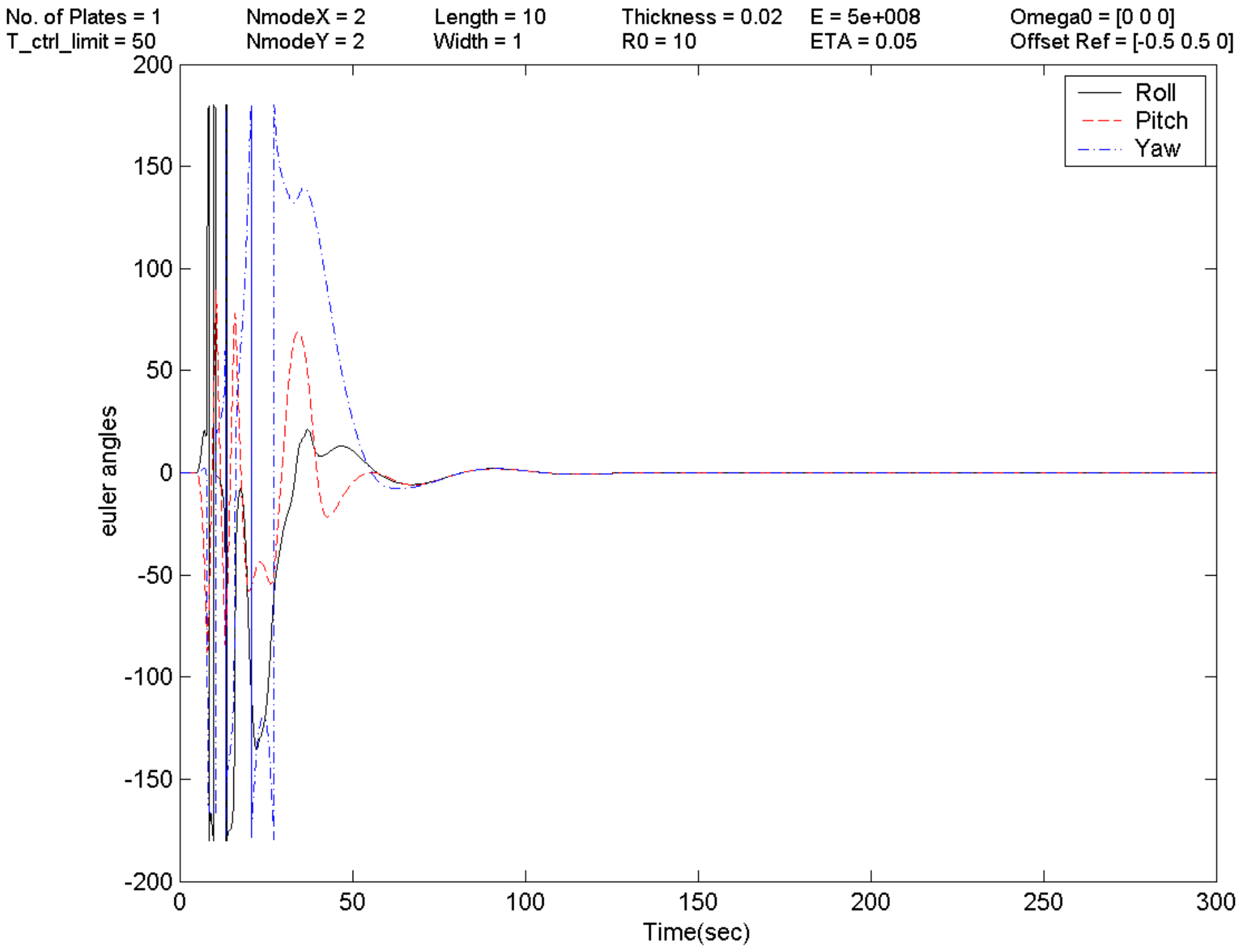

Figure 5.41: ARM case 13 - Spacecraft attitude (Euler angles)

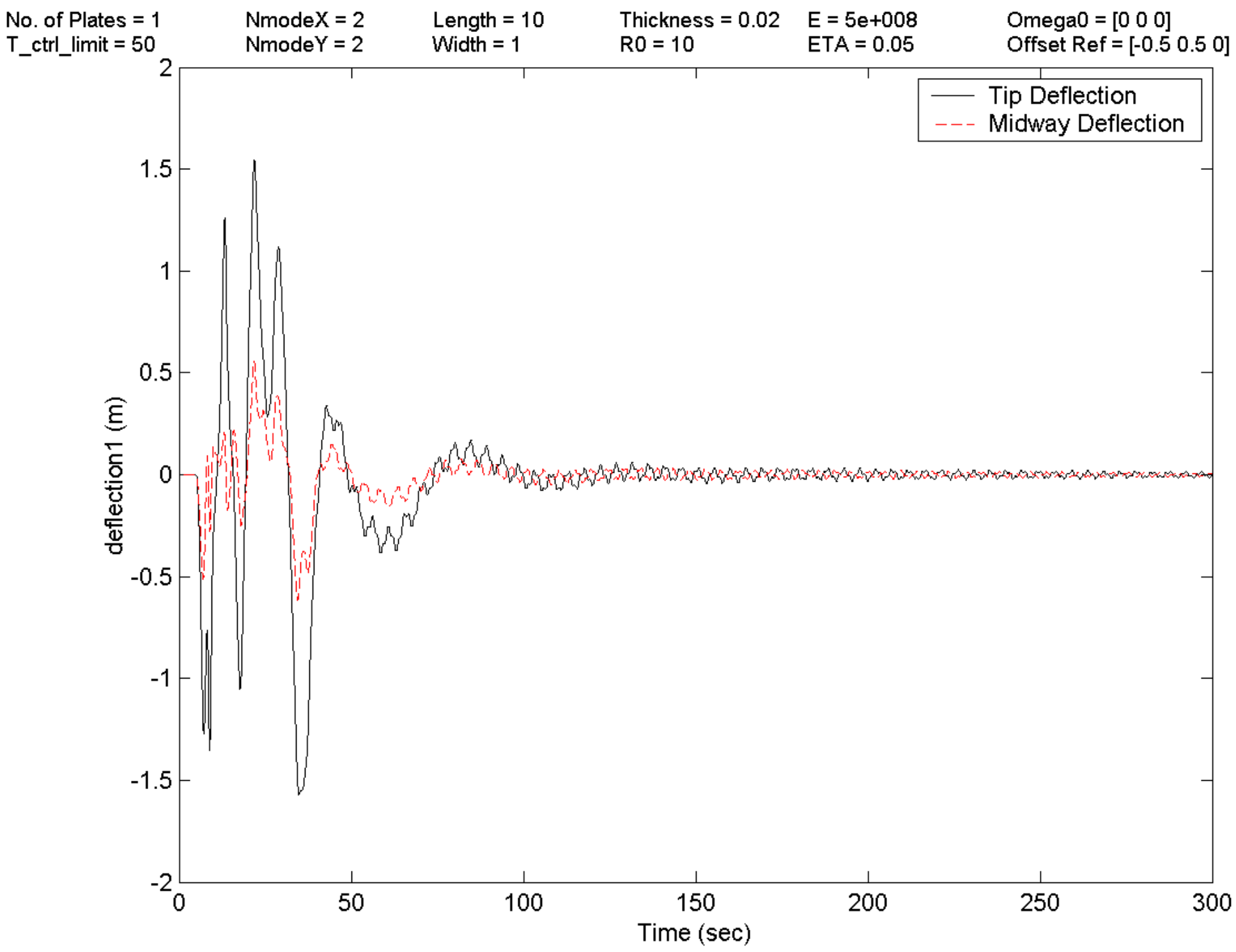

Figure 5.42: ARM case 13 - Deflection ( $w$ ) at tip $x = .5; y = 10$ and midway $x = .5; y = 5$

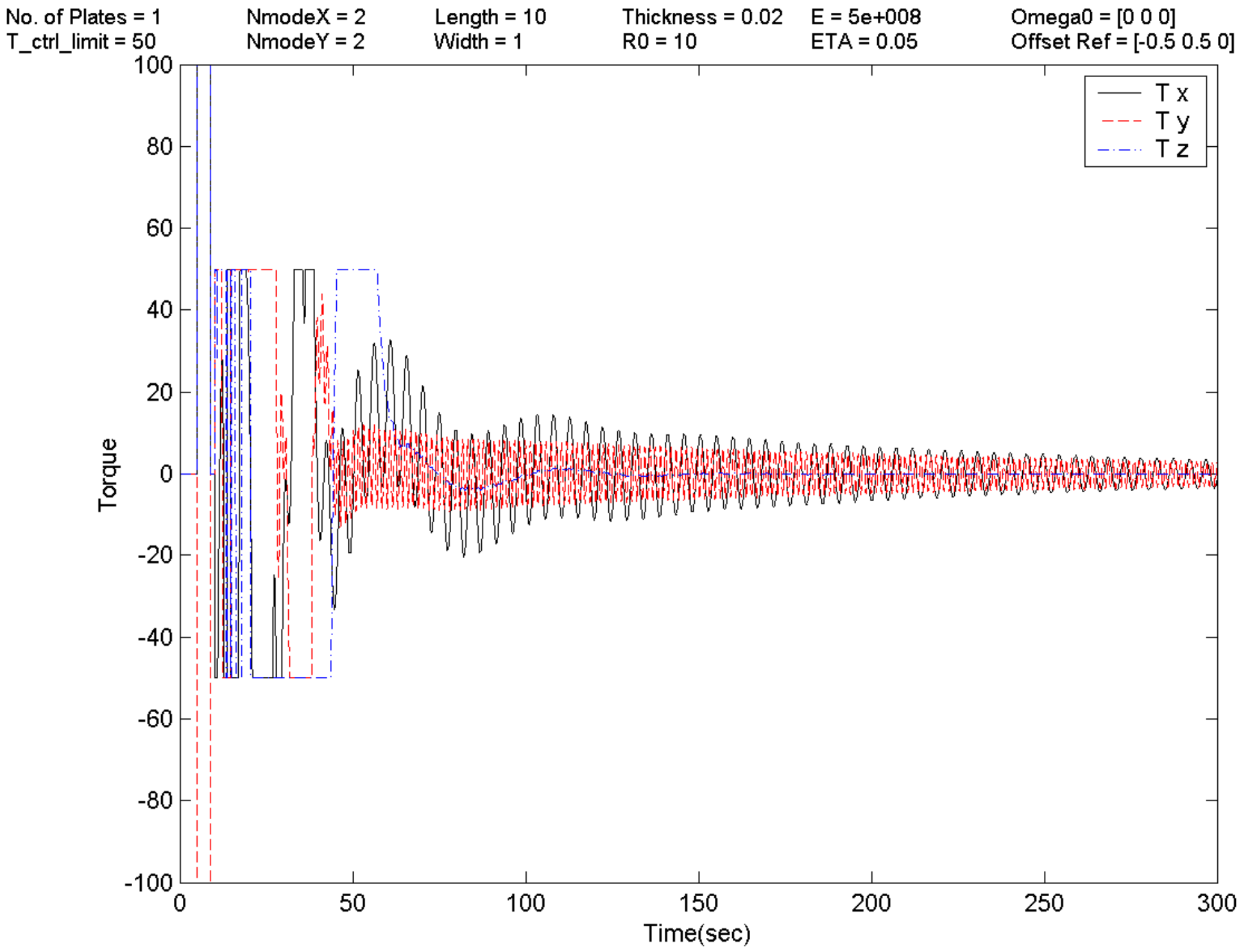

Figure 5.43: ARM case 13 - Initial disturbance and attitude recovery torques ( $\vec{\tau}$ )

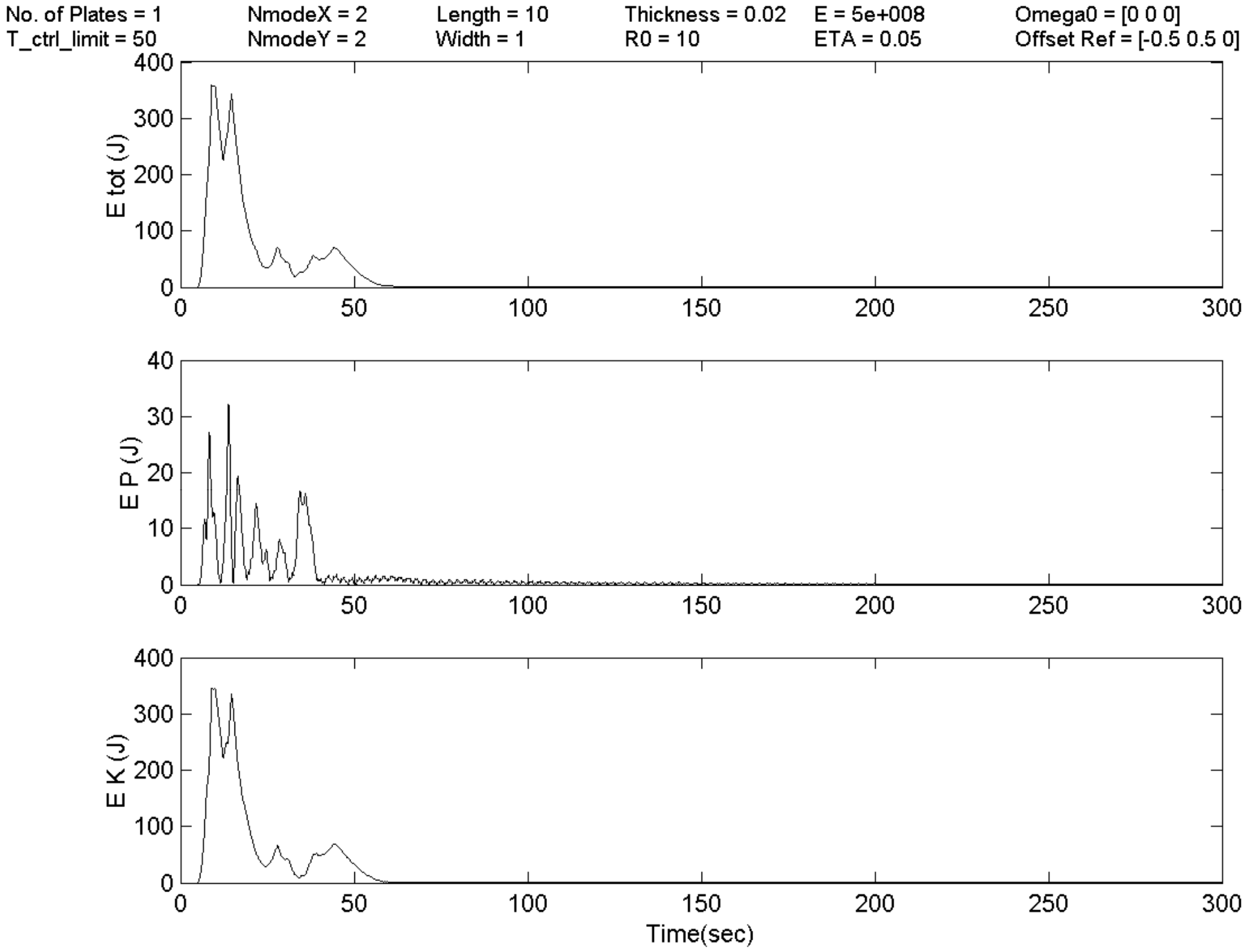

Figure 5.44: ARM case 13 - Spacecraft total, potential, and kinetic energies

## 5.4 Concluding Remarks

Nine different attitude recovery simulation cases were presented in this chapter based on two different malfunctions, namely, (i) a severe on-board malfunction affecting mainly the *x*-axis; and (ii) a large disturbance torque affecting all three attitude axes. For all the simulation cases where the vibration suppression gains in attitude recovery controller were active (e.g. cases 6-13 where $\underline{K}_{P\chi} \neq \underline{0}$; $\underline{K}_{D\chi} \neq \underline{0}$), the system performance requirements of Section 1.3 were met and of course, the controller gains can be further tuned to decrease the attitude recovery maneuver time, if needed. It is also noted that the controller is fairly insensitive to parameter and model uncertainties in the particular cases discussed and that the first mode of vibration is the dominant mode and hence a low-order controller will perform adequately under most conditions.

Finally, it is important to also realize that the attitude recovery controller presented in this thesis, for both the rigid and flexible spacecraft, can also work as effectively for a slew maneuver which usually induces less excitations on the system.

# CHAPTER 6

## SPACECRAFT SYSTEM STABILITY

In this chapter, the focus will be on the important topic of closed-loop stability for the flexible spacecraft. Indeed, given the spacecraft attitude recovery controller, described in Chapters 5, the first and foremost important question regarding its various properties is whether or not it is stable, as an unstable controller is useless and will most likely jeopardize the mission. Therefore, a careful and rigorous mathematical proof of stability is essential and, towards this objective, our primary analysis tool will be the Lyapunov theory or otherwise known as Lyapunov's *direct method.* In essence, Lyapunov's direct method allows the determination of the stability properties of our nonlinear system by construction of a scalar function (e.g. a quadratic function) for the system states and examining the function's time evolution. For asymptotic stability, we require that this Lyapunov function be always positive and its rate of change be always negative. For the spacecraft attitude recovery system, it will be sufficient to insure that the system is at least locally asymptotically stable. Many existing theorems and results are used in this chapter that can be found, along with proofs, in references [88]-[90] and [148].

## 6.1 Internal and Zero Dynamics

We recall that our system has well-defined relative degrees, $r_i$, for $i = 1, 2, 3$ at $\vec{x}_o$ such that the conditions (6.1) are satisfied and input-output feedback linearization implementation is possible, namely

$(i)$ The scalar functions $L_{\vec{g}_j} L_{\vec{f}}^k h_i = 0$ for $\forall i, j, k$ ; $1 \le i, j \le 3$ ; $0 \le k \le r_i - 1$

$$(ii) \text{ The decoupling matrix } E = \begin{bmatrix} L_{\vec{g}_1} L_{\vec{f}}^{r_1-1} h_1 & L_{\vec{g}_2} L_{\vec{f}}^{r_1-1} h_1 & L_{\vec{g}_3} L_{\vec{f}}^{r_1-1} h_1 \\ L_{\vec{g}_1} L_{\vec{f}}^{r_2-1} h_2 & L_{\vec{g}_2} L_{\vec{f}}^{r_2-1} h_2 & L_{\vec{g}_3} L_{\vec{f}}^{r_2-1} h_2 \\ L_{\vec{g}_1} L_{\vec{f}}^{r_3-1} h_3 & L_{\vec{g}_2} L_{\vec{f}}^{r_3-1} h_3 & L_{\vec{g}_3} L_{\vec{f}}^{r_3-1} h_3 \end{bmatrix} \tag{6.1}$$

is invertible in the neighborhood $\Omega$ of $x_o$

In equations (5.12)-(5.14), we have shown that using feedback linearization, we are able to convert the system dynamics (2.59)-(2.60) into two parts

(i) an external observable linear system (5.13) for which a controller based on linear control theory was devised in (5.16);
(ii) an internal unobservable nonlinear dynamics given by (5.14).

We need to show that both parts of the closed-loop system dynamics are stable together. The proof of stability of part (ii) will indicate that our closed-loop system is minimum phase and hence there is at least a possibility that the overall system is asymptotically stable. However, before analyzing part (ii), there is a need to simplify it by finding appropriate coordinate transformations (i.e. diffeomorphisms) to rearrange the system in the *normal form*. This internal dynamics of the system is examined next.

### 6.1.1 Internal Dynamics of the Flexible Spacecraft

Using the outputs $\vec{y}$ of the system (2.60), their derivatives, the relative degrees $r_i = 2$ for $i = 1, 2, 3$ and the total relative degree of $r = \sum_{i=1}^{3} r_i = 6$, one can devise a non-linear coordinate transformation $T$ to map $\vec{x}$ in (2.60) to $\vec{\mu}$, as given by

$$\vec{\mu}=T(\vec{x})=T\begin{pmatrix} q_i \\ \dot{q}_i \\ --- \\ \chi_1 \\ \vdots \\ \dot{\chi}_{npq} \end{pmatrix}=\begin{bmatrix} \Psi_i(\vec{x})=h_i \\ \Psi_{i+3}(\vec{x})=L_{\vec{f}}h_i \\ --- \\ \Phi_1(\vec{x}) \\ \vdots \\ \Phi_{m-8}(\vec{x}) \end{bmatrix}=\begin{bmatrix} q_i \\ \dot{q}_i \\ --- \\ \Phi_1(\vec{x}) \\ \vdots \\ \Phi_{m-8}(\vec{x}) \end{bmatrix} \triangleq \begin{bmatrix} \vec{\zeta} \\ --- \\ \vec{\eta} \end{bmatrix} \quad \text{where } i=1,2,3 \qquad (6.2)$$

with $\vec{\zeta}$ representing the new rigid states that stay the same as before (i.e. $\vec{q}$ *and* $\dot{\vec{q}}$ ) and $\vec{\eta}$ representing the remaining states which no longer have the same clear physical meaning as the elastic states $\vec{\chi}$ *and* $\dot{\vec{\chi}}$. The coordinate transformation $T(\vec{x})$ must be a diffeomorphism which requires $T(\vec{x})$ to be a continuous, differentiable and bijective (i.e. onto and one-to-one) mapping and the inverse mapping $T^{-1}(\vec{x})$ (i.e. $\Psi^{-1}(\vec{x}),\Phi^{-1}(\vec{x})$) must exist. Using our new states, our closed-loop system (5.13)-(5.14) can be converted into the following *normal form*

$$\begin{bmatrix} \dot{\vec{\zeta}} \\ --- \\ \dot{\vec{\eta}} \end{bmatrix} \triangleq \begin{bmatrix} \dot{\zeta}_i \\ \dot{\zeta}_{i+3} \\ --- \\ \dot{\vec{\eta}} \end{bmatrix} = \begin{bmatrix} \zeta_{i+3} \\ v_i \\ --- \\ \vec{\alpha}(\vec{\zeta},\vec{\eta})+\vec{\beta}(\vec{\zeta},\vec{\eta})\vec{v} \end{bmatrix} \qquad (6.3)$$

where the $\dot{\vec{\zeta}}$ *and* $\dot{\vec{\eta}}$ represent the external and internal dynamics of the closed-loop system. If the distribution $\Gamma$, spanned by smooth vector fields $\vec{g}_j$, $j=1,2,3$ is involutive near $\vec{x}_o$ then a diffeomorphism $\Phi(\vec{x})$ exists such that $\vec{\beta}(\vec{\zeta},\vec{\eta})=0$, and therefore the internal dynamics has the simpler form of $\dot{\vec{\eta}}=\vec{\alpha}(\vec{\zeta},\vec{\eta})$ which is now independent of the input $\vec{v}$. The zero dynamics is then defined as [88]

$$\dot{\vec{\eta}}=\vec{\alpha}(0,\vec{\eta}) \qquad (6.4)$$

with output $y_i = q_i = \zeta_i$, $i = 1, 2, 3$ constrained to be equal to zero (also then $\dot{q}_i = \zeta_{i+3} = 0$) for all times by proper choice of initial conditions $\vec{\zeta}(t_o)$ and feedback $\vec{v}(0, \vec{\eta})$.

By definition, an involutive distribution implies that it is closed under the Lie bracket defined as $\left[\vec{g}_i, \vec{g}_j\right] = L_{\vec{g}_i}\vec{g}_j - L_{\vec{g}_j}\vec{g}_i$. In our case, it means that the dimension of the distribution $\Gamma = \text{span}\left\{g_j\right\}$, $j = 1,2,3$ is always 3 and does not change when a new vector field $\vec{r}$, which is generated by the Lie bracket of any combination of two of the vector fields $\vec{g}_j$, is included in the set of the vector field generating the distribution $\Gamma$. This is in fact the case for our system, in other words, every new Lie bracket generated vector field $\vec{r}$ is always dependent on vector fields $\vec{g}_j$. Given that our distribution $\Gamma$ is involutive, then by the Frobenius Theorem [88], the set of differential equations

$$L_{\vec{g}_j}\Phi_k = \frac{\partial \Phi_k}{\partial \vec{x}}\vec{g}_j = 0, \quad k = 1,2,\ldots,m-8; \quad j = 1,2,3 \tag{6.5}$$

has a full set of solutions. These solutions are the appropriate diffeomorphisms, $\Phi(\vec{x})$, which result in $\vec{\beta}(\vec{\zeta}, \vec{\eta}) = 0$. Also note that we further require the following two conditions be met for the appropriate $\Phi(\vec{x})$, namely

$$\begin{array}{l} (i) \quad \Phi_k(0) = 0 \quad k = 1,2,\ldots,m-8, \text{ and} \\ (ii) \quad \nabla\Phi_k \text{ are linearly independent.} \end{array} \tag{6.6}$$

### 6.1.2 General Formulation for Internal Dynamics of a class on Nonlinear Systems

Consider the familiar $n$-dimensional nonlinear system of the form

$$\begin{array}{l} \dot{\vec{x}} = \vec{f}(\vec{x}) + \underline{G}(\vec{x})\,\vec{\tau} \\ \vec{y} = \vec{h}(\vec{x}) \end{array} \tag{6.7}$$

where $\vec{x} \in \mathbb{R}^n$ is the state vector given by

$$\vec{x} = \begin{bmatrix} x_1 & \cdots & x_p & | & x_{p+1} & \cdots & x_r & \cdots & x_n \end{bmatrix}^T = \begin{bmatrix} \vec{x}_\alpha^T & | & \vec{x}_\beta^T \end{bmatrix}^T \quad (6.8)$$

where $\vec{\tau} \in \mathbb{R}^p$ is the control input, $\vec{y} \in \mathbb{R}^s$ is the system output, $\vec{f} \in \mathbb{R}^n$ and $\vec{h} \in \mathbb{R}^s$ are smooth vector fields (i.e. belong to class $C^\infty$). Matrix $\underline{G} \in \mathbb{R}^{n\times p}$ is composed of columns that are each smooth vector fields $\vec{g}_j$, $j = 1,\ldots,p$ defined on $\mathbb{R}^n$ and note that matrix $\underline{G}$ reduces to a column vector in case of a single-input system where $p = 1$. Also note that problems of interest will have $p \le s \le r < n$, where $r$ is the well-defined relative degree of the system.

We assume that the $\vec{f}$ and $\underline{G}$ terms in (6.7) can be given as

$$\begin{aligned} \vec{f} &= \underline{M}^{-1}(\vec{x}_\beta)\,\vec{l}(\vec{x}) \\ \underline{G} &= \underline{M}^{-1}_{\{n,p\}}(\vec{x}_\beta) \end{aligned} \quad (6.9)$$

where $\underline{M}$ is the system matrix (e.g. Mass matrix in case of a mechanical system) and $\vec{l}$ is the combination of all linear and non-linear terms. The matrix $\underline{M}^{-1}_{\{n,p\}}$ is obtained by taking the first $p$ columns of matrix $\underline{M}^{-1}$. We have assumed that $\underline{M}$ is only a function of a subset of the state variables (i.e. $\vec{x}_\beta$) which is a reasonable assumption for many classes of physical systems such as mechanical systems, like our flexible spacecraft under consideration.

The system matrix $\underline{M}$ can be partitioned into four block matrices as

$$\underline{M} = \begin{bmatrix} \underline{M}_{11} & \underline{M}_{12} \\ \underline{M}_{21} & \underline{M}_{22} \end{bmatrix} \quad (6.10)$$

For most physical systems, we have a symmetric matrix $\underline{M}$ such that $\underline{M}_{21} = \underline{M}_{12}^T$. The block matrices are specified as $\underline{M}_{11} \in \mathbb{R}^{p\times p}$, $\underline{M}_{12} \in \mathbb{R}^{p\times(n-p)}$, $\underline{M}_{21} \in \mathbb{R}^{(n-p)\times p}$ and $\underline{M}_{22} \in \mathbb{R}^{(n-p)\times(n-p)}$. Using the block matrix inversion formula [127], we can obtain

$$\underline{M}^{-1} = \begin{bmatrix} \underline{F}_{11}^{-1} & -\underline{F}_{11}^{-1}\underline{M}_{12}\underline{M}_{22}^{-1} \\ -\underline{M}_{22}^{-1}\underline{M}_{12}^{T}\underline{F}_{11}^{-1} & \underline{M}_{22}^{-1} + \underline{M}_{22}^{-1}\underline{M}_{12}^{T}\underline{F}_{11}^{-1}\underline{M}_{12}\underline{M}_{22}^{-1} \end{bmatrix} \tag{6.11}$$
$$\textit{where} \quad \underline{F}_{11} = \underline{M}_{11} - \underline{M}_{12}\underline{M}_{22}^{-1}\underline{M}_{12}^{T}$$

For the sake of our subsequent development, the matrices $\underline{M}_{11}$, $\underline{M}_{22}$ and $\underline{F}_{11}$ must be nonsingular so that $\underline{M}^{-1}$ is invertible. The definition and specification of zero dynamics require the *normal form* of system (6.7) which can be obtained easily once we have found the appropriate mappings (i.e. diffeomorphisms). In general there are $n-p$ diffeomorphisms, $\Phi_k$, that can be solved for, of which only $n-r$ are non-trivial and of interest. All $\Phi_k$'s must respect the conditions set forth in (6.5)-(6.6) with $j=1,\ldots,p$ and $k=1,\ldots,n-p$. It's condition (6.5) that can be used to our advantage to formulate the problem of finding $\Phi_k$ in a new light. We express $L_{\vec{g}_j}\Phi_k = 0$ in its matrix form

$$L_{\vec{g}_j}\Phi_k = \nabla\Phi_k\,\vec{g}_j = 0, \qquad j=1,\ldots,p \quad ,k=1,\ldots,n-p$$
$$\begin{bmatrix} \dfrac{\partial\Phi_k}{\partial x_1} & \cdots & \dfrac{\partial\Phi_k}{\partial x_n} \end{bmatrix} \begin{bmatrix} | & & | \\ \vec{g}_1 & | \cdots | & \vec{g}_p \\ | & & | \end{bmatrix} = \vec{0}^T \tag{6.12}$$

and now we take the transpose of (6.12) to obtain

$$\begin{bmatrix} \vec{g}_1^T \\ - \; - \; - \\ \vdots \\ - \; - \; - \\ \vec{g}_p^T \end{bmatrix} \begin{bmatrix} \dfrac{\partial\Phi_k}{\partial x_1} \\ \vdots \\ \dfrac{\partial\Phi_k}{\partial x_n} \end{bmatrix} = \vec{0} \quad or \quad \underline{G}^T(\nabla\Phi_k)^T = \vec{0} \tag{6.13}$$

Hence $(\nabla\Phi_k)^T = \mathcal{N}(\underline{G}^T)$ where $\mathcal{N}(\underline{G}^T)$ is the *Null Space* of $\underline{G}^T$. In essence, we can now partly solve the problem by finding the Null Space of $\underline{G}^T$ using linear algebraic techniques. The matrix $\underline{G}^T$ is the first $p$ rows of $\underline{M}^{-1}$ obtained from (6.11) as

$$\underline{G}^T = \begin{bmatrix} \underline{F}_{11}^{-1} & -\underline{F}_{11}^{-1}\underline{M}_{12}\underline{M}_{22}^{-1} \end{bmatrix} \tag{6.14}$$

Now assume that $\mathcal{N}(\underline{G}^T) = \underline{X}$ so that

$$\underline{G}^T\underline{X} = \underline{0} \tag{6.15}$$

and let $\underline{X} = \begin{bmatrix} \underline{N} \\ \underline{I} \end{bmatrix}$ where $\underline{I}$ is the identity matrix. Using (6.14) and (6.15), we solve for $\underline{X} \in \mathbb{R}^{n\times(n-p)}$

$$\underline{G}^T\underline{X} = \begin{bmatrix} \underline{F}_{11}^{-1} & -\underline{F}_{11}^{-1}\underline{M}_{12}\underline{M}_{22}^{-1} \end{bmatrix} \begin{bmatrix} \underline{N} \\ \underline{I} \end{bmatrix} = \underline{0} \tag{6.16}$$

Expanding (6.16) gives

$$\underline{F}_{11}^{-1}\underline{N} - \underline{F}_{11}^{-1}\underline{M}_{12}\underline{M}_{22}^{-1}\underline{I} = \underline{0} \tag{6.17}$$

By inspecting (6.17), we find that

$$\underline{N} = \underline{M}_{12}\underline{M}_{22}^{-1} \tag{6.18}$$

and hence

$$\mathcal{N}(\underline{G}^T) = \underline{X} = \begin{bmatrix} \underline{M}_{12}\underline{M}_{22}^{-1} \\ \underline{I} \end{bmatrix} \tag{6.19}$$

The rank of $\underline{X}$ provides the number of independent solutions for $\Phi_k$, and as $\underline{X}$ includes the $(n-p)\times(n-p)$ identity matrix then there are no zero columns in $\underline{X}$ and $Rank(\underline{X}) = n - p$. From (6.19) we can solve for $\Phi_k, k = 1, \ldots, n-p$, which will be used to transform system (6.7) into its *normal form*. Each column in matrix (6.19) represents the gradient of $\Phi_k$ (i.e. $(\nabla\Phi_k)^T$) and expanding (6.19) gives

$$\begin{bmatrix} \frac{\partial \Phi_k}{\partial x_1} \\ \vdots \\ \frac{\partial \Phi_k}{\partial x_n} \end{bmatrix} = \underbrace{\begin{bmatrix} \overbrace{\begin{matrix} N_{11} & \cdots & N_{1(n-p)} \\ \vdots & \ddots & \vdots \\ N_{p1} & \cdots & N_{p(n-p)} \end{matrix}}^{p\times(n-p)} \\ \begin{matrix} -- & -- & --- \end{matrix} \\ \underbrace{\begin{matrix} 1 & \cdots & 0 \\ \vdots & \ddots & \vdots \\ 0 & \cdots & 1 \end{matrix}}_{(n-p)\times(n-p)} \end{bmatrix}}_{n\times(n-p)} \tag{6.20}$$

Recalling that $N_{jk}$ terms in (6.20) are not a function of $\vec{x}_\alpha = \begin{bmatrix} x_1 & \cdots & x_p \end{bmatrix}$, we can obtain the closed-form expression for $\Phi_k$ as

$$\Phi_k = x_1 N_{1k} + x_2 N_{2k} + \cdots + x_p N_{pk} + x_{p+k}.1 + c_k \qquad \mathit{for} \quad 1 \le k \le n-p, \quad c_k \in \mathbb{R} \tag{6.21}$$

Let's define a vector containing all the $n-p$ diffeomorphisms $\Phi_k$ as $\vec{\Phi}^T = \begin{bmatrix} \Phi_1 & \Phi_2 & \cdots & \Phi_{n-p} \end{bmatrix}$, we can then express the analytical solution for all $\Phi_k$, which are in essence the new *normal variables* $\vec{\eta}$, as

$$\vec{\eta}^T = \vec{\Phi}^T = \begin{bmatrix} x_1 & \cdots & x_p \end{bmatrix} \underline{N} + \begin{bmatrix} x_{p+1} & \cdots & x_n \end{bmatrix} + \vec{c}^T \tag{6.22}$$

or simply as

$$\vec{\eta}^T = \begin{bmatrix} \vec{x}_\alpha^T & | & \vec{x}_\beta^T \end{bmatrix} \begin{bmatrix} \underline{M}_{12} \underline{M}_{22}^{-1} \\ \underline{I} \end{bmatrix} + \vec{c}^T = \vec{x}^T \underline{X} + \vec{c}^T \tag{6.23}$$

where $\vec{c} \in \mathbb{R}^{(n-p)\times 1}$ is a vector of arbitrary constants, which could be zero.

Having obtained $\vec{\eta} = \vec{\Phi}(\vec{x})$, we can solve for $\vec{x} = \vec{\Phi}^{-1}(\vec{\eta})$, knowing that the inverse map, $\vec{\Phi}^{-1}$, exists by definition. Once this is done, both $\underline{M}$ and $\vec{l}$ can be expressed as a function of $\vec{\eta}$: $\underline{M}(\vec{\eta})$ and $\vec{l}(\vec{\zeta},\vec{\eta})$ where $\vec{\zeta} = \Psi(\vec{x})$ are the transformed $\vec{x}_\alpha$ states (refer to (6.2) as an example).

The corresponding *normal form* of (6.7) can then be obtained by differentiating (6.23)

$$\dot{\vec{\eta}}^T = \dot{\vec{x}}^T \underline{X} + \vec{x}^T \dot{\underline{X}} \tag{6.24}$$

Using equations (6.7), (6.9), (6.15) and (6.19) with (6.24), we can expand (6.24)

$$\dot{\vec{\eta}}^T = \left(\underline{M}^{-1}\vec{l} + \underline{G}\,\vec{\tau}\right)^T \underline{X} + \vec{x}^T \begin{bmatrix} \dot{\underline{N}} \\ \underline{0} \end{bmatrix} = \vec{\tau}^T \underbrace{\underline{G}^T \underline{X}}_{=\underline{0}} + \vec{l}^T \underline{M}^{-1} \underline{X} + \vec{x}_\alpha^T \dot{\underline{N}} \tag{6.25}$$

Substituting (6.11) and (6.19) into (6.25) and simplifying, we obtain

$$\begin{aligned} \dot{\vec{\eta}}^T &= \vec{l}^T \begin{bmatrix} \underline{F}_{11}^{-1} & -\underline{F}_{11}^{-1}\underline{M}_{12}\underline{M}_{22}^{-1} \\ -\underline{M}_{22}^{-1}\underline{M}_{12}^T\underline{F}_{11}^{-1} & \underline{M}_{22}^{-1} + \underline{M}_{22}^{-1}\underline{M}_{12}^T\underline{F}_{11}^{-1}\underline{M}_{12}\underline{M}_{22}^{-1} \end{bmatrix} \begin{bmatrix} \underline{M}_{12}\underline{M}_{22}^{-1} \\ \underline{I} \end{bmatrix} + \vec{x}_\alpha^T \dot{\underline{N}} \\ &= \vec{l}^T \begin{bmatrix} \underline{0} \\ \underline{M}_{22}^{-1} \end{bmatrix} + \vec{x}_\alpha^T \left( \dot{\underline{M}}_{12}\underline{M}_{22}^{-1} + \underline{M}_{12}\dot{\underline{M}}_{22}^{-1} \right) \end{aligned} \tag{6.26}$$

The final compact form of (6.26) as a function of the new states $\vec{\zeta}$ and $\vec{\eta}$ is simply

$$\dot{\vec{\eta}} = \underline{M}_{22}^{-1}(\vec{\eta})\vec{l}_\beta(\vec{\zeta},\vec{\eta}) + \left( \underline{M}_{22}^{-1}(\vec{\eta})\dot{\underline{M}}_{12}^T(\vec{\eta}) + \dot{\underline{M}}_{22}^{-1}(\vec{\eta})\underline{M}_{12}^T(\vec{\eta}) \right)\vec{\zeta} \tag{6.27}$$

where $\vec{l}_\beta$ refers to the subset of vector $\vec{l}$ composed of $(p+1)^{th}$ to $n^{th}$ elements.

**Theorem 6.1:** Given a nonlinear system of the form

$$\dot{\vec{x}} = \vec{f}(\vec{x}) + \underline{G}(\vec{x})\,\vec{\tau} = \vec{f} + \sum_{j=1}^{p} \vec{g}_j \tau_j$$

$$\vec{y} = \vec{h}(\vec{x})$$

where $\vec{x} \in \mathbb{R}^n$, $\vec{\tau} \in \mathbb{R}^p$ and $\vec{y} \in \mathbb{R}^s$ are the system state, input and output, respectively, and $\vec{f} \in \mathbb{R}^n$, $\vec{h} \in \mathbb{R}^s$ and $\vec{g}_j \in \mathbb{R}^n$, $j = 1,\ldots,p$, are smooth vector fields belonging to class $C^\infty$. Assuming the following properties hold:

(i) $r$ is the relative degree of the system and is well defined.

(ii) $\vec{f}$ can be expressed as $\underline{M}^{-1}(\vec{x}_\beta)\,\vec{l}(\vec{x})$ where $\underline{M} \in \mathbb{R}^{n\times n}$ is the invertible system matrix which is a function of only the state variables $\vec{x}_\beta = \begin{bmatrix} x_{p+1} & \cdots & x_n \end{bmatrix}^T$ and $\vec{l} \in \mathbb{R}^n$ contains all linear and non-linear terms.

(iii) Distribution $\Gamma$, spanned by smooth vector fields $\vec{g}_j$, $j = 1,\ldots,p$ is involutive.

(iv) $p \le s \le r < n$

then the new *normal variables* $\vec{\eta} \in \mathbb{R}^{n-p}$, corresponding to $\vec{x}_\beta$ and given by the diffeomorphisms $\Phi_k$, $k = 1,\ldots,n-p$, are

$$\vec{\eta} = \underline{M}_{22}^{-1}\underline{M}_{12}^{T}\vec{x}_\alpha + \vec{x}_\beta$$

where $\vec{x}_\alpha = \begin{bmatrix} x_1 & \cdots & x_p \end{bmatrix}^T$,

and the rate of change of $\vec{\eta}$, which constitutes the key part of the *normal form* of the original nonlinear system, is given by

$$\dot{\vec{\eta}} = \underline{M}_{22}^{-1}\vec{l}_\beta + \left(\underline{M}_{22}^{-1}\dot{\underline{M}}_{12}^{T} + \dot{\underline{M}}_{22}^{-1}\underline{M}_{12}^{T}\right)\vec{\zeta}$$

where $\vec{l}_\beta$ refers to the subset of vector $\vec{l}$ composed of $(p+1)^{th}$ to $n^{th}$ elements and $\vec{\zeta}$ is the new *normal variables* corresponding to $\vec{x}_\alpha$ obtained using trivial mappings.

**Corollary 6.2:** The zero dynamics of system (6.7) is simply the normal equation (6.27) with all outputs kept at zero. Assuming that the output $\vec{y}$ in (6.7) is a linear function of the first $s$ elements of the state vector, $\vec{y} = \underline{C}\,\vec{x}_{\{1..s\}}$, where $\underline{C} \in \mathbb{R}^{s\times s}$ is any non-singular, constant real matrix, then the zero dynamics will have the simple form

$$\dot{\vec{\eta}} = \underline{M}_{22}^{-1}\vec{l}_\beta \tag{6.28}$$

We need the non-singularity assumption since otherwise, $\underline{C}$ could have a null space and hence $\vec{x}_{\{1..s\}}$ need not be zero when $\vec{y} = 0$. We have also assumed that new normal variables $\vec{\zeta}$ remain the same as $\vec{x}_{\{1..s\}}$, after trivial mappings $\Psi(\vec{x})$. Note that $\vec{x}_{\{1..s\}}$ refers to the sub-vector of $\vec{x}$ composed of the first $s$ elements.

### 6.1.3 Zero Dynamics for Two Representative Cases

In what follows, we have considered two representative cases for constructing and analyzing the zero dynamics of our flexible spacecraft composed of a rigid bus and one single appendage. Specifically,

Case (i): assuming only the first vibration modes in each direction of the appendage (i.e. $p = q = 1$ *and* $m = 10$);

Case (ii): assuming two vibration modes in each direction of the appendage (i.e. $p = q = 2$ *and* $m = 16$).

In both cases, we assume that the appendage parameters (e.g. $a, b, E, etc...$) are unknown so that we can obtain a general result as a function of these design parameters.

Using Theorem (6.1), the diffeomorphisms $\Phi(\vec{x})$ for the 2 cases were obtained to be

$$case\,(i): \quad \Phi_1(\vec{x}) = \chi_1;$$

$$\Phi_2(\vec{x}) = \dot{\chi}_1 + \begin{Bmatrix} (0.6b^2 + 0.8bd_y + 1.6\chi_1 d_z)\omega_x(\tilde{q},\dot{\tilde{q}}) \\ -(0.4a + 0.8d_x)b\omega_y(\tilde{q},\dot{\tilde{q}}) \\ -(0.8a + 1.6d_x)\chi_1\omega_z(\tilde{q},\dot{\tilde{q}}) \end{Bmatrix}$$

$$\begin{aligned} case\,(ii): \quad & \Phi_1(\vec{x}) = \chi_1; \quad \Phi_2(\vec{x}) = \chi_2; \quad \Phi_3(\vec{x}) = \chi_3; \quad \Phi_4(\vec{x}) = \chi_4 \\ & \Phi_5(\vec{x}) = \dot{\chi}_1 + 6.1\omega_x(\tilde{q},\dot{\tilde{q}}) + \left(0.05\chi_3 - 0.01\chi_4\right)\omega_z(\tilde{q},\dot{\tilde{q}}) \\ & \Phi_6(\vec{x}) = \dot{\chi}_2 + 1.1\omega_x(\tilde{q},\dot{\tilde{q}}) + \left(-0.01\chi_3 + 0.2\chi_4\right)\omega_z(\tilde{q},\dot{\tilde{q}}) \\ & \Phi_7(\vec{x}) = \dot{\chi}_3 + 0.2\omega_y(\tilde{q},\dot{\tilde{q}}) + \left(0.05\chi_1 - 0.01\chi_2\right)\omega_z(\tilde{q},\dot{\tilde{q}}) \\ & \Phi_8(\vec{x}) = \dot{\chi}_4 + 0.1\omega_y(\tilde{q},\dot{\tilde{q}}) + \left(-0.01\chi_1 + 0.2\chi_2\right)\omega_z(\tilde{q},\dot{\tilde{q}}) \end{aligned} \tag{6.29}$$

where $\omega_x(\tilde{q},\dot{\tilde{q}})$, $\omega_y(\tilde{q},\dot{\tilde{q}})$ and $\omega_z(\tilde{q},\dot{\tilde{q}})$ were defined in (2.17). Hence using (6.29) to obtain the new coordinates $\vec{\eta}^T = \left[\Phi_1(\vec{x})\ldots\Phi_{m-8}(\vec{x})\right] = \left[\vec{\eta}_a^T \; \vec{\eta}_b^T\right]$, and setting the outputs to zero as per (6.4), the zero dynamics can then be obtained for each case as follows:

Case (i):

In this case, the new coordinates are $\vec{\eta}^T = \left[\Phi_1(\vec{x})\,\Phi_2(\vec{x})\right] = \left[\eta \; \dot{\eta}\right]$, and the zero dynamics is given by

$$\ddot{\eta} + \frac{\xi}{ab}\dot{\eta} + \frac{1.03Eh^3}{\rho b^4(1-\gamma^2)}\eta = 0 \tag{6.30}$$

This is an interesting result showing that the zero dynamics is globally exponentially stable for all design parameters (e.g. $a, b, E, etc.$).

Case (ii):

In this case, the new coordinates for the zero dynamics are $\vec{\eta}^T = \left[\Phi_1(\vec{x})\ldots\Phi_8(\vec{x})\right] = \left[\vec{\eta}_a^T \; \vec{\eta}_b^T\right]$, and using Corollary (6.2), the zero dynamics can be

derived to be

$$\begin{bmatrix} \dot{\vec{\eta}}_a \\ \dot{\vec{\eta}}_b \end{bmatrix} = \begin{bmatrix} \underline{0} & \underline{1} \\ -c_2\underline{C} & -c_1\underline{1} \end{bmatrix} \begin{bmatrix} \vec{\eta}_a \\ \vec{\eta}_b \end{bmatrix} \tag{6.31}$$

where $\vec{\eta}_a = \begin{bmatrix} \eta_1 & \eta_2 & \eta_3 & \eta_4 \end{bmatrix}^T$, $\vec{\eta}_b = \begin{bmatrix} \dot{\eta}_1 & \dot{\eta}_2 & \dot{\eta}_3 & \dot{\eta}_4 \end{bmatrix}^T$ and

$$c_1 = \frac{\xi}{ab} \quad and \quad c_2 = \frac{Eh^3}{12\rho\left(1\text{-}\gamma^2\right)} \tag{6.32}$$

$$\underline{C} = \begin{bmatrix} \lambda_1\alpha_1 & 0 & 0 & 0 \\ 0 & \lambda_1\alpha_2 & 0 & 0 \\ 0 & 0 & \lambda_1\alpha_3 + \lambda_2\beta_1 & -\lambda_2\beta_3 \\ 0 & 0 & -\lambda_2\beta_3 & \lambda_1\alpha_4 + \lambda_2\beta_2 \end{bmatrix} \tag{6.33}$$

and $\alpha_1 = \alpha_3 = 0.1236, \alpha_2 = \alpha_4 = 4.8552, \ \beta_1 = 0.5577, \beta_2 = 3.8901, \beta_3 = 0.8856$,

$\lambda_1 = \dfrac{100}{b^4}$ *and* $\lambda_2 = \dfrac{200(1-\gamma)}{a^2b^2}$ are all positive real numbers.

To analyze the stability properties of the zero dynamics (6.31), we use a Lyapunov function candidate $L_\eta$

$$L_\eta = \vec{\eta}^T \underline{P}\, \vec{\eta} \tag{6.34}$$

It is clear that $L_\eta > 0$ when $\underline{P} \succ 0$ (i.e. $\underline{P}$ is positive definite) except at the origin when $L_\eta = 0$, but we also need $\dot{L}_\eta < 0$ for global asymptotical stability

$$\dot{L}_\eta = \dot{\vec{\eta}}^T \underline{P}\, \vec{\eta} + \vec{\eta}^T \underline{P}\, \dot{\vec{\eta}} \tag{6.35}$$

Equation (6.31) can be re-written as

$$\dot{\vec{\eta}} = \begin{bmatrix} \underline{0} & \underline{1} \\ -c_2\underline{C} & -c_1\underline{1} \end{bmatrix} \vec{\eta} \tag{6.36}$$

Hence, equation (6.35) becomes

$$\dot{L}_\eta = \vec{\eta}^T \underline{A}^T \underline{P}\,\vec{\eta} + \vec{\eta}^T \underline{P}\,\underline{A}\,\vec{\eta} = \vec{\eta}^T \left(\underline{A}^T \underline{P} + \underline{P}\,\underline{A}\right) \vec{\eta} \tag{6.37}$$

where

$$\underline{A} = \begin{bmatrix} \underline{0} & \underline{1} \\ -c_2\underline{C} & -c_1\underline{1} \end{bmatrix} \tag{6.38}$$

In order to insure asymptotic stability of the equilibrium point, we must have $\left(\underline{A}^T \underline{P} + \underline{P}\,\underline{A}\right) = -\underline{Q}$ where $\underline{Q} \succ 0$. This is the standard Lyapunov equation and needs to be solved for $\underline{P}$, given for example that $\underline{Q} = \underline{1}$.

The matrix $\underline{P}$ can be written in block form and hence we can expand the Lyapunov equation as

$$\begin{bmatrix} \underline{0} & -c_2\underline{C} \\ \underline{1} & -c_1\underline{1} \end{bmatrix} \begin{bmatrix} \underline{p}_{11} & \underline{p}_{12} \\ \underline{p}_{21} & \underline{p}_{22} \end{bmatrix} + \begin{bmatrix} \underline{p}_{11} & \underline{p}_{12} \\ \underline{p}_{21} & \underline{p}_{22} \end{bmatrix} \begin{bmatrix} \underline{0} & \underline{1} \\ -c_2\underline{C} & -c_1\underline{1} \end{bmatrix} = \begin{bmatrix} -\underline{1} & \underline{0} \\ \underline{0} & -\underline{1} \end{bmatrix} \tag{6.39}$$

and since we want $\underline{P} \succ 0$, then $\underline{P} = \underline{P}^T$ which requires $\underline{p}_{11} = \underline{p}_{11}^T$, $\underline{p}_{22} = \underline{p}_{22}^T$ and $\underline{p}_{21} = \underline{p}_{12}^T$. To further simplify the problem, we can also make the assumption that $\underline{p}_{12} = \underline{p}_{12}^T$. Now, equation (6.39) can be re-written as 4 matrix equations that need to be solved for $\underline{p}_{11}, \underline{p}_{12}$ and $\underline{p}_{22}$, that is

$$\begin{aligned} -c_2\underline{C}\,\underline{p}_{12} - c_2\underline{p}_{12}\underline{C} &= -\underline{1} \\ 2\underline{p}_{12} - 2c_1\underline{p}_{22} &= -\underline{1} \\ -c_2\underline{C}\,\underline{p}_{22} + \underline{p}_{11} - c_1\underline{p}_{12} &= \underline{0} \\ -c_2\underline{p}_{22}\underline{C} + \underline{p}_{11} - c_1\underline{p}_{12} &= \underline{0} \end{aligned} \tag{6.40}$$

After necessary algebraic manipulations, the solutions can be found to be

$$
\begin{aligned}
\underline{p}_{11} &= \frac{c_1\underline{C}^{-1}}{2c_2} + \frac{c_2\underline{C}}{2c_1} + \frac{1}{2c_1} \\
\underline{p}_{12} &= \frac{\underline{C}^{-1}}{2c_2} \\
\underline{p}_{22} &= \frac{\underline{C}^{-1}}{2c_1c_2} + \frac{1}{2c_1}
\end{aligned}
\tag{6.41}
$$

Hence we find that a solution, $\underline{P}$, satisfying the Lyapunov equation exists and is given by (6.41). In order to ensure that $\underline{P} \succ 0$ as required in (6.34), the following conditions must be met [127]

$$
\begin{aligned}
&(i) \quad \underline{p}_{11} \succ 0 \\
&(ii) \quad \underline{p}_{12}\underline{p}_{11}^{-1}\underline{p}_{12} - \underline{p}_{22} \prec 0
\end{aligned}
\tag{6.42}
$$

Condition (i) is satisfied if $\underline{C} \succ 0$, since if $\underline{C} \succ 0$ then $\underline{C}^{-1} \succ 0$ and given that both $c_1, c_2 > 0$, then all the three terms in $\underline{p}_{11}$ are positive definite matrices, the sum of which is also positive definite.

One way to determine if a symmetric *nxn* matrix is positive definite is to insure that all the leading principle minors $\Delta_i, i = 1,2,...n$ are positive. For $\underline{C}$, this results in the following conditions:

$$
\begin{aligned}
&|\Delta_1| = \lambda_1\alpha_1 > 0, \quad |\Delta_2| = \lambda_1\lambda_2\alpha_1\alpha_2 > 0, \quad |\Delta_3| = \lambda_1\lambda_2\alpha_1\alpha_2(\lambda_1\alpha_3 + \lambda_2\beta_1) > 0, \quad \textit{and} \\
&|\Delta_4| = \lambda_1\lambda_2\alpha_1\alpha_2[(\lambda_1\alpha_3 + \lambda_2\beta_1)(\lambda_1\alpha_4 + \lambda_2\beta_2) - \lambda_2^2\beta_3^2)] > 0
\end{aligned}
$$

The first three conditions are trivially satisfied since all variables involved are positive. For the fourth condition, we need to verify that

$$
(\lambda_1\alpha_3 + \lambda_2\beta_1)(\lambda_1\alpha_4 + \lambda_2\beta_2) > \lambda_2^2\beta_3^2 \tag{6.43}
$$

Condition (6.43) can be expanded in all its terms as:

$$(\lambda_1^2\alpha_3\alpha_4 + \lambda_1\lambda_2\alpha_4\beta_1 + \lambda_1\lambda_2\alpha_3\beta_2 + \lambda_2^2\beta_1\beta_2) > \lambda_2^2\beta_3^2 \tag{6.44}$$

Now since $\beta_1\beta_2 > \beta_3^2$ and the other three terms of the left hand side of (6.44) are all positive, this proves that condition (6.43) is always satisfied and hence $\underline{C}$ is positive definite, which in turn implies that $\underline{p}_{11} \succ 0$.

Now we turn our attention to condition (ii) in (6.42) which, using $\underline{p}_{12}$ and $\underline{p}_{22}$ from (6.41), can be written as

$$\underbrace{\frac{1}{4c_2^2}\overbrace{\left(\underline{C}\,\underline{p}_{11}\underline{C}\right)^{-1}}^{\underline{a}} - \frac{\underline{C}^{-1}}{2c_1c_2} - \frac{\underline{1}}{2c_1}}_{\underline{N}} \prec 0 \tag{6.45}$$

In order for matrix $\underline{N} \prec 0$, we must have $\vec{x}^T\underline{N}\,\vec{x} < 0$ for all $\vec{x} \in \left\{\mathbb{R}^4 \setminus \vec{0}\right\}$. Noting that $\underline{a}$ and $\underline{C}^{-1}$ are positive definite matrices, we can use the following inequalities for a positive definite matrix $\underline{m}$ [89]

$$\lambda_{\min}\left(\underline{m}\right)\left\|\vec{x}\right\|^2 \le \vec{x}^T\underline{m}\,\vec{x} \le \lambda_{\max}\left(\underline{m}\right)\left\|\vec{x}\right\|^2 \tag{6.46}$$

to re-write $\vec{x}^T\underline{N}\,\vec{x} < 0$ condition as

$$\frac{1}{4c_2^2}\lambda_{\max}\left(\underline{a}\right)\left\|\vec{x}\right\|^2 - \frac{1}{2c_1c_2}\lambda_{\min}\left(\underline{C}^{-1}\right)\left\|\vec{x}\right\|^2 - \frac{1}{2c_1}\lambda_{\min}\left(\underline{1}\right)\left\|\vec{x}\right\|^2 < 0 \tag{6.47}$$

Equation (6.47) can be simplified to

$$\lambda_{\max}\left(\underline{a}\right) < 4c_2^2\left(\frac{0.17238}{2c_1c_2} + \frac{1}{2c_1}\right) \tag{6.48}$$

where $\underline{a} = \left(\underline{C}\ \underline{p}_{11}\underline{C}\right)^{-1}$.

Equation (6.48) imposes a condition on $c_1$ and $c_2$ (i.e. design parameters $a, b, E, etc...$) which needs to be met so that (6.45) is satisfied and hence the system will have a linear time invariant zero dynamics which is asymptotically stable. For our example spacecraft, described in Section 5.2 and with appendage parameters $a = 1, b = 10, \rho = 10, E = 5.0e8$, $h = 0.02, \gamma = 0.3$ and $\xi = 0.05$, condition (6.48) is met as $\lambda_{\max}\left(\underline{a}\right) = 0.12$ and the right hand-side of (6.48) is a very large number ($\approx 5.4e6$).

This is an interesting result showing that the linear time invariant zero dynamics is globally asymptotically stable for realistic and common design parameters (e.g. $a, b, E, etc.$), for a representative case where two vibration modes in each direction of the appendage were active.

## 6.2 Stability Analysis for the Flexible Spacecraft

We have so far established that the zero dynamics of the flexible spacecraft is asymptotically stable for two representative cases. We now turn our attention to the proof of stability of the overall close loop system (5.13)-(5.14) for the case when $p = q = 2\ and\ m = 16$. We will again make use of the Lyapunov theory to determine the stability properties of the closed loop system and will use the new coordinates $\vec{\eta}^T = \left[\Phi_1(\vec{x})\ldots\Phi_8(\vec{x})\right] = \left[\vec{\eta}_a^T\ \vec{\eta}_b^T\right]$ obtained from (6.29). This specific coordinate selection simplifies the problem considerably since in this representation, the internal dynamics (5.14) will be independent of input $\vec{v}$ (i.e. as shown earlier $\dot{\vec{\eta}} = \vec{\alpha}(\tilde{q}, \dot{\tilde{q}}, \vec{\eta})$). Using the actual diffeomorphisms (6.29), we now rewrite the system (5.13)-(5.14) as

$$\begin{bmatrix} \dot{\vec{q}}_e \\ \ddot{\vec{q}}_e \\ \cdots \\ \dot{\vec{\eta}}_a \\ \dot{\vec{\eta}}_b \end{bmatrix} = \underbrace{\begin{bmatrix} \underline{0} & \underline{1} & \vdots & \underline{0} & \underline{0} \\ -\underline{K}_{Pq} & -\underline{K}_{Dq} & \vdots & \underline{K}_{P\chi} & \underline{K}_{D\chi} \\ \cdots & \cdots & + & \cdots & \cdots \\ \underline{0} & \underline{0} & \vdots & \underline{0} & \underline{1} \\ \underline{0} & \underline{0} & \vdots & c_2\underline{C} & c_1\underline{1} \end{bmatrix}}_{\underline{A}} \begin{bmatrix} \vec{q}_e \\ \dot{\vec{q}}_e \\ \cdots \\ \vec{\eta}_a \\ \vec{\eta}_b \end{bmatrix} + \underbrace{\begin{bmatrix} 0 \\ \vec{f}_{\dot{\vec{q}}_e}(\tilde{q},\dot{\tilde{q}},\vec{\eta}_a) \\ \cdots \\ \vec{f}_{\vec{\eta}}(\tilde{q},\dot{\tilde{q}},\vec{\eta}_a,\vec{\eta}_b) \\ 0 \end{bmatrix}}_{\vec{f}_{\vec{\mu}}} \tag{6.49}$$

or simply as

$$\dot{\vec{\mu}} = \underline{A}\,\vec{\mu} + \vec{f}_{\vec{\mu}} \tag{6.50}$$

where $\vec{\mu}^T = \begin{bmatrix} \vec{q}_e^T & \dot{\vec{q}}_e^T & \vec{\eta}_a^T & \vec{\eta}_b^T \end{bmatrix}$ and the nonlinear terms in the system equation (6.50) are all contained in the $\vec{f}_{\vec{\mu}}$ term which is given by

$$\vec{f}_{\vec{\mu}} = \begin{bmatrix} 0 \\ \vec{f}_{\dot{\vec{q}}_e}(\tilde{q},\dot{\tilde{q}},\vec{\eta}_a) \\ \cdots \\ \vec{f}_{\vec{\eta}}(\tilde{q},\dot{\tilde{q}},\vec{\eta}_a,\vec{\eta}_b) \\ 0 \end{bmatrix} \tag{6.51}$$

where

$$\vec{f}_{\dot{\vec{q}}_e} = -\underline{K}_{D\chi} \begin{bmatrix} 6.1\omega_x + \left(0.05\eta_3 - 0.01\eta_4\right)\omega_z \\ 1.1\omega_x + \left(-0.01\eta_3 + 0.2\eta_4\right)\omega_z \\ 0.2\omega_y + \left(0.05\eta_1 - 0.01\eta_2\right)\omega_z \\ 0.1\omega_y + \left(-0.01\eta_1 + 0.2\eta_2\right)\omega_z \end{bmatrix} \tag{6.52}$$

and

$$\vec{f}_{\vec{\eta}} = \begin{bmatrix} \begin{pmatrix} 0.003\omega_x - 0.3\omega_x^2\eta_1 + 0.7\omega_x^2\eta_2 + -0.05\omega_x\omega_y\eta_3 + 0.01\omega_x\omega_y\eta_4 \\ +\omega_y^2\eta_1 + 0.05\omega_z\dot{\eta}_3 - 0.01\omega_z\dot{\eta}_4 - 6.1\omega_y\omega_z - 1.3\omega_z^2\eta_1 + 0.7\omega_z^2\eta_2 \end{pmatrix} \\ \begin{pmatrix} 0.7\omega_x^2\eta_1 - 5.9\omega_x^2\eta_2 + 0.01\omega_x\omega_y\eta_3 - 0.25\omega_x\omega_y\eta_4 + \omega_y^2\eta_2 \\ -0.01\omega_z\dot{\eta}_3 + 0.25\omega_z\dot{\eta}_4 - 1.2\omega_y\omega_z + 0.7\omega_z^2\eta_1 - 7.0\omega_z^2\eta_2 \end{pmatrix} \\ \begin{pmatrix} -0.3\omega_x^2\eta_3 + 0.7\omega_x^2\eta_4 - 0.05\omega_x\omega_y\eta_1 + 0.01\omega_x\omega_y\eta_2 + \omega_y^2\eta_3 \\ +0.05\omega_z\dot{\eta}_1 - 0.01\omega_z\dot{\eta}_2 - 0.04\omega_x\omega_z - 1.3\omega_z^2\eta_3 + 0.7\omega_z^2\eta_4 \end{pmatrix} \\ \begin{pmatrix} 0.7\omega_x^2\eta_3 - 5.9\omega_x^2\eta_4 - 0.25\omega_x\omega_y\eta_2 + \omega_y^2\eta_4 + 0.001\omega_z\eta_2 \\ -0.01\omega_z\dot{\eta}_1 + 0.25\omega_z\dot{\eta}_2 - 0.08\omega_x\omega_z + 0.7\omega_z^2\eta_3 - 7.0\omega_z^2\eta_4 \end{pmatrix} \end{bmatrix} \tag{6.53}$$

and where $\omega_x$, $\omega_y$ and $\omega_z$ are functions of $\tilde{q}$ and $\dot{\tilde{q}}$ given by (2.17). By visual inspection, it is noted that both $\vec{f}_{\vec{\mu}}$ and $\partial\vec{f}_{\vec{\mu}} / \partial\vec{\mu}$ are continuous and hence $\vec{f}_{\vec{\mu}}$ is locally Lipschitz.

To analyze the stability properties of the closed-loop system dynamics (6.50), we use a Lyapunov function candidate $L_\mu$ given by

$$L_\mu = \vec{\mu}^T \underline{U} \vec{\mu} \tag{6.54}$$

Clearly $L_\mu > 0$ when $\underline{U} \succ 0$, except at the origin when $L_\mu = 0$. To show that $\dot{L}_\mu < 0$ for asymptotical stability of the equilibrium point of the closed loop system, we have

$$\dot{L}_\mu = \vec{\mu}^T \underline{U}\, \dot{\vec{\mu}} + \dot{\vec{\mu}}^T \underline{U}\, \vec{\mu} \tag{6.55}$$

Using (6.50), (6.55) now becomes

$$\begin{aligned} \dot{L}_\mu &= \vec{\mu}^T \underline{U}\,\underline{A}\, \vec{\mu} + \vec{\mu}^T \underline{U}\, \vec{f}_{\vec{\mu}} + \vec{\mu}^T\, \underline{A}^T \underline{U}\, \vec{\mu} + \vec{f}_{\vec{\mu}}^{\,T} \underline{U}\, \vec{\mu} \\ &= \vec{\mu}^T \underbrace{\left( \underline{U}\,\underline{A} + \underline{A}^T \underline{U} \right)}_{-\underline{Q}} \vec{\mu} + 2\vec{\mu}^T \underline{U}\, \vec{f}_{\vec{\mu}} \end{aligned} \tag{6.56}$$

The Lyapunov equation obtained from the first term on the right-hand side of (6.56) must be solved for $\underline{U}$ such that the first term of (6.56) becomes negative. The following inequality holds [89]

$$-\vec{\mu}^T \underline{Q}\, \vec{\mu} \leq -\lambda_{\min}(\underline{Q}) \left\| \vec{\mu} \right\|^2 \tag{6.57}$$

where $\lambda_{\min}(\underline{Q})$ is the smallest eigenvalue of $\underline{Q}$ and $\left\| \vec{\mu} \right\|$ is the vector 2-norm of $\vec{\mu}$. The norm of the second term on the left-hand side of (6.56) is given by

$$\left\| 2\vec{\mu}^T \underline{U}\, \vec{f}_{\vec{\mu}} \right\| \leq 2 \left\| \vec{\mu} \right\| \left\| \underline{U} \right\| \left\| \vec{f}_{\vec{\mu}} \right\| \tag{6.58}$$

where $\left\| \underline{U} \right\|$ is the matrix 2-norm of $\underline{U}$. Now since $\vec{f}_{\vec{\mu}}$ is locally Lipschitz, the following inequality holds

$$\left\| \vec{f}_{\vec{\mu}} \right\| \leq l \left\| \vec{\mu} \right\|, \quad \vec{\mu} \in \Omega_{\mu} \tag{6.59}$$

where $l \in \mathbb{R}$ is the Lipschitz constant.

Finally using (6.57)-(6.59) in (6.56) we obtain

$$\dot{L}_{\mu} \leq -\lambda_{\min}(\underline{Q}) \left\| \vec{\mu} \right\|^2 + 2l \left\| \underline{U} \right\| \left\| \vec{\mu} \right\|^2, \quad \vec{\mu} \in \Omega_{\mu} \tag{6.60}$$

We require that $\dot{L}_{\mu} < 0$ and using (6.60), this imposes the following condition

$$l < \frac{\lambda_{\min}(\underline{Q})}{2 \left\| \underline{U} \right\|} \tag{6.61}$$

Condition (6.61) ensures that $\dot{L}_{\mu}$ is always negative definite and hence the origin is an asymptotically stable equilibrium point. For our example of the flexible spacecraft with 2 flexible modes (case ii), with $\underline{Q} = \underline{1}$, we obtain $\lambda_{\min}(\underline{Q}) = 1$ and $\|\underline{U}\| = 2.12 \times 10^5$. Thus condition (6.61) can be satisfied in a small neighborhood $\Omega_{\mu}$, where the proof of asymptotic stability is valid and which nevertheless guarantees that the origin is indeed an asymptotically stable equilibrium point. Hence, we have shown that our transformed closed loop system (6.50) is locally asymptotically stable which implies that our original system (5.13)-(5.14) is asymptotically stable as well.

One can trade off system performance for a proof resulting in a larger region of attraction, $\Omega_{\mu}$, by making use of the following theorem (see [89] for the proof):

**Theorem 6.3**: *The origin of (5.13)-(5.14) is asymptotically stable if the origin of the zero dynamics (6.4) is asymptotically stable.*

It was shown that the origin of the zero dynamics (6.4) is asymptotically stable and hence the closed loop system is asymptotically stable, however the above theorem is only valid when $\underline{K}_{P\chi}$ *and* $\underline{K}_{D\chi}$ are set to zero which increases the appendage vibration suppression time considerably as was shown in attitude recovery maneuver case 7, presented in Section 5.2. Therefore, in a real life application, one would prefer our proof giving a smaller region of attraction but with a much better closed loop performance.

## 6.3 Concluding Remarks

This chapter was devoted to the important topic of stability analysis for the flexible spacecrafts. The internal nonlinear and unobservable system was shown to have an associated zero dynamics which is asymptotically stable, and this was proved for two representative cases. The overall closed-loop stability of the flexible spacecraft was also analyzed rigorously and shown to be locally asymptotically stable using Lyapunov's method, for the more generic of the two cases.

# CHAPTER 7

# CONCLUSIONS
# AND DIRECTIONS FOR FUTURE WORK

In this thesis, the important problem of automated attitude recovery for rigid and flexible spacecrafts was investigated using nonlinear control. The thesis can be broken down in three major parts, namely spacecraft dynamics (Chapter 2), simulation (Chapter 3) and control (Chapters 4-6).

In the dynamics section, a general closed-form model for attitude and flexible dynamics for a class of spacecrafts was derived in detail. The spacecraft considered had a star topology with a rigid bus and Flexible Plate-type Appendages (FPAs). One of the main motivating factors for the use of flexible plate-type appendages in this thesis was that they have not been treated extensively in the attitude dynamics and control literature. The flexible dynamics derivations were based on the hybrid coordinates approach and resulted in system equations which are nonlinear and highly coupled due to the addition of flexible appendages. There is indeed a strong coupling effect between the attitude and the elastic motion of the appendages.

In the simulation part, a high-fidelity, user-friendly modeling and control software environment was designed and developed which has been validated through extensive numerical simulations using analytical test cases or the principles of energy and angular momentum conservation. The flexible spacecraft simulator test-bed which is based on the dynamic formulation derived in this thesis, is not only a valuable asset for investigation of the attitude recovery problem, but can also be used for other research in the area of flexible attitude dynamics and control.

Finally in the control section, work was undertaken to develop suitable nonlinear controllers based on feedback linearization approach. Given that the flexible spacecraft is under-actuated, the input-output linearization technique was specifically used to break up the system into two distinct parts, namely (1) an external linearizable system for which a linear controller was easily implemented and (2) an internal nonlinear unobservable system for which the associated zero dynamics was shown to be globally asymptotically stable for two representative cases. The overall closed-loop stability of the flexible spacecraft was also analyzed rigorously and shown to be locally asymptotically stable using Lyapunov's method. The controllers were shown to be robust against modeling and parametric uncertainties for a typical flexible spacecraft through simulations. In this section of the thesis, the simulator was used extensively for the investigation of the attitude recovery of spacecraft using the input-output linearization technique and simulation results were presented to demonstrate the capability and advantages of the proposed modeling and control techniques for attitude recovery under severe malfunctions affecting the spacecraft.

Also, as part of the research performed under the control topic, a new explicit analytical solution for the construction of normal form and the zero dynamics of a general class of nonlinear systems was developed and applied specifically to the flexible spacecraft system, but this solution has applications in other areas of research.

The following points identify specific improvements or additional work that could be undertaken as part of future activities in this area of research. These have been categorized under the three main headings of dynamics, simulation and control, and under each heading the items are listed by order of priority:

Future Work in Dynamics:

- To enhance the dynamics formulation, one could include flexible beam-type appendages and implement this new part of the formulation in the spacecraft simulator.

- Presently, the formulation assumes that the spacecraft is in a fully deployed configuration which is the case for almost the entire duration of a spacecraft's mission. However, for short periods of several minutes, many spacecrafts must deploy their appendages (e.g. solar panels, antennas, scientific instruments, etc.) which can cause disturbing torques and also change the dynamics of the system during the deployment. This dynamics can be added to the formulation so that it can be simulated and analyzed in special cases.

- Another improvement is to include the necessary modifications to allow the possibility of adding a point mass anywhere along or at the tip of a flexible appendage (e.g. sometimes a magnetometer is placed at the tip of a flexible beam, in order to distance it from the magnetic field generated by the spacecraft).

Future Work in Simulation:

- The spacecraft simulator can be further refined and validated, through comparative studies with other commercially or otherwise available simulators, and eventually using actual flight data obtained from a flexible spacecraft. This will increase the simulator's value and dependability for future research in the area of flexible spacecraft dynamic and control.

- Simulation models for ACS sensors and actuators can be included in the overall simulation architecture which will increase the overall fidelity of the simulation and further validate the capabilities of the controller.

- Presently the simulator is setup with one kind of appendage as far as shape (i.e. rectangular), boundary condition (i.e. cantilevered), and mode shape functions (i.e. eigenfunctions) are concerned. However, the dynamics formulation is already generic and can handle any appendage shapes (e.g. circular, elliptical, etc.) with different boundary conditions (e.g. pinned) and any feasible mode shape functions (e.g. polynomials). This addition will require the evaluation of integrals of appropriate mode shapes either analytically, as is done for the rectangular case, or numerically.

- Currently, the 3D animation tool can display up to two appendages in two separate windows with the attitude of the spacecraft being displayed in a third window. However, a future improvement could incorporate all displays into one, and to allow the animation of more appendages than just two.

Future Work in Control:

- The proof of the zero dynamics of the flexible system can be extended to a generic $m$ dimensional system using Corollary 6.2. Preliminary work in this area has been done and indicates that the zero dynamics of the general $m$ dimensional system is indeed asymptotically stable.

- The dynamics of the actuators (e.g. on-off dynamics of the thrusters) can be taken into account in order to investigate their effects on the system closed-loop response and robustness.

- In this thesis, we made the assumption that all state variables were available for use by the controller, however for many spacecrafts, the states will not be directly measured and an observer must be included to estimate the missing states. We can use a nonlinear observer to accomplish this task.

- Finally, with the addition of the observers, one would need to redo the proof for the closed-loop stability of the overall system including the observers.

In conclusion, we note that at the time of writing of this thesis, part of the work performed has already been published in three conference papers [149]-[151], one journal paper has been submitted for publication [152] and two other journal papers are being prepared for submission [153]-[154].

# Appendix A

## Derivation of the Attitude and Flexible Dynamics of a Star Topology Spacecraft

## A.1 System Configuration and Frames of Reference

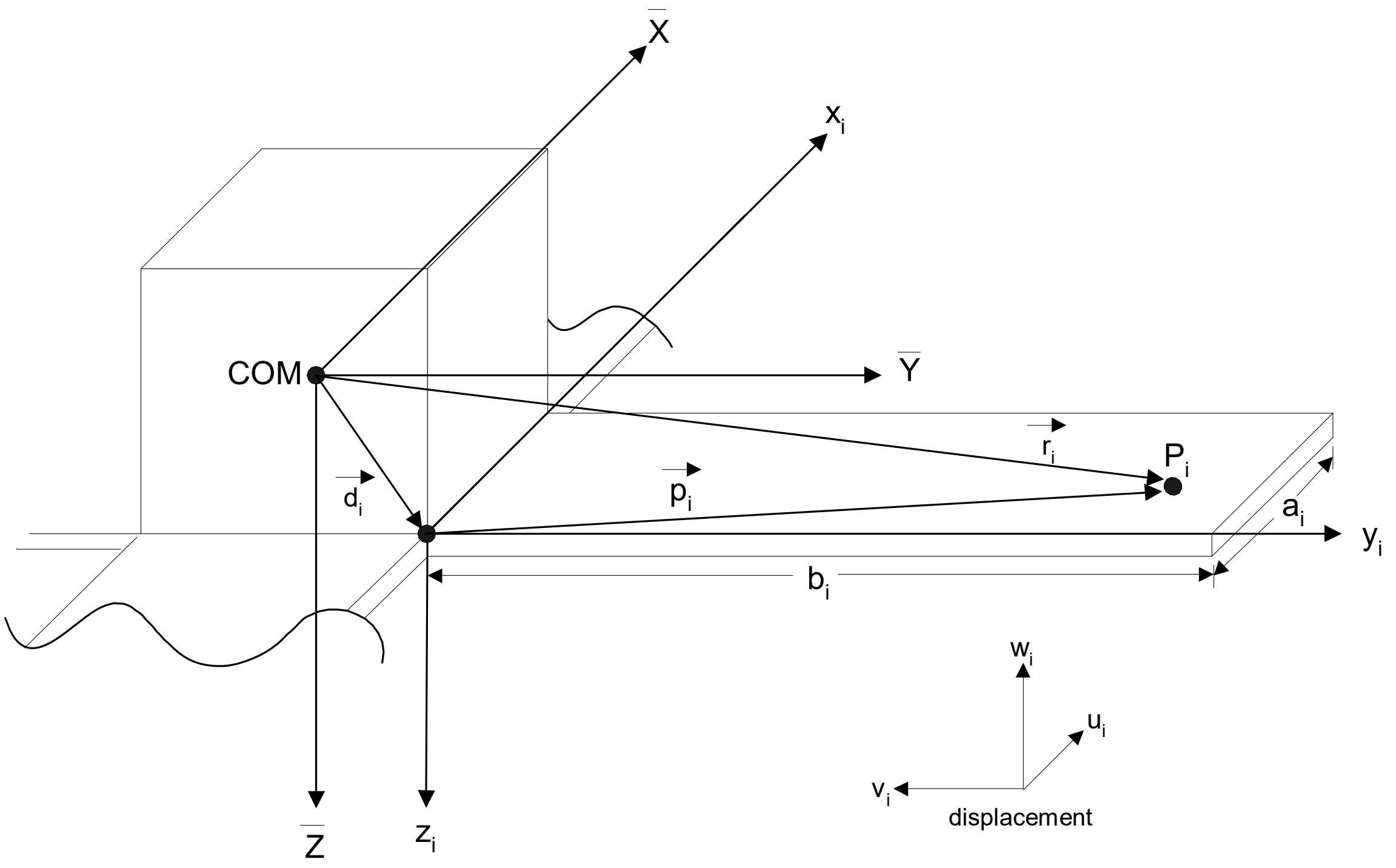

*Notations:*

| | |
|---|---|
| $\vec{v}_c$ | Velocity vector of the Center Of Mass (COM) |
| $\vec{\omega}_c$ | Angular body rate about the center of mass |
| $(\vec{r})_g$ or $(\underline{M})_g$ | Vector $\vec{r}$ or matrix $\underline{M}$ expressed in celestial inertial reference frame |
| $(\vec{r})_c$ or $(\underline{M})_c$ | Vector $\vec{r}$ or matrix $\underline{M}$ expressed in spacecraft/body reference frame |
| $(\vec{r})_p$ or $(\underline{M})_p$ | Vector $\vec{r}$ or matrix $\underline{M}$ expressed in appendage reference frame |

The components of a vector $\vec{r}$ are expressed as $r_x$, $r_y$ and $r_z$.

The rotation matrices that transform vectors and matrices from appendage reference frame to Spacecraft reference frame are denoted by ${}_c(R_i)_p$ for the $i^{th}$ appendage. The rotation matrices that transform vectors and matrices from Spacecraft reference frame to appendage reference frame are denoted by ${}_p(R_i)_c$ for the $i^{th}$ appendage.

The two rotation matrices above are related as follows: ${}_p(R_i)_c = {}_c(R_i)_p^{-1}$.

## A.2 Kinetic Energy

The *system* under consideration is a spacecraft composed of a rigid bus with *n* Flexible Plate-type Appendages (FPAs) and is assumed to be in a fully deployed configuration. The system's total kinetic energy is the summation of the kinetic energies associated with

- the rigid bus rotation ($T_c$);
- the appendage rotation and vibration ($T_{\text{FPA}}$);
- the orbital motion of the system's center of mass ($T_o$); and
- the internal momentum/reaction wheels' rotation ($T_w$), if any.

That is,

$$\begin{aligned} T_{\text{total}} &= T_{(\text{attitude+vibration})} + T_o + T_w \\ T_{\text{total}} &= T_c + T_{\text{FPA}} + T_o + T_w \end{aligned} \tag{A1}$$

From (A1) and $T_{\text{FPA}} = \sum_{i=1}^{n} T_i$ :

$$T_{\text{total}} = T_w + T_o + T_c + \sum_{i=1}^{n} T_i \tag{A2}$$

where $n$ is the number of appendages and the subscript $i$ designates the $i^{th}$ appendage.

The orbital kinetic energy is given by:

$$T_o = \frac{1}{2} m_t \left( \vec{v}_c^T \vec{v}_c \right) \tag{A3}$$

where $\vec{v}_c$ is the orbital velocity of the system's center of mass and $m_t$ is the total mass of the system defined as:

$$m_t = \sum_{i=1}^{n} m_i + m_c \tag{A4}$$

where $m_c$ is the rigid bus mass and $m_i$ is the mass of the $i^{\text{th}}$ appendage.

The bus rotational kinetic energy is given by (A5) where $\underline{I}_c$ is the bus inertia matrix:

$$T_c = \frac{1}{2}\vec{\omega}_c^T \underline{I}_c \vec{\omega}_c \tag{A5}$$

The rotational plus vibrational kinetic energies of the appendage are obtained by integrating the square of the velocity of a point on the appendage over the entire area, $A_i$, of the appendage, that is:

$$T_i = \frac{1}{2}\int_{A_i} \rho_i \vec{v}_i^T \vec{v}_i \mathrm{d}A_i \tag{A6}$$

where $\rho_i$ is the area density of the $i^{th}$ appendage given by:

$$\rho_i = \frac{m_i}{A_i} \tag{A7}$$

For a rectangular appendage, (A6) becomes:

$$T_i = \frac{1}{2}\int_0^{a_i}\int_0^{b_i} \rho_i \vec{v}_i^T \vec{v}_i \mathrm{d}y_i \mathrm{d}x_i$$

where $a_i$ and $b_i$ are the width and length of the $i^{th}$ appendage, respectively.

The velocity vector of point $P_i$ (on the $i^{th}$ appendage) is expressed as (in celestial reference frame):

$$\left(\vec{v}_i\right)_g = \left(\vec{v}_c\right)_g + \left(\dot{\vec{r}}_i\right)_g \tag{A8}$$

or as (in appendage reference frame):

$$\left(\vec{v}_i\right)_p = \left(\vec{v}_c\right)_p + \left(\dot{\vec{r}}_i\right)_p + \left(\vec{\omega}_i\right)_p \times \left(\vec{r}_i\right)_p \tag{A9}$$

where $\vec{\omega}_i$ is the angular velocity of the $i^{\text{th}}$ appendage which is the same as the rigid bus $\vec{\omega}_c$ expressed in the appendage reference frame (i.e. $\vec{\omega}_i = {}_p(R_i)_c\, \vec{\omega}_c$). Also note that the orbital velocity term, ($\vec{v}_c$), of point $P_i$ velocity, ($\vec{v}_i$), in equation (A9) will be ignored since the contribution of $\vec{v}_c$ has already been accounted for in the system's orbital kinetic energy in (A3).

The position vector of the point $P_i$ on the appendage, $\vec{r}_i$, is given by:

$$\left(\vec{r}_i\right)_p = \left(\vec{d}_i\right)_p + \left(\vec{p}_i\right)_p \tag{A10}$$

Taking into account the displacement of the point $P_i$ on the appendage (displacement in the $u$ direction is neglected), we have:

$$\left(\vec{r}_i\right)_p = \begin{bmatrix} d_{ix} + x_i \\ d_{iy} + y_i - v_i \\ d_{iz} + w_i \end{bmatrix} \tag{A11}$$

Note that $x_i, y_i \in \mathbb{R}$ and $v_i$ and $w_i$ are functions of $x_i, y_i$ and time $t$.

Expanding $\left(\dot{\vec{r}}_i\right)_p + \left(\vec{\omega}_i\right)_p \times \left(\vec{r}_i\right)_p$ in (A9), we obtain:

$$\left(\dot{\vec{r}}_i\right)_p + \left(\vec{\omega}_i\right)_p \times \left(\vec{r}_i\right)_p = \begin{bmatrix} 0 \\ -\dot{v}_i \\ \dot{w}_i \end{bmatrix} + \begin{vmatrix} \hat{i} & \hat{j} & \hat{k} \\ \omega_{ix} & \omega_{iy} & \omega_{iz} \\ \left(d_{ix}+x_i\right) & \left(d_{iy}+y_i-v_i\right) & \left(d_{iz}+w_i\right) \end{vmatrix} \tag{A12}$$

Simplifying further gives

$$\left(\dot{\vec{r}}_i\right)_p + \left(\vec{\omega}_i\right)_p \times \left(\vec{r}_i\right)_p = \begin{bmatrix} \omega_{iy}\left(d_{iz}+w_i\right) - \omega_{iz}\left(d_{iy}+y_i-v_i\right) \\ \omega_{iz}\left(d_{ix}+x_i\right) - \omega_{ix}\left(d_{iz}+w_i\right) - \dot{v}_i \\ \omega_{ix}\left(d_{iy}+y_i-v_i\right) - \omega_{iy}\left(d_{ix}+y_i\right) + \dot{w}_i \end{bmatrix} \tag{A13}$$

The square of the velocity is given by equations (A14)-(A17):

$$\left[\left(\dot{\vec{r}}_i\right)_p + \left(\vec{\omega}_i\right)_p \times \left(\vec{r}_i\right)_p\right]^T \left[\left(\dot{\vec{r}}_i\right)_p + \left(\vec{\omega}_i\right)_p \times \left(\vec{r}_i\right)_p\right] = A + B + C \tag{A14}$$

$$A = \omega_{iy}^2\left(d_{iz}+w_i\right)^2 + \omega_{iz}^2\left(d_{iy}+y_i-v_i\right)^2 - 2\omega_{iy}\omega_{iz}\left(d_{iz}+w_i\right)\left(d_{iy}+y_i-v_i\right) \tag{A15}$$

$$\begin{aligned} B = {} & \omega_{ix}^2\left(d_{iy}+w_i\right)^2 + \omega_{iz}^2\left(d_{ix}+x_i\right)^2 + \dot{v}_i^2 - 2\omega_{ix}\omega_{iz}\left(d_{iz}+w_i\right)\left(d_{ix}+x_i\right) \\ & - 2\omega_{iz}\left(d_{ix}+x_i\right)\dot{v}_i + 2\omega_{ix}\left(d_{iz}+w_i\right)\dot{v}_i \end{aligned} \tag{A16}$$

$$\begin{aligned} C = {} & \omega_{ix}^2\left(d_{iy}+y_i-v_i\right)^2 + \omega_{iy}^2\left(d_{ix}+x_i\right)^2 + \dot{w}_i^2 - 2\omega_{ix}\omega_{iy}\left(d_{iy}+y_i-v_i\right)\left(d_{ix}+x_i\right) \\ & + 2\omega_{iz}\left(d_{i2}+y_i-v_i\right)\dot{w}_i - 2\omega_{iy}\left(d_{ix}+x_i\right)\dot{w}_i \end{aligned} \tag{A17}$$

By rearranging the terms in (A14), we obtain:

$$\left[\left(\dot{\vec{r}}_i\right)_p+\left(\vec{\omega}_i\right)_p\times\left(\vec{r}_i\right)_p\right]^T\left[\left(\dot{\vec{r}}_i\right)_p+\left(\vec{\omega}_i\right)_p\times\left(\vec{r}_i\right)_p\right]=\omega_{ix}^2\left[\left(d_{iz}+w_i\right)^2+\left(d_{iy}+y_i-v_i\right)^2\right]$$
$$+\omega_{iy}^2\left[\left(d_{iz}+w_i\right)^2+\left(d_{ix}+x_i\right)^2\right]$$
$$+\omega_{iz}^2\left[\left(d_{ix}+x_i\right)^2+\left(d_{iy}+y_i-v_i\right)^2\right]$$
$$+\omega_{ix}2\left[\left(d_{iz}+w_i\right)\dot{v}_i+\left(d_{iy}+y_i-v_i\right)\dot{w}_i\right]$$
$$+\omega_{iy}\left[-2\left(d_{ix}+x_i\right)\dot{w}_i\right] \tag{A18}$$
$$+\omega_{iz}\left[-2\left(d_{ix}+x_i\right)\dot{v}_i\right]$$
$$+\omega_{ix}\omega_{iy}\left[-2\left(d_{iy}+y_i-v_i\right)\left(d_{ix}+x_i\right)\right]$$
$$+\omega_{ix}\omega_{iz}\left[-2\left(d_{iz}+w_i\right)\left(d_{ix}+x_i\right)\right]$$
$$+\omega_{iy}\omega_{iz}\left[-2\left(d_{iz}+w_i\right)\left(d_{iy}+y_i-v_i\right)\right]$$
$$+\dot{w}^2+\dot{v}^2$$

We now rewrite (A18) in matrix form where the square of the time derivative of the $v$ displacement, $\dot{v}^2$, is neglected:

$$\left[\left(\dot{\vec{r}}_i\right)_p+\left(\vec{\omega}_i\right)_p\times\left(\vec{r}_i\right)_p\right]^T\left[\left(\dot{\vec{r}}_i\right)_p+\left(\vec{\omega}_i\right)_p\times\left(\vec{r}_i\right)_p\right]=$$
$$\begin{bmatrix}\omega_{ix} & \omega_{iy} & \omega_{iz}\end{bmatrix}\begin{bmatrix}a_1 & \frac{a_7}{2} & \frac{a_8}{2}\\ \frac{a_7}{2} & a_2 & \frac{a_9}{2}\\ \frac{a_8}{2} & \frac{a_9}{2} & a_3\end{bmatrix}\begin{bmatrix}\omega_{ix}\\ \omega_{iy}\\ \omega_{iz}\end{bmatrix}+\begin{bmatrix}a_4 & a_5 & a_6\end{bmatrix}\begin{bmatrix}\omega_{ix}\\ \omega_{iy}\\ \omega_{iz}\end{bmatrix}+\dot{w}^2 \tag{A19}$$

In (A19) the values of $a_k$ , $k=1,2,\ldots,9$ , are given by:

$$
\begin{aligned}
a_1 &= \left[ \left( d_{iz} + w_i \right)^2 + \left( d_{iy} + y_i - v_i \right)^2 \right] \\
a_2 &= \left[ \left( d_{iz} + w_i \right)^2 + \left( d_{ix} + x_i \right)^2 \right] \\
a_3 &= \left[ \left( d_{ix} + x_i \right)^2 + \left( d_{iy} + y_i - v_i \right)^2 \right] \\
a_4 &= 2\left[ \left( d_{iz} + w_i \right) \dot{v}_i + \left( d_{iy} + y_i - v_i \right) \dot{w}_i \right] \\
a_5 &= \left[ -2\left( d_{ix} + x_i \right) \dot{w}_i \right] \\
a_6 &= \left[ -2\left( d_{ix} + x_i \right) \dot{v}_i \right] \\
a_7 &= \left[ -2\left( d_{iy} + y_i - v_i \right)\left( d_{ix} + x_i \right) \right] \\
a_8 &= \left[ -2\left( d_{iz} + w_i \right)\left( d_{ix} + x_i \right) \right] \\
a_9 &= \left[ -2\left( d_{iz} + w_i \right)\left( d_{iy} + y_i - v_i \right) \right]
\end{aligned}
\qquad \text{(A20)}
$$

We can now rewrite (A19) in a condensed form:

$$
\left[ \left( \dot{\vec{r}}_i \right)_p + \left( \vec{\omega}_i \right)_p \times \left( \vec{r}_i \right)_p \right]^T \left[ \left( \dot{\vec{r}}_i \right)_p + \left( \vec{\omega}_i \right)_p \times \left( \vec{r}_i \right)_p \right] = \vec{\omega}_i^T \underline{I}_{i_f} \vec{\omega}_i + \vec{g}_{i_f}^T \vec{\omega}_i + \dot{w}_i^2 \qquad \text{(A21)}
$$

where

$$
\underline{I}_{i_f} = \frac{1}{2} \begin{bmatrix} 2a_1 & a_7 & a_8 \\ a_7 & 2a_2 & a_9 \\ a_8 & a_9 & 2a_3 \end{bmatrix} \qquad \vec{g}_{i_f}^T = \left[ a_4 \quad a_5 \quad a_6 \right] \qquad \text{(A22)}
$$

Using (A2), the attitude and vibrational kinetic energies are combined to form a new term:

$$
T_{\omega\chi} = T_c + \sum_{i=1}^{n} T_i \qquad \text{(A23)}
$$

Using (A5), (A6) and (A9) in (A23), we get

$$T_{\omega\chi} = \frac{1}{2}\vec{\omega}_c^T \underline{I}_c \vec{\omega}_c + \frac{1}{2}\sum_{i=1}^{n}\int_{A_i} \rho_i \left[\left(\dot{\vec{r}}_i\right)_p + \left(\vec{\omega}_i\right)_p \times \left(\vec{r}_i\right)_p\right]^T \left[\left(\dot{\vec{r}}_i\right)_p + \left(\vec{\omega}_i\right)_p \times \left(\vec{r}_i\right)_p\right] \mathrm{d}A_i \qquad \text{(A24)}$$

and substituting (A21) in (A24) gives

$$T_{\omega\chi} = \frac{1}{2}\vec{\omega}_c^T \underline{I}_c \vec{\omega}_c + \frac{1}{2}\sum_{i=1}^{n}\rho_i \int_{A_i} \vec{\omega}_i^T \underline{I}_{i_f} \vec{\omega}_i \mathrm{d}A_i + \frac{1}{2}\sum_{i=1}^{n}\rho_i \int_{A_i} \vec{g}_{i_f}^T \vec{\omega}_i \mathrm{d}A_i + \frac{1}{2}\sum_{i=1}^{n}\rho_i \int_{A_i} \dot{w}_i^2 \mathrm{d}A_i \qquad \text{(A25)}$$

By rearranging the terms in (A25), one obtains:

$$T_{\omega\chi} = \frac{1}{2}\vec{\omega}_c^T \underline{I}_c \vec{\omega}_c + \frac{1}{2}\vec{\omega}_i^T \sum_{i=1}^{n}\left[\int_{A_i} \rho_i \underline{I}_{i_f} \mathrm{d}A_i\right]\vec{\omega}_i + \frac{1}{2}\sum_{i=1}^{n}\left[\int_{A_i} \rho_i \vec{g}_{i_f}^T \mathrm{d}A_i\right]\vec{\omega}_i + \frac{1}{2}\sum_{i=1}^{n}\int_{A_i} \rho_i \dot{w}_i^2 \mathrm{d}A_i \qquad \text{(A26)}$$

Knowing the following formulas for derivatives of expressions involving vectors and matrices:

| $y$ (scalar or vector)<br>$\vec{\mathbf{x}}$ = vector<br>$\underline{\mathbf{A}}$ = matrix | $\frac{\partial y}{\partial \vec{\mathbf{x}}}$ |
|---|---|
| $\underline{\mathbf{A}}\vec{\mathbf{x}}$ | $\underline{\mathbf{A}}^T$ |
| $\vec{\mathbf{x}}^T \underline{\mathbf{A}}$ | $\underline{\mathbf{A}}$ |
| $\vec{\mathbf{x}}^T \vec{\mathbf{x}}$ | $2\vec{\mathbf{x}}$ |
| $\vec{\mathbf{x}}^T \underline{\mathbf{A}}\vec{\mathbf{x}}$ | $\underline{\mathbf{A}}\vec{\mathbf{x}} + \underline{\mathbf{A}}^T \vec{\mathbf{x}}$ |

the partial derivative of kinetic energy $T_{\omega\chi}$ with respect to $\vec{\omega}$ is derived below and will be used in the next section to formulate the attitude dynamics equation:

$$\frac{\partial T_{\omega\chi}}{\partial \vec{\omega}} = \frac{1}{2}\left(\underline{I}_c + \underline{I}_c^T\right)\vec{\omega}_c + \frac{1}{2}\sum_{i=1}^{n}\left[\int_{A_i} \rho_i \underline{I}_{i_f} \mathrm{d}A_i + \int_{A_i} \rho_i \underline{I}_{i_f}^T \mathrm{d}A_i\right]\vec{\omega}_i + \frac{1}{2}\sum_{i=1}^{n}\left[\int_{A_i} \rho_i \vec{g}_{i_f}^T \mathrm{d}A_i\right] \qquad \text{(A27)}$$

which can be simplified to:

$$\frac{\partial T_{\omega\chi}}{\partial \vec{\omega}} = \underline{I}_c \vec{\omega}_c + \sum_{i=1}^{n} \underline{I}_i \vec{\omega}_i + \sum_{i=1}^{n} \vec{\kappa}_i \qquad \text{(A28)}$$

where $\underline{I}_i$ is the inertia tensor of the appendage expressed in appendage reference frame:

$$\underline{I}_i = \int_{A_i} \rho_i \underline{I}_{i_f} \, \mathrm{d}A_i \qquad \text{(A29)}$$

and $\vec{\kappa}_i$ the angular momentum due to appendage flexibility expressed in appendage reference frame:

$$\vec{\kappa}_i = \frac{1}{2} \int_{A_i} \rho_i \vec{g}_{i_f}^{T} \, \mathrm{d}A_i \qquad \text{(A30)}$$

## A.3 Attitude Dynamics Equation

We start with the attitude dynamic equation expressed in the spacecraft reference frame:

$$\frac{\mathrm{d}\vec{l}}{\mathrm{d}t} + \vec{\omega} \times \vec{l} = \vec{\tau} \qquad \text{(A31)}$$

where $\vec{\tau}$ is the external torque and $\vec{l}$ is the total angular momentum of the system.

Considering our specific case:

$$\vec{l} = \frac{\partial T_{\omega\chi}}{\partial \vec{\omega}} + \vec{h}_w \qquad \text{(A32)}$$

where $\vec{h}_w$ is the internal angular momentum due to rotating mechanical wheels, if any (i.e. momentum wheels, reaction wheels or control moment gyros).

The attitude dynamics equation then becomes:

$$\frac{\mathrm{d}}{\mathrm{d}t}\left(\frac{\partial T_{\omega\chi}}{\partial\vec{\omega}}+\vec{h}_w\right)+\vec{\omega}\times\left(\frac{\partial T_{\omega\chi}}{\partial\vec{\omega}}+\vec{h}_w\right)=\vec{\tau} \tag{A33}$$

or

$$\frac{\mathrm{d}}{\mathrm{d}t}\left(\frac{\partial T_{\omega\chi}}{\partial\vec{\omega}}+\vec{h}_w\right)+\underline{\Omega}\left(\frac{\partial T_{\omega\chi}}{\partial\vec{\omega}}+\vec{h}_w\right)=\vec{\tau} \tag{A34}$$

with

$$\underline{\Omega}=\begin{bmatrix}0 & -\omega_z & \omega_y\\ \omega_z & 0 & -\omega_x\\ -\omega_y & \omega_x & 0\end{bmatrix} \tag{A35}$$

By considering the complete system with all the appendages included in (A34), we get:

$$\frac{\mathrm{d}}{\mathrm{d}t}\left(\frac{\partial T_{\omega\chi}}{\partial\vec{\omega}}+\vec{h}_w\right)=\underline{I}_c\dot{\vec{\omega}}+\frac{1}{2}\sum_{i=1}^{n}\left(\underline{I}_i+\underline{I}_i^T\right)\dot{\vec{\omega}}+\frac{1}{2}\sum_{i=1}^{n}\left(\dot{\underline{I}}_i+\dot{\underline{I}}_i^T\right)\vec{\omega}+\sum_{i=1}^{n}\dot{\vec{\kappa}}_i+\dot{\vec{h}}_w \tag{A36}$$

We can now define the system inertia matrix, derivative of system inertia matrix, angular momentum due to all appendage flexibility and derivative of angular momentum due to all appendage flexibility (expressed in spacecraft reference frame), respectively as:

$$\underline{I}_t=\underline{I}_c+\sum_{i=1}^{n}(\underline{I}_i)_c$$
$$(\underline{I}_i)_c={}_c(R_i)_p^T\,\underline{I}_i\,{}_c(R_i)_p$$

$$\dot{\underline{I}}_t = \sum_{i=1}^{n} (\dot{\underline{I}}_i)_c$$

$$(\dot{\underline{I}}_i)_c = {}_c(R_i)_p^T \, \dot{\underline{I}}_i \, {}_c(R_i)_p$$

$$\vec{\kappa}_t = \sum_{i=1}^{n} (\vec{\kappa}_i)_c \qquad \text{(A37)}$$

$$(\vec{\kappa}_i)_c = {}_c(R_i)_p \, \vec{\kappa}_i$$

$$\dot{\vec{\kappa}}_t = \sum_{i=1}^{n} (\dot{\vec{\kappa}}_i)_c$$

$$(\dot{\vec{\kappa}}_i)_c = {}_c(R_i)_p \, \dot{\vec{\kappa}}_i$$

and therefore (A36) can be re-written as:

$$\frac{\mathrm{d}}{\mathrm{d}t}\left( \frac{\partial T_{\omega\chi}}{\partial \vec{\omega}} + \vec{h}_w \right) = \underline{I}_t \dot{\vec{\omega}} + \dot{\underline{I}}_t \vec{\omega} + \dot{\vec{\kappa}}_t + \dot{\vec{h}}_w \qquad \text{(A38)}$$

Substituting (A38) and (A28) in (A34), we obtain the final form of the attitude dynamics equation in spacecraft reference frame:

$$\underline{I}_t \dot{\vec{\omega}} + \dot{\underline{I}}_t \vec{\omega} + \underline{\Omega}\left( \underline{I}_t \vec{\omega} + \vec{\kappa}_t + \vec{h}_w \right) + \dot{\vec{\kappa}}_t = \vec{\tau} \qquad \text{(A39)}$$

Note that $\dot{\vec{h}}_w$ in (A38) represents the torque generated by the wheels which can be included in $\vec{\tau}$ and hence does not appear explicitly in (A39).

## A.4 Appendage Discretization

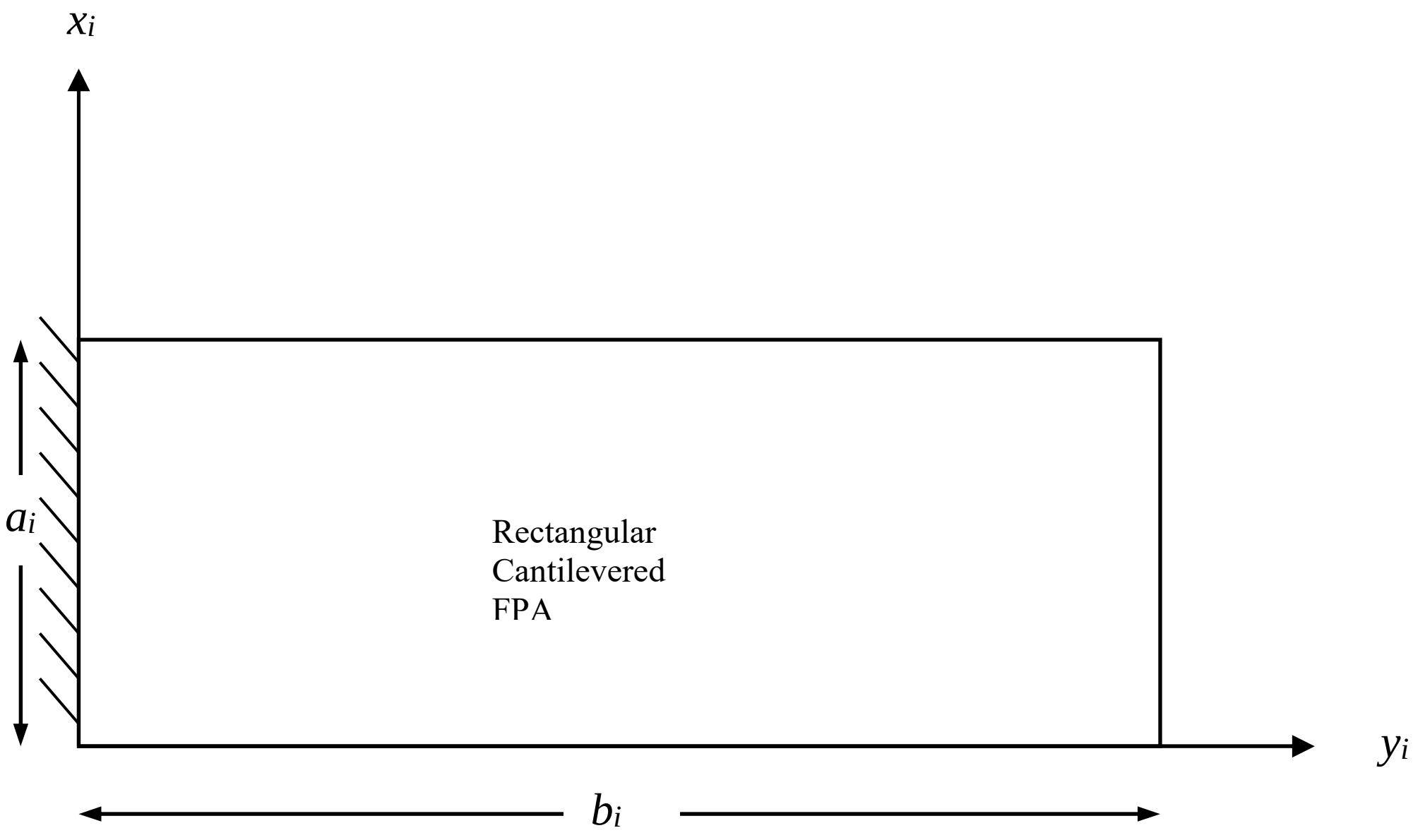

$p$ mode in $x_i$ direction

$q$ mode in $y_i$ direction

The Raleigh-Ritz method is used to discretize the continuous appendage. Displacement in the $w$ direction (perpendicular to the page) for the $i^{th}$ appendage is given by:

$$w_i = w_i\left(x_i, y_i, t\right) = \sum_{r=1}^{p}\sum_{s=1}^{q} \chi_{i_{rs}}\left(t\right)\phi_r\left(x_i\right)\ \psi_s\left(y_i\right) \tag{A40}$$

where $\chi_{i_{rs}}$ is a non-dimensional vector and $\phi_r$ and $\psi_s$ are vectors of orthogonal admissible functions (mode shapes) satisfying all geometric boundary conditions. Using a vector notation, equation (A40) becomes:

$$w_i\left(x_i, y_i, t\right) = \vec{\phi}_r^T\ \underline{\chi}_{i_{rs}}\ \vec{\psi}_s = \vec{C}_{\phi\psi}^T\ \vec{\chi}_{i_{rs}} \tag{A41}$$

where the matrix $\underline{\chi}_{i_{rs}}$ has been changed into a vector form. For clarity the subscript $i$, identifying the $i^{th}$ appendage, will be dropped in the following equations:

$$\vec{C}_{\phi\psi}^{T} = \left[ \phi_1\psi_1 \quad \phi_1\psi_2 \quad \cdots \quad \phi_1\psi_q \mid \phi_2\psi_1 \quad \phi_2\psi_2 \quad \cdots \quad \phi_2\psi_q \mid \cdots \mid \phi_p\psi_1 \quad \phi_p\psi_2 \quad \cdots \quad \phi_p\psi_q \right] \quad \text{(A42)}$$

$$\vec{\chi}^{T} = \left[ \chi_{11} \quad \chi_{12} \quad \cdots \quad \chi_{1q} \mid \chi_{21} \quad \chi_{22} \quad \cdots \quad \chi_{2q} \mid \cdots \mid \chi_{p1} \quad \chi_{p2} \quad \cdots \quad \chi_{pq} \right] \quad \text{(A43)}$$

Note that $\vec{C}_{\phi\psi}$ is a function of $x$ and $y$ (spatial function), and $\vec{\chi}$ is only a function of time $t$ (temporal function).

The boundary conditions for the cantilevered appendage are: clamped for *(x,0)*, and free for *(x,b)*, *(0,y)* and *(a,y)*. The mode shape functions for the free-free and clamped-free beams, $\phi_r(x)$ and $\psi_s(y)$ are given by:

$$\phi_1(x) = \frac{1}{\sqrt{a}}$$

$$\phi_2(x) = \sqrt{\frac{12}{a^3}}\left(x - \frac{a}{2}\right) \qquad \text{(free-free)} \qquad \text{(A44)}$$

for $r = 3, 4, 5, \ldots, p$

$$\phi_r(x) = \frac{1}{\sqrt{a}}\left\{\left[\cosh\left(\lambda_r \frac{x}{a}\right) + \cos\left(\lambda_r \frac{x}{a}\right)\right] - \sigma_r\left[\sinh\left(\lambda_r \frac{x}{a}\right) + \sin\left(\lambda_r \frac{x}{a}\right)\right]\right\}$$

for $s = 1, 2, 3, \ldots, q$ (clamped-free) (A45)

$$\psi_s(y) = \frac{1}{\sqrt{b}}\left\{\left[\cosh\left(\lambda_s \frac{y}{b}\right) - \cos\left(\lambda_s \frac{y}{b}\right)\right] - \sigma_s\left[\sinh\left(\lambda_s \frac{y}{b}\right) - \sin\left(\lambda_s \frac{y}{b}\right)\right]\right\}$$

where $\sigma_r$ and $\sigma_s$ are given by

$$\sigma_r = \frac{\cosh(\lambda_r) - \cos(\lambda_r)}{\sinh(\lambda_r) - \sin(\lambda_r)} \qquad \text{(A46)}$$

$$\sigma_s = \frac{\sinh(\lambda_s) - \sin(\lambda_s)}{\cosh(\lambda_s) + \cos(\lambda_s)} \qquad \text{(A47)}$$

and $\lambda_r$ and $\lambda_s$ are the roots of

$$1 - \cosh(\lambda_r)\cos(\lambda_r) = 0 \qquad \text{(A48)}$$

$$1 + \cosh(\lambda_s)\cos(\lambda_s) = 0 \qquad \text{(A49)}$$

Note that $\phi_1$ and $\phi_2$ represent the rigid modes of a free-free beam for translation and rotation, respectively.

Some of the terms that will be used throughout the formulation are provided next:

$$w_i = w_i(x_i, y_i, t) = \vec{C}_{\phi\psi}^T \ \vec{\chi}_i \qquad \text{(A50)}$$

$$w_i^2 = w_i^T w_i \qquad \text{(A51)}$$

$$w_i^2 = \vec{\chi}_i^T \vec{C}_{\phi\psi} \vec{C}_{\phi\psi}^T \vec{\chi}_i \qquad \text{(A52)}$$

$$\frac{\partial w_i}{\partial \vec{\chi}_i} = \vec{C}_{\phi\psi} \qquad \text{(A53)}$$

$$\frac{\partial w_i^2}{\partial \vec{\chi}_i} = \vec{C}_{\phi\psi} \vec{C}_{\phi\psi}^T \vec{\chi} + \left(\vec{C}_{\phi\psi} \vec{C}_{\phi\psi}^T\right)^T \vec{\chi} = 2\vec{C}_{\phi\psi} \vec{C}_{\phi\psi}^T \vec{\chi} \qquad \text{(A54)}$$

$$\frac{\partial w_\mathrm{i}}{\partial x_\mathrm{i}} = \vec{C}_{\phi'\psi}^T \vec{\chi}_i \qquad \text{(A55)}$$

$$\frac{\partial^2 w_i}{\partial x_i^2} = \vec{C}_{\phi''\psi}^{T} \vec{\chi}_i \tag{A56}$$

$$\frac{\partial w_i}{\partial y_i} = \vec{C}_{\phi\psi'}^{T} \vec{\chi}_i \tag{A57}$$

$$\frac{\partial^2 w_i}{\partial y_i^2} = \vec{C}_{\phi\psi''}^{T} \vec{\chi}_i \tag{A58}$$

$$\frac{\partial^2 w_i}{\partial x_i \, \partial y_i} = \vec{C}_{\phi'\psi'}^{T} \vec{\chi}_i \tag{A59}$$

$$\dot{w}_i = \vec{C}_{\phi\psi}^{T} \dot{\vec{\chi}}_i \tag{A60}$$

$$\dot{w}_i^2 = \left(\vec{C}_{\phi\psi}^{T} \dot{\vec{\chi}}_i\right)^T \left(\vec{C}_{\phi\psi}^{T} \dot{\vec{\chi}}_i\right) = \left(\dot{\vec{\chi}}_i^T \vec{C}_{\phi\psi}\right)\left(\vec{C}_{\phi\psi}^{T} \dot{\vec{\chi}}_i\right) = \dot{\vec{\chi}}_i^T \vec{C}_{\phi\psi} \vec{C}_{\phi\psi}^{T} \dot{\vec{\chi}}_i \tag{A61}$$

The displacement in the $v$ direction can be found to be (see Section A.9 in this appendix for the derivation)

$$v_i = \frac{1}{2}\int_0^y \left(\frac{\partial \; w_i(x,\eta,t)}{\partial \eta}\right)^2 \, \mathrm{d}\eta \tag{A62}$$

Note that a change of variable (i.e. $y$ has been replaced by $\eta$) has taken place in the $w_i$ term, to avoid confusions with terms involving both $v_i$ and d$y$ integrals (e.g. appendage inertia terms in Section A.5). We can now rewrite (A62) using the vector notation:

$$v = \frac{1}{2}\int_0^y \left(\vec{C}_{\phi(x)\psi'(\eta)}^{T} \vec{\chi}_i\right)^2 \, \mathrm{d}\eta$$

$$v = \frac{1}{2}\int_0^y \left( \vec{\chi}_i^T \vec{C}_{\phi\psi'} \vec{C}_{\phi\psi'}^T \vec{\chi}_i \right) \mathrm{d}\eta \tag{A63}$$

The time derivative of (A63) is given by:

$$\dot{v} = \int_0^y \left( \dot{\vec{\chi}}_i^T \vec{C}_{\phi\psi'} \vec{C}_{\phi\psi'}^T \vec{\chi}_i \right) \mathrm{d}\eta \tag{A64}$$

and finally by:

$$\dot{v} = \dot{\vec{\chi}}_i^T \left[ \int_0^y \vec{C}_{\phi\psi'} \vec{C}_{\phi\psi'}^T \; \mathrm{d}\eta \right] \vec{\chi}_i \tag{A65}$$

Also recall that the last term in the kinetic energy (A26) can now be expressed as

$$\frac{1}{2}\int_{A_i} \rho_i \dot{w}_i^2 \mathrm{d}A_i = \frac{1}{2}\int_{A_i} \rho_i \left( \dot{\vec{\chi}}_i^T \vec{C}_{\phi\psi} \vec{C}_{\phi\psi}^T \dot{\vec{\chi}}_i \right) \mathrm{d}A_i = \frac{1}{2}\rho_i \; \dot{\vec{\chi}}_i^T \left[ \int_{A_i} \left( \vec{C}_{\phi\psi} \vec{C}_{\phi\psi}^T \right) \mathrm{d}A_i \right] \dot{\vec{\chi}}_i \tag{A66}$$

## A.5 $\underline{I}_i$ Inertia Terms

We now choose the rectangular geometrical shape for the appendage and expand the appendage's inertia matrix $\underline{I}_i$ for this specific case. From equations (A22) and (A29), the first element of the appendage inertia matrix (expressed in appendage reference frame) is:

$$I_{ixx} = \int_{A_i} \rho_i a_1 \mathrm{d}A_i \tag{A67}$$

Substituting the first term of (A20) in (A67) gives:

$$I_{ixx} = \rho_i \int_0^{a_i}\int_0^{b_i} \left[ \left(d_{iz} + w_i\right)^2 + \left(d_{iy} + y_i - v_i\right)^2 \right] \mathrm{d}y_i \mathrm{d}x_i \tag{A68}$$

and expanding (A68), we get

$$I_{ixx} = \rho_i \int_0^{a_i}\int_0^{b_i} \left( d_{iz}^2 + w_i^2 + 2d_{iz}w_i + d_{iy}^2 + y_i^2 + v_i^2 + 2d_{iy}y_i - 2d_{iy}v_i - 2y_i v_i \right) \mathrm{d}y_i \mathrm{d}x_i \tag{A69}$$

Now by regrouping terms, we obtain

$$\begin{aligned} I_{ixx} = {} & \rho_i \int_0^{a_i}\int_0^{b_i} (d_{iy}^2 + d_{iz}^2) \mathrm{d}y_i \mathrm{d}x_i + \rho_i \int_0^{a_i}\int_0^{b_i} (y_i^2 + 2d_{iy}y_i) \mathrm{d}y_i \mathrm{d}x_i - 2d_{iy}\rho_i \int_0^{a_i}\int_0^{b_i} v_i \mathrm{d}y_i \mathrm{d}x_i \\ & + 2d_{iz}\rho_i \int_0^{a_i}\int_0^{b_i} w_i \mathrm{d}y_i \mathrm{d}x_i + \rho_i \int_0^{a_i}\int_0^{b_i} w_i^2 \mathrm{d}y_i \mathrm{d}x_i - 2\rho_i \int_0^{a_i}\int_0^{b_i} (y_i v_i) \mathrm{d}y_i \mathrm{d}x_i \end{aligned} \tag{A70}$$

and by integrating the first two terms of (A70) and using (A63), (A50) and (A52), we get:

$$I_{ixx} = \rho_i \left\{ \begin{array}{l} a_i b_i \left( d_{iy}^2 + d_{iz}^2 \right) + a_i \left( \dfrac{b_i^3}{3} \right) + a_i b_i^2 d_{iy} - 2 d_{iy} \int_0^{a_i} \int_0^{b_i} \left[ \dfrac{1}{2} \int_0^{y_i} \left[ \vec{\chi}_i^T \vec{C}_{\phi\psi'} \vec{C}_{\phi\psi'}^T \vec{\chi}_i \right] \mathrm{d}y \right] \mathrm{d}y_i \mathrm{d}x_i \\ +2 d_{iz} \int_0^{a_i} \int_0^{b_i} \vec{C}_{\phi\psi}^T \vec{\chi}_i \mathrm{d}y_i \mathrm{d}x_i + \int_0^{a_i} \int_0^{b_i} \vec{\chi}_i^T \vec{C}_{\phi\psi} \vec{C}_{\phi\psi}^T \vec{\chi}_i \mathrm{d}y_i \mathrm{d}x_i \\ -2 \int_0^{a_i} \int_0^{b_i} y_i \left[ \dfrac{1}{2} \int_0^{y_i} \left[ \vec{\chi}_i^T \vec{C}_{\phi\psi'} \vec{C}_{\phi\psi'}^T \vec{\chi}_i \right] \mathrm{d}\eta \right] \mathrm{d}y_i \mathrm{d}x_i \end{array} \right\} \quad \text{(A71)}$$

or

$$I_{ixx} = \rho_i \left\{ \begin{array}{l} \left[ a_i b_i \left( d_{iy}^2 + d_{iz}^2 \right) \right] + a_i \left( \dfrac{b_i^3}{3} \right) + a_i b_i^2 d_{iy} - d_{iy} \vec{\chi}_i^T \left[ \int_0^{a_i} \int_0^{b_i} \int_0^{y_i} \vec{C}_{\phi\psi'} \vec{C}_{\phi\psi'}^T \mathrm{d}\eta \mathrm{d}y_i \mathrm{d}x_i \right] \vec{\chi}_i \\ +2 d_{iz} \left[ \int_0^{a_i} \int_0^{b_i} \vec{C}_{\phi\psi}^T \mathrm{d}y_i \mathrm{d}x_i \right] \vec{\chi}_i + \vec{\chi}_i^T \left[ \int_0^{a_i} \int_0^{b_i} \vec{C}_{\phi\psi} \vec{C}_{\phi\psi}^T \mathrm{d}y_i \mathrm{d}x_i \right] \vec{\chi}_i \\ -\vec{\chi}_i^T \left[ \int_0^{a_i} \int_0^{b_i} \int_0^{y_i} y_i \left[ \vec{C}_{\phi\psi'} \vec{C}_{\phi\psi'}^T \right] \mathrm{d}\eta \mathrm{d}y_i \mathrm{d}x_i \right] \vec{\chi}_i \end{array} \right\} \quad \text{(A72)}$$

Defining the following mode shape integrals:

$$\underline{M}_{\phi\psi'_\eta} = \int_0^{a_i} \int_0^{b_i} \int_0^{y_i} \vec{C}_{\phi\psi'} \vec{C}_{\phi\psi'}^T \mathrm{d}\eta \ \mathrm{d}y_i \mathrm{d}x_i \quad \text{(A73)}$$

$$\vec{m}_{\phi\psi} = \int_0^{a_i} \int_0^{b_i} \vec{C}_{\phi\psi}^T \ \mathrm{d}y_i \mathrm{d}x_i \quad \text{(A74)}$$

$$\underline{M}_{\phi\psi} = \int_0^{a_i} \int_0^{b_i} \vec{C}_{\phi\psi} \vec{C}_{\phi\psi}^T \ \mathrm{d}y_i \mathrm{d}x_i \quad \text{(A75)}$$

$$\underline{M}_{y\phi\psi'_\eta} = \int_0^{a_i} \int_0^{b_i} y_i \int_0^{y_i} \left( \vec{C}_{\phi\psi'} \vec{C}_{\phi\psi'}^T \right) \mathrm{d}\eta \ \mathrm{d}y_i \mathrm{d}x_i \quad \text{(A76)}$$

where $\phi = \phi(x)$, $\psi = \psi(x)$ and $\psi' = \dfrac{\mathrm{d}\psi(\eta)}{\mathrm{d}\eta}$.

and substituting (A73)-(A76) in (A72), the final value of $I_{ixx}$ is obtained to be:

$$I_{ixx} = \rho_i \begin{pmatrix} \left[a_i b_i \left(d_{iy}^2 + d_{iz}^2\right)\right] + a_i \left(\dfrac{b_i^3}{3}\right) + a_i b_i^2 d_{iy} - d_{iy} \vec{\chi}_i^T \underline{M}_{\phi\psi'_\eta} \vec{\chi}_i + 2 d_{iz} \vec{m}_{\phi\psi}^T \vec{\chi}_i \\ + \vec{\chi}_i^T \underline{M}_{\phi\psi} \vec{\chi}_i - \vec{\chi}_i^T \underline{M}_{y\phi\psi'_\eta} \vec{\chi}_i \end{pmatrix} \tag{A77}$$

The same process is repeated for the remaining elements of $\underline{I}_i$ where the subscript $i$ is dropped for simplicity:

$$I_{xy} = I_{yx} = \int_A \rho \frac{a_7}{2} \mathrm{d}A$$

$$I_{xy} = -\rho \int_0^a \int_0^b \left(d_y + y - v\right)\left(d_x + x\right) \mathrm{d}y\mathrm{d}x$$

$$I_{xy} = -\rho \int_0^a \int_0^b \left(d_y d_x + d_y x + d_x y + xy - d_x v - xv\right) \mathrm{d}y\mathrm{d}x$$

$$I_{xy} = -\rho_i \left[ d_y d_x \left(ab\right) + d_y \frac{a^2 b}{2} + d_x \frac{ab^2}{2} + \frac{a^2 b^2}{4} - d_x \int_0^a \int_0^b v \,\mathrm{d}y\mathrm{d}x - \int_0^a \int_0^b xv \,\mathrm{d}y\mathrm{d}x \right]$$

$$I_{xy} = -\rho \left[ d_y d_x \left(ab\right) + d_y \frac{a^2 b}{2} + d_x \frac{ab^2}{2} + \frac{a^2 b^2}{4} - \frac{d_x}{2} \vec{\chi}^T \underline{M}_{\phi\psi'_\eta} \vec{\chi} - \frac{1}{2} \vec{\chi}^T \underline{M}_{x\phi\psi'_\eta} \vec{\chi} \right] \tag{A78}$$

where

$$\underline{M}_{x\phi\psi'_\eta} = \int_0^a \int_0^b x \int_0^y \vec{C}_{\phi\psi'} \vec{C}_{\phi\psi'}^T \,\mathrm{d}\eta\mathrm{d}y\mathrm{d}x \tag{A79}$$

$$I_{xz} = I_{zx} = \int_A \rho \frac{a_8}{2} \mathrm{d}A$$

$$I_{xz} = -\rho \int_0^a \int_0^b \left(d_z + w\right)\left(d_x + x\right)\,\mathrm{d}y\mathrm{d}x$$

$$I_{xz} = -\rho \int_0^a \int_0^b \left(d_z d_x + d_z x + d_x w + xw\right)\,\mathrm{d}y\mathrm{d}x$$

$$I_{xz} = -\rho \left[ d_z d_x \left(ab\right) + d_z \frac{a^2 b}{2} + d_x \int_0^a \int_0^b w\,\mathrm{d}y\mathrm{d}x - \int_0^a \int_0^b xw\,\mathrm{d}y\mathrm{d}x \right]$$

$$I_{xz} = -\rho \left[ d_z d_x \left(ab\right) + d_z \frac{a^2 b}{2} + d_x \vec{m}_{\phi\psi}^T \vec{\chi} + \vec{m}_{x\phi\psi}^T \vec{\chi} \right] \tag{A80}$$

where

$$\vec{m}_{x\phi\psi} = \int_0^a \int_0^b x\vec{C}_{\phi\psi}\,\mathrm{d}y\mathrm{d}x \tag{A81}$$

$$I_{yy} = \int_A \rho\, a_2 \mathrm{d}A$$

$$I_{yy} = \rho \int_0^a \int_0^b \left[ \left(d_z + w\right)^2 + \left(d_x + x\right)^2 \right] \mathrm{d}y\mathrm{d}x$$

$$I_{yy} = \rho \int_0^a \int_0^b \left[ \left(d_z^2 + w^2 + 2d_z w\right) + \left(d_{ix}^2 + x^2 + 2d_x x\right) \right] \mathrm{d}y\mathrm{d}x$$

$$I_{yy} = \rho \left[ \left(d_z^2 + d_x^2\right) ab + d_x b a^2 + b\frac{a^3}{3} + \vec{\chi}^T \underline{M}_{\phi\psi} \vec{\chi} + 2d_z \vec{m}_{\phi\psi}^T \vec{\chi} \right] \tag{A82}$$

$$I_{yz} = I_{zy} = \int_A \rho \frac{a_9}{2} \mathrm{d}A$$

$$I_{yz} = \rho \int_0^a \int_0^b \left[ \left( d_z + w \right) \left( d_y + y - v \right) \right] \mathrm{d}y \mathrm{d}x$$

$$I_{yz} = -\rho \int_0^a \int_0^b \left( d_z d_y + d_z y - d_z v + d_y w + yw - \cancel{vw} \right) \mathrm{d}y \mathrm{d}x \qquad (vw \text{ is negligible})$$

$$I_{yz} = -\rho \left[ \left( d_y d_z \right) ab + d_z \frac{b^2 a}{2} - \frac{d_z}{2} \vec{\chi}^T \underline{M}_{\phi\psi'_\eta} \vec{\chi} + d_y \vec{m}^T_{\phi\psi} \vec{\chi} + \vec{m}^T_{y\phi\psi} \vec{\chi} \right] \qquad \text{(A83)}$$

where

$$\vec{m}_{y\phi\psi} = \int_0^a \int_0^b y \vec{C}_{\phi\psi} \; \mathrm{d}y \mathrm{d}x \qquad \text{(A84)}$$

$$I_{zz} = \int_A \rho \, a_3 \mathrm{d}A$$

$$I_{zz} = \rho \int_0^a \int_0^b \left[ \left( d_x + x \right)^2 + \left( d_y + y - v \right)^2 \right] \mathrm{d}y \mathrm{d}x$$

$$I_{zz} = \rho \int_0^a \int_0^b \left( \left( d_x^2 + 2 d_x x + x^2 \right) + d_y^2 + y^2 + \cancel{v^2} + 2 d_y y - 2 d_y v - 2 yv \right) \mathrm{d}y \mathrm{d}x \qquad (v^2 \text{ is negligible})$$

$$I_{zz} = \rho \left[ \left( d_x^2 + d_y^2 \right) ab + 2 d_x \frac{a^2 b}{2} + 2 d_y \frac{ab^2}{2} + \frac{ab^3}{3} + \frac{a^3 b}{3} - 2 d_y \int_0^a \int_0^b v \; \mathrm{d}y \mathrm{d}x - 2 \int_0^a \int_0^b yv \; \mathrm{d}y \mathrm{d}x \right]$$

$$I_{zz} = \rho \left[ \left( d_x^2 + d_y^2 \right) ab + d_x a^2 b + d_y ab^2 + \frac{ab^3}{3} + \frac{a^3 b}{3} - d_y \vec{\chi}^T \underline{M}_{\phi\psi'_\eta} \vec{\chi} - \vec{\chi}^T \underline{M}_{y\phi\psi'_\eta} \vec{\chi} \right] \qquad \text{(A85)}$$

By differentiating $\underline{I}_i$ terms with respect to time, we obtain the following set of equations:

$$\dot{I}_{xx} = \rho\begin{bmatrix} -d_y \dot{\vec{\chi}}^T \left(\underline{M}_{\phi\psi'_\eta} + \underline{M}^T_{\phi\psi'_\eta}\right)\vec{\chi} + 2d_z \vec{m}^T_{\phi\psi}\dot{\vec{\chi}} + \dot{\vec{\chi}}^T\left(\underline{M}_{\phi\psi} + \underline{M}^T_{\phi\psi}\right)\vec{\chi} \\ -\dot{\vec{\chi}}^T\left(\underline{M}_{y\phi\psi'_\eta} + \underline{M}^T_{y\phi\psi'_\eta}\right)\vec{\chi} \end{bmatrix} \tag{A86}$$

$$\dot{I}_{yy} = \rho\left[\dot{\vec{\chi}}^T\left(\underline{M}_{\phi\psi} + \underline{M}^T_{\phi\psi}\right)\vec{\chi} + 2d_z\vec{m}^T_{\phi\psi}\dot{\vec{\chi}}\right] \tag{A87}$$

$$\dot{I}_{zz} = \rho\left[-d_y\dot{\vec{\chi}}^T\left(\underline{M}_{\phi\psi'_\eta} + \underline{M}^T_{\phi\psi'_\eta}\right)\vec{\chi} - \dot{\vec{\chi}}^T\left(\underline{M}_{y\phi\psi'_\eta} + \underline{M}^T_{y\phi\psi'_\eta}\right)\vec{\chi}\right] \tag{A88}$$

$$\dot{I}_{xy} = \dot{I}_{yx} = \frac{\rho}{2}\left[d_x\dot{\vec{\chi}}^T\left(\underline{M}_{\phi\psi'_\eta} + \underline{M}^T_{\phi\psi'_\eta}\right)\vec{\chi} + \dot{\vec{\chi}}^T\left(\underline{M}_{x\phi\psi'_\eta} + \underline{M}^T_{x\phi\psi'_\eta}\right)\vec{\chi}\right] \tag{A89}$$

$$\dot{I}_{xz} = \dot{I}_{zx} = -\rho\left[d_x\vec{m}^T_{\phi\psi}\dot{\vec{\chi}} + \vec{m}^T_{x\phi\psi}\dot{\vec{\chi}}\right] \tag{A90}$$

$$\dot{I}_{yz} = \dot{I}_{zy} = -\rho\left[-\frac{d_z}{2}\dot{\vec{\chi}}^T\left(\underline{M}_{\phi\psi'_\eta} + \underline{M}^T_{\phi\psi'_\eta}\right)\vec{\chi} + d_y\vec{m}^T_{\phi\psi}\dot{\vec{\chi}} + \vec{m}^T_{y\phi\psi}\dot{\vec{\chi}}\right] \tag{A91}$$

## A.6 Angular Momentum due to Flexibility, $\vec{\kappa}$

We now expand the angular momentum terms due to flexibility, $\vec{\kappa}$, for the $i^{\text{th}}$ appendage expressed in appendage reference frame. Here again, we drop the subscript $i$ for simplicity of presentation. From equations (A22) and (A30), the $x$ component of the angular momentum due to flexibility is:

$$\kappa_x = \frac{1}{2}\rho\int_0^a\int_0^b a_4 \mathrm{d}y\mathrm{d}x \tag{A92}$$

Substituting the fourth term of (A20) in (A92) gives:

$$\kappa_x = \frac{1}{2}\rho\int_0^a\int_0^b 2\left[\left(d_z + w\right)\dot{v} + \left(d_y + y - v\right)\dot{w}\right] \mathrm{d}y\mathrm{d}x \tag{A93}$$

and by expanding (A93), we get:

$$\kappa_x = \frac{1}{2}\rho\int_0^a\int_0^b 2\left[\left(d_z\dot{v} + \cancel{w\dot{v}}\right) + \left(d_y\dot{w} + y\dot{w} - \cancel{v\dot{w}}\right)\right] \mathrm{d}y\mathrm{d}x \quad (w\dot{v} \text{ and } v\dot{w} \text{ are negligible}) \tag{A94}$$

Substituting (A73), (A74) and (A84) in (A94), we obtain the final value of $\kappa_x$ as:

$$\kappa_x = \rho\left[d_z\ \dot{\vec{\chi}}^T \underline{M}_{\phi\psi'_\eta}\vec{\chi} + d_y\vec{m}^T_{\phi\psi}\dot{\vec{\chi}} + \vec{m}^T_{y\phi\psi}\dot{\vec{\chi}}\right] \tag{A95}$$

The same derivation process is repeated for the $y$ and $z$ components of $\vec{\kappa}$:

$$\kappa_y = \frac{1}{2}\rho\int_0^a\int_0^b a_5 \,\mathrm{d}y\mathrm{d}x$$

$$\kappa_y = \frac{1}{2}\rho\int_0^a\int_0^b\left[-2\left(d_x + x\right)\dot{w}\right]\ \mathrm{d}y\mathrm{d}x$$

$$\int_0^{b_i}\int_0^{a_i}\left(\frac{\partial^2 w}{\partial y^2}\right)^2 \mathrm{d}y\mathrm{d}x = \int_0^{a_i}\int_0^{b_i}\left(\vec{C}^T_{\phi\psi''}\vec{\chi}\right)^2\ \mathrm{d}y\mathrm{d}x \tag{A96}$$

$$\kappa_z = \frac{1}{2}\rho\int_0^a\int_0^b a_6\ \mathrm{d}y\mathrm{d}x$$

$$\kappa_z = \rho\int_0^a\int_0^b\left[\left(d_x + x\right)\dot{v}\right]\ \mathrm{d}y\mathrm{d}x$$

$$\kappa_z = -\rho\left[d_x\dot{\vec{\chi}}^T\underline{M}_{\phi\psi'_\eta}\vec{\chi} + \dot{\vec{\chi}}^T\underline{M}_{x\phi\psi'_\eta}\vec{\chi}\right] \tag{A97}$$

By differentiating $\vec{\kappa}$ with respect to time, we obtain the following set of equations:

$$\dot{\kappa}_x = \rho\left[d_z\ \vec{\chi}^T\underline{M}^T_{\phi\psi'_\eta}\ddot{\vec{\chi}} + d_z\ \dot{\vec{\chi}}^T\underline{M}_{\phi\psi'_\eta}\dot{\vec{\chi}} + d_y\vec{m}^T_{\phi\psi}\ddot{\vec{\chi}} + \vec{m}^T_{y\phi\psi}\ddot{\vec{\chi}}\right] \tag{A98}$$

$$\dot{\kappa}_y = -\rho\left[d_x\vec{m}^T_{\phi\psi}\ddot{\vec{\chi}} + \vec{m}^T_{x\phi\psi}\ddot{\vec{\chi}}\right] \tag{A99}$$

$$\dot{\kappa}_z = -\rho\left[d_x\vec{\chi}^T\underline{M}^T_{\phi\psi'_\eta}\ddot{\vec{\chi}} + d_x\dot{\vec{\chi}}^T\underline{M}_{\phi\psi'_\eta}\dot{\vec{\chi}} + \vec{\chi}^T\underline{M}^T_{x\phi\psi'_\eta}\ddot{\vec{\chi}} + \dot{\vec{\chi}}^T\underline{M}_{x\phi\psi'_\eta}\dot{\vec{\chi}}\right] \tag{A100}$$

We write $\dot{\kappa}_j$ as a function of $\dot{\kappa}_{j\dot{\chi}}$ and $\dot{\vec{\kappa}}_{j\ddot{\chi}}$ where *j* refers to a vector component, $x, y\ or\ z$

$$\dot{\kappa}_j = \dot{\kappa}_{j\dot{\chi}} + \dot{\vec{\kappa}}^T_{j\ddot{\chi}}\ddot{\vec{\chi}} \tag{A101}$$

where the $\dot{\kappa}_{j\dot{\chi}}$ components are scalars and are given as (for *i*^th^ appendage):

$$\dot{\kappa}_{x\dot{\chi}} = \rho\left[d_z\ \dot{\vec{\chi}}^T \underline{M}_{\phi\psi'_\eta}\dot{\vec{\chi}}\right] \tag{A102}$$

$$\dot{\kappa}_{y\dot{\chi}} = 0 \tag{A103}$$

$$\dot{\kappa}_{z\dot{\chi}} = -\rho\left[d_x\dot{\vec{\chi}}^T \underline{M}_{\phi\psi'_\eta}\dot{\vec{\chi}} + \dot{\vec{\chi}}^T \underline{M}_{x\phi\psi'_\eta}\dot{\vec{\chi}}\right] \tag{A104}$$

and $\dot{\vec{\kappa}}_{j\ddot{\chi}}$ components are vectors with *pq* components and are given as (for *i*$^{th}$ appendage):

$$\dot{\vec{\kappa}}^T_{x\ddot{\chi}} = \rho\left[d_z\ \vec{\chi}^T \underline{M}_{\phi\psi'_\eta} + d_y\vec{m}^T_{\phi\psi} + \vec{m}^T_{y\phi\psi}\right] \tag{A105}$$

$$\dot{\vec{\kappa}}^T_{y\ddot{\chi}} = -\rho\left[d_x\vec{m}^T_{\phi\psi} + \vec{m}^T_{x\phi\psi}\right] \tag{A106}$$

$$\dot{\vec{\kappa}}^T_{z\ddot{\chi}} = -\rho\left[d_x\vec{\chi}^T \underline{M}_{\phi\psi'_\eta} + \vec{\chi}^T \underline{M}_{x\phi\psi'_\eta}\right] \tag{A107}$$

Therefore, for the *i*$^{th}$ appendage, we have:

$$\underbrace{\begin{bmatrix}\dot{\kappa}_x\\ \dot{\kappa}_y\\ \dot{\kappa}_z\end{bmatrix}}_{\dot{\vec{\kappa}}} = \underbrace{\begin{bmatrix}\dot{\kappa}_{x\dot{\chi}}\\ \dot{\kappa}_{y\dot{\chi}}\\ \dot{\kappa}_{z\dot{\chi}}\end{bmatrix}}_{\dot{\vec{\kappa}}_{\dot{\chi}}} + \underbrace{\begin{bmatrix}\dot{\vec{\kappa}}^T_{x\ddot{\chi}}\\ \dot{\vec{\kappa}}^T_{y\ddot{\chi}}\\ \dot{\vec{\kappa}}^T_{z\ddot{\chi}}\end{bmatrix}}_{\underline{\dot{\kappa}}^T_{\ddot{\chi}}}\begin{bmatrix}\ddot{\vec{\chi}}\end{bmatrix} \tag{A108}$$

Using (A108), we can re-write the attitude dynamics equation (A39) in a compact matrix form (in spacecraft reference frame) as:

$$\begin{bmatrix}\underline{I}_t & \underline{\dot{\kappa}}^T_{\ddot{\chi}}\end{bmatrix}\begin{bmatrix}\dot{\vec{\omega}}\\ \ddot{\vec{\chi}}\end{bmatrix} = \left[-\left(\dot{\underline{I}}_t\vec{\omega} + \underline{\Omega}\left(\underline{I}_t\vec{\omega} + \vec{\kappa}_t + \vec{h}_w\right) + \dot{\vec{\kappa}}_{\dot{\chi}} + \dot{\vec{h}}_w\right)\right] + \vec{\tau} \tag{A109}$$

## A.7 Potential Energy

The total potential energy is obtain by summing the orbital, attitude and flexible potential energies:

$$K_{\text{total}} = K_{\text{orbital}} + K_{\text{attitude}} + K_{\text{flexible}} \tag{A110}$$

The only term in (A110) which drives the vibrational equations is the potential (flexible) energy stored in *n* deflected appendages and is given by the following equation:

$$K_{flexible} = \frac{1}{2}\sum_{i=1}^{n} D_i \int_0^{a_i}\int_0^{b_i}\left[\left(\frac{\partial^2 w_i}{\partial x_i^2}\right)^2 + \left(\frac{\partial^2 w_i}{\partial y_i^2}\right)^2 + 2\gamma\frac{\partial^2 w_i}{\partial x_i^2}\frac{\partial^2 w_i}{\partial y_i^2} + 2(1-\gamma)\left(\frac{\partial^2 w_i}{\partial x_i \partial y_i}\right)^2\right] \mathrm{d}y_i \mathrm{d}x_i \tag{A111}$$

where $D_i$ is the flexural rigidity and is defined by:

$$D_i = \frac{E_i \; h_i^3}{12\;(1\text{-}\gamma_i^2)} \tag{A112}$$

In (A112), $h_i$ is the thickness of the $i^{th}$ appendage, $\gamma_i$ and $E_i$ are the Poisson ratio and the modulus of elasticity of the material, respectively.

Expanding the first term of equation (A111) using (A56) gives:

$$\int_0^{a_i}\int_0^{b_i}\left(\frac{\partial^2 w_i}{\partial x_i^2}\right)^2 \mathrm{d}y_i \mathrm{d}x_i = \int_0^{a_i}\int_0^{b_i}\left(\vec{C}^T_{\phi''\psi}\vec{\chi}_i\right)^2 \mathrm{d}y_i \mathrm{d}x_i \tag{A113}$$

$$\int_0^{a_i}\int_0^{b_i}\left(\frac{\partial^2 w_i}{\partial x_i^2}\right)^2 \mathrm{d}y_i \mathrm{d}x_i = \int_0^{a_i}\int_0^{b_i}\left(\vec{\chi}_i^T \; \vec{C}_{\phi''\psi} \; \vec{C}^T_{\phi''\psi} \; \vec{\chi}_i\right) \mathrm{d}y_i \mathrm{d}x_i \tag{A114}$$

$$\int_0^{a_i}\int_0^{b_i}\left(\frac{\partial^2 w_i}{\partial \, x_i^2}\right)^2 \mathrm{d}y_i \mathrm{d}x_i = \vec{\chi}_i^T \left[\int_0^{a_i}\int_0^{b_i}\left(\vec{C}_{\phi''\psi} \; \vec{C}^T_{\phi''\psi}\right) \mathrm{d}y_i \mathrm{d}x_i\right] \vec{\chi}_i \tag{A115}$$

Expanding the second term of equation (A111) using (A58) gives:

$$\int_0^{a_i}\int_0^{b_i}\left(\frac{\partial^2 w_i}{\partial y_i^2}\right)^2 \mathrm{d}y_i \mathrm{d}x_i = \int_0^{a_i}\int_0^{b_i}\left(\vec{C}^T_{\phi\psi''}\vec{\chi}_i\right)^2 \mathrm{d}y_i \mathrm{d}x_i \tag{A116}$$

$$\int_0^{a_i}\int_0^{b_i}\left(\frac{\partial^2 w_i}{\partial y_i^2}\right)^2 \mathrm{d}y_i \mathrm{d}x_i = \int_0^{a_i}\int_0^{b_i}\left(\vec{\chi}_i^T\ \vec{C}_{\phi\psi''}\ \vec{C}^T_{\phi\psi''}\ \vec{\chi}_i\right) \mathrm{d}y_i \mathrm{d}x_i \tag{A117}$$

$$\int_0^{a_i}\int_0^{b_i}\left(\frac{\partial^2 w_i}{\partial y_i^2}\right)^2 \mathrm{d}y_i \mathrm{d}x_i = \vec{\chi}_i^T\left[\int_0^{a_i}\int_0^{b_i}\left(\vec{C}_{\phi\psi''}\ \vec{C}^T_{\phi\psi''}\right) \mathrm{d}y_i \mathrm{d}x_i\right]\vec{\chi}_i \tag{A118}$$

Expanding the third term of equation (A111) using (A56) and (A58) gives:

$$\int_0^{a_i}\int_0^{b_i}\frac{\partial^2 w_i}{\partial x_i^2}\frac{\partial^2 w_i}{\partial y_i^2}\ \mathrm{d}y_i \mathrm{d}x_i = \int_0^{a_i}\int_0^{b_i}\ \left(\vec{C}^T_{\phi''\psi}\vec{\chi}_i\right)^T\left(\vec{C}^T_{\phi\psi''}\vec{\chi}_i\right) \mathrm{d}y_i \mathrm{d}x_i \tag{A119}$$

$$\int_0^{a_i}\int_0^{b_i}\frac{\partial^2 w_i}{\partial x_i^2}\frac{\partial^2 w_i}{\partial y_i^2}\ \mathrm{d}y_i \mathrm{d}x_i = \int_0^{a_i}\int_0^{b_i}\ \left(\vec{\chi}_i^T\ \vec{C}_{\phi''\psi}\ \vec{C}^T_{\phi\psi''}\ \vec{\chi}_i\right) \mathrm{d}y_i \mathrm{d}x_i \tag{A120}$$

$$\int_0^{a_i}\int_0^{b_i}\frac{\partial^2 w_i}{\partial x_i^2}\frac{\partial^2 w_i}{\partial y_i^2}\ \mathrm{d}y_i \mathrm{d}x_i = \vec{\chi}_i^T\left[\int_0^{a_i}\int_0^{b_i}\ \left(\vec{C}_{\phi''\psi}\ \vec{C}^T_{\phi\psi''}\right) \mathrm{d}y_i \mathrm{d}x_i\right]\vec{\chi}_i \tag{A121}$$

Expanding the last term of equation (A111) using (A59) gives:

$$\int_0^{b_i}\int_0^{a_i}\left(\frac{\partial^2 \omega_i}{\partial x_i \partial y_i}\right)^2\ dx_i dy_i = \int_0^{a_i}\int_0^{b_i}\ \left(\vec{C}^T_{\phi'\psi'}\vec{\chi}_i\right)^2\ dy_i\ dx_i \tag{A122}$$

$$\int_0^{a_i}\int_0^{b_i}\left(\frac{\partial^2 \omega_i}{\partial x_i \partial y_i}\right)^2\ \mathrm{d}y_i \mathrm{d}x_i = \int_0^{a_i}\int_0^{b_i}\ \left(\vec{\chi}_i^T\ \vec{C}_{\phi'\psi'}\ \vec{C}^T_{\phi'\psi'}\ \vec{\chi}_i\right) \mathrm{d}y_i \mathrm{d}x_i \tag{A123}$$

$$\int_0^{a_i}\int_0^{b_i}\left(\frac{\partial^2\omega_i}{\partial x_i \partial y_i}\right)^2 \mathrm{d}y_i\mathrm{d}x_i = \vec{\chi}_i^T\left[\int_0^{a_i}\int_0^{b_i}\left(\vec{C}_{\phi'\psi'}\ \vec{C}_{\phi'\psi'}^T\right)\mathrm{d}y_i\mathrm{d}x_i\right]\vec{\chi}_i \tag{A124}$$

By defining the following mode shape integrals:

$$\underline{M}_{\phi''\psi} = \int_0^{a_i}\int_0^{b_i}\left(\vec{C}_{\phi''\psi}\ \vec{C}_{\phi''\psi}^T\right)\mathrm{d}y_i\mathrm{d}x_i \tag{A125}$$

$$\underline{M}_{\phi\psi''} = \int_0^{a_i}\int_0^{b_i}\left(\vec{C}_{\phi\psi''}\ \vec{C}_{\phi\psi''}^T\right)\mathrm{d}y_i\mathrm{d}x_i \tag{A126}$$

$$\underline{M}_{\phi''\psi''} = \int_0^{a_i}\int_0^{b_i}\left(\vec{C}_{\phi''\psi}\ \vec{C}_{\phi\psi''}^T\right)\mathrm{d}y_i\mathrm{d}x_i \tag{A127}$$

$$\underline{M}_{\phi'\psi'} = \int_0^{a_i}\int_0^{b_i}\left(\vec{C}_{\phi'\psi'}\ \vec{C}_{\phi'\psi'}^T\right)\mathrm{d}y_i\mathrm{d}x_i \tag{A128}$$

we rewrite (A111) using (A115), (A118), (A121), and (A124)-(A128) as follows:

$$K_{flexible} = \frac{1}{2}\sum_{i=1}^{n} D_i\ \vec{\chi}_i^T\ \left[\underline{M}_{\phi''\psi} + \underline{M}_{\phi\psi''} + 2\gamma_i\ \underline{M}_{\phi''\psi''} + 2\left(1-\gamma_i\right)\underline{M}_{\phi'\psi'}\right]\vec{\chi}_i \tag{A129}$$

and finally by defining:

$$\underline{V}_i = D_i\left[\underline{M}_{\phi''\psi} + \underline{M}_{\phi\psi''} + 2\gamma_i\ \underline{M}_{\phi''\psi''} + 2\left(1-\gamma_i\right)\underline{M}_{\phi'\psi'}\right] \tag{A130}$$

we obtain the final form of the potential energy stored in the *n* deflected appendages as:

$$K_{\text{flexible}} = \frac{1}{2}\sum_{i=1}^{n}\ \vec{\chi}_i^T\underline{V}_i\ \vec{\chi}_i \tag{A131}$$

## A.8 Vibrational (Flexural) Equations

The vibrational equations of motion are obtained by using the conventional form of Lagrange's equation:

$$\frac{\mathrm{d}}{\mathrm{d}t}\left(\frac{\partial L}{\partial \dot{q}_k}\right) - \frac{\partial L}{\partial q_k} = Q_k \qquad k = 1, 2, 3\ldots, npq \tag{A132}$$

where

$$L = T(\vec{\chi}_1, \vec{\chi}_2, ..., \vec{\chi}_{npq}, \dot{\vec{\chi}}_1, \dot{\vec{\chi}}_2, ..., \dot{\vec{\chi}}_{npq}) - K(\vec{\chi}_1, \vec{\chi}_2, ..., \vec{\chi}_{npq}) \tag{A133}$$

In scalar form (A132) becomes:

$$\frac{\mathrm{d}}{\mathrm{d}t}\left(\frac{\partial T}{\partial \dot{\chi}_{irs}}\right) - \frac{\partial T}{\partial \chi_{irs}} + \frac{\partial K}{\partial \chi_{irs}} = Q_{irs} \qquad i = 1,2, ..., n\ ;\ \ r = 1,2, ..., p\ ;\ \ s = 1,2, ..., q \tag{A134}$$

where $n$ is the number of appendage and $p$ and $q$ are the number of modes in the Free-Free and the Clamped-Free directions, respectively. The term $Q$ denotes the generalized forces due to environmental disturbances.

In vector form (A134) becomes:

$$\frac{\mathrm{d}}{\mathrm{d}t}\left(\frac{\partial T}{\partial \dot{\vec{\chi}}_i}\right) - \frac{\partial T}{\partial \vec{\chi}_i} + \frac{\partial K}{\partial \vec{\chi}_i} = \vec{Q}_i \tag{A135}$$

Using equations (A2), (A3), (A5), (A26), (A29) and (A30), we obtain the equation for the kinetic energy of the satellite expressed in appendage reference frame:

$$T = \frac{1}{2} m_t v_c^2 + \frac{1}{2}\vec{\omega}^T \underline{I}_c \vec{\omega} + \frac{1}{2}\vec{\omega}^T \sum_{i=1}^{n} \underline{I}_i\ \vec{\omega} + \sum_{i=1}^{n} \vec{\kappa}_i^T \vec{\omega} + \frac{1}{2}\sum_{i=1}^{n} \rho_i \int_{A_i} \dot{w}_i^2 \mathrm{d}A_i \tag{A136}$$

By differentiating (A136) with respect to the time derivative of $\vec{\chi}_i$, we obtain:

$$\frac{\partial T}{\partial \dot{\vec{\chi}}_i} = \frac{\partial}{\partial \dot{\vec{\chi}}_i} \begin{pmatrix} \frac{1}{2} m_c v_c^2 + \frac{1}{2} \vec{\omega}^T \underline{I}_c \vec{\omega} + \frac{1}{2} \vec{\omega}^T \sum_{i=1}^{n} \underline{I}_i \ \vec{\omega} \\ + \sum_{i=1}^{n} \begin{bmatrix} \kappa_{ix} & \kappa_{iy} & \kappa_{iz} \end{bmatrix} \begin{bmatrix} \omega_x \\ \omega_y \\ \omega_z \end{bmatrix} + \frac{1}{2} \sum_{i=1}^{n} \rho_i \left( \dot{\vec{\chi}}_i^T \underline{M}_{\phi\psi} \dot{\vec{\chi}}_i \right) \end{pmatrix} \tag{A137}$$

Using previously derived mode shape integrals and dropping the subscript *i*, we write:

$$\begin{aligned} \frac{\partial T}{\partial \dot{\vec{\chi}}} &= \rho \omega_x \left( d_z \underline{M}_{\phi\psi'_\eta} \vec{\chi} + d_y \vec{m}_{\phi\psi} + \vec{m}_{y\phi\psi} \right) - \rho \omega_y \left( d_x \vec{m}_{\phi\psi} + \vec{m}_{x\phi\psi} \right) \\ &\quad - \rho \omega_z \left( d_x \underline{M}_{\phi\psi'_\eta} \vec{\chi} + \underline{M}_{x\phi\psi'_\eta} \vec{\chi} \right) + \rho \ \underline{M}_{\phi\psi} \ \dot{\vec{\chi}} \end{aligned} \tag{A138}$$

and differentiating (A138) with respect to time gives:

$$\begin{aligned} \frac{\mathrm{d}}{\mathrm{d}t}\left( \frac{\partial T}{\partial \dot{\vec{\chi}}} \right) &= \rho \left[ \dot{\omega}_x d_z \underline{M}_{\phi\psi'_\eta} \ \vec{\chi} + \omega_x d_z \underline{M}_{\phi\psi'_\eta} \dot{\vec{\chi}} \right] + \rho \ \underline{M}_{\phi\psi}^T \ddot{\vec{\chi}} \\ &\quad + \rho \dot{\omega}_x \left( d_y \vec{m}_{\phi\psi} + \vec{m}_{y\phi\psi} \right) - \rho \dot{\omega}_y \left( d_x \vec{m}_{\phi\psi} + \vec{m}_{x\phi\psi} \right) \\ &\quad - \rho \left[ \dot{\omega}_z \left( d_x \underline{M}_{\phi\psi'_\eta} + \underline{M}_{x\phi\psi'_\eta} \right) \vec{\chi} + \omega_z \left( d_x \underline{M}_{\phi\psi'_\eta} + \underline{M}_{x\phi\psi'_\eta} \right) \dot{\vec{\chi}} \ \right] \end{aligned} \tag{A139}$$

Next we differentiate (A136), with respect to $\vec{\chi}$, to obtain:

$$\frac{\partial T}{\partial \vec{\chi}} = \frac{\partial}{\partial \vec{\chi}} \left[ \frac{1}{2} m_t v_c^2 + \frac{1}{2} \vec{\omega}^T \underline{I}_c \vec{\omega} + \frac{1}{2} \vec{\omega}^T \sum_{i=1}^{n} \underline{I}_i \ \vec{\omega} + \sum_{i=1}^{n} \vec{\kappa}_i^T \vec{\omega} + \frac{1}{2} \sum_{i=1}^{n} \rho_i \int_{A_i} \dot{w}_i^2 \mathrm{d}A_i \right] \tag{A140}$$

and by keeping the terms contributing to the differentiation of *T* with respect to $\vec{\chi}$, we are only left with:

$$\frac{\partial T}{\partial \vec{\chi}} = \frac{\partial}{\partial \vec{\chi}} \left[ \frac{1}{2} \vec{\omega}^T \underline{I} \ \vec{\omega} + \vec{\kappa}^T \vec{\omega} \right] \tag{A141}$$

Expanding the terms of (A141) gives:

$$\frac{\partial T}{\partial \vec{\chi}}=\frac{\partial}{\partial \vec{\chi}}\left[\frac{1}{2}\left\{\begin{bmatrix}\omega_x & \omega_y & \omega_z\end{bmatrix}\begin{bmatrix}I_{xx} & I_{xy} & I_{xz}\\ I_{yx} & I_{yy} & I_{yz}\\ I_{zx} & I_{zy} & I_{zz}\end{bmatrix}\begin{bmatrix}\omega_x\\ \omega_y\\ \omega_z\end{bmatrix}\right\}+\begin{bmatrix}\kappa_x & \kappa_y & \kappa_z\end{bmatrix}\begin{bmatrix}\omega_x\\ \omega_y\\ \omega_z\end{bmatrix}\right] \tag{A142}$$

and simplifying (A142) in the next two steps, results in:

$$\frac{\partial T}{\partial \vec{\chi}}=\frac{\partial}{\partial \vec{\chi}}\left[\frac{1}{2}\left\{\begin{bmatrix}\omega_x & \omega_y & \omega_z\end{bmatrix}\begin{bmatrix}I_{xx}\omega_x+I_{xy}\omega_y+I_{xz}\omega_z\\ I_{yx}\omega_x+I_{yy}\omega_y+I_{yz}\omega_z\\ I_{zx}\omega_x+I_{zy}\omega_y+I_{zz}\omega_z\end{bmatrix}\right\}+\begin{bmatrix}\kappa_x\omega_x+\kappa_y\omega_y+\kappa_z\omega_z\end{bmatrix}\right]$$

$$\frac{\partial T}{\partial \vec{\chi}}=\frac{\partial}{\partial \vec{\chi}}\begin{bmatrix}\frac{1}{2}\begin{Bmatrix}I_{xx}\omega_x^2+I_{xy}\omega_x\omega_y+I_{xz}\omega_x\omega_z\\ +I_{yx}\omega_x\omega_y+I_{yy}\omega_y^2+I_{yz}\omega_y\omega_z\\ +I_{zx}\omega_x\omega_z+I_{zy}\omega_y\omega_z+I_{zz}\omega_z^2\end{Bmatrix}\\ +\kappa_x\omega_x+\kappa_y\omega_y+\kappa_z\omega_z\end{bmatrix}$$

$$\frac{\partial T}{\partial \vec{\chi}}=\frac{\partial}{\partial \vec{\chi}}\begin{bmatrix}\frac{1}{2}I_{xx}\omega_x^2+I_{xy}\omega_x\omega_y+I_{xz}\omega_x\omega_z+\frac{1}{2}I_{yy}\omega_y^2+I_{yz}\omega_y\omega_z\\ +\frac{1}{2}I_{zz}\omega_z^2+\kappa_x\omega_x+\kappa_y\omega_y+\kappa_z\omega_z\end{bmatrix} \tag{A143}$$

Carrying out the differentiation in (A143) and substituting the mode shape integrals, we have:

$$\frac{\partial T}{\partial \vec{\chi}} = \rho \begin{bmatrix} \omega_x^2 \left( -d_y \tilde{M}_{\phi\psi'_\eta} \ \vec{\chi} + d_z \vec{m}_{\phi\psi} + \tilde{M}_{\phi\psi} \ \vec{\chi} - \tilde{M}_{y\phi\psi'_\eta} \ \vec{\chi} \right) \\ +\omega_x \omega_y \left( d_x \tilde{M}_{\phi\psi'_\eta} \ \vec{\chi} + \tilde{M}_{x\phi\psi'_\eta} \ \vec{\chi} \right) + \omega_x \omega_z \left( -d_x \vec{m}_{\phi\psi} \ - \vec{m}_{x\phi\psi} \ \right) \\ +\omega_y^2 \left( \tilde{M}_{\phi\psi} \ \vec{\chi} + d_z \vec{m}_{\phi\psi} \right) + \omega_y \omega_z \left( d_z \tilde{M}_{\phi\psi'_\eta} \ \vec{\chi} - d_y \vec{m}_{\phi\psi} - \vec{m}_{y\phi\psi} \right) \\ +\omega_z^2 \left( -d_y \tilde{M}_{\phi\psi'_\eta} \ \vec{\chi} - \tilde{M}_{y\phi\psi'_\eta} \ \vec{\chi} \right) + \omega_x \left( d_z \ \underline{M}^T_{\phi\psi'_\eta} \dot{\vec{\chi}} \right) \\ +\omega_z \left( -d_x \ \underline{M}^T_{\phi\psi'_\eta} \dot{\vec{\chi}} - \underline{M}^T_{x\phi\psi'_\eta} \dot{\vec{\chi}} \right) \end{bmatrix} \tag{A144}$$

where

$$\tilde{M}_{\phi\psi'_\eta} = \frac{1}{2} \left( \underline{M}_{\phi\psi'_\eta} + \underline{M}^T_{\phi\psi'_\eta} \right) \tag{A145}$$

$$\tilde{M}_{x\phi\psi'_\eta} = \frac{1}{2} \left( \underline{M}_{x\phi\psi'_\eta} + \underline{M}^T_{x\phi\psi'_\eta} \right) \tag{A146}$$

$$\tilde{M}_{\phi\psi} = \frac{1}{2} \left( \underline{M}_{\phi\psi} + \underline{M}^T_{\phi\psi} \right) \tag{A147}$$

$$\tilde{M}_{y\phi\psi'_\eta} = \frac{1}{2} \left( \underline{M}_{y\phi\psi'_\eta} + \underline{M}^T_{y\phi\psi'_\eta} \right) \tag{A148}$$

Noting that for our particular system, all mode shape matrices $\underline{M}$ , except for $\underline{M}_{\phi''\psi''}$, are symmetric (i.e. $\underline{M}^T = \underline{M}$ ), then all matrices $\tilde{M} = \underline{M}$ .

Using (A131), we can obtain:

$$\frac{\partial K}{\partial \vec{\chi}} = \frac{\partial K_{\text{flexible}}}{\partial \vec{\chi}} = \frac{1}{2} \left( \underline{V} + \underline{V}^T \right) \vec{\chi} \tag{A149}$$

and after substitution of (A139), (A144) and (A149) in Lagrange's equation (A135), we get the vibrational equation for the $i^{th}$ appendage as:

$$\begin{bmatrix} \rho\left(\dot{\omega}_x d_z \underline{M}_{\phi\psi'_\eta}\ \vec{\chi} + \omega_x d_z \underline{M}_{\phi\psi'_\eta}\ \dot{\vec{\chi}}\right) + \rho\dot{\omega}_x\left(d_y \vec{m}_{\phi\psi} + \vec{m}_{y\phi\psi}\right) - \rho\dot{\omega}_y\left(d_x \vec{m}_{\phi\psi} + \vec{m}_{x\phi\psi}\right) \\ -\rho\left[\dot{\omega}_z\left(d_x \underline{M}_{\phi\psi'_\eta} + \underline{M}_{x\phi\psi'_\eta}\right)\vec{\chi} + \omega_z\left(d_x \underline{M}_{\phi\psi'_\eta} + \underline{M}_{x\phi\psi'_\eta}\right)\dot{\vec{\chi}}\ \right] + \rho\ \underline{M}_{\phi\psi}\ddot{\vec{\chi}} \end{bmatrix} -$$

$$\rho\begin{bmatrix} \omega_x^2\left(-d_y \underline{M}_{\phi\psi'_\eta}\ \vec{\chi} + d_z \vec{m}_{\phi\varphi} + \underline{M}_{\phi\psi}\ \vec{\chi} - \underline{M}_{y\phi\psi'_\eta}\ \vec{\chi}\right) \\ +\omega_x\omega_y\left(d_x \underline{M}_{\phi\psi'_\eta}\ \vec{\chi} + \underline{M}_{x\phi\psi'_\eta}\ \vec{\chi}\right) + \omega_x\omega_z\left(-d_x \vec{m}_{\phi\psi} - \vec{m}_{x\phi\psi}\right) \\ +\omega_y^2\left(\underline{M}_{\phi\psi}\ \vec{\chi} + d_z \vec{m}_{\phi\psi}\right) + \omega_y\omega_z\left(d_z \underline{M}_{\phi\psi'_\eta}\ \vec{\chi} - d_y \vec{m}_{\phi\psi} - \vec{m}_{y\phi\psi}\right) \\ +\omega_z^2\left(-d_y \underline{M}_{\phi\psi'_\eta}\ \vec{\chi} - \underline{M}_{y\phi\psi'_\eta}\ \vec{\chi}\right) + \omega_x\left(d_z\ \underline{M}_{\phi\psi'_\eta}\dot{\vec{\chi}}\right) \\ +\omega_z\left(-d_x\ \underline{M}_{\phi\psi'_\eta}\dot{\vec{\chi}} - \underline{M}_{x\phi\psi'_\eta}\dot{\vec{\chi}}\right) \end{bmatrix} + \frac{1}{2}\left(\underline{V} + \underline{V}^T\right)\vec{\chi} = \vec{Q} \qquad \text{(A150)}$$

and rearranging the terms in (A150) gives:

$$\rho\underline{M}_{\phi\psi}\ddot{\vec{\chi}} + \rho\left[d_z \underline{M}_{\phi\psi'_\eta}\vec{\chi} + d_y \vec{m}_{\phi\psi} + \vec{m}_{y\phi\psi}\right]\dot{\omega}_x$$

$$+\rho\left[-d_x \vec{m}_{\phi\psi} - \vec{m}_{x\phi\psi}\right]\dot{\omega}_y + \rho\left[-d_x \underline{M}_{\phi\psi'_\eta}\vec{\chi} - \underline{M}_{x\phi\psi'_\eta}\vec{\chi}\right]\dot{\omega}_z =$$

$$\rho\begin{bmatrix} -\frac{\left(\underline{V} + \underline{V}^T\right)}{2\rho} + \omega_x^2\left(-d_y \underline{M}_{\phi\psi'_\eta} + \underline{M}_{\phi\psi} - \underline{M}_{y\phi\psi'_\eta}\right) \\ +\omega_y^2 \underline{M}_{\phi\psi} + \omega_z^2\left(-d_y \underline{M}_{\phi\psi'_\eta} - \underline{M}_{y\phi\psi'_\eta}\right) + \\ \omega_x\omega_y\left(d_x \underline{M}_{\phi\psi'_\eta} + \underline{M}_{x\phi\psi'_\eta}\right) + \omega_y\omega_z\left(d_z \underline{M}_{\phi\psi'_\eta}\right) \\ \end{bmatrix}\vec{\chi} + \qquad \text{(A151)}$$

$$\rho\left[-\omega_x d_z \underline{M}_{\phi\psi'_\eta} + \omega_z\left(d_x \underline{M}_{\phi\psi'_\eta} + \underline{M}_{x\phi\psi'_\eta}\right) + \omega_x d_z \underline{M}_{\phi\psi'_\eta} - \omega_z\left(d_x \underline{M}_{\phi\psi'_\eta} + \underline{M}_{x\phi\psi'_\eta}\right)\right]\dot{\vec{\chi}} +$$

$$\rho\left[\omega_x^2 d_z \vec{m}_{\phi\psi} + \omega_y^2 d_z \vec{m}_{\phi\psi} - \omega_x\omega_z\left(d_x \vec{m}_{\phi\psi} + \vec{m}_{x\phi\psi}\right) - \omega_y\omega_z\left(d_y \vec{m}_{\phi\psi} + \vec{m}_{y\phi\psi}\right)\right] + \vec{Q}$$

Note that the coefficients of the $\dot{\vec{\chi}}$ term in (A151) cancel out and hence the $\dot{\vec{\chi}}$ term drops to zero but since there is some damping in any flexible structure, we will add a damping term to (A151). We have assumed viscous damping and no modal coupling as far as damping is concerned which results in a diagonal damping matrix, $\underline{D}$, with entries $\xi$ for the damping parameter. The $\vec{Q}$ term representing the generalized forces due to environmental disturbances is set to zero for our present purposes.

The final form of the vibrational equation for the $i^{th}$ appendage can be then given by:

$$\underline{M}_{\phi\psi}\ddot{\vec{\chi}}+\dot{\vec{\kappa}}_{x\ddot{\chi}}\,\dot{\omega}_x+\dot{\vec{\kappa}}_{y\ddot{\chi}}\,\dot{\omega}_y+\dot{\vec{\kappa}}_{z\ddot{\chi}}\,\dot{\omega}_z=\underline{K}\,\vec{\chi}-\underline{D}\,\dot{\vec{\chi}}+\vec{C} \tag{A152}$$

where

$$\underline{K}=\begin{bmatrix}-\dfrac{\left(\underline{V}+\underline{V}^T\right)}{2\,\rho}+\omega_x^2\left(-d_y\underline{M}_{\phi\psi'_\eta}+\underline{M}_{\phi\psi}-\underline{M}_{y\phi\psi'_\eta}\right)\\+\omega_y^2\underline{M}_{\phi\psi}+\omega_z^2\left(-d_y\underline{M}_{\phi\psi'_\eta}-\underline{M}_{y\phi\psi'_\eta}\right)+\\\omega_x\omega_y\left(d_x\underline{M}_{\phi\psi'_\eta}+\underline{M}_{x\phi\psi'_\eta}\right)+\omega_y\omega_z\left(d_z\underline{M}_{\phi\psi'_\eta}\right)\\ \end{bmatrix} \tag{A153}$$

$$\vec{C}=\begin{bmatrix}\omega_x^2 d_z\vec{m}_{\phi\psi}+\omega_y^2 d_z\vec{m}_{\phi\psi}-\omega_x\omega_z\left(d_x\vec{m}_{\phi\psi}+\vec{m}_{x\phi\psi}\right)\\-\omega_y\omega_z\left(d_y\vec{m}_{\phi\psi}+\vec{m}_{y\phi\psi}\right)\end{bmatrix} \tag{A154}$$

and the $\dot{\vec{\kappa}}_{j\ddot{\chi}}$ terms are the same as (A105)-(A107) without the multiplicative $\rho$ term:

$$\begin{aligned}\dot{\vec{\kappa}}_{x\ddot{\chi}}&=d_z\underline{M}_{\phi\psi'_\eta}\vec{\chi}+d_y\vec{m}_{\phi\psi}+\vec{m}_{y\phi\psi}\\\dot{\vec{\kappa}}_{y\ddot{\chi}}&=-d_x\vec{m}_{\phi\psi}-\vec{m}_{x\phi\psi}\\\dot{\vec{\kappa}}_{z\ddot{\chi}}&=-d_x\underline{M}_{\phi\psi'_\eta}\vec{\chi}-\underline{M}_{x\phi\psi'_\eta}\vec{\chi}\end{aligned} \tag{A155}$$

Finally, the vibrational equations of the $i^{th}$ appendage can be written using a compact matrix form, as

$$\begin{bmatrix}\dot{\underline{\kappa}}_{\ddot{\chi}} & \underline{M}_{\phi\psi}\end{bmatrix}\begin{bmatrix}\dot{\vec{\omega}}\\\ddot{\vec{\chi}}\end{bmatrix}=\begin{bmatrix}\underline{K}\,\vec{\chi}-\underline{D}\,\dot{\vec{\chi}}+\vec{C}\end{bmatrix} \tag{A156}$$

where $\dot{\underline{\kappa}}_{\ddot{\chi}}=\left[\dot{\vec{\kappa}}_{x\ddot{\chi}}\;\dot{\vec{\kappa}}_{y\ddot{\chi}}\;\dot{\vec{\kappa}}_{z\ddot{\chi}}\right]$.

The complete set of the attitude and flexural equations of the system, (A109) with no reaction/momentum wheels and (A156), can be combined and rewritten as:

$$\begin{bmatrix} \underline{I}_t & \dot{\underline{\kappa}}^T_{\ddot{\vec{\chi}}} \\ \dot{\underline{\kappa}}_{\ddot{\vec{\chi}}} & \underline{M}_{\phi\psi} \end{bmatrix} \begin{bmatrix} \dot{\vec{\omega}} \\ \ddot{\vec{\chi}} \end{bmatrix} = \begin{bmatrix} -\left(\dot{\underline{I}}_t\vec{\omega} + \underline{\Omega}\left(\underline{I}_t\vec{\omega} + \vec{\kappa}_t + \vec{h}_w\right) + \dot{\vec{\kappa}}_{\dot{\vec{\chi}}}\right) \\ \underline{K}\,\vec{\chi} - \underline{D}\,\dot{\vec{\chi}} + \vec{C} \end{bmatrix} + \begin{bmatrix} \vec{\tau} \\ 0 \end{bmatrix} \qquad \text{(A157)}$$

or equivalently,

$$\begin{bmatrix} \dot{\vec{\omega}} \\ \ddot{\vec{\chi}} \end{bmatrix} = \begin{bmatrix} \underline{I}_t & \dot{\underline{\kappa}}^T_{\ddot{\vec{\chi}}} \\ \dot{\underline{\kappa}}_{\ddot{\vec{\chi}}} & \underline{M}_{\phi\psi} \end{bmatrix}^{-1} \begin{bmatrix} -\left(\dot{\underline{I}}_t\vec{\omega} + \underline{\Omega}\left(\underline{I}_t\vec{\omega} + \vec{\kappa}_t + \vec{h}_w\right) + \dot{\vec{\kappa}}_{\dot{\vec{\chi}}}\right) \\ \underline{K}\,\vec{\chi} - \underline{D}\,\dot{\vec{\chi}} + \vec{C} \end{bmatrix} + \begin{bmatrix} \underline{I}_t & \dot{\underline{\kappa}}^T_{\ddot{\vec{\chi}}} \\ \dot{\underline{\kappa}}_{\ddot{\vec{\chi}}} & \underline{M}_{\phi\psi} \end{bmatrix}^{-1} \begin{bmatrix} \vec{\tau} \\ 0 \end{bmatrix} \qquad \text{(A158)}$$

The mass matrix inversion in (A158) can be obtained by using the block matrix inversion formula [127], and hence we can decouple the system of equations

$$\begin{aligned} \begin{bmatrix} \dot{\vec{\omega}} \\ \ddot{\vec{\chi}} \end{bmatrix} = & \begin{bmatrix} \underline{F}_{11}^{-1} & -\underline{I}_t^{-1}\dot{\underline{\kappa}}^T_{\ddot{\vec{\chi}}}\underline{F}_{22}^{-1} \\ -\underline{F}_{22}^{-1}\dot{\underline{\kappa}}_{\ddot{\vec{\chi}}}\underline{I}_t^{-1} & \underline{F}_{22}^{-1} \end{bmatrix} \begin{bmatrix} -\left(\dot{\underline{I}}_t\vec{\omega} + \underline{\Omega}\left(\underline{I}_t\vec{\omega} + \vec{\kappa}_t + \vec{h}_w\right) + \dot{\vec{\kappa}}_{\dot{\vec{\chi}}}\right) \\ \underline{K}\,\vec{\chi} - \underline{D}\,\dot{\vec{\chi}} + \vec{C} \end{bmatrix} + \\ & \begin{bmatrix} \underline{F}_{11}^{-1} & -\underline{I}_t^{-1}\dot{\underline{\kappa}}^T_{\ddot{\vec{\chi}}}\underline{F}_{22}^{-1} \\ -\underline{F}_{22}^{-1}\dot{\underline{\kappa}}_{\ddot{\vec{\chi}}}\underline{I}_t^{-1} & \underline{F}_{22}^{-1} \end{bmatrix} \begin{bmatrix} \vec{\tau} \\ 0 \end{bmatrix} \end{aligned} \qquad \text{(A159)}$$

or equivalently,

$$\begin{bmatrix} \dot{\vec{\omega}} \\ \ddot{\vec{\chi}} \end{bmatrix} = \begin{bmatrix} -\underline{F}_{11}^{-1}\vec{l}_{\vec{\omega}} - \underline{I}_t^{-1}\dot{\underline{\kappa}}^T_{\ddot{\vec{\chi}}}\underline{F}_{22}^{-1}\vec{l}_{\vec{\chi}} \\ -\underline{F}_{22}^{-1}\dot{\underline{\kappa}}_{\ddot{\vec{\chi}}}\underline{I}_t^{-1}\vec{l}_{\vec{\omega}} + \underline{F}_{22}^{-1}\vec{l}_{\vec{\chi}} \end{bmatrix} + \begin{bmatrix} \underline{F}_{11}^{-1} \\ -\underline{F}_{22}^{-1}\dot{\underline{\kappa}}_{\ddot{\vec{\chi}}}\underline{I}_t^{-1} \end{bmatrix} \vec{\tau} \qquad \text{(A160)}$$

where

$$
\begin{aligned}
\underline{F}_{11} &= \underline{I}_t - \underline{\dot{\kappa}}_{\ddot{\chi}}^T \underline{M}_{\phi\psi}^{-1} \underline{\dot{\kappa}}_{\ddot{\chi}} \\
\underline{F}_{22} &= \underline{M}_{\phi\psi} - \underline{\dot{\kappa}}_{\ddot{\chi}} \underline{I}_t^{-1} \underline{\dot{\kappa}}_{\ddot{\chi}}^T \\
\vec{l}_{\dot{\vec{\omega}}} &= \dot{\underline{I}}_t \vec{\omega} + \underline{\Omega}\left(\underline{I}_t \vec{\omega} + \vec{\kappa}_t + \vec{h}_w\right) + \dot{\vec{\kappa}}_{\dot{\vec{\chi}}} \\
\vec{l}_{\ddot{\vec{\chi}}} &= \underline{K}\, \vec{\chi} - \underline{D}\, \dot{\vec{\chi}} + \vec{C}
\end{aligned}
\tag{A161}
$$

By defining $\vec{\delta}^T = \left[ \vec{\chi}^T \ \dot{\vec{\chi}}^T \right]$, (A160) can be expressed in a concise state space representation as,

$$
\begin{aligned}
\dot{\vec{x}} &= \begin{bmatrix} \dot{\vec{\omega}} \\ \dot{\vec{\delta}} \end{bmatrix} = \begin{bmatrix} \vec{f}_{\vec{\omega}}(\vec{\omega}, \vec{\delta}) \\ \vec{f}_{\vec{\delta}}(\vec{\omega}, \vec{\delta}) \end{bmatrix} + \begin{bmatrix} \underline{G}_{\vec{\omega}}(\vec{\delta}) \\ \underline{G}_{\vec{\delta}}(\vec{\delta}) \end{bmatrix} \vec{\tau} \ \square \ \vec{f}(\vec{x}) + \underline{G}(\vec{x})\, \vec{\tau} \\
\vec{y} &= \vec{h}(\vec{x})
\end{aligned}
\tag{A162}
$$

where the $\vec{f}$ and $\underline{G}$ terms can be obtained directly from (A160) and the state and output vectors are given as

$$
\begin{aligned}
\vec{x}^T &= [\vec{\omega}^T \quad \vec{\delta}^T] = [\omega_x \ \omega_y \ \omega_z \ \chi_1 \ \chi_2 \cdots \chi_{npq} \ \dot{\chi}_1 \ \dot{\chi}_2 \cdots \dot{\chi}_{npq}] \\
\vec{y} &= \vec{h}(\vec{x}) = \begin{bmatrix} x_1 & x_2 & x_3 \end{bmatrix}^T = \begin{bmatrix} \omega_x & \omega_y & \omega_z \end{bmatrix}^T
\end{aligned}
\tag{A163}
$$

This represents a *m*-dimensional system with $m = 3 + 2npq$, where $n, p, q$ were defined previously.

## A.9 Derivation of the $v$ displacement

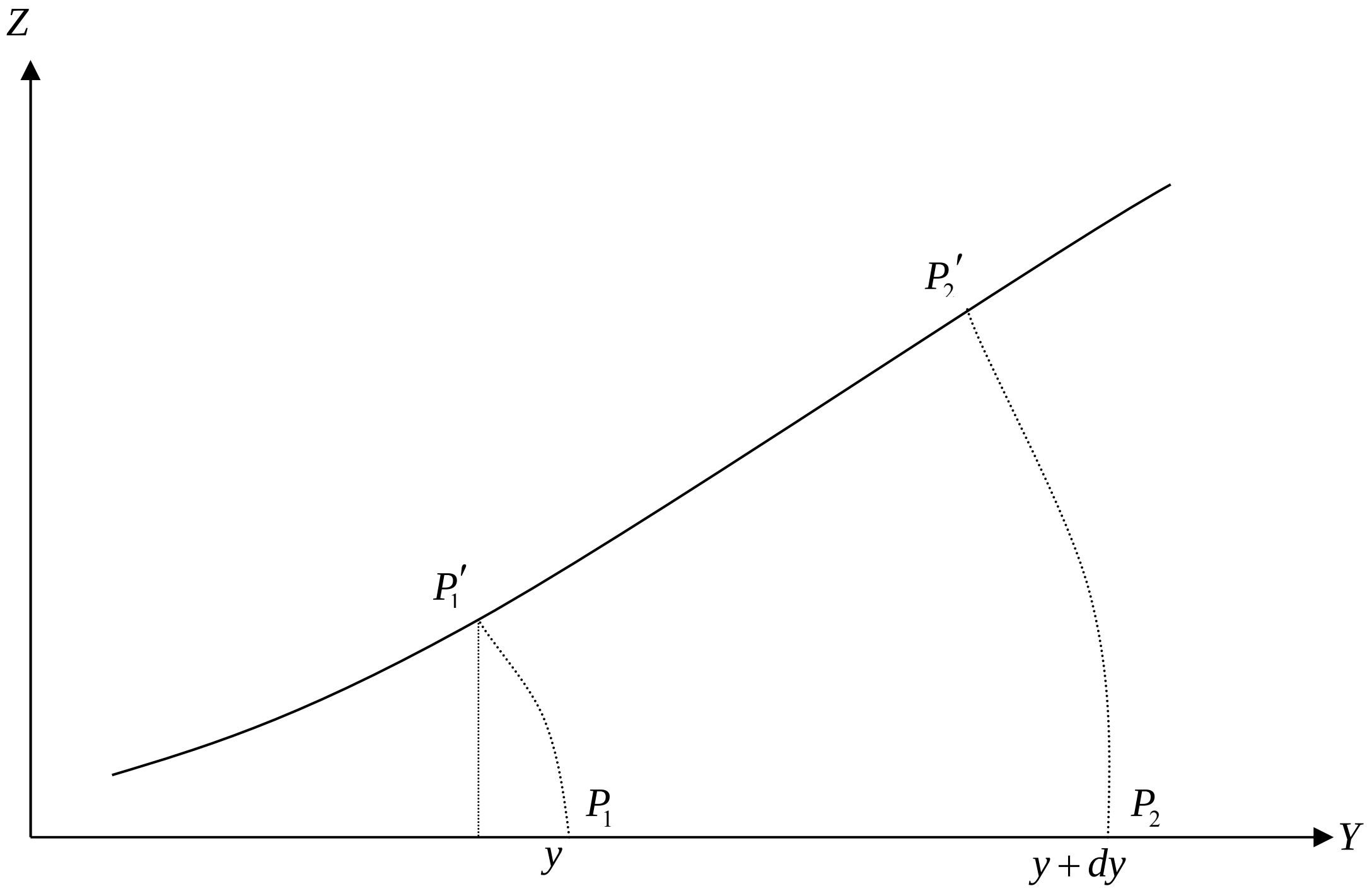

Two fixed points on the undeflected appendage are selected as $P_1:(y,0)$ and $P_2:(y+dy,0)$ as shown in the figure above. After a deflection of the appendage, the new coordinates of these points are given as $P_1':\left(y-v(y),w(y)\right)$ and $P_2':\left(y+dy-v(y+dy),w(y+dy)\right)$.

Using the Pythagorian theorem, we have

$$\mathrm{d}y^2=\left[y+\mathrm{d}y-v(y+\mathrm{d}y)-y+v(y)\right]^2+\left[w(y+\mathrm{d}y)-w(y)\right]^2 \qquad \text{(A164)}$$

and by defining

$$v(y+\mathrm{d}y)=v(y)+\frac{\mathrm{d}v}{\mathrm{d}y}\mathrm{d}y \qquad \text{(A165)}$$

and

$$w(y+\mathrm{d}y) = w(y) + \frac{dw}{\mathrm{d}y}\mathrm{d}y \tag{A166}$$

equation (A164) can then be reduced to:

$$\mathrm{d}y^2 = \left(\mathrm{d}y - \mathrm{d}v\right)^2 + \left(\mathrm{d}w\right)^2 \tag{A167}$$

By expanding (A167), we obtain

$$\mathrm{d}y^2 = \mathrm{d}y^2 - 2\mathrm{d}y\mathrm{d}v + \mathrm{d}v^2 + \mathrm{d}w^2$$

$$\mathrm{d}v^2 - 2\mathrm{d}y\mathrm{d}v - \mathrm{d}w^2 = 0 \tag{A168}$$

and solving for the roots of (A168):

$$\mathrm{d}v = \frac{-(-2\mathrm{d}y)}{2} \pm \frac{\sqrt{4\mathrm{d}y^2 - 4\mathrm{d}w^2}}{2} \tag{A169}$$

The positive solution of (A169) is trivial and leads to the case where $\mathrm{d}y = 0$, which is not of interest. We can see this by allowing $\mathrm{d}w = 0$ and hence $\mathrm{d}v = 0$ in (A169):

$$\mathrm{d}y + \sqrt{\mathrm{d}y^2 - 0} = 0$$

$$2\mathrm{d}y = 0$$

$$\mathrm{d}y = 0$$

The negative solution of (A169) is of interest and is given by:

$$\mathrm{d}v = \mathrm{d}y - \sqrt{\mathrm{d}y^2 - \mathrm{d}w^2}$$

$$\mathrm{d}v = \mathrm{d}y - \mathrm{d}y\sqrt{1-\frac{\mathrm{d}w^2}{\mathrm{d}y^2}} \tag{A170}$$

We now integrate both sides of (A170) to find the expression for $v$:

$$\int_0^y \mathrm{d}v = \int_0^y \left(1-\sqrt{1-\left(\frac{\mathrm{d}w}{\mathrm{d}y}\right)^2}\right)\mathrm{d}y$$

$$v = \int_0^y \left(1-\underbrace{\sqrt{1-\underbrace{\left(\frac{\mathrm{d}w}{\mathrm{d}y}\right)}_{x}^2}}_{f(x)}\right)\mathrm{d}y \tag{A171}$$

Let's set $x = \frac{\mathrm{d}w}{\mathrm{d}y}$ and $f(x) = \sqrt{1-x^2}$ and find the first and second derivatives of $f(x)$ as:

$$f'(x) = \frac{-a}{\sqrt{1-a^2}} \tag{A172}$$

and

$$f''(x) = \frac{-1}{\sqrt{1-a^2}} + \frac{-a^2}{\sqrt{(1-a^2)^3}} \tag{A173}$$

We now use (A172) and (A173) to express $f(x)$ using a Maclaurin series, as:

$$f(x) = f(0) + f'(0)x + f''(0)\frac{x^2}{2!} + h.o.t.$$

$$f(x) = 1 + \frac{0x}{1} + \frac{-1x^2}{2} + h.o.t \tag{A174}$$

and by substituting (A174) into (A171) and ignoring the higher order terms (h.o.t.), we obtain the final form of the approximate $v$ displacement as:

$$\begin{aligned} v &= \int_0^y \left(1 - \left(1 - \frac{x^2}{2}\right)\right) \mathrm{d}y = \int_0^y \left(\frac{x^2}{2}\right) \mathrm{d}y \\ &= \frac{1}{2}\int_0^y \left(\frac{\mathrm{d}w}{\mathrm{d}y}\right)^2 \mathrm{d}y \end{aligned} \tag{A175}$$

# Appendix B

## Integrals of Mode Shapes

Mode shape integrals resulting from the Lagrangian derivation of the system equations were derived analytically in closed-form expressions (except for one case, $\underline{M}_4(y)$) and are provided here. Note that $\otimes$ is the tensor or Kronecker product.

Mode Shape Integrals – Matrices $\underline{M}_k = \underline{M}_k(x) \otimes \underline{M}_k(y)$

| $k$ | $\underline{M}_k(x,y)$ | $\underline{M}_k(x)$ | $\underline{M}_k(y)$ |
|---|---|---|---|
| 1 | $\underline{M}_{\phi\psi'_\eta}$ | $\underline{M}_{\phi\phi} = \int_0^a \vec{\phi}\vec{\phi}^T \mathrm{d}x$ | $\underline{M}_{\psi'_\eta\psi'_\eta} = \int_0^b \int_0^y \vec{\psi}'_\eta \vec{\psi}'^T_\eta \mathrm{d}\eta \mathrm{d}y$ |
| 2 | $\underline{M}_{x\phi\psi'_\eta}$ | $\underline{M}_{x\phi\phi} = \int_0^a x\vec{\phi}\vec{\phi}^T \mathrm{d}x$ | $\underline{M}_{\psi'_\eta\psi'_\eta} = \int_0^b \int_0^y \vec{\psi}'_\eta \vec{\psi}'^T_\eta \mathrm{d}\eta \mathrm{d}y$ |
| 3 | $\underline{M}_{\phi\psi}$ | $\underline{M}_{\phi\phi} = \int_0^a \vec{\phi}\vec{\phi}^T \mathrm{d}x$ | $\underline{M}_{\psi\psi} = \int_0^b \vec{\psi}\vec{\psi}^T \mathrm{d}y$ |
| 4 | $\underline{M}_{y\phi\psi'_\eta}$ | $\underline{M}_{\phi\phi} = \int_0^a \vec{\phi}\vec{\phi}^T \mathrm{d}x$ | $\underline{M}_{y\psi'_\eta\psi'_\eta} = \int_0^b y \int_0^y \vec{\psi}'_\eta \vec{\psi}'^T_\eta \mathrm{d}\eta \mathrm{d}y$ |
| 5 | $\underline{M}_{\phi''\psi}$ | $\underline{M}_{\phi''\phi''} = \int_0^a \vec{\phi}''\vec{\phi}''^T \mathrm{d}x$ | $\underline{M}_{\psi\psi} = \int_0^b \vec{\psi}\vec{\psi}^T \mathrm{d}y$ |
| 6 | $\underline{M}_{\phi\psi''}$ | $\underline{M}_{\phi\phi} = \int_0^a \vec{\phi}\vec{\phi}^T \mathrm{d}x$ | $\underline{M}_{\psi''\psi''} = \int_0^b \vec{\psi}''\vec{\psi}''^T \mathrm{d}y$ |
| 7 | $\underline{M}_{\phi''\psi''}$ | $\underline{M}_{\phi''\phi} = \int_0^a \vec{\phi}''\vec{\phi}^T \mathrm{d}x$ | $\underline{M}_{\psi\psi''} = \int_0^b \vec{\psi}\vec{\psi}''^T \mathrm{d}y$ |
| 8 | $\underline{M}_{\phi'\psi'}$ | $\underline{M}_{\phi'\phi'} = \int_0^a \vec{\phi}'\vec{\phi}'^T \mathrm{d}x$ | $\underline{M}_{\psi'\psi'} = \int_0^b \vec{\psi}'\vec{\psi}'^T \mathrm{d}y$ |

| $k$ | $\underline{M}_k(x)$ | $\underline{M}_k(y)$ |
|---|---|---|
| 1 | $\underline{M}_{\phi\phi} = \begin{cases} a & i = j \\ 0 & i \neq j \end{cases}$ | $\underline{M}_{\psi'_\eta\psi'_\eta} = \begin{cases} 2 + \frac{\sigma_i^2 \lambda_i^2}{2} - \sigma_i\lambda_i & i = j \\ \frac{8\lambda_i^2\lambda_j^2}{(\lambda_j^2 + (-1)^{i+j}\lambda_i^2)^2} - \frac{4\lambda_i^2\lambda_j^2(\sigma_j\lambda_j - \sigma_i\lambda_i)}{\lambda_j^4 - \lambda_i^4} & i \neq j \end{cases}$ |
| 2 | $\underline{M}_{x\phi\phi} = \begin{cases} 0 & (1,i) \text{ and } (i,1),\ i \geq 3 \\ -8\sqrt{3}a^2 \frac{\sigma_i}{\lambda_i^3}[1-(-1)^i] & (2,i) \text{ and } (i,2), i \geq 3 \\ -\frac{a^2}{2\sqrt{3}} & (1,2) \text{ and } (2,1) \\ \frac{a^2}{2} & i = j \\ 8a^2 \frac{\sigma_i\lambda_i\sigma_j\lambda_j(\lambda_i^4 + \lambda_j^4)}{(\lambda_i^4 - \lambda_j^4)^2}[(-1)^{i+j} - 1] & i \neq j,\ i \geq 3,\ j \geq 3 \end{cases}$ | Same as $\underline{M}_{\psi'_\eta\psi'_\eta} = \underline{M}_1(y)$ |
| 3 | Same as $\underline{M}_{\phi\phi} = \underline{M}_1(x)$ | $\underline{M}_{\psi\psi} = \begin{cases} b & i = j \\ 0 & i \neq j \end{cases}$ |
| 4 | Same as $\underline{M}_{\phi\phi} = \underline{M}_1(x)$ | Values obtained numerically using MATHCAD.<br>For example, for *p=q=3*<br>$\underline{M}_{y\psi'_\eta\psi'_\eta} = b\begin{bmatrix} 1.193 & -0.686 & -0.792 & -0.546 & -0.454 & -0.372 & -0.322 & -0.280 & -0.250 \\ -0.686 & 6.478 & 0.169 & -2.911 & -1.889 & -1.943 & -1.613 & -1.518 & -1.341 \\ -0.792 & 0.169 & 17.86 & 3.271 & 6.154 & -3.326 & -4.170 & -3.334 & 3.409 \\ -0.546 & -2.912 & 3.274 & 36.05 & 8.570 & -9.841 & -4.363 & -6.582 & -4.996 \\ -0.454 & -1.889 & -6.154 & 8.570 & 60.80 & 15.92 & -14.01 & -5.059 & -9.125 \\ -0.372 & -1.943 & -3.326 & -9.841 & 15.92 & 92.13 & 25.30 & -18.67 & 5.446 \\ -0.322 & -1.613 & -4.170 & -4.363 & 14.04 & 25.30 & 130.0 & 36.69 & -23.80 \\ -0.280 & -1.518 & -3.334 & -6.582 & -5.059 & -18.67 & 36.69 & 174.5 & 50.10 \\ -0.250 & -1.340 & -3.409 & -4.996 & 9.125 & -5.446 & -23.80 & 50.10 & 225.6 \end{bmatrix}$ |
| 5 | $\underline{M}_{\phi''\phi''} = \begin{cases} \dfrac{\lambda_i^4}{a^3} & i = j,\ i \geq 3,\ j \geq 3 \\ 0 & \text{otherwise} \end{cases}$ | Same as $\underline{M}_{\psi\psi} = \underline{M}_3(y)$ |
| 6 | Same as $\underline{M}_{\phi\phi} = \underline{M}_1(x)$ | $\underline{M}_{\psi''\psi''} = \begin{cases} \dfrac{\lambda_i^4}{b^3} & i = j \\ 0 & \text{otherwise} \end{cases}$ |
| 7 | $\underline{M}_{\phi''\phi} = \begin{cases} \dfrac{2\sigma_i\lambda_i}{a}[1-(-1)^i] & (i,\ 1),\ i \geq 3 \\ \{\ \dfrac{2\sqrt{3}\sigma_i\lambda_i}{a}[1-(-1)^i] & \\ \quad -\dfrac{4\sqrt{3}}{a}[1+(-1)^i-(-1)^i\sigma_i\lambda_i]\ \} & (i,2),\ i \geq 3 \\ \dfrac{\sigma_i\lambda_i}{a}(2-\sigma_i\lambda_i) & i = j,\ i \geq 3,\ j \geq 3 \\ \dfrac{4\lambda_i^4(\sigma_j\lambda_j-\sigma_i\lambda_i)}{a(\lambda_j^4-\lambda_i^4)}[1+(-1)^{i+j}] & i \neq j,\ i \geq 3,\ j \geq 3 \\ 0 & \text{otherwise} \end{cases}$ | $\underline{M}_{\psi\psi''} = \begin{cases} \dfrac{\sigma_i\lambda_i}{b}(2-\sigma_i\lambda_i) & i = j \\ \dfrac{4\lambda_j^2(\sigma_i\lambda_i-\sigma_j\lambda_j)}{b(\lambda_i^4-\lambda_j^4)}[\lambda_i^2+(-1)^{i+j}\lambda_j^2] & i \neq j \end{cases}$ |
| 8 | $\underline{M}_{\phi'\phi'} = \begin{cases} \dfrac{12}{a} & (2,2) \\ \dfrac{4\sqrt{3}}{\text{a}}[(-1)^i+1] & (2,i) \text{ and } (i,2),\ i \geq 3 \\ \dfrac{\lambda_i\sigma_i}{a}[6+\lambda_i\sigma_i] & i = j,\ i \geq 3, j \geq 3 \\ \dfrac{4\lambda_j\lambda_i(\sigma_j\lambda_i^3-\sigma_i\lambda_j^3)}{a\left(\lambda_i^4-\lambda_j^4\right)}[(-1)^{i+j}+1] & i \neq j,\ i \geq 3, j \geq 3 \\ 0 & \text{otherwise} \end{cases}$ | $\underline{M}_{\psi'\psi'} = \begin{cases} \dfrac{\sigma_i\lambda_i}{b}(2+\sigma_i\lambda_i) & i = j \\ \dfrac{4\lambda_j\lambda_i}{b\left(\lambda_i^4-\lambda_j^4\right)}[(-1)^{i+j}(\sigma_j\lambda_i^3-\sigma_i\lambda_j^3) & \\ \quad -\lambda_j\lambda_i\left(\sigma_i\lambda_i-\sigma_j\lambda_j\right)] & i \neq j \end{cases}$ |

Note that all mode shape matrices above (except $\underline{M}_7$) are symmetric.

Mode Shape Integrals – Vectors $\underline{m}_k = \underline{m}_k(x) \otimes \underline{m}_k(y)$

| $k$ | $\vec{m}_k(x,y)$ | $\vec{m}_k(x)$ | $\vec{m}_k(y)$ |
|---|---|---|---|
| 1 | $\vec{m}_{\phi\psi}$ | $\vec{m}_\phi = \int_0^a \vec{\phi}\ \mathrm{d}x$ | $\vec{m}_\psi = \int_0^b \vec{\psi}\ \mathrm{d}y$ |
| 2 | $\vec{m}_{y\phi\psi}$ | $\vec{m}_\phi = \int_0^a \vec{\phi}\ \mathrm{d}x$ | $\vec{m}_{y\psi} = \int_0^b y\vec{\psi}\ \mathrm{d}y$ |
| 3 | $\vec{m}_{x\phi\psi}$ | $\vec{m}_{x\phi} = \int_0^a x\vec{\phi}\ \mathrm{d}x$ | $\vec{m}_\psi = \int_0^b \vec{\psi}\ \mathrm{d}y$ |

| $k$ | $\vec{m}_k(x)$ | $\vec{m}_k(y)$ |
|---|---|---|
| 1 | $\vec{m}_\phi = \begin{cases} \mathrm{a} & i=1 \\ 0 & \text{otherwise} \end{cases}$ | $\vec{m}_\psi = \frac{2b\sigma_i}{\lambda_i}$ |
| 2 | Same as $\vec{m}_\phi = \vec{m}_1(x)$ | $\vec{m}_{y\psi} = \frac{2b^2}{\lambda_i^2}$ |
| 3 | $\vec{m}_{x\phi} = \begin{cases} \frac{\mathrm{a}^2}{2} & i=1 \\ -\frac{\mathrm{a}^2}{2\sqrt{3}} & i=2 \\ 0 & \text{otherwise} \end{cases}$ | Same as $\vec{m}_\psi = \vec{m}_1(y)$ |

# Appendix C

## Environmental Disturbance Torques

The following four sub-sections discuss briefly each of the four environmental disturbance torques. Hughes [3] provides a more detailed discussion of the environmental disturbances affecting the spacecraft.

## C.1 Gravity Gradient Torque

The gravity gradient torque arises because the gravitational force varies over the unsymmetrical mass distribution of the spacecraft body. This gravity gradient torque expressed in the spacecraft reference frame varies throughout the orbit as well since the radius vector from the centre of the Earth to the centre of mass of the spacecraft varies in the spacecraft reference frame.

The following assumptions are usually made to simplify the gravity gradient torque model:

i) Only the effect of the Earth is considered.
ii) Earth is considered to have a spherical shape and a symmetrical mass distribution.
iii) The spacecraft is considered as a single, small body compared to the Earth.

The instantaneous gravity gradient torque is given by

$$\vec{\tau}_g = \frac{3\mu_E}{\|\vec{r}_c\|^3}\left[\vec{r}_c \times (\underline{I}_t \vec{r}_c)\right] \tag{C1}$$

where $\mu_E$ is the Earth's gravitational constant, $\vec{r}_c$ is the unit vector defining the spacecraft center of mass position from the center of the Earth, expressed in the spacecraft reference frame, and $\underline{I}_t$ is the spacecraft total inertia matrix which includes the flexibility effects.

## C.2 Magnetic Torque

The magnetic torque is often not determined accurately due to the inaccuracy of the spacecraft's residual magnetic dipole vector arising from permanent magnets and current loops within the spacecraft (excluding the magnetorquers, if any). On the other hand, the Earth's magnetic field has been defined accurately in the form of a series of spherical harmonics. However, due to the inaccuracy of the spacecraft's residual magnetic dipole, there is no incentive for considering a complicated expression for the Earth's magnetic field; hence a dipole model was used where all but the first term of the spherical harmonics are ignored.

The magnetic disturbance torque is given by

$$\vec{\tau}_m = \vec{m}_c \times \vec{b}_c \tag{C2}$$

where $\vec{m}_c$ is the spacecraft residual magnetic dipole vector, and $\vec{b}_c$ is the Earth's magnetic field vector at the spacecraft's position which can be computed , using a simple magnetic dipole model [6], as follows

$$\vec{b}_c = \frac{a^3 H_0}{\left\|\vec{r}_c\right\|^3}\left[3\left(\vec{m}^T \vec{r}_c\right)\vec{r}_c - \vec{m}\right] \tag{C3}$$

where $a^3 H_0$ and $\vec{m}$ are the Earth's magnetic dipole strength and unit vector, respectively.

## C.3 Solar Radiation Torque

The solar pressure torque is produced due to the accumulative force imparted by the sun on the spacecraft body and the offset of the spacecraft's optical center of pressure from the spacecraft's center of mass. In order to simplify the calculation of solar radiation torque, the reflected and emitted radiation from the Earth is ignored, leaving the solar radiation as the only radiation source to consider.

The solar radiation torque is given by

$$\vec{\tau}_s = \vec{r}_s \times \vec{f}_s \tag{C4}$$

where $\vec{r}_s$ is the vector from the spacecraft center of mass to the optical center of pressure and the solar radiation force, $\vec{f}_s$, is calculated using

$$\vec{f}_s = \frac{F_s}{c} A_s (1 + \mathrm{q}) \vec{u}_s \tag{C5}$$

where $F_s$ is the solar flux constant, $c$ is the speed of light, $A_s$ is the exposed solar surface area, $\mathrm{q}$ is the reflectance factor (ranging from 0 to 1) and $\vec{u}_s$ is the unit vector defining the direction of the solar radiation, pointing away form the Sun towards the spacecraft surface, expressed in the spacecraft reference frame.

## C.4 Aerodynamic Torque

The aerodynamics disturbance torque is due to the accumulative force imparted by the molecules found in the upper atmosphere and the offset of the spacecraft's aerodynamic center of pressure from the spacecraft's center of mass.

A number of assumptions are usually made which are consistent with the calculation of drag for aircraft and can only be used as rough estimates for spacecraft drag force. The drag force was calculated using aerodynamic theory which assumes continuum flow. However, at the usual orbital altitudes, the spacecraft encounters only free-molecular flow where the average separation distance of air molecules is large compared to the size of the spacecraft. As the names imply, the treatments of these two flows are significantly different, however, as the calculation of the drag force due to free-molecular flow is challenging and requires rigorous mathematical modelling, it was deemed unnecessary for the purposes of this thesis. Also, for precise results, the variation of the Earth's atmosphere would have to be modelled. This variation has two causes: solar activity and the difference in atmospheric height of the day and night sides of the Earth.

The aerodynamic torque is given by

$$\vec{\tau}_a = \vec{r}_a \times \vec{f}_a \tag{C6}$$

where $\vec{r}_a$ is the vector from the spacecraft center of mass to the aerodynamic center of pressure and the aerodynamic drag force, $\vec{f}_a$, is calculated using

$$\vec{f}_a = \frac{1}{2}\rho c_d A_a \left\|\vec{v}_t\right\| \vec{v}_t \tag{C7}$$

where $\rho$ is the atmospheric density, $c_d$ is the drag coefficient, $A_a$ is the exposed aerodynamic surface area and $\vec{v}_t$ is the spacecraft velocity vector expressed in the spacecraft reference frame.